# Towards Effective Research-Paper Recommender Systems and User Modeling based on Mind Maps

**Dissertation**
zur Erlangung des akademischen Grades
Doktoringenieur (Dr.-Ing.)

Angenommen durch die Fakultät für Informatik
der Otto-von-Guericke-Universität Magdeburg

von Diplom Wirtschaftsinformatiker Jöran Beel, MSc
geboren am 19.02.1980 in Herdecke

GutachterInnen:
Prof. Dr. Andreas Nürnberger
Prof. Dr. Klaus Turowski
Prof. Dr. Alesia Zuccala

Magdeburg, den 17. März 2015

# Abstract

User modeling and recommender systems are often seen as key success factors for companies such as Google, Amazon, and Netflix. However, while user-modeling and recommender systems successfully utilize items like emails, news, social tags, and movies, they widely neglect mind-maps as a source for user modeling. We consider this a serious shortcoming since we assume user modeling based on mind maps to be equally effective as user modeling based on other items. Hence, millions of mind-mapping users could benefit from user-modeling applications such as recommender systems.

The objective of this doctoral thesis is to develop an effective user-modeling approach based on mind maps. To achieve this objective, we integrate a research-paper recommender system in our mind-mapping and reference-management software *Docear*. The recommender system builds user models based on the users' mind maps, and recommends research papers based on the user models. As part of our research, we identify several variables relating to mind-map-based user modeling, and evaluate the variables' impact on user-modeling effectiveness with an offline evaluation, a user study, and an online evaluation based on 430,893 recommendations displayed to 4,700 users.

We find, among others, that the number of analyzed nodes, the time when nodes were modified, the visibility of nodes, the relations between nodes, and the number of children and siblings of a node affect the effectiveness of user modeling. When all variables are combined in a favorable way, this novel user-modeling approach achieves click-through rates of 7.20%, which is nearly twice as effective as the best baseline. In addition, we show that user modeling based on mind maps performs about as well as user modeling based on other items, namely the research articles users downloaded or cited. Our findings let us to conclude that user modeling based on mind maps is a promising research field, and that developers of mind-mapping applications should integrate recommender systems into their applications. Such systems could create additional value for millions of mind-mapping users.

As part of our research, we also address the question of how to evaluate recommender systems adequately. This question is highly discussed in the recommender-system community, and we provide some new results and arguments. Among others, we show that offline evaluations often cannot predict results of online evaluations and user studies in the field of research-paper recommender systems. We also show that click-through rate and user rating correlate well ($r$=0.78). We discuss these findings, including some inherent problems of offline evaluations, and conclude that offline evaluations are probably unsuitable for evaluating research-paper recommender systems, while both user studies and online evaluations are adequate evaluation methods.

We also introduce a new weighting scheme, TF-ID*u*F, which could be relevant for recommender systems in general. In addition, we are first to compare the weighting scheme CC-IDF against CC only, and we research concept drift in the context of research-paper recommender systems, with the result that interests of researchers seem to shift after about four months. Last, but not least, we publish the architecture of Docear's recommender system, as well as four datasets relating to the users, recommendations, and document corpus of Docear and its recommender system.

# Zusammenfassung

Empfehlungsdienste und Nutzermodellierungssysteme sind wichtige Erfolgsfaktoren für Unternehmen wie Google, Amazon und Netflix. Während solche Systeme die Interessen von Nutzern erfolgreich an Hand von verfassten Emails, gelesenen Nachrichten, geschauten Filmen, etc. ableiten, ignorieren Unternehmen bisher weitestgehend Mind-Maps als Quelle für Nutzermodellierung. Dies sehen wir als Problem, denn Millionen von Mind-Mapping Nutzern könnten von Mind-Map basierten Nutzermodellierungssystemen wie beispielsweise Empfehlungsdiensten profitieren.

Das Ziel der vorliegenden Doktorarbeit ist es, ein effektives Nutzermodellierungs-verfahren zu entwickeln zur Realisierung von Empfehlungsdiensten in Mind-Mapping-Tools. Hierzu entwickeln wir ein Empfehlungsdienst für unsere Mind-Mapping und Referenzmanagementsoftware *Docear*. Wir identifizieren Variablen, welche die Effektivität der Nutzermodellierung beeinflussen. Den Einfluss evaluieren wir mit einer Offline Evaluation, einer Nutzerstudie, und einer Online Evaluation basierend auf 430,893 Empfehlungen für 4,700 Nutzer.

Die Evaluation zeigt, dass, unter anderem, die Anzahl der analysierten Knoten, der Zeitpunkt wann Knoten modifiziert werden, die Beziehung zwischen Knoten, und die Anzahl von Kinder- und Geschwisterknoten die Effektivität der Nutzermodellierung beeinflussen. Werden alle Faktoren in einem mind-map-spezifischen Nutzermodellierungsverfahren berücksichtigt, ist dieses Verfahren nahezu doppelt so effektiv wie Standardverfahren. Wir zeigen außerdem, dass Mind-Map-basierte Nutzermodellierung ähnlich effektiv ist, wie Nutzermodellierung in anderen Bereichen. Die Ergebnisse lassen uns schlussfolgern, dass Nutzermodellierung basierend auf Mind-Maps ein vielversprechendes Forschungsgebiet ist, und dass Entwickler von Mind-Mapping Tools Empfehlungsdienste integrieren sollten. Solche Empfehlungsdienste würden ähnlichen Mehrwert für die Anwender und Entwickler schaffen, wie Empfehlungsdienste in anderen Bereichen.

Im Rahmen unserer Arbeit beschäftigen wir uns auch mit der Frage, wie Empfehlungsdienste angemessen evaluiert werden können. Diese Frage wird in der wissenschaftlichen Gemeinschaft derzeit intensiv diskutiert, und wir tragen mit neuen Erkenntnissen zu dieser Diskussion bei. Unter anderem zeigen wir, dass Offline Evaluationen häufig nicht in der Lage sind, die Effektivität von Empfehlungsverfahren in der Praxis vorherzusagen. Wir zeigen außerdem, dass Click-Through Rates (CTR) und Nutzerzufriedenheit eine starke Korrelation aufweisen ($r$=0,78). Wir diskutieren diese Ergebnisse, einschließlich einiger inhärenter Probleme von Offline Evaluationen, und schlussfolgern, dass Offline Evaluationen vermutlich ungeeignet sind für die Evaluation von (Literatur-) Empfehlungsdiensten. Nutzerstudien und Online Evaluationen hingegen, erscheinen beide gleichermaßen geeignet.

Wir stellen in unserer Arbeit außerdem ein neues Verfahren zur Gewichtung von Wörtern und Zitationen für Nutzermodellierung vor (TF-ID*u*F). Zusätzlich evaluieren wir das Gewichtungsverfahren CC-IDF mit dem Ergebnis, dass CC-IDF vermutlich nicht effektiv ist. Wir untersuchen auch den Einfluss von ‚Concept-Drift' in Literaturempfehlungs-diensten, und finden heraus, dass nur die Arbeiten der letzten vier Monate für Nutzermodellierung verwendet werden sollten. Schließlich publizieren wir die Architektur und vier Datensets von Docear's Empfehlungsdienst.

# Table of Contents

# List of Figures

# List of Tables

# Glossary

| | |
|---|---|
| ATR | Annotation-Through Rate |
| CBF | Content-Based Filtering |
| CC-IDF | Weighting scheme analog to TF-IDF but based on citations instead of terms |
| CiTR | Cite-Through Rate |
| CF | Collaborative Filtering |
| CTR | Click-Through Rate |
| DTR | Download-Through Rate |
| IDF | Inverse Document Frequency |
| IF | Information Filtering |
| IR | Information Retrieval |
| LTR | Link-Through Rate |
| MRR | Mean Reciprocal Rank |
| nDCG | normalized Discounted Cumulative Gain |
| P@N | Precision at position n |
| *r* | Pearson correlation coefficient |
| TF | Term Frequency |
| TF-IDF | Weighting scheme based on Term Frequency and Inverse Document Frequency |
| TF-ID*u*F | Novel weighting scheme like TF-IDF but based on a user's personal document corpus |
| VSM | Vector Space Model |

# Acknowledgements

This doctoral thesis would not exist without the support of various individuals and institutions to whom I would like to express my sincere thanks.

I am particularly indebted to Claus Rautenstrauch, who mentored me during my undergraduate and graduate studies, and was my PhD supervisor, until he tragically passed away. Claus always encouraged me to explore the world, to broaden my horizons, and to question the status quo. Without him, I probably would have never studied at Macquarie University Sydney, Australia, Lancaster University, UK, and most certainly, I would never have begun the pursuit of my PhD.

Deepest gratitude is also due to my supervisor Andreas Nürnberger. He greatly dedicated himself to my work, particularly with respect to the development of Docear. Without Andreas, Docear – the foundation for my research – would not exist. Furthermore, Andreas was a great motivator throughout the time I was working on my thesis and his guidance and knowledge was of great value.

I thank SAP and the Very-Large-Business-Applications Lab (VLBA-Lab) – particularly the directors, Hans Henning Arndt and Klaus Turowski – for their support, which allowed me to start my PhD. The time at the VLBA-Lab has been largely inspiring and laid the foundation for my current research. Working at the VLBA-lab also allowed me to teach at the WADI University in Syria. I experienced Syria as a beautiful country with welcoming people and students eager to gain knowledge. I wish my former students and colleagues all the best, and that Syria may soon overcome the civil war and find the peace its people deserve.

Bela Gipp was one of the best friends and colleagues one could ask for. He never spared constructive feedback, he challenged my ideas, and he always motivated me to go the extra mile. I appreciated his diligence and I am grateful for his ideas and clarity of thought, which has proven to be very valuable to my work.

Stefan Langer and Marcel Genzmehr have stood along my side for more than three years, day-by-day, to work on Docear and its recommender system. I am sincerely thankful for their passion, their knowledge, and the sacrifice they accepted to make Docear the powerful and unique tool it has become. It has been a pleasure working with them, and I hope that many more years will follow.

I thank Erik Wilde and Jim Pitman from the University of California, Berkeley, for their generosity, time, and knowledge. My work and research at UC Berkeley has been an experience of a lifetime that has shaped my understanding of science.

I also thank the University of California, Berkeley, and the Otto-von-Guericke University, as well as the German Academic Exchange Service (DAAD), the European Union, and the federal state of Lower-Saxony for their support, both financially and in terms of resources. Their support was key to the development of Docear and for conducting my research. I also thank the DAAD, ACM, and IEEE for several travel grants that allowed me to participate in conferences, such as the ACM/IEEE Joint Conference on Digital Libraries (JCDL) and the ACM Recommender Systems Conference (RecSys).

Georgia Kapitsaki has been a great research fellow for many months, and I am greatly indebted to her for the invitation to the University of Cyprus (UCY). The work-environment at UCY was amazing, and Georgia's dedication was invaluable to my research.

I would like to thank Debora Weber-Wulff for providing me with the opportunity to teach and work at HTW Berlin. The students at HTW Berlin were highly motivated and exceptionally talented – working with them was a pleasure and has contributed to the success of Docear and my research.

Gratitude is also owed to the many students, volunteers, and research fellows that aided in the development of Docear and its recommender system as well as in the research. I would like to thank particularly Mario Lipinski, Corinna Breitinger, Norman Meuschke, Christoph Müller, Simon Hewitt, Ammar Shaker, Nick Friedrich, Cheng Xie, Julius Seltenheim, Alexander Schwank, Florian Wokurka, Michael Schleichard, Paul Stüber, Sebastian Götte, Patrick Lühne, Mathias Silbermann, the developers of the mind-mapping software Freeplane (Dimitry Polivaev, Volker Boercher, and many others), and the developers of the reference management software JabRef (Morten Omholt Alver, Oliver Kopp, and many others). My gratitude goes also to the 4,700 Docear users who gave permission to use their data for my research.

Finally, yet importantly, I would like to thank my girlfriend Carina, who gave me support and backup throughout my PhD. Particularly during the past months, which were characterized by long working hours and little sleep, Carina was of invaluable help and a great motivator.

# 1. Introduction

## 1.1 Problem Setting

Items such as emails, social tags, or research articles are often utilized beyond their original purpose. The goal of this “extended use” is typically to enhance existing services, provide new services, or generate additional revenue. For instance, social tags are intended to organize private webpage collections, but search engines utilize them to enhance webpage indexing [441]. Emails are intended as a means of communication, but Google utilizes them for generating user models and displaying personalized advertisements [150], and research articles are intended to communicate research results, but the articles, or more precisely their references, are utilized to measure the impact of researchers and academic institutions [189].

We propose that mind-maps could also be utilized beyond their original purpose, similar to social tags, emails, and research articles. In a preliminary study, we developed eight ideas of how mind maps could be utilized to provide new services, enhance existing services, and generate additional revenues (cf. Appendix B.4, p. 184). We explored the feasibility of the ideas and concluded that user modeling was the most promising.

User modeling is the process of inferring information about users by analyzing the users’ items or actions [104, 442]. User models are required by many applications such as personalized search engines, adaptive graphical user interfaces, and recommender systems. Particularly, recommender systems are often seen as key factors to the increase of user satisfaction and revenue generation. For instance, *Amazon* considers its recommender system as a “key differentiation factor” [350], *Google*’s business model (i.e. personalized advertisement) is heavily dependent on user modeling [151], and the movie rental and streaming service *Netflix* offered one million US dollars to whomever could improve its recommender system by 10% [303]. Given the popularity of recommender systems in general, we find it surprising, that researchers and developers of mind-mapping tools showed little interest in user modeling or recommender systems based on mind maps.

In the research community, we are first to explore the field of mind-map-based user modeling, to the best of our knowledge. In practice, two companies made the first experiences: Both *MindMeister* and *Mindomo*, utilized mind maps for user modeling in the context of personalized advertisement. MindMeister extracted terms from the node that a user had created or edited most recently, and used these terms as a user model. MindMeister then sent the user model, i.e. the terms, to Amazon's Web Service as a search query. Amazon returned book recommendations that matched the query, and MindMeister displayed the recommendations in a window next to the mind map (Figure 1). Mindomo applied a similar concept using *Google AdSense* instead of Amazon. Both companies have since abandoned their user-modeling systems, although they still actively maintain their mind-mapping tools in general. In an email, Mindomo explained, "people were not really interested" in the advertisement[1]. We were surprised about Mindomo's statement because it contradicted our expectations about the usefulness of user modeling based on mind maps.

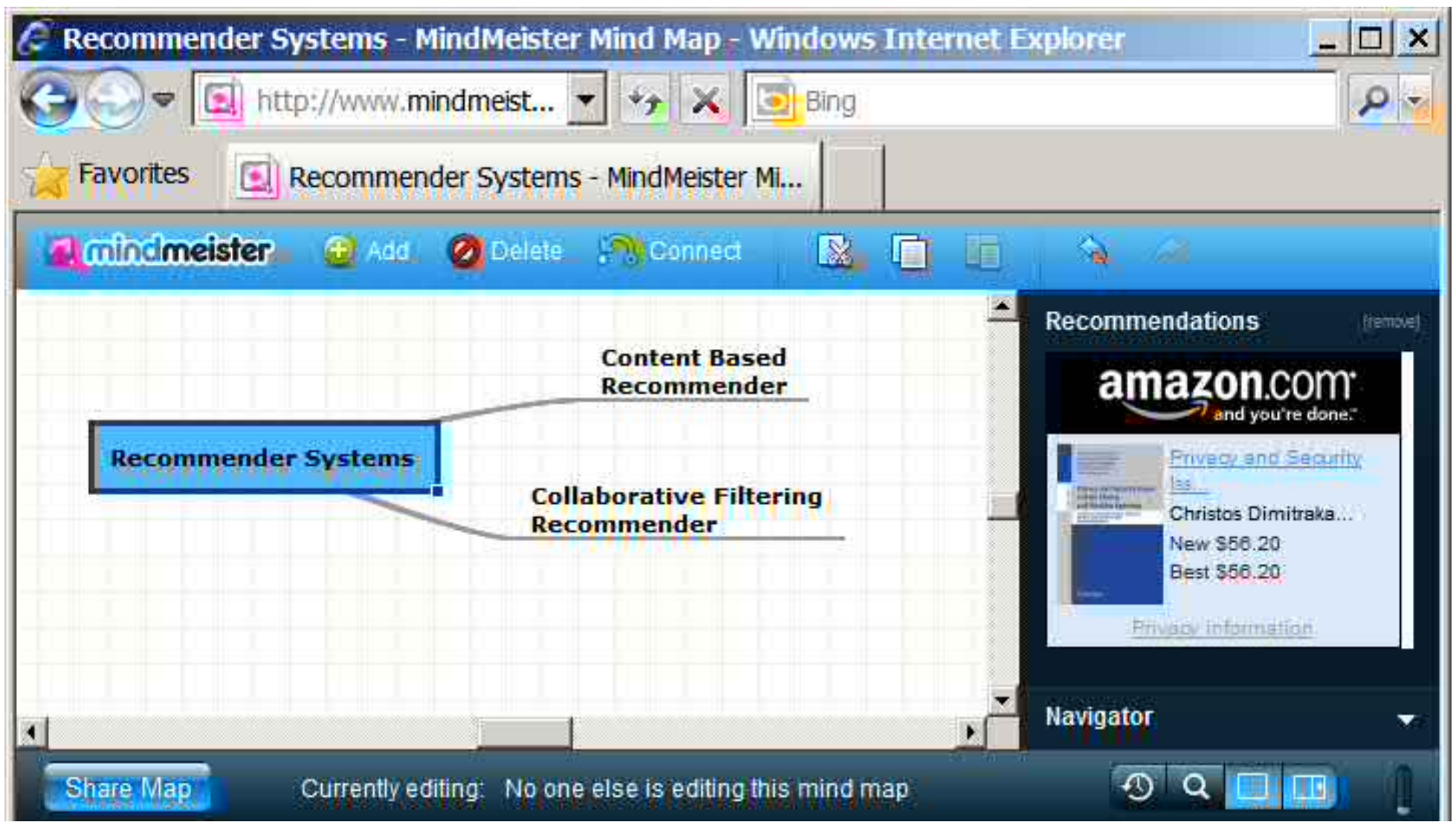

Figure 1: Personalized advertisement in MindMeister

To explore the effectiveness of mind-map-based user modeling in more detail, and to find reasons why Mindomo and MindMeister might have abandoned personalized advertisements, we conducted a preliminary study (Appendix B, p. 179). We re-implemented MindMeister's approach and used it in our mind mapping software, *Docear*, to recommend research papers. Instead of using Amazon's Web Service or Google AdSense, we built our own corpus of recommendation candidates and used Apache Lucene to match candidates with user models. In Docear, MindMeister's approach, i.e. utilizing the terms from the most recently edited or created node, achieved click-through rates (CTR) between

[1] Email by Daniel Sima of the Mindomo team, October 3, 2011. Permission for publication was granted.

0.2% and around 1%. Compared to other recommender systems [253, 291, 345], such a CTR is disappointing, which might explain why Mindomo and MindMeister abandoned their recommender systems.

Besides MindMeister's user-modeling approach, there are two more approaches, both following the popular idea of content-based filtering (CBF), that we consider rather obvious to use with mind-maps. One approach is to build user models based on terms contained in all nodes of a users' current mind map. The next approach is to utilize terms from all mind maps ever created by a user. As part of the preliminary study, we implemented these approaches in Docear and both achieved CTRs of around 6% (Appendix B, p. 179). Such a CTR is reasonable and significantly better than MindMeister's approach. We were surprised that rather similar user-modeling approaches differed in their effectiveness by a factor of six. Apparently, small differences in the algorithms – such as whether to utilize terms from a single node or from the entire mind map – have a significant impact on user modeling performance.

## 1.2 Motivation

Given the popularity of recommender systems for movies, e-commerce, etc., and the encouraging results of our preliminary study, we wanted to explore the potential of mind-map-based user modeling further. Our motivation was twofold.

First, we have a self-interest in mind-map-based user modeling and recommender systems. Since 2009, we have been developing the reference management software *Docear*, previously known as *SciPlore MindMapping* [33, 35]. Docear has around 40,000 users[2] who manage their academic literature and references with mind maps. These users would benefit from a user-modeling application such as a recommender system.

Second, a significant number of mind-mapping users could benefit from recommender systems, as well as the developers of the mind-mapping tools (cf. Appendix B, p. 179). The website *Mind-Mapping.org* lists 142 actively maintained mind-mapping tools[3]. The mind-mapping tools are used by an estimated two million active users who create around five millions mind maps every year (cf. Appendix B, p. 179). Developing an effective user-modeling approach should encourage developers of mind-mapping applications to integrate recommender systems in their applications and thereby provide additional value to their users.

[2] More than 20,000 users registered, and based on the number of webpage visitors and update requests, we estimate that a similar number of researchers uses Docear without registration.

## 1.3 Research Objective, Questions, and Tasks

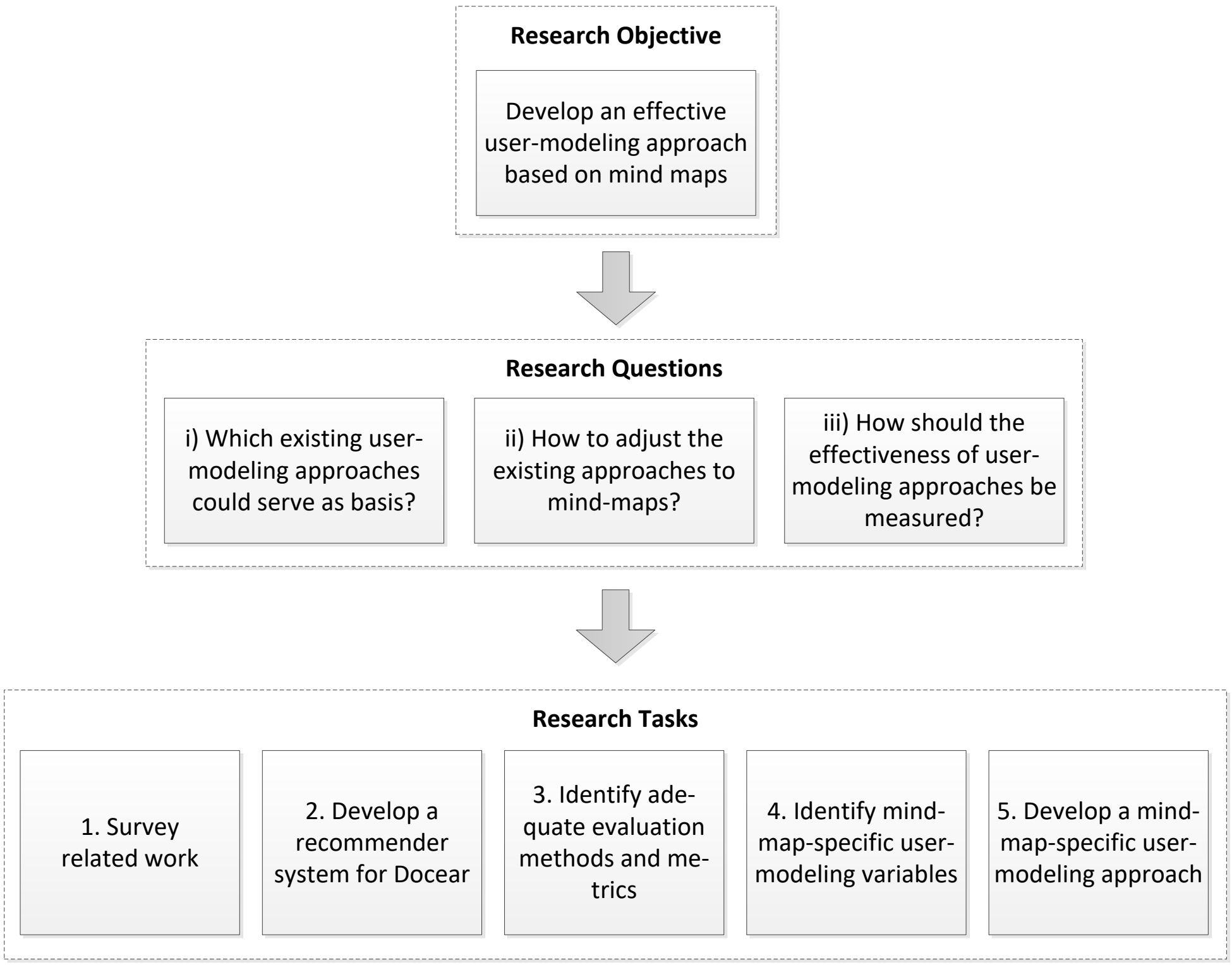

Figure 2: Research objective, questions, and tasks

Given the many users that could benefit from mind-map-based user modeling and recommender systems, we defined our research objective as follows (cf. Figure 2):

***Develop an effective user-modeling approach based on mind maps***

[3] Some tools offer mind-mapping only as secondary feature in addition to other visualization techniques, such as concept maps or Gantt charts

The research objective should be seen in the context of research-paper recommender systems, because we conducted our research based on Docear, and most of Docear’s users are researchers, hence our decision to use user models for a research-paper recommender systems.

The research objective leads us to ask three research questions:

i. Which existing user-modeling approaches could serve as a basis for mind-map-based user modeling?

ii. Could the effectiveness of existing approaches be increased by adjusting them to the special characteristics of mind maps?

iii. How should the effectiveness of user-modeling approaches be measured?

To answer the research questions, we defined five tasks:

**Task 1: Survey related work**

*There are a significant number of publications about mind maps, user modeling, recommender systems, and recommender-systems evaluation. We assumed that the existing work would help to identify user-modeling approaches that could serve the basis for a mind-map-specific user-modeling approach. In addition, we expected that the literature could help identify adequate methods and metrics to evaluate user-modeling approaches. Therefore, the first task is to conduct a thorough review of the corresponding literature.*

Chapter 3, p. 29

**Task 2: Develop a recommender system for Docear**

*The second task is to develop a recommender system and integrate it into Docear, to being able to conduct our research in a real-world scenario. The recommender system should be capable of applying different user-modeling approaches and evaluation methods.*

Chapter 5.1, p. 73

**Task 3: Identify adequate evaluation methods and metrics**

*The literature survey revealed that there is uncertainty about the adequacy of evaluation methods and metrics. Therefore, the third task is to conduct additional research to find adequate evaluation methods and metrics for our particular scenario (user modeling and research-paper recommendations based on mind maps).*

Chapter 5.2, p. 87

**Task 4: Identify mind-map-specific user-modeling variables**

*As a preliminary step toward the research goal, the fourth task is to identify variables that affect user modeling based on mind maps, and to assess the variables' impact on user-modeling effectiveness.*

Chapter 5.3, p. 105

**Task 5: Develop a mind-map-specific user-modeling approach**

*The fifth task is to combine the variables from Task 4 in a single algorithm to obtain an effective mind-map-specific user-modeling approach. The approach should be compared against adequate baselines.*

Chapter 5.4, p. 121

## 1.4 Outline

Chapter 2 presents basic information that is crucial for understanding this doctoral thesis. This includes an introduction to mind mapping, user modeling, recommender systems, and some definitions. The chapter only covers the basics – readers familiar with the topics are advised to skip to Chapter 3.

Chapter 3 presents the results of our literature survey (Research Task 1). Since there is no literature about user modeling and recommender systems based on mind maps, the review focuses on recommender systems in general, more precisely on *research-paper* recommender systems, and their evaluation.

Chapter 4 presents the methodology. This includes information about how Docear's recommender system was built, information about Docear’s users, an explanation of how recommendations are generated, and how user studies, offline evaluations, and online evaluations were conducted.

Chapter 5 presents the results of our research and splits it into four parts, one for each research task (Tasks 2-5): In Part One, Docear's research-paper recommender system is presented (Task 2), including its architecture and four datasets (5.1, p. 73). The datasets contain information about Docear's users, the recommendation corpus, and delivered recommendations. Both the architecture and datasets help understanding and reproducing our research. In addition, the datasets allow further analyses that go beyond our own research. In Part Two, results of Research Task 3 are presented, i.e. different methods for recommender-systems evaluation are compared, and discussed (5.2 p. 87). The discussion concludes that click-through rate is the most appropriate evaluation metric for Docear's scenario. In addition, it is concluded that offline evaluations, the most common evaluation method for recommender systems, are probably inappropriate for evaluating (research-paper) recommender systems. In Part Three, the results of Task 4 are presented, i.e. the effect of several variables on user modeling based on mind maps (5.3, p. 105). Among others, it is shown how the number of analyzed nodes and the visibility of nodes affecting user-modeling effectiveness. In Part Four, results of task 5 are presented, i.e. the variables are combined in a novel mind-map-specific user modeling and recommendation approach (5.4, p. 121). A comparison with standard user-modeling approaches shows that the mind-map-specific approach is around twice as effective.

This thesis concludes with a summary of the contributions (Chapter 6), and provides an outlook for further research (Chapter 7). The Appendix contains additional information, e.g. our preliminary study (Appendix B), details on Docear’s users (Appendix H), the individual recommendation approaches that we surveyed (Appendix F), and the patent application that we filed for our mind-map-specific user-modeling approach (Appendix K).

# 2. Fundamentals

This chapter introduces mind mapping (p. 9), Docear (p. 12), some definitions (p. 14), user modeling (p. 16), recommender systems (p. 18), and related research fields (p. 27). A general understanding of these topics is important to following our research and discussion. Readers familiar with the topics may skip this chapter.

## 2.1 Mind Mapping

Mind mapping is a technique for recording and organizing information and for the development of new ideas with special types of documents called "mind maps" [176]. The structure of mind maps is similar to outlines and consists of three elements: *nodes*, *connections*, and *visual clues*. When users create a new mind map, they start with a *root node,* in which they write the central concept that the mind map is about [103]. To detail the central concept, users create sub nodes, i.e. child nodes, branching from the root node. To detail the child nodes, users create further sub nodes, and so on. This process is similar to creating an outline with a title, chapters, paragraphs, and sentences.

Mind maps are often used for tasks like brainstorming, knowledge management, note taking, project planning, decision making, and career planning [123]. Originally, mind mapping was done with pen and paper, but since the 1980s, more than one-hundred software tools for aiding users in creating mind maps have been developed [125].

Figure 3 shows a mind map that we created with Docear to organize academic conferences and journals. In the root node, we wrote the central topic ("Conferences and Journals"), and then created child nodes representing categories (e.g. "Information Retrieval" and "User Modeling"). Below each category, we created further child nodes with hyperlinks to websites of corresponding journals and conferences (red arrows indicate a hyperlink). For the "UMAP conference" node, we created an additional note with information about a paper that we submitted to the conference. A circle at the end of a node indicates that the node has child nodes, which are hidden ("folded"). Clicking the circle would unfold the node, i.e. make its child nodes visible again.

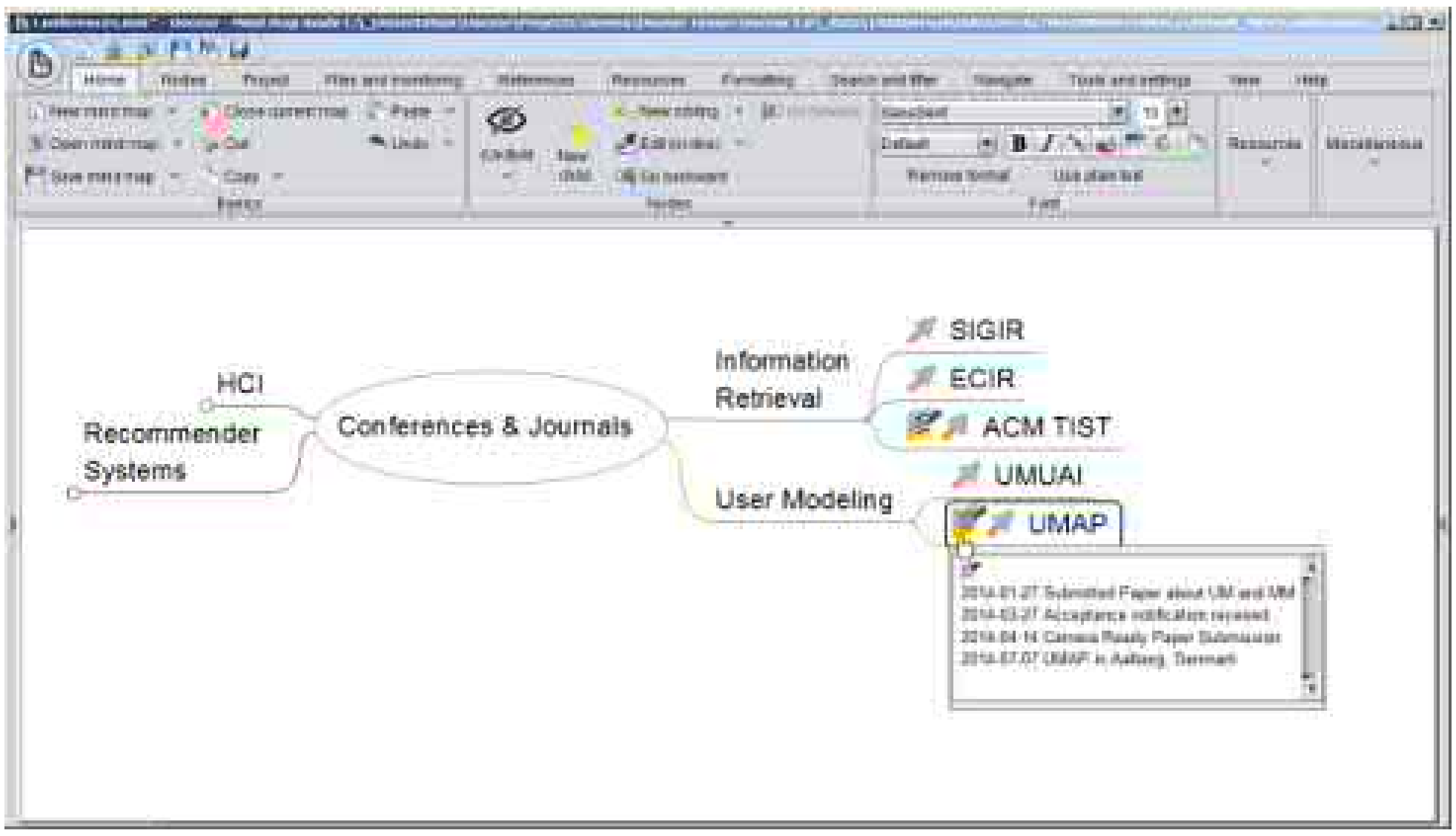

Figure 3: Mind map for managing conferences and journals

Figure 4 shows a mind map created for career planning. For some nodes, we attached icons that indicate the progress of certain tasks.

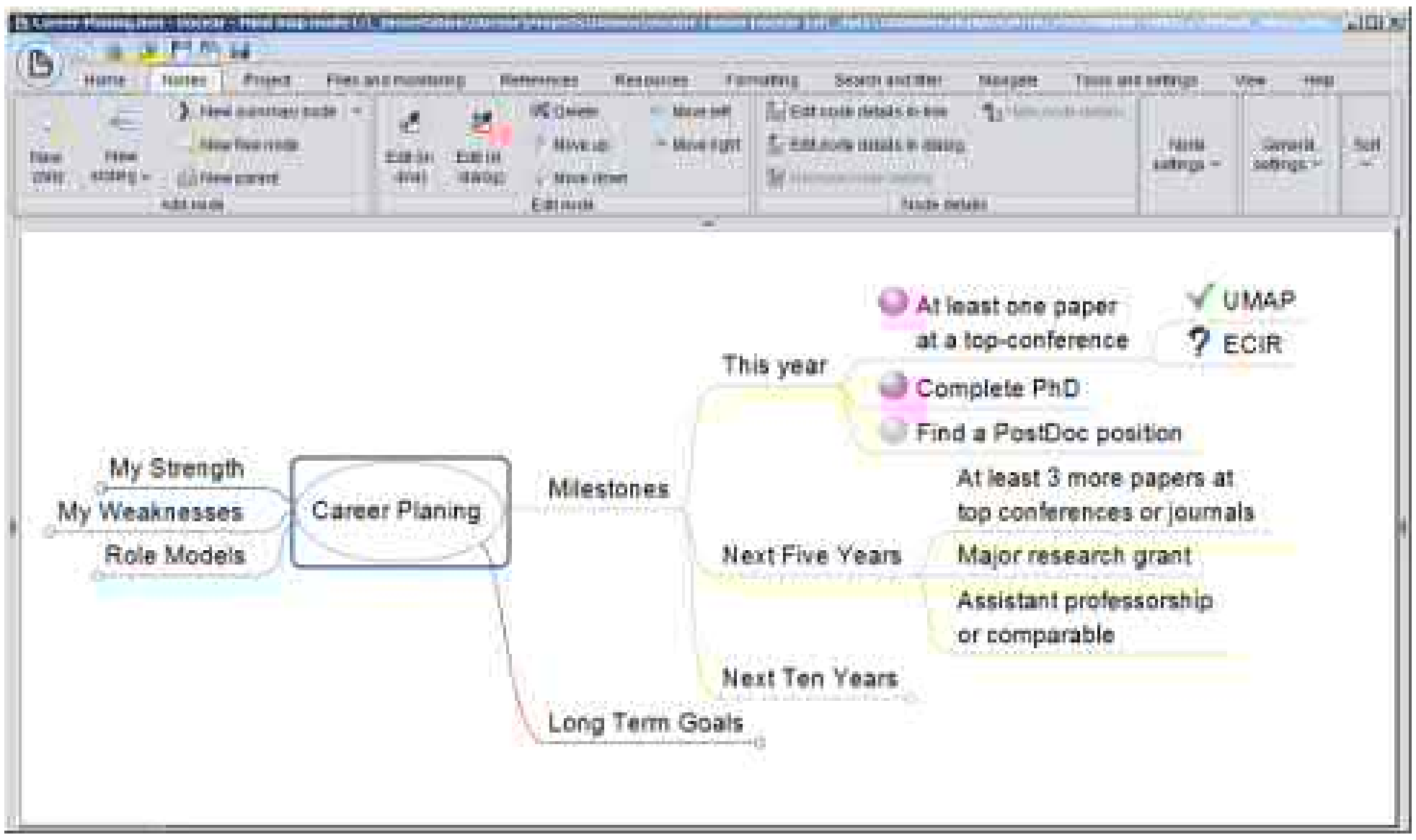

Figure 4: Mind map for career planning

While mind maps share many characteristics with outlines, as well as with other document types such as emails or research articles, mind maps also possess several unique characteristics.

First, mind maps are often personal, while most other document types such as emails and research papers are meant to be seen by at least one other person. As such, we assume that mind maps might not be formulated as well as content in other documents. For instance, mind-maps might contain more abbreviations or spelling errors, because users do not worry about those people who might not be able to understand the mind maps.

Second, mind maps might be less “standardized.” For example, while the structure of research articles is rather standard (title, abstract, headings, body text, etc.), mind maps are used for various tasks. We would assume that mind maps used for project planning differ in structure and content from a mind map used to plan a vacation. This might lead to challenges when it comes to selecting and weighting certain features, e.g. terms, of a mind map. While terms in a title of a research paper are obviously more descriptive than words in the paper’s appendix, such obviousness does not exist for different nodes in mind maps.

Third, mind maps evolve over time, and mind-map-based user-modeling systems might consider this evolution. In contrast, other user-modeling applications typically get access to content when the items are finally published. However, the evolution of mind maps is likely very dissimilar. A brainstorming mind map might have a lifespan of a few hours. A mind map for planning one’s next vacation probably has a lifespan of a few weeks. A mind map for managing literature might be used over several years. This could lead to challenges when the evolution of the mind map is to be considered by a user-modeling system.

Further differences relate to formatting and layout options that are often different for mind maps from other document types. For instance, it is common to fold nodes in mind maps that are not needed at a particular time (cf. Figure 5). Such an option usually does not exist when, for example, writing emails.

The unique characteristics of mind maps led us to the assumption that by considering these characteristics in the user modeling process, user-modeling effectiveness can be improved when compared to the standard user-modeling approaches neglecting the characteristics.

## 2.2 Docear

Docear is an open-source JAVA application for managing PDF files, annotations, and references with mind maps. Figure 5 shows an example of mind maps. In that mind map, we created categories reflecting our research interests ("Academic Search Engines"), subcategories ("Google Scholar"), and sorted PDFs by category and subcategory. Docear imported annotations that we made in the PDFs with a third party PDF editor (comments, highlighted text, and bookmarks). Clicking a PDF icon in the mind map opens the linked PDF file. Docear also extracts metadata from PDF files (e.g. title and journal name), and displays the metadata when the cursor hovers over a PDF icon. Overall, the information is organized similarly to other reference managers such as Endnote, Zotero, or Mendeley (Figure 6), with the difference being that Docear uses mind maps while other reference managers use tables or social tags.

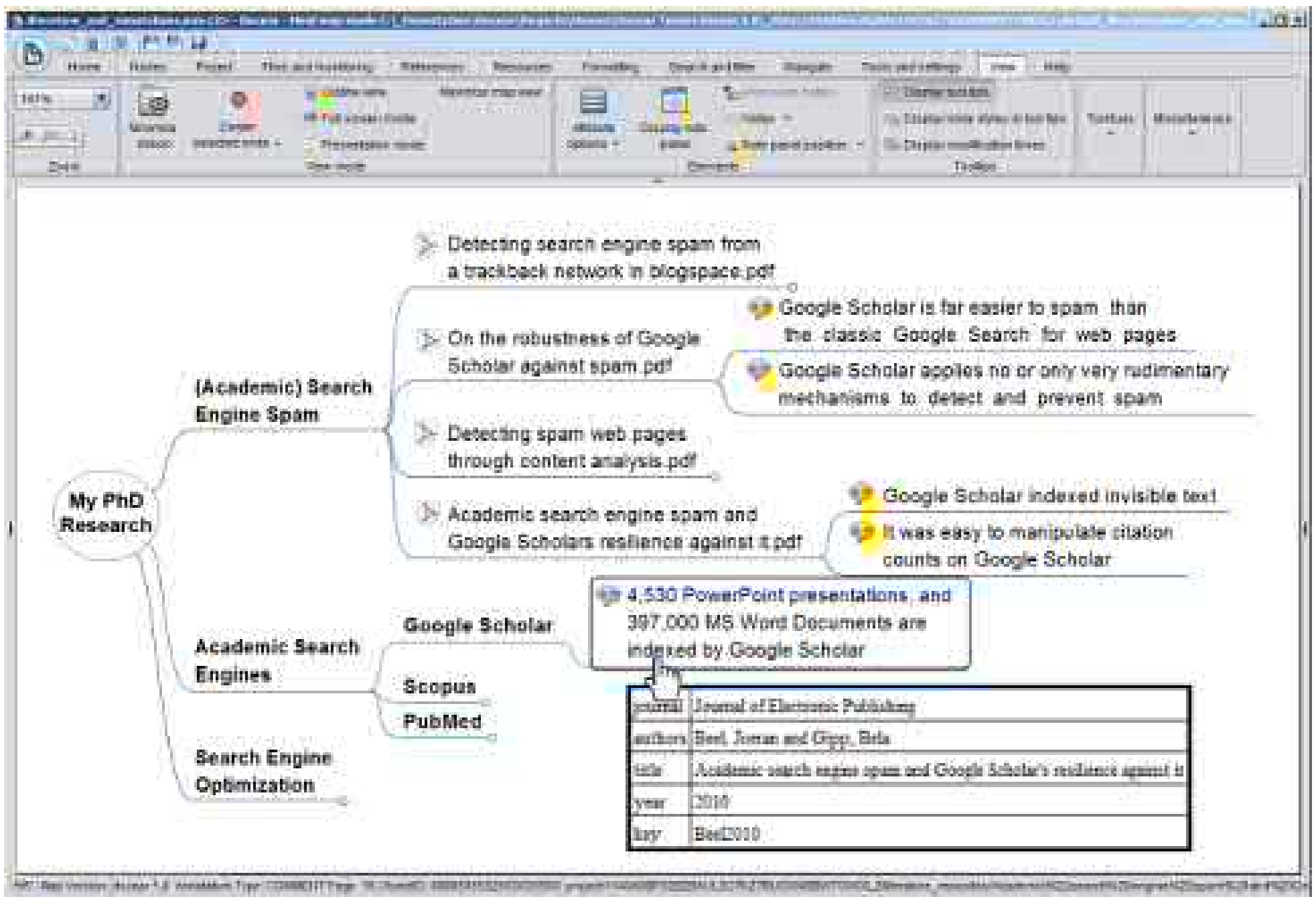

Figure 5: Mind map for organizing academic literature and notes

In addition to the organization of PDFs and references, Docear enables users to draft their own assignments, papers, books, etc. in mind maps. Figure 7 shows a mind map that represents a draft for a new research paper. In the mind map, we outlined the paper that we wanted to write and included LaTeX formulas and images, as well as some of the PDFs and citations from the mind map in Figure 5. We use the term "citation" to refer to a reference or link in a mind map to a research paper. For instance, in Figure 5, nodes with a PDF icon link to a PDF file, typically a research article. If such a link exists, this is seen as a citation for the linked research article. A citation is also made when a user added bibliographic data, such as title and author, to a node (even if the node did not link a PDF).

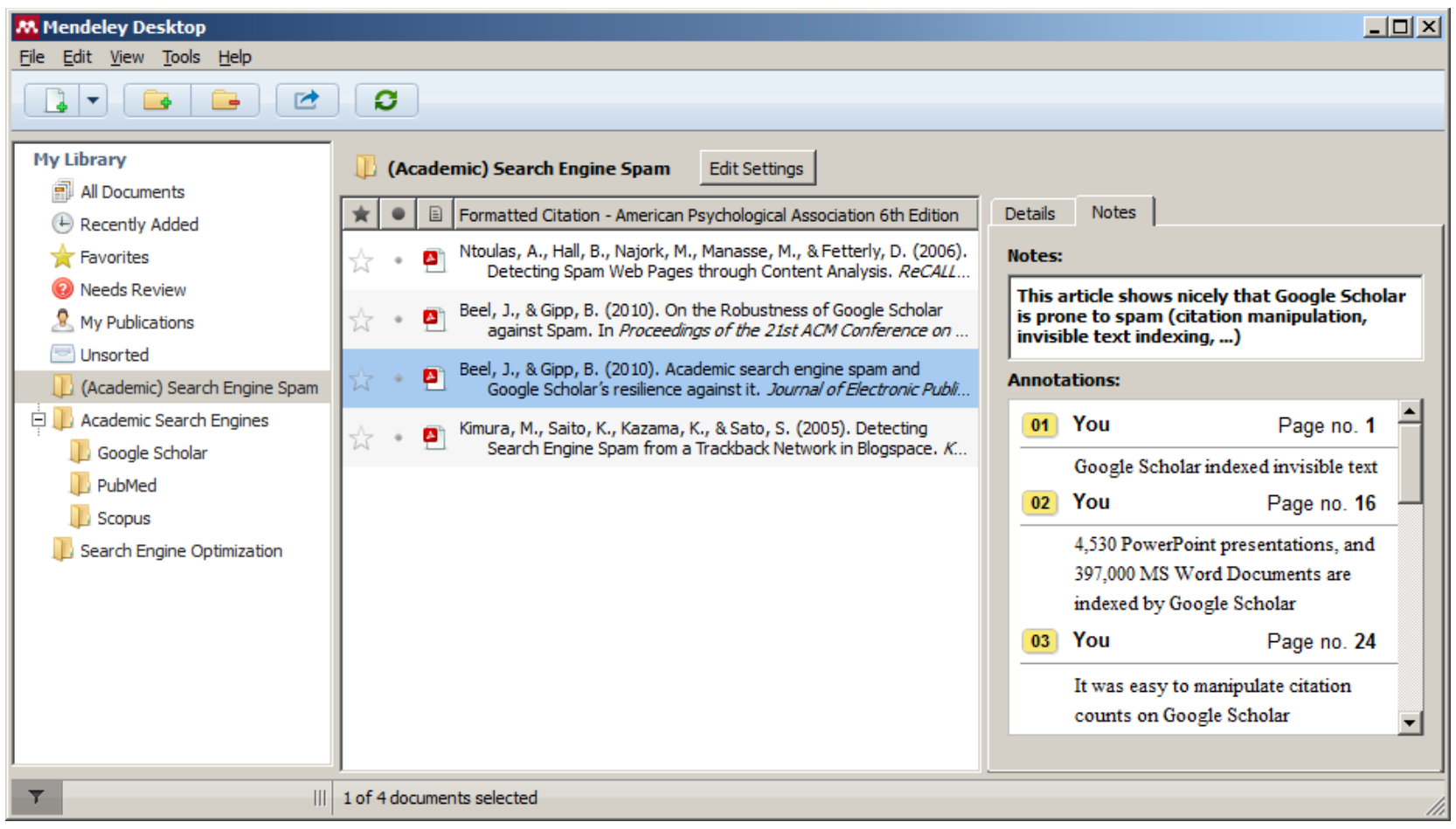

Figure 6: Reference management in Mendeley

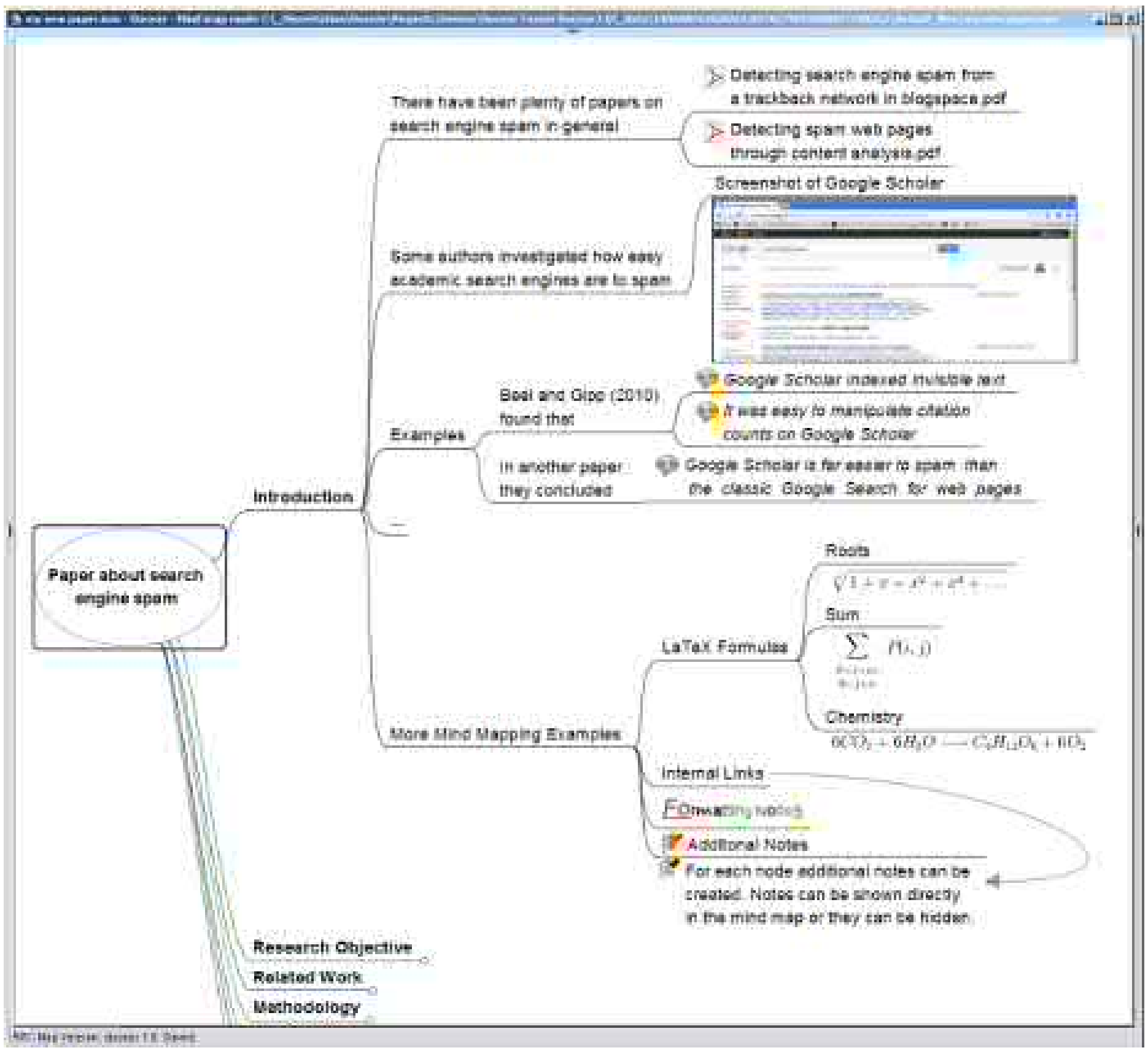

Figure 7: Mind map as a draft for a new research paper

In addition, we developed Docear4Word, an add-on for Microsoft Word for managing references (Figure 8). Docear4Word is based on BibTeX and the Citation Style Language (CSL), features over 1,700 citation styles (Harvard, IEEE, ACM, etc.), is published as open source, and runs with Microsoft Word 2002 (and later) on Windows XP (and later). Docear4Word is similar to the MS-Word add-ons that reference managers like Endnote, Zotero, or Citavi offer, with the difference that it is being developed to work with the de-facto standard BibTeX and thereby work with almost any reference manager. For more details about Docear4Word, refer to Appendix E and http://docear.org.

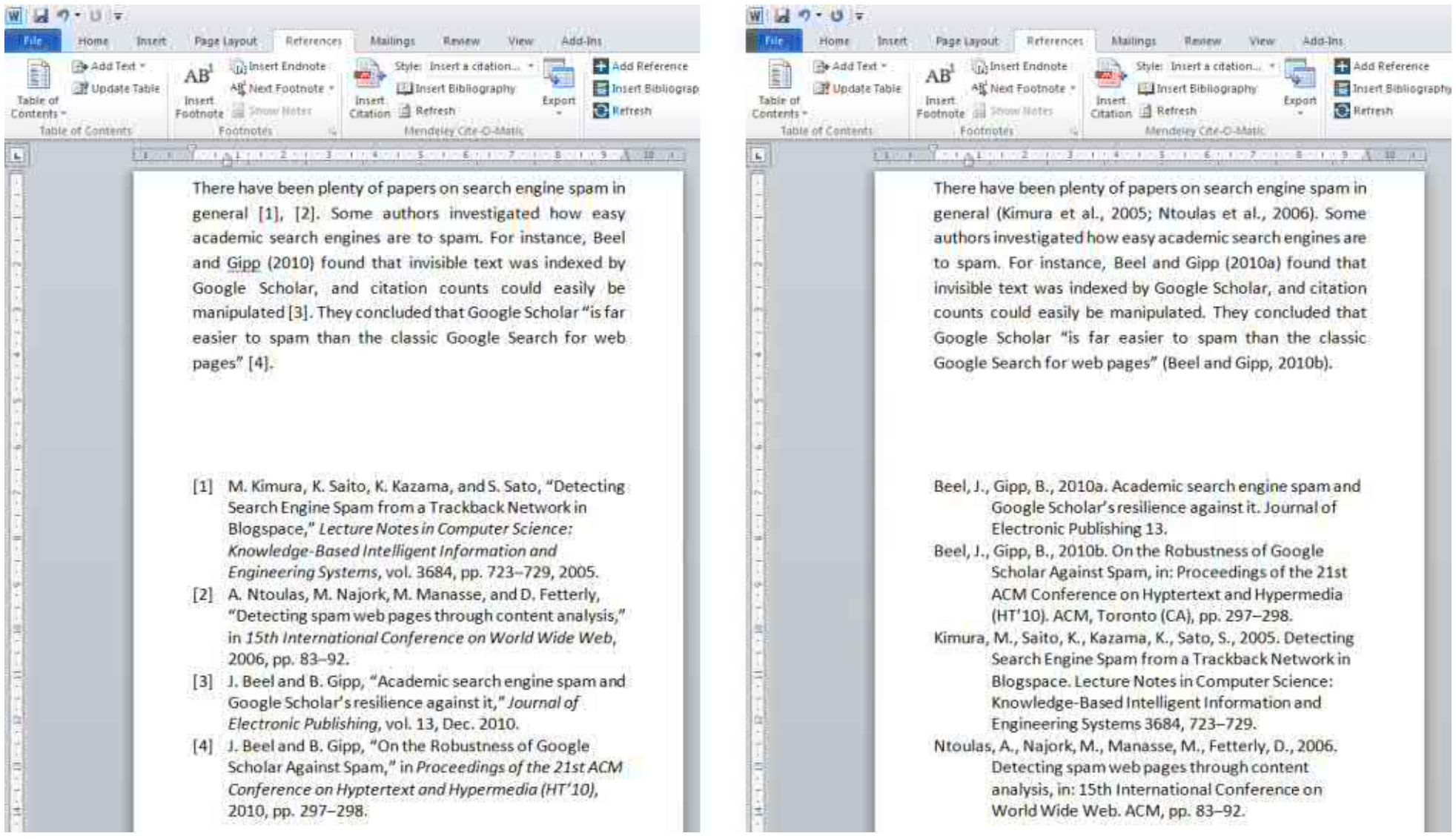

Figure 8: Word document with IEEE (left) and Harvard (right) citation style

## 2.3 Definitions

We use the term "idea" to refer to a hypothesis about how recommendations could be effectively generated. To differentiate how specific the idea is, we distinguish between recommendation classes, approaches, algorithms, and implementations (Figure 9).

We define a "recommendation class" as the least specific idea, namely a broad concept that vaguely describes how recommendations might be given. For instance, collaborative filtering (CF) and content-based filtering (CBF) fundamentally differ in their underlying ideas: the underlying idea of CBF is that users are interested in items that are similar to items the users previously liked. In contrast, the idea of CF is that users like items that the users' peers liked. However, these ideas are rather vague and leave room for speculation about how the idea is actually realized.

A "recommendation approach" is a model of how to bring a recommendation class into practice. For instance, the idea behind CF can be realized with user-based CF [335], content-boosted CF [280], and various other approaches [356]. These approaches are quite different but are each consistent with the central idea of CF. Nevertheless, approaches are rather vague, leaving room for speculation about how recommendations are precisely generated.

A "recommendation algorithm" describes in detail the idea behind a recommendation approach. For instance, an algorithm of a CBF approach would specify whether terms were extracted from the title of a document or from the body of the text, and how terms are processed (e.g. stop-word removal or stemming) and weighted (e.g. TF-IDF). Algorithms are not necessarily complete. For instance, pseudo-code might contain only the most important information and ignore basics such as weighting schemes. This means that for a particular recommendation approach there might be several (slightly) different algorithms.

Finally, an "implementation" is the actual source code of an algorithm that can be compiled and applied in a recommender system. It fully details how recommendations are generated and leaves no room for speculation. It is therefore the most specific idea about how recommendations might be generated.

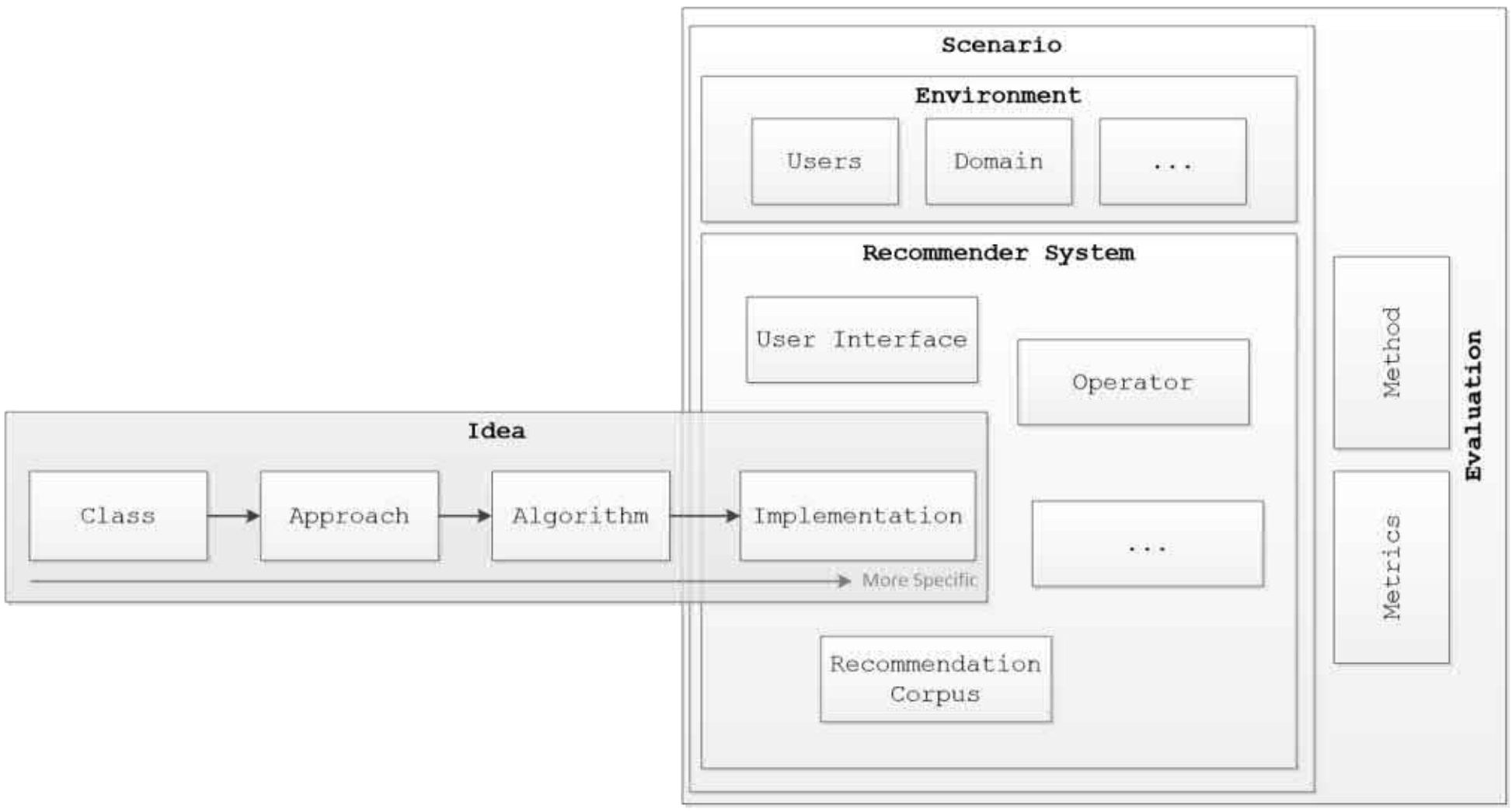

Figure 9: Illustration of recommendation ideas, systems, etc.

A "recommender system" is a fully functional software system that applies at least one implementation to give recommendations. In addition, recommender systems feature several other components, such as a user interface, a corpus of recommendation candidates, and an operator that owns/runs the system. Some recommender systems also apply two or more recommendation approaches. For instance, CiteULike lets their users choose between two approaches to use [57, 89].

The "recommendation scenario" describes the entire setting of a recommender system, including the recommender system and the recommendation environment, i.e. the domain and user characteristics.

"Evaluation" describes any kind of assessment that measures the effectiveness of ideas. There are different methods to evaluate recommender systems, which will be later introduced.

By “effectiveness,” we refer to the degree to which a recommender system achieves its objective. Generally, the objective of a recommender system is to provide “good” [158] and “useful” [168] recommendations that make users “happy” [124] by satisfying their “needs” [275]. The needs of users vary, and consequently, different items might make users happy. For instance, some users might be interested in *novel* research-paper recommendations; others might be interested in *authoritative* research-paper recommendations; and, of course, users require recommendations specific to their fields of research [393]. When we use the term “effectiveness,” we refer to whatever objective the evaluator might have wanted to measure. In addition, we use the terms “performance” and “effectiveness” interchangeably.

## 2.4 User Modeling

In daily life, many situations require user modeling, which is, in a broad sense, the ability to understand someone’s needs and to adapt to them [104, 360]. *Rich* provided the following example of a librarian who needs to know basic information about a library visitor to being able to recommend books to this visitor:

> *“Someone walks into a large library, tells the librarian that he is interested in China, and asks for some books. What sort of books does the librarian recommend? That depends. Is the person a small child who just saw a TV show about China and wants to see more pictures of such an exotic place? Is the person a high school student doing a term paper? Or maybe a prospective tourist? Or a scholar interested in Eastern thought? Can the person read Chinese? The librarian*

> *needs to know these things before he can point the reader to the right books. Some of what he needs to know he'll know before he even thinks about it, such as the approximate age of the person. Some things he'll assume until he has evidence to the contrary, such as that the person does not read Chinese. To find out other things, he'll ask a few specific questions. Only after he has a rough model of the person he's talking to can he answer the question."* [337]

There are different types of computer systems adapting to the needs of their users. *Kobsa* lists ten fields including intelligent interfaces, cognitive engineering, intelligent tutoring, expert systems, and guidance systems [208]. Others add more fields to that list, for instance, educational hypermedia [68], navigation support [127], and dialog strategy [442]. However, we see three main categories in user modeling and that is human computer interaction, user prediction & plan recognition, and information filtering/retrieval.

Human computer interaction (HCI) includes the adaption of software's interfaces [208], navigation [127], and presentation [127] to the users' needs. For instance, for elderly users, larger fonts might be used to compensate vision impairments [161] or menu items may be re-sorted depending on how often they are used [119]. Plan and goal prediction focuses on the actions of a user. For instance, *Thai-Nghe et al.* build user models of students to predict the students' future performance [389]. *Germanakos et al.* predict future purchases from customers [127]. *Hirsh et al.* predict which command line users will enter next [175]. *Macskassy et al.* predict which emails of a user are worth being forwarded to the user's mobile device [260]. However, HCI and plan and goal recognition are out of scope of this doctoral thesis.

The focus of this thesis lies on user modeling for recommender systems, which is a sub-discipline of information filtering, which is a sub-discipline of information retrieval, and *a "research area that offers tools for discriminating between relevant and irrelevant information by providing personalized assistance for continuous retrieval of information*" [224].

The probably two most important questions in user modeling and recommender systems are 1) how to identify the user's information needs, and 2) how to find items satisfying the users' needs? Potential answers to these questions are covered in the next section.

## 2.5 Recommender Systems

### 2.5.1 Introduction

Ideally, a recommender system identifies the users' needs automatically by inferring the needs from the user's item interactions. Alternatively, the recommender system asks users to specify their needs by providing a list of keywords or through some other method. However, in this case a recommender system becomes very much like a search engine and loses one of its main features, namely the capability to recommend items even if users do not know exactly what they need.

To identify users' information needs and match these needs with items, researchers proposed several recommendation classes such as collaborative filtering and content-based filtering, as well as feature-based, knowledge-based, behavior-based, citation-based, context-based, and rule-based recommendations, and many more [70, 79, 244, 322, 330, 392, 425]. We consider the following seven classes to be most appropriate for distinguishing the approaches in the field of research-paper recommender systems:

1. Stereotyping
2. Content-based Filtering
3. Collaborative Filtering
4. Co-Occurrence
5. Graph-based
6. Global Relevance
7. Hybrid

In the following sections, stereotypes, content-based filtering, collaborative filtering, and co-occurrence recommendations are introduced. The other classes are briefly introduced later as they are not that commonly used.

### 2.5.2 Recommendation Classes

#### *2.5.2.1 Stereotyping*

Stereotyping is one of the earliest user modeling and recommendation classes. It was introduced by *Rich* in the recommender system *Grundy*, which recommended novels to its users [337]. *Rich* was inspired by stereotypes from psychology that allowed psychologists to quickly judge people based on a few characteristics. *Rich* defined stereotypes – which she called "facets" – as collections of characteristics. For instance, Grundy assumed that male users have "a fairly high tolerance for violence and suffering, as well as a preference for thrill, suspense, fast plots, and a negative interest in romance" [337]. Consequently, Grundy recommended books that had been manually classified to suit the facets.

One major problem with stereotypes is that they may pigeonhole users. While many men have a negative interest in romance, certainly not all do. Similarly, a recommender system that recommends sausages to users because they are German might please those who actually like sausages, but Germans who are vegetarian or Muslim might feel uncomfortable [210]. In addition, building stereotypes is often labor intensive, as the items typically need to be manually classified for each facet. This limits the number of e.g. books that could reasonably be personalized [15].

Advocates of stereotypes argue that once the stereotypes are created the recommender system needs little computing power and may perform quite well in practice. For instance, *Weber and Castillo* observed that female users were usually searching for the composer Richard Wagner when they entered the search query 'Wagner' on *Yahoo!* [410]. In contrast, male users entering the same query usually were looking for the Wagner paint sprayer. *Weber and Castillo* modified the search algorithm to show the Wikipedia page for Richard Wagner to female users, and the homepage of the Wagner paint sprayer company to male users searching for 'Wagner.' As a result, user satisfaction increased. Similarly, the travel agency *Orbitz* observed that Macintosh users were "40% more likely to book a four- or five-star hotel than PC users" and when booking the same hotel, Macintosh users booked the more expensive rooms [271]. Consequently, Orbitz assigned their website visitors to either the "Mac User" or "PC user" stereotype, and Mac users received recommendations for pricier hotels than PC users. All parties benefited – users received more relevant search results, and Orbitz received higher commissions.

#### *2.5.2.2 Content-based filtering*

Content-based filtering (CBF) is one of the most widely used and researched recommendation approaches [256]. One central component of CBF is the user modeling process, in which the interests of users are inferred from the items that users interacted with. "Items" are usually textual, for instance emails [313] or webpages [5]. "Interaction" is typically established through actions such as downloading, buying, authoring, or tagging an item. Items are represented by a content model containing the items' features. Features are typically word-based, i.e. single words, phrases, n-grams, etc. Some recommender systems also use non-textual features such as writing style [352, 353], layout information [112, 357], and XML tags [74]. Typically, only the most descriptive features are used to model an item and users and these features are typically weighted. Once the most discriminative features are identified, they are stored, typically as vector that contains the features and their weights. The user model typically consists of the features of a user's items. To find recommendations, the user model and recommendation candidates are compared in e.g. the vector space model and similarities are calculated e.g. with Cosine.

CBF has a number of advantages compared to stereotypes. CBF allows a more individual personalization so the recommender system can determine the best recommendations for each user individually, rather than be limited by stereotypes. CBF also requires less labor since user models can be created automatically.

On the downside, content-based filtering requires more computing power than stereotyping. Each item must be analyzed for its features, user models need to be built, and similarity calculations need to be performed. If there are many users and many items, these calculations require significant resources. Content-based filtering is also criticized for low serendipity and overspecialization because it recommends items as similar as possible to the ones a user already knows [256]. Content-based filtering also ignores quality and popularity of items [106]. For instance, two research papers may be considered equally relevant by a CBF recommender system because the papers share the same terms with the user model. However, one paper might be written by an authority in the field, well structured, and presenting original results, while the other paper might be penned by a student, poorly written and just paraphrasing other research papers. Ideally, a recommender system should recommend only the first candidate but a CBF system would fail to do so. Another criticism of content-based filtering is that it is dependent on access to the item's features [106]. For research-paper recommendations, usually PDFs must be processed and converted to text, document fields must be identified, and features such as terms must be extracted. None of these tasks is trivial and they may introduce errors in the recommendations [46, 95, 266].

#### 2.5.2.3 *Collaborative filtering*

The term "collaborative filtering" (CF) was coined in 1992 by *Goldberg et al.*, who proposed that "information filtering can be more effective when humans are involved in the filtering process" [148]. The concept of collaborative filtering as it is understood today was introduced two years later by *Resnick et al.* [335]. Their theory was that users like what like-minded users like, whereas two users were considered like-minded when they rated items alike. When like-minded users were identified, items that one user rated positively were recommended to the other user, and vice versa. Compared to CBF, CF offers three advantages. First, CF is content independent, i.e. no error-prone item processing is required [166, 348, 393]. Second, because the ratings are done by humans, CF takes into account real quality assessments [106]. Finally, CF is supposed to provide serendipitous recommendations because recommendations are not based on *item* similarity but on *user* similarity [166, 275].

A general problem of CF is the "cold start problem," which may occur in three situations [348]: new users, new items, and new communities or disciplines. If a new user rates few or no items, the system cannot find like-minded users and therefore cannot provide recommendations. If an item is new in the system and has not been rated yet by at least one user, it cannot be recommended. In a new community, no users have rated items, so no recommendations can be made and as a result, the incentive for users to rate items is low.

There are further critiques of CF. Computing time for CF tends to be higher than for content-based filtering [348]. Collaborative filtering in general is less scalable and requires more offline data processing than CBF [362]. *Torres et al.* note that collaborative filtering creates similar users [393] and *Sundar et al.* criticize that collaborative filtering dictates opinions [375]. *Lops* makes the criticism that collaborative filtering systems are black boxes that cannot explain why an item is recommended except that other users liked it [256]. Manipulation is also considered a problem: since collaborative filtering is based on user opinions, blackguards might try to manipulate ratings to promote their products so they are recommended more often [277–279].

#### 2.5.2.4 *Co-occurrence recommendations*

To give *co-occurrence* recommendations, those items are recommended that frequently co-occur with some source items. One of the first co-occurrence application was co-citation analysis introduced by *Small* in 1973 [359]. *Small* proposed that two papers are the more related to each other, the more often they are co-cited. This concept was adopted by many others, the most popular example being Amazon's "*Customers Who Bought This Item Also Bought….*" Amazon analyzes which items are frequently bought together, and when a customer browses a product, items frequently bought with that item are recommended.

One advantage of co-occurrence recommendations is the focus on relatedness instead of similarity. Similarity expresses how many features two items have in common. Recommending similar items, as CBF is doing, is often not ideal because similar items are not serendipitous [370]. In contrast, *relatedness* expresses how closely coupled two items are, not necessarily dependent on their features. For instance, two papers sharing the same features (words) are similar. In contrast, paper and pen are not similar but related, because both are required for writing letters. Hence, co-occurrence recommendations provide more serendipitous recommendations and are comparable to collaborative filtering. In addition, no access to content is needed and complexity is rather low. It is also rather easy to generate anonymous recommendations, and hence to assure users' privacy. On the downside, recommendations are not that highly personalized and items can be recommended only if they co-occur at least once with another item.

### 2.5.3 Recommender-Systems Evaluation

To evaluate recommender systems, some researchers distinguish between "offline" and "online evaluations" [436], between "data-centric" and "user-centric" evaluations [344], and between "live user experiments" and "offline analyses" [168]. We adopt the classification by *Ricci et al.* [336] and offer further sub-classification, somewhat inspired by [191], i.e. we distinguish between user studies, online evaluations, and offline evaluations. Our classification is illustrated in Figure 10 and explained in the following sections

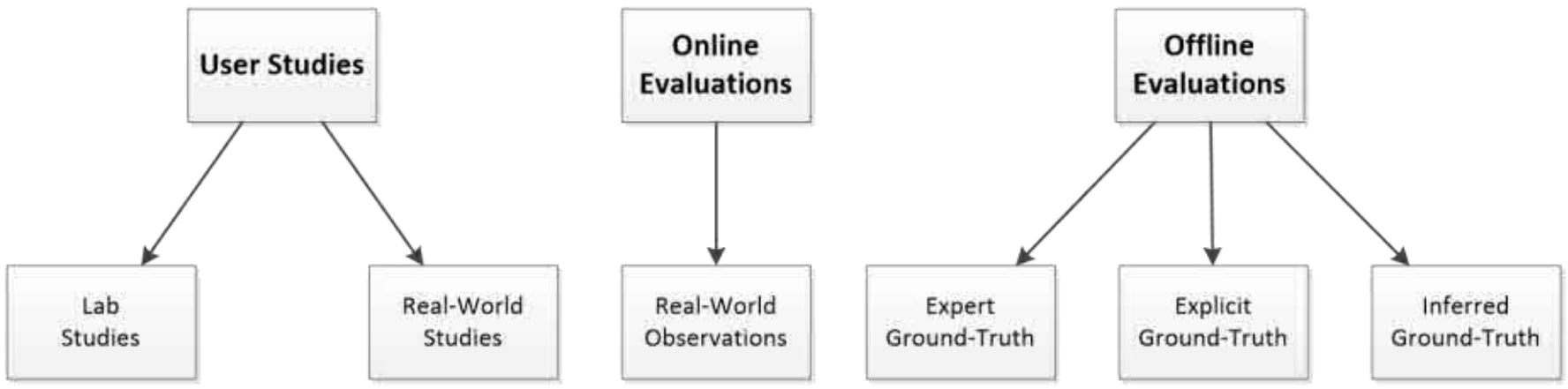

Figure 10: Classification of evaluation methods

#### *2.5.3.1 User studies*

User studies typically measure user satisfaction through explicit ratings. Users receive recommendations generated by different recommendation approaches, rate the recommendations, and the community considers the approach with the highest average rating most effective [336]. Study participants are typically asked to quantify their overall satisfaction with the recommendations. However, they might also be asked to rate individual aspects of a recommender system, for instance, how novel or authoritative recommendations are [393], or how suitable they are for non-experts [219]. A user study can also collect qualitative feedback, but this is rarely done in the field of (research-paper) recommender systems [366, 367].

We distinguish between “lab” and “real-world” user studies. In lab studies, participants are aware that they are part of a user study, which, as well as several other factors, might affect their behavior and thereby the evaluation's results [154, 245]. In real-world studies, participants are not aware of the study and rate recommendations for their own benefit, for instance because the recommender system improves recommendations based on the ratings (i.e. relevance feedback [256]), or ratings are required to generate recommendations (i.e. collaborative filtering [335]).

Often, user studies are considered the optimal evaluation method [346]. However, the outcomes of user studies may depend on the questions users are asked. *Cremonesi et al.* found that it makes a difference whether users are asked for the "perceived relevance" or "global satisfaction" of recommendations [97]. Similarly, it made a difference whether users were asked to rate the *novelty* or *relevance* of recommendations [96]. A large number of participants are also crucial to user study validity, which makes user studies relatively expensive to conduct. The number of required participants, to receive statistically significant results, depends on the number of approaches being evaluated, the number of recommendations being displayed, and the variations in the results [73, 255]. However, as rough estimate, at least a few dozen participants are required, often more.

#### *2.5.3.2 Online evaluations*

Online evaluations originated from online advertising and e-commerce. They measure the acceptance rates of recommendations in real-world recommender systems. Acceptance rates are often measured by click-through rates (CTR), i.e. the ratio of clicked recommendations to displayed recommendations. For instance, if a recommender system displays 10,000 recommendations and 120 are clicked, the CTR is 1.2%. Other metrics include the ratio of downloaded or bought items to the items displayed. Acceptance rate is typically interpreted as an implicit measure for user satisfaction. The assumption is that when a user clicks, downloads, or buys a recommended item, the user liked the recommendation. Of course, this assumption is not always reliable because users might buy a book but rate it negatively after reading it. However, in some cases, metrics such as CTR can be an *explicit* measures of effectiveness, namely when the operator receives money, e.g. for clicks on recommendations.

Online evaluations are not without criticism. *Zheng et al.* showed that CTR and relevance do not always correlate and concluded that "CTR may not be the optimal metric for online evaluation of recommender systems" and "CTR should be used with precaution" [436]. In addition, conducting online evaluations requires significantly more time than offline evaluations, they are more expensive, and they can only be conducted by researchers who have access to a real-world recommender system.

#### *2.5.3.3 Offline evaluations*

Offline evaluations typically measure the *accuracy* of a recommender system based on a ground-truth [200, 212]. To measure accuracy, precision at position *n* (P@n) is often used to express how many items of the ground-truth are recommended within the top *n* recommendations. Other common evaluation metrics include recall, F-measure, mean reciprocal rank (MRR), normalized discounted cumulative gain (nDCG), mean absolute error, and root mean square error. Offline evaluations are also sometimes used to evaluate aspects such as novelty or serendipity of recommendations [124].

We define three types of ground-truths.

‘*Explicit ground-truths*’ contain explicit information about how much users liked certain items, whereas *liked* typically means how well users *rated* an item. To evaluate a recommendation approach, some ratings are removed from the dataset and the recommendation approach predicts the ratings for the removed items. The closer the predicted ratings are to the original ratings, the more accurate the recommender approach is. Figure 11 illustrates the idea of an explicit ground-truth. User *u* has watched five movies and rated how much she liked them on a scale of 1 to 5. Movies D and E are removed from the collection. The recommendation approach predicts the ratings for movies D and E. The prediction for movie D was “perfect” (4) and the prediction for movie E was close (2 instead of 1). Consequently, the evaluated approach would be quite accurate.

*“Inferred ground-truths”* are typically based on personal item collections of users, for instance a list of papers a user cited, or a list of books a user bought. The assumption is that the items in the users' personal collection – and only these items – would have been good recommendations. To evaluate a recommender system based on such a ground-truth, random items are removed from the collections, and recommendations are created based on the remaining items. The more of the removed items are recommended, the more accurate the approach is. At first glance, this concept seems similar to explicit ground-truths, but it is not as Figure 12 illustrates. User *u* has three research papers in her collection (Paper A, B, and C). The recommendation approach recommends three papers (Paper C, D, and E), only one of which is in *u*'s collection (Paper C). Only paper C is considered a "good" recommendation. We propose that this concept is fundamentally flawed, because also Paper D and E might have been relevant recommendations. We will elaborate on this criticism later.

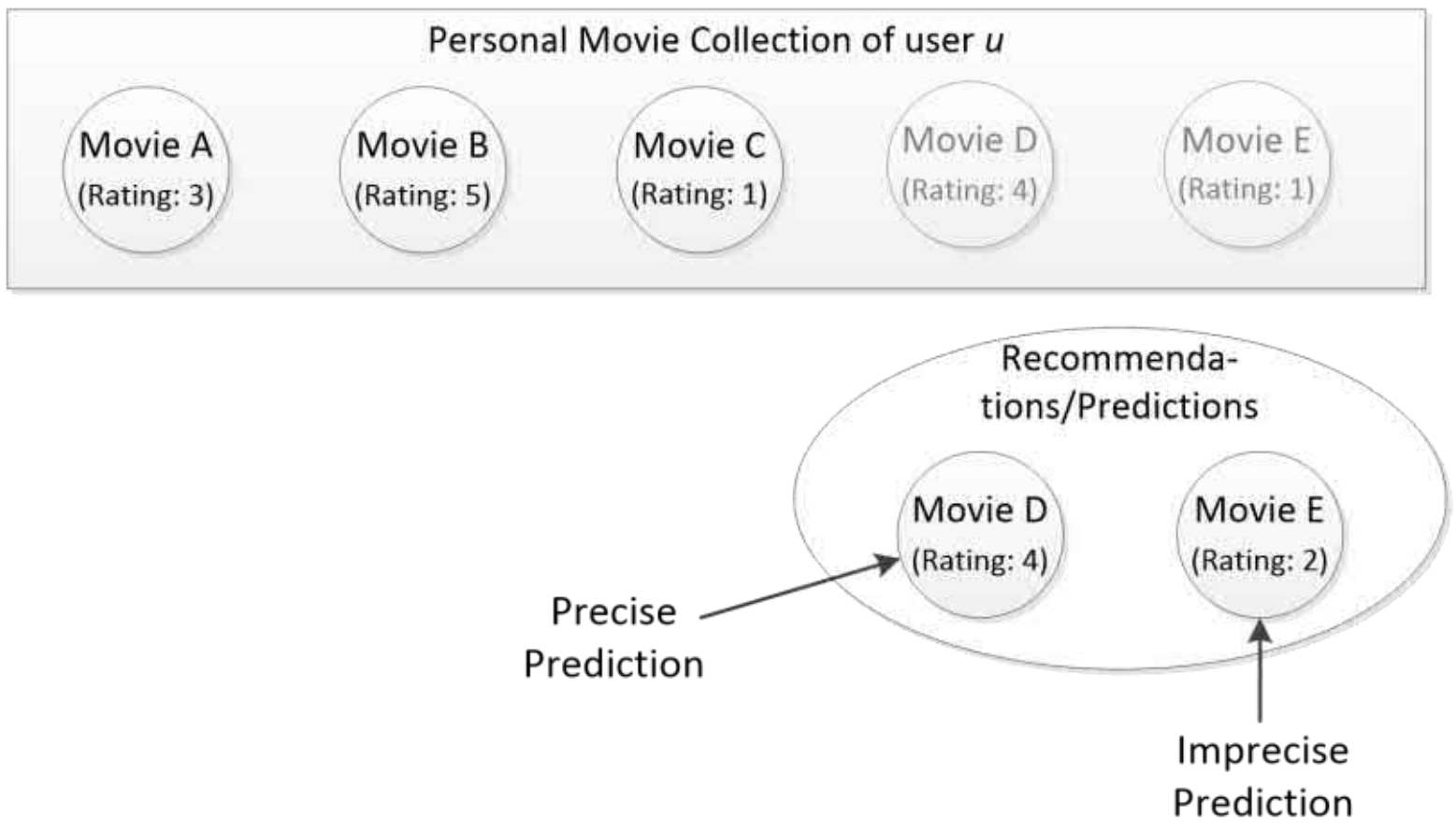

Figure 11: Illustration of an explicit ground-truth

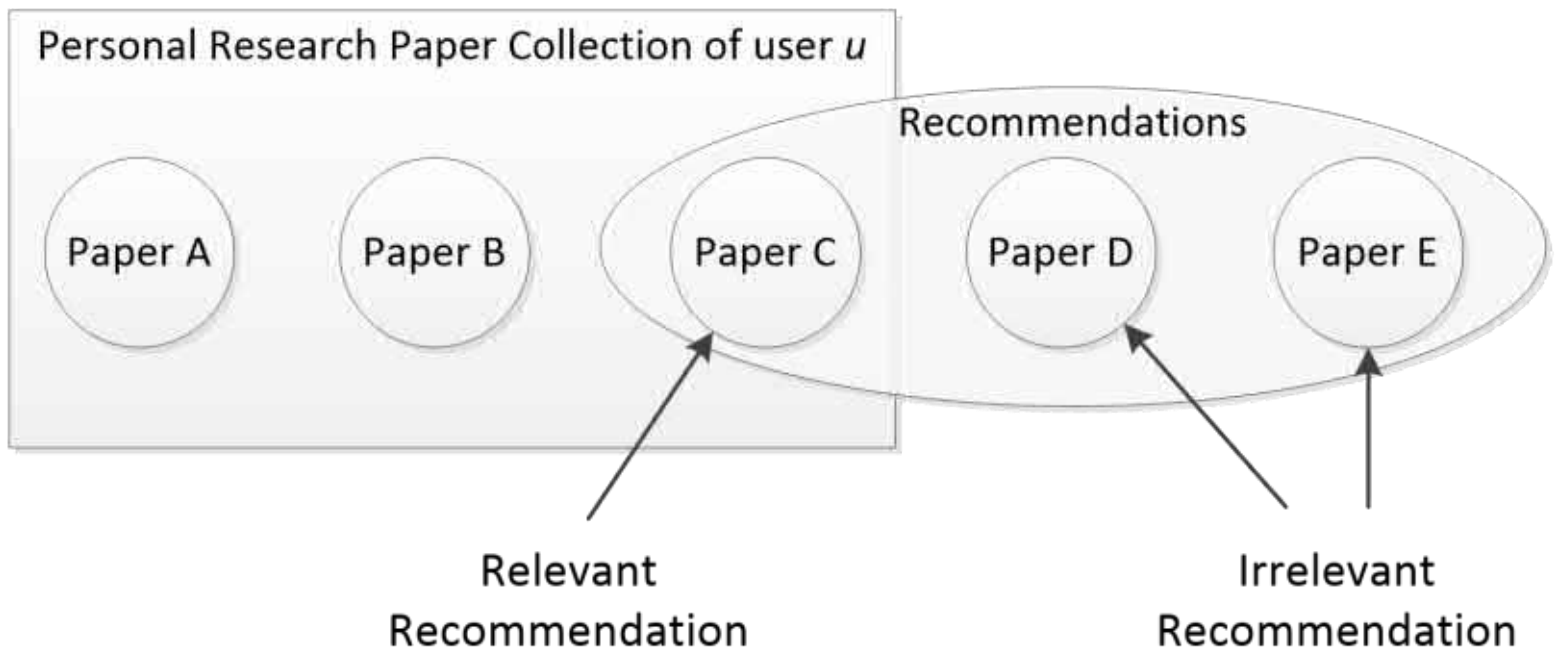

Figure 12: Illustration of inferred ground-truth

*"Expert ground-truths"* contain item classifications that are manually compiled by topical experts. Examples of such datasets include the TREC or MeSH classification. In these datasets, documents such as webpages or research papers are classified according to the information needs they satisfy. In MeSH, for instance, terms from a controlled vocabulary are assigned to research papers. Papers with the same MeSH terms are considered similar. For an evaluation, some papers of one MeSH category are taken as input and the more papers of the same category are recommended, the more accurate the algorithm is.

Offline evaluations were originally meant to identify a number of promising recommendation approaches [158, 269, 331, 336]. These approaches should then be evaluated in detail with a user study or online evaluation to identify the most effective approaches. However, criticism has been raised on the assumption that offline evaluation could predict an algorithm's effectiveness in online evaluations or user studies. More precisely, several researchers have shown that results from offline evaluations do not necessarily correlate with results from user studies or online evaluations [96, 97, 169, 170, 274, 344, 394]. This means that approaches that are effective in offline evaluations are not necessarily effective in real-world recommender systems. Therefore, *McNee et al.* criticized that

> *"the research community's dependence on offline experiments [has] created a disconnect between algorithms that score well on accuracy metrics and algorithms that users will find useful."* [275]

Several more researchers voiced criticism of offline evaluations. *Jannach et al.* stated that "the results of offline [evaluations] may remain inconclusive or even misleading" and "real-world evaluations and, to some extent, lab studies represent probably the best methods to evaluate systems" [191]. *Knijnenburg et al.* reported that "the presumed link between algorithm accuracy […] and user experience […] is all but evident" [207]. *Said et al.* consider "on-line evaluation [as] the only technique able to measure the true user satisfaction" [346]. *Rashid et al.* criticize that biases in the offline datasets may cause bias in the evaluation [331]. The main reason for the criticism in the literature is that offline evaluations ignore human factors; yet human factors strongly affect overall user satisfaction with recommendations.

Despite the criticism, offline evaluations are the predominant evaluation method in the recommender community [192] and "surprisingly few studies [evaluate] algorithms in live experiments with real users" [207].

#### *2.5.3.4 The operator's perspective*

It is commonly assumed that the objective of a recommender system is to make users "happy" (cf. 2.3, p. 14). However, there is another important stakeholder who is often ignored in the general recommender literature: the operator of a recommender system [158].

Operators of recommender systems often are assumed to be satisfied when their users are satisfied, but this is not always the case. Operators may also want to keep down costs for labor, disk storage, memory, CPU power, and traffic [336]. Therefore, for operators, an effective recommender system may be one that can be developed, operated, and maintained at a low cost. Operators may also want to generate a profit from the recommender system [158]. Such operators might prefer to recommend items with higher profit margins, even if user satisfaction was not optimal. For instance, publishers might be more interested in recommending papers the user would have to pay for than papers the user could freely download.

## 2.6 Related Research Fields

Several research fields are related to user modeling and (research-paper) recommender systems. While we did not survey these fields, we are introducing them, so interested readers may broaden their research into these directions.

Research on *academic search engines* deals with calculating relevancies between research papers and search queries [63, 338, 339]. The techniques are often similar to those used by research-paper recommender systems. In some cases, recommender systems and academic search engines are even identical. As shown later in detail, some of the recommender systems require their users to provide keywords that represent their interests. In such cases, research-paper recommender systems do not differ from academic search engines where users provide keywords to retrieve relevant papers. Consequently, these fields are highly related and most approaches for academic search engines are relevant for research-paper recommender systems.

The *reviewer assignment problem* targets using information-retrieval and information-filtering techniques to automate the assignment of conference papers to reviewers [109]. The differences from research-paper recommendations are minimal: in the reviewer assignment problem a relatively small number of paper submissions *must* be assigned to a small number of users, i.e. reviewers; research-paper recommender systems recommend a few papers out of a large corpus to a relatively large number of users. However, the techniques are usually identical. The reviewer assignment problem was first addressed by *Dumais and Nielson* in 1992 [109], six years before *Giles et al.* introduced the first research-paper recommender system [139]. A good survey on the reviewer assignment problem was published by *Wang et al.* [406].

*Scientometrics* deals with analyzing the impact of researchers, research articles and the links between them. Scientometrics researchers use several techniques to calculate document relatedness or to rank a collection of articles, and some of them – h-index [174], co-citation strength [359] and bibliographic coupling strength [203] – have also been applied by research-paper recommender systems [54, 413, 427]. However, there are many more metrics in scientometrics that might be relevant for research-paper recommender systems [443].

Other related research fields include book recommender systems [292], educational recommender systems [69], academic alerting services [114], expert search [100], automatic summarization of academic articles [194, 289, 388], academic news feed recommenders [91, 318], academic event recommenders [206], venue recommendations [421], citation recommenders for patents [308], recommenders for academic datasets [358], and plagiarism detection. The latter, like many research-paper recommenders, utilizes text and citation analysis to identify similar documents [141, 431, 439]. In addition, research that relates to crawling the web and analyzing academic articles can be useful for building research-paper recommender systems, for instance, author name extraction and disambiguation [246], title extraction [36, 46, 160, 178, 319], or citation extraction and matching [237]. Finally, most of the research about content-based [256] or collaborative filtering [336, 348] from other domains (e.g. movies or news) is relevant for research-paper recommender systems as well.

# 3. Related Work[4]

This chapter presents related work on mind mapping, research-paper recommender systems, and recommender-systems evaluation. The primary goal of the review was to identify promising user-modeling approaches to apply with mind maps, as well as to identify adequate evaluation methods and metrics to measure the effectiveness of recommendation approaches (cf. research questions *i* and *iii*, p. 4). Apart from answering research question *i* and *iii*, the review aimed at providing a comprehensive and critical overview of available research-paper recommender systems, and the approaches and techniques they apply, as well as to identify potential problems that require further research. This enables researchers and developers to (a) learn about the most important aspects of research-paper recommender systems, (b) identify promising fields of research, and (c) motivate the community to solve the most urgent problems that currently hinder the effective use of research-paper recommender systems.

The focus of the survey lies on 70 recommendation approaches that were presented in 127 research articles. We analyze the use of recommendation classes such as collaborative filtering, the use of document fields such as title, abstract, or citation context, and the use of weighting schemes such as TF-IDF. We review the approaches' evaluations, including which evaluation methods were applied (e.g. user-studies or offline evaluations), which evaluation metrics were used (e.g. precision or recall), how many participants the user studies had, and how strong datasets were pruned. A discussion and critical analysis of the most serious limitations in the research field follows, exploring inadequate evaluations, sparse information on algorithms, neglecting the user modeling process and overall user satisfaction, and not transferring research results into practice. A review of the individual recommendation approaches can be found in Appendix F (p. 219).

## 3.1 Introduction

Mind maps received significant attention in various research fields. In the field of human computer interaction (HCI), *Faste and Lin* evaluated the effectiveness of mind mapping tools and developed a framework for mind-map-based collaboration [116]. In the field of document engineering and text mining, *Kudelic et al.* created mind maps from texts automatically [223], and *Bia et al.* utilized mind maps to model semi-structured documents, i.e. XML files and the corresponding DTDs, schemas, and XML instances [56]. In the field of education, *Jamieson* researched how graph analysis techniques could be used with mind maps to quantify the learning of students [190], and *Somers et al.* used mind maps to research how knowledgeable business school students are [361]. Furthermore, mind maps have been used to implement a lambda calculator [83], to filter search results from Google [440], to conduct peer-review [349], present software requirements [92], and there are numerous studies about the effectiveness of mind maps as learning tool [84, 98, 99, 123, 176, 193, 226, 302, 305, 364, 405, 407].

---

[4] Parts of this chapter have been published as: J. Beel, S. Langer, M. Genzmehr, B. Gipp, C. Breitinger, and A. Nürnberger, "Research Paper Recommender System Evaluation: A Quantitative Literature Survey," in Proceedings of the Workshop on Reproducibility and Replication in Recommender Systems Evaluation (RepSys) at the ACM Recommender System Conference (RecSys), 2013, pp. 15–22.

However, the research on mind maps is not helpful for developing a user-modeling and recommender system based on mind maps. Therefore, we shifted the focus of our literature review from mind maps to recommender systems. Since the body of literature in the field of recommender systems is huge, we decided to narrow down our review to *research-paper recommender system*s, as we wanted to apply our mind-map-based user modeling in the context of such systems.

The first research-paper recommender system was presented in 1998, by *Giles et al.* as part of the *CiteSeer* project [139]. Since then, at least 216 more articles about research-paper recommender systems were published [1, 2, 4, 6, 9, 10, 12, 14, 33, 35, 42, 44, 45, 47, 48, 52, 54, 57–61, 75, 78–81, 85–90, 94, 101, 102, 107, 110, 111, 113, 115, 117, 118, 121, 122, 126, 128–138, 140, 142, 145, 152, 153, 155–157, 162–165, 171–173, 177, 179–183, 185–187, 196–199, 202, 204, 209, 211, 213–222, 227, 228, 231–236, 238, 239, 241, 244, 247, 249–251, 258, 264, 265, 268, 270, 274, 275, 281–286, 288, 290, 291, 293–301, 304, 307, 309–312, 315, 320, 321, 323, 324, 326–328, 332, 334, 340–342, 347, 354, 360, 366–369, 371–374, 376–387, 390, 391, 393, 395–401, 404, 408, 409, 411–420, 422–424, 426–429, 432, 434, 435, 437][5]. The yearly number of publications steadily increases: 66 of the 217 surveyed articles (30%) were published just in the past two years (Figure 13 & Table 1). The few existing literature surveys in the field cover only a fraction of the published articles [156, 244, 360]. Hence, they do not help in obtaining an overview of the research field, and identifying the most promising approaches (neither generally, nor for our specific purpose of generating recommendations based on mind maps).

[5] Numbers are based on our literature search. Although, we believe our survey to be the most comprehensive survey about research-paper recommender systems, we may have missed a few articles. In addition, most likely, more than 40 papers were published in 2013 since we conducted the literature search in January 2014. Articles presented at conferences in late 2013 most likely had not been published in conferences proceedings by January 2014, and hence were not found through our search. Hence, the total number of papers published is probably somewhat more than 217.

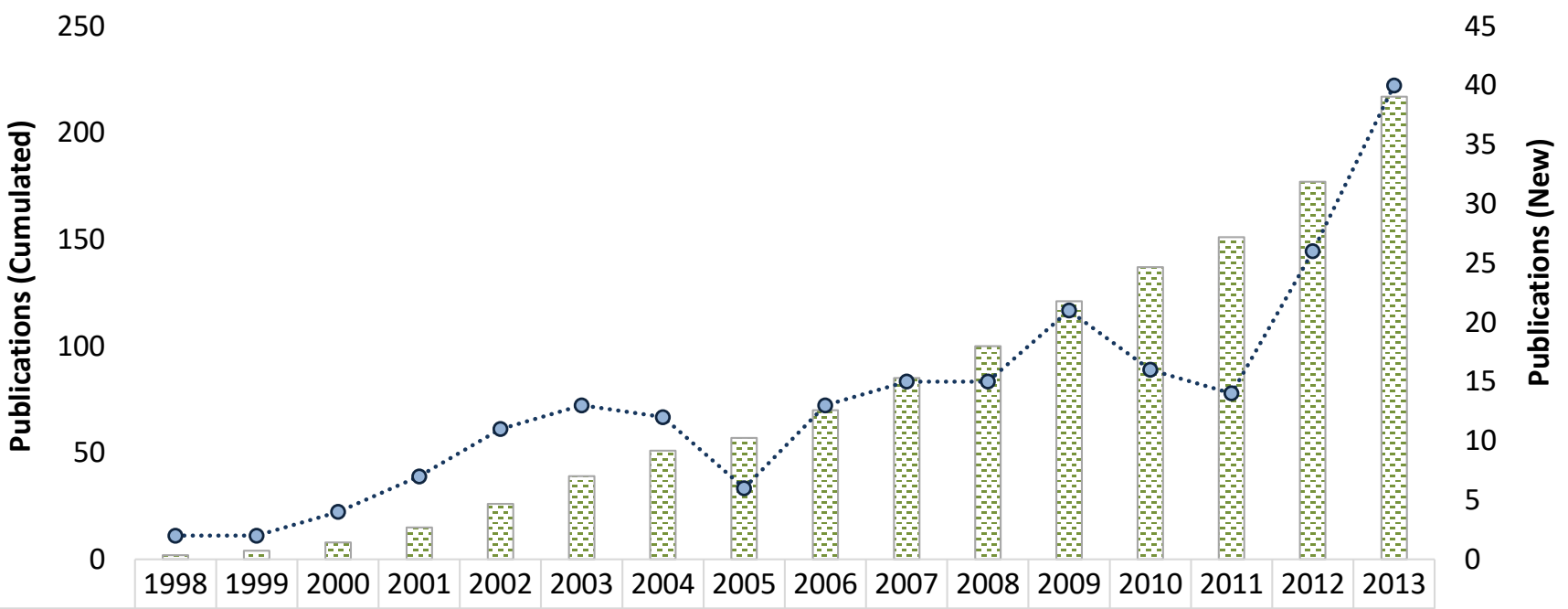

| | 1998 | 1999 | 2000 | 2001 | 2002 | 2003 | 2004 | 2005 | 2006 | 2007 | 2008 | 2009 | 2010 | 2011 | 2012 | 2013 |
|---|---|---|---|---|---|---|---|---|---|---|---|---|---|---|---|---|
| Cumulated | 2 | 4 | 8 | 15 | 26 | 39 | 51 | 57 | 70 | 85 | 100 | 121 | 137 | 151 | 177 | 217 |
| New (per year) | 2 | 2 | 4 | 7 | 11 | 13 | 12 | 6 | 13 | 15 | 15 | 21 | 16 | 14 | 26 | 40 |

Year

Figure 13: Published papers per year[5]

The 217 surveyed articles were obtained through a literature search in June 2013 and January 2014. We conducted the search via *Google Scholar*, *ACM Digital Library*, *Springer Link*, and *ScienceDirect*, and searched for *[paper | article | citation] [recommender | recommendation] [system | systems]*. All articles that had relevance for research-paper recommender systems were downloaded. Our relevance judgment made use of the title, and the abstract where the title was not clear. We examined the bibliography of each article. When an entry in the bibliography pointed to a relevant article not yet downloaded, we also downloaded that article. We expanded our search to websites, blogs, patents, and presentations on major academic recommender systems. These major academic services include the academic search engines *CiteSeer*(x)[6], *Google Scholar* (*Scholar Update*)[7], and *PubMed*[8]; the social network *ResearchGate*[9]; and the reference managers *CiteULike*[10] and *Mendeley*[11]. While these systems offer recommender systems along with their main services, there are also a few stand-alone recommender systems, namely *BibTip*[12], *bX*[13], *RefSeer*[14], *TheAdvisor*[15] and an experimental system called *Sarkanto*[16]. For clarity, we use the term "article" to refer to the 217 reviewed documents, and the term "paper" to refer to documents being recommended by research-paper recommender systems[17].

[6] http://citeseerx.ist.psu.edu
[7] http://scholar.google.com/scholar?sciupd=1&hl=en&as_sdt=0,5
[8] http://www.ncbi.nlm.nih.gov/pubmed
[9] http://www.researchgate.net/
[10] http://www.citeulike.org/
[11] http://www.mendeley.com/

Table 1: List of papers by year

| **Year** | **1998** | **1999** | **2000** | **2001** | **2002** | **2003** | **2004** | **2005** | **2006** | **2007** | **2008** | **2009** | **2010** | **2011** | **2012** | **2013** |
|---|---|---|---|---|---|---|---|---|---|---|---|---|---|---|---|---|
| **Reference** | [59, 139] | [236, 340] | [60, 117, 320, 413] | [126, 131, 133, 239, 284, 341, 398] | [129, 130, 132, 182, 274, 282, 285, 297, 309, 310, 397] | [58, 128, 134–138, 157, 183, 368, 381, 383, 386] | [111, 179, 238, 286, 312, 323, 324, 377, 378, 382, 393, 396] | [2, 78, 79, 211, 360, 409] | [1, 4, 61, 85, 94, 122, 153, 171, 173, 198, 247, 275, 311] | [11, 121, 181, 199, 251, 265, 270, 326, 327, 332, 369, 385, 401, 418, 423] | [57, 80, 115, 165, 172, 228, 290, 291, 296, 298, 328, 379, 411, 432, 437] | [9, 35, 88, 90, 101, 102, 107, 142, 145, 209, 283, 293–295, 304, 367, 376, 380, 384, 404, 417] | [54, 86, 110, 140, 164, 197, 202, 235, 268, 315, 354, 372, 399, 408, 419, 434] | [12, 33, 89, 118, 155, 156, 163, 232, 250, 258, 301, 334, 390, 395] | [14, 87, 152, 162, 177, 180, 185–187, 196, 213, 215, 217–219, 231, 233, 234, 244, 288, 387, 412, 414, 415, 426, 427] | [6, 42, 44, 45, 47, 48, 52, 75, 81, 113, 204, 214, 216, 220–222, 227, 241, 249, 264, 281, 299, 300, 307, 321, 342, 347, 366, 371, 373, 374, 391, 400, 416, 420, 422, 424, 428, 429, 435] |

The 217 articles consist of peer reviewed conference articles (71%), journal articles (14%), pre-prints (4%), and other formats such as PhD theses, patens, presentations and web pages (Table 2). When referring to a large number of recommender systems with certain properties, we cite only three exemplary articles. For instance, when we report how many recommender systems apply content-based filtering, we report the number or percentage and provide three exemplary references [47, 187, 236].

---

[12] http://www.bibtip.com/
[13] http://www.exlibrisgroup.com/category/bXUsageBasedServices
[14] http://refseer.ist.psu.edu/
[15] http://theadvisor.osu.edu/
[16] http://lab.cisti-icist.nrc-cnrc.gc.ca/Sarkanto/
[17] Some recommender systems also recommended "citations" but in our opinion, differences between recommending papers and citations are marginal, which is why we do not distinguish between these two terms in the remainder.

Table 2: Article types

| Journal articles | Conference papers | PhD Theses | Master's Theses | Patents | Pre-prints/unpublished | Other |
|---|---|---|---|---|---|---|
| 14% | 71% | 1% | 1% | 1% | 4% | 7% |

We used all 217 articles for some quantitative analyses about, e.g. page counts, citation counts, and number of authors. Citation counts were retrieved from Google Scholar in early 2014. Some researchers have reservations about using Google Scholar as source for citation counts [23, 184, 306], but the numbers should give a sufficient idea of a paper's popularity.

Of the 217 articles, we reviewed 127 articles about 70 recommendation approaches in detail [4, 9, 12, 33, 35, 42, 47, 48, 52, 54, 57–59, 61, 75, 80, 86–90, 94, 107, 110, 111, 113, 118, 121, 128–140, 142, 145, 152, 153, 155, 162–165, 171–173, 179, 180, 182, 183, 185–187, 196, 197, 199, 202, 209, 211, 213, 215–220, 228, 231–236, 238, 239, 247, 250, 251, 258, 274, 275, 282–286, 290, 291, 294–296, 298, 301, 304, 309–312, 320, 323, 324, 326, 327, 334, 342, 347, 369, 372, 390, 393, 395, 399, 401, 408, 409, 413, 417, 427, 428, 437]. We read those 127 articles thoroughly and present their main ideas and results in this survey. The remaining 90 articles were excluded for one of the following reasons:

- 58 articles were excluded because we considered them to be of little significance [1, 2, 11, 14, 60, 78, 79, 85, 101, 102, 115, 117, 126, 157, 177, 181, 198, 265, 268, 270, 288, 293, 297, 315, 328, 332, 340, 341, 354, 366–368, 376–387, 396–398, 404, 411, 412, 414, 415, 418, 419, 423, 426, 432, 434]. We judged articles to be of little significance when they provided neither an evaluation nor an interesting new approach; when they were not understandable due to serious English errors; or when they were out of scope (although the article's title suggested some relevance to research-paper recommender systems). One example of an article out of scope is 'Research Paper Recommender Systems - A Subspace Clustering Approach' [2]. The title seems relevant for this survey, but the article presents a collaborative filtering approach that does not focus on research-paper recommender systems. Instead, it is based on the *Movielens* dataset, which contains ratings of movies.
- 28 articles were excluded because they were found during the second round of literature search in January 2014 [6, 45, 81, 122, 204, 214, 221, 222, 227, 241, 249, 264, 281, 299, 300, 307, 321, 371, 373, 374, 391, 400, 416, 420, 422, 424, 429, 435] when we were researching the number of articles published in 2013 so we

could create Figure 13. It would have been interesting to include these articles in the in-depth review, but in the time it would have taken to review them, some more articles would have been published, and we would never have finished the survey.

- Four articles were literature surveys on research-paper recommender systems; hence they did not presented any new approaches and were not relevant for our survey [44, 156, 244, 360].

Overall, the reviewed articles were comprehensive, with a median page count of nine. Almost half of the articles (45.78%) had 10 or more pages (Figure 15). Another 16.2% had eight or nine pages. Only 21.8% of the articles had four or fewer pages. Citation counts follow a typical power-law distribution: a few articles gained many citations (maximum was 528 [139]) and many articles had few citations (Figure 16 and Figure 17). Mean citation count was 39, and median was nine. From the reviewed articles, 19.80% had no citations, 32.67% had less than 10 citations. Not surprisingly, the earlier an article was published, the more citations it tended to have (Figure 14).

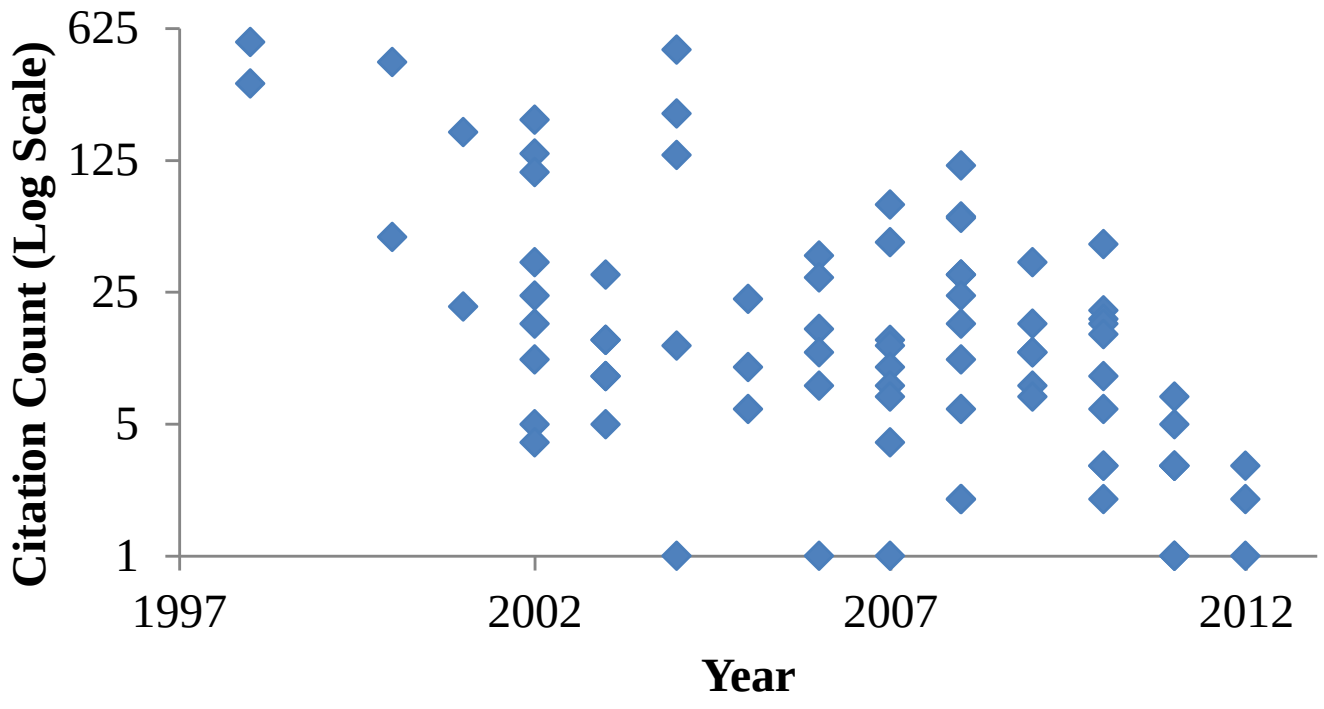

Figure 14: Citation counts by year[18]

[18] Articles with no citations are not plotted due to the log scale

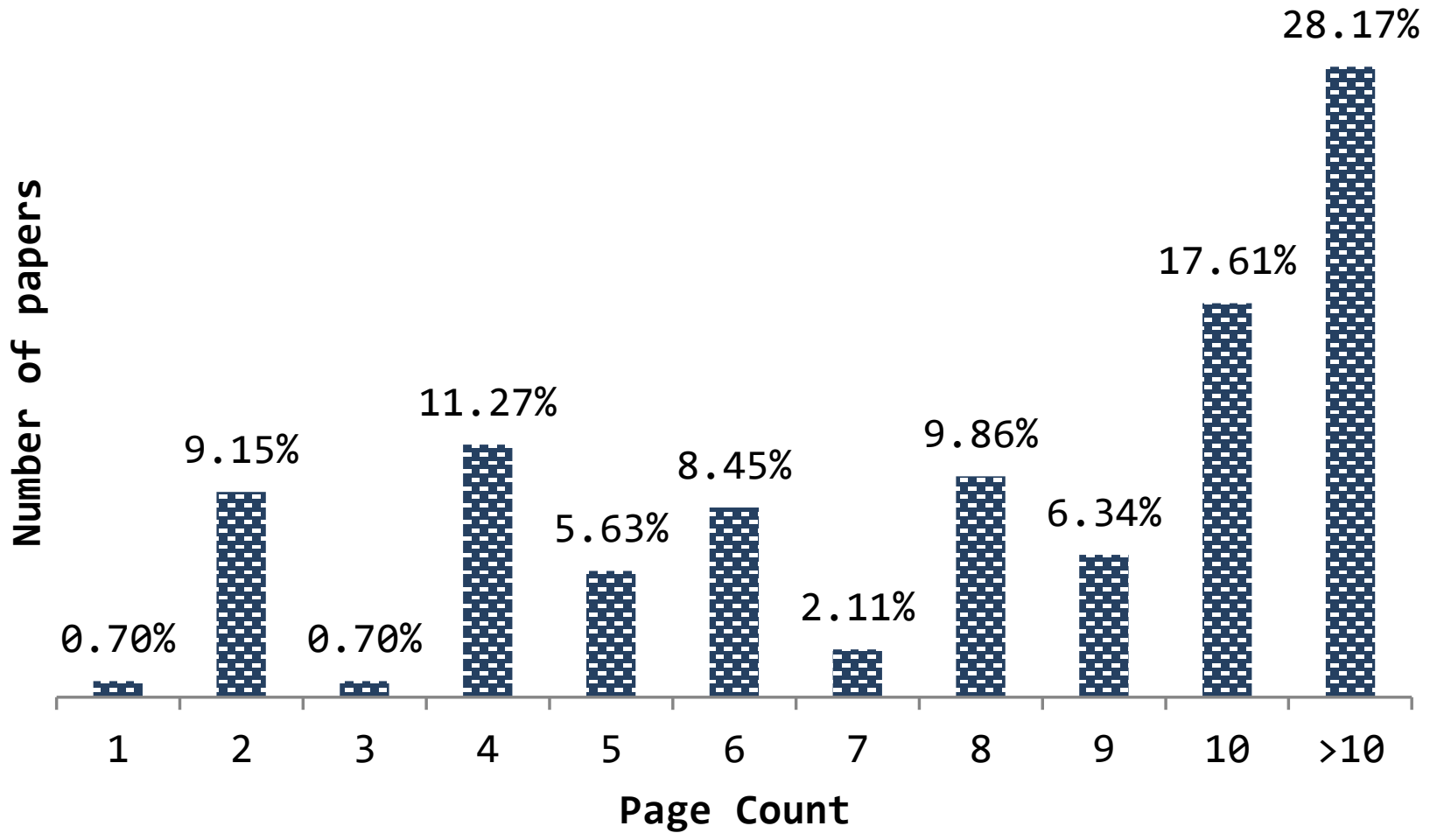

Figure 15: Page count of reviewed articles

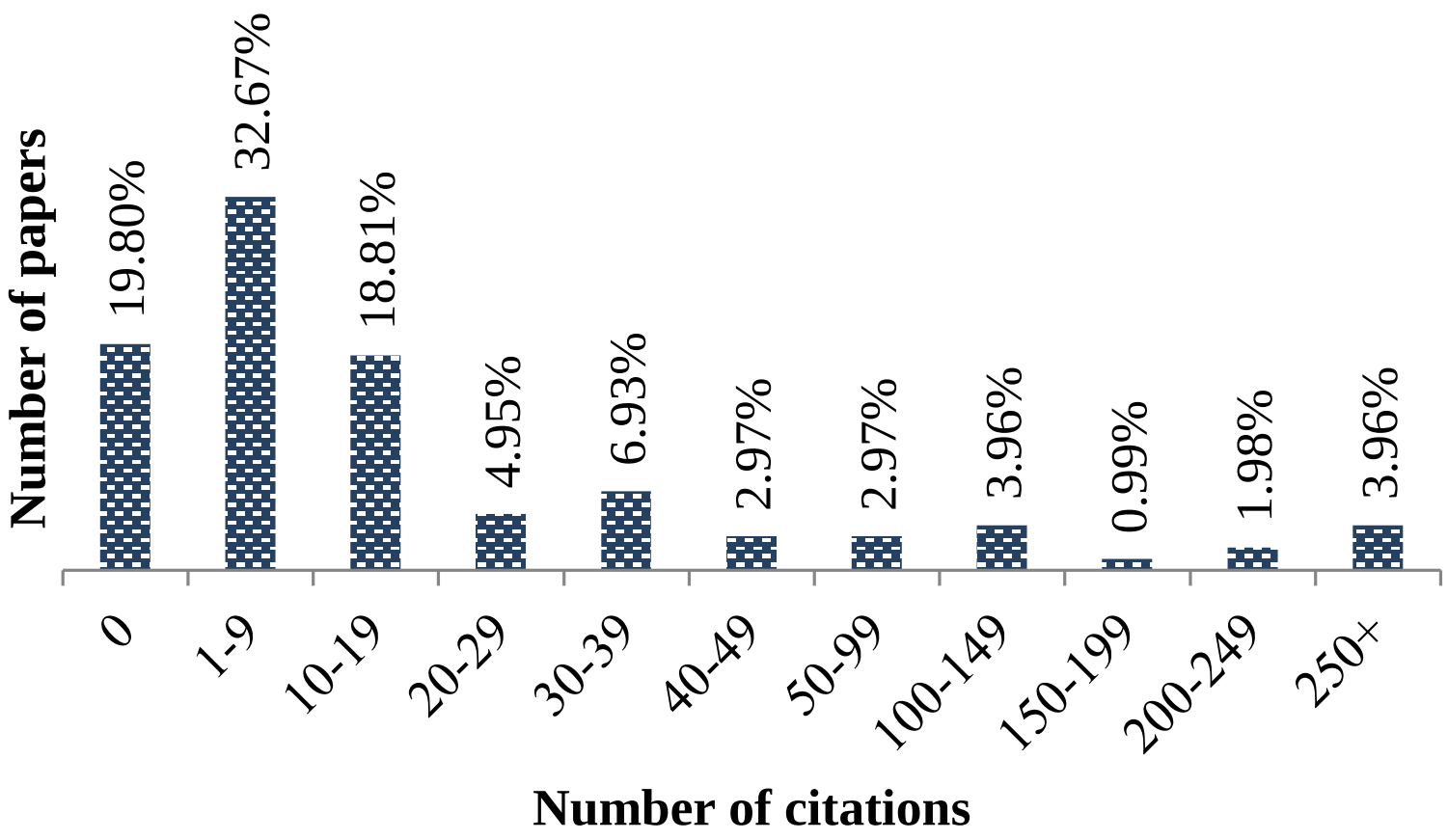

Figure 16: Citation Counts Overview

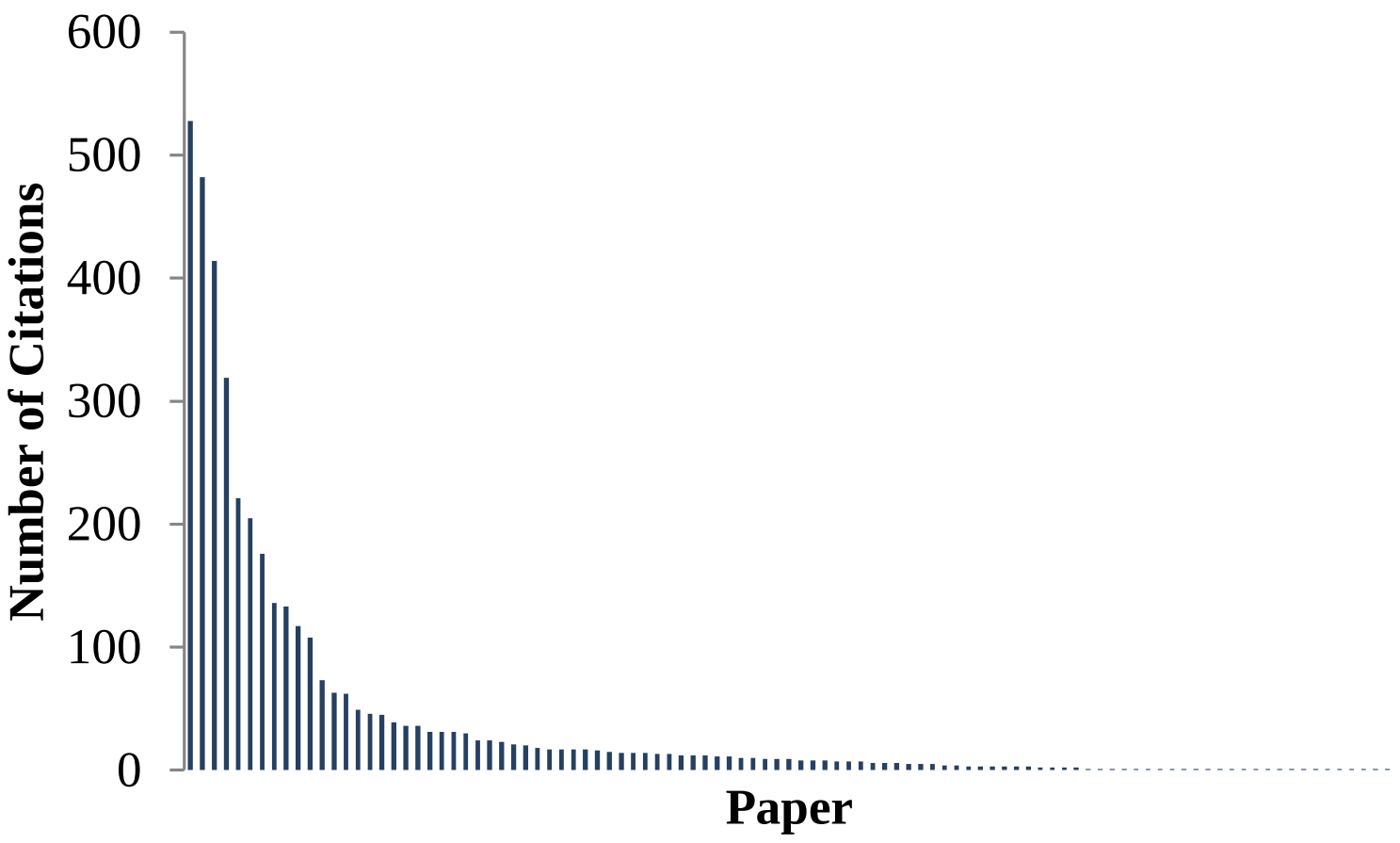

Figure 17: Citation counts of the reviewed papers

## 3.2 Survey of the Recommendation Classes

### 3.2.1 Content-Based Filtering

In the research-paper recommender-system community, CBF is the predominant approach: of the 70 reviewed approaches, 34 (49%) apply the idea of CBF [47, 164, 284], not including the hybrid approaches, which also mostly apply CBF. For the reviewed approaches, "interaction" between users and items was typically established through authorship [152, 372, 395], having papers in one's personal collection [47, 88, 185], adding social tags [118], or downloading [320], reading [417], and browsing papers [61, 183, 291].

Most approaches use plain words as features, although some use *n-grams* [118, 301], *topics* (words and word combinations that occurred as social tags on CiteULike) [196], and *concepts* that were inferred from the Anthology Reference Corpus (ACL) corpus via Latent Dirichlet Allocation [54], and assigned to papers through machine learning. Only a few approaches utilize non-textual features, and if they do then these non-textual features are typically utilized *in addition* to words. *Giles et al.* used citations in the same way as words were used and weighted the citations with the standard TF-IDF measure (they called this method *CC-IDF*) [139]. Others adopted the idea of CC-IDF or used it as baseline [47, 59, 111]. *Zarrinkalam and Kahani* considered *authors* as features and determined similarities by the number of authors two items share [428].

The approaches extracted words from the title [235, 251, 342], abstract [110, 164, 196], header [139], introduction [182], foreword [182], author-provided keywords [110, 182, 187], and bibliography [111], as well as from the papers' body text [202, 301, 342]. The approaches further extracted words from external sources such as social tags [187, 197], ACM classification tree and DMOZ categories [284, 286], and citation context [164, 179, 202]. Utilizing citation context is similar to the way search engines use anchor analysis for webpage indexing since the 1990's [65, 272]. Citation context analysis was also used in academic search [355] before it was used by research-paper recommender systems.

It is well known that words from different document fields have different discriminative powers [263]. For instance, a word occurring in the title is usually more meaningful than a word occurring in the body text. *Nascimento et al.* accounted for this and weighted terms from the title three times stronger than terms from the body-text, and text from the abstract twice as strong [301]. This weighting scheme was arbitrarily selected and not based on empirical evidence.

*Huang et al.* experimented with different weights for papers' content and citation context [179]. They found that an equal weight for both fields achieved the highest precision. The other reviewed approaches that used text from different fields did not report on any field weighting.

The most popular model to store item representations and user models was the Vector Space Model (VSM), which was used by 93% of those approaches that reported the utilized model. Other approaches modeled their users as graph [309, 310, 409], as list of topics that were assigned through machine learning [284], or as an ACM hierarchy [209]. Of those who used VSM, all but one used the cosine measure to calculate similarities between user models and recommendation candidates. In 1998, *Giles et al.* compared headers of documents with a string distance measure [59], but neither they nor others mentioned that technique again, which leads us to the assumption that the string edit distance was not effective.

TF-IDF was the most popular weighting scheme (83%) among those approaches for which the scheme was specified. Other weighting schemes included plain term frequency (TF) [118, 301, 372], and techniques that the authors called "phrase depth" and "life span" [118].

### 3.2.2 Collaborative Filtering

From the reviewed approaches, only nine (13%) apply collaborative filtering, and none uses explicit ratings [274, 320, 399]. *Yang et al.* intended to let users rate research papers, but users were "too lazy to provide ratings" [417]. This illustrates one of the main problems of CF: it requires user participation, but often the motivation to participate is low. To overcome this problem, *Yang et al.* inferred implicit ratings from the number of pages the users read: the more pages users read, the more the users were assumed to like the documents [417]. *Pennock et al.* interpreted interactions such as downloading a paper, adding it to ones' profile, editing paper details, and viewing its bibliography as positive votes [320]. *McNee et al.* assumed that an author's citations indicate a positive vote for a paper [274]. They postulated that when two authors cite the same papers, they are like-minded. Similar, if a user reads or cites a paper the citations of the cited paper are supposed to be liked by the user.

Using inferred ratings annihilates CF's advantage of being based on real quality assessments. This criticism applies to citations as well as to other types of implicit ratings [66, 254, 259]. For instance, we reference papers in this survey that had inadequate evaluations, or were written in barely understandable English. Hence, interpreting these citations as positive vote would be misguiding. Similarly, when a user spends lots of time reading a paper this *could* mean that the paper contains interesting information, and the user would rate the paper positively; but it could also mean that the paper is just difficult to understand and requires a lot of effort to read. Consequently, CF's advantage of explicit human quality assessments mostly vanishes when implicit ratings are used.

Using citations might also annihilate CF's second advantage of being content-independent. Typically, reliable citation data is not widely available. Therefore, access to the papers' content is required to build a citation network, but this process is even more fault-prone than word extraction in CBF. In CBF, "only" the text of the papers must be extracted, and maybe fields such as title or abstracts must be identified. For citation-based CF the text must also be extracted but in this text, the bibliography and its individual references must be identified, including their various fields (such as title and author). This is usually an error-prone task [95].

A general problem of collaborative filtering in the domain of research-paper recommender systems is sparsity. *Vellino* compared the (implicit) ratings on Mendeley (research papers) and Netflix (movies), and found that sparsity on Mendeley was three orders of magnitude higher than on Netflix [400]. This is caused by the different ratio of users and items. In domains such as movie recommendations, there are typically few items and many users. For instance, the movie recommender MovieLens has 65,000 users and 5,000 movies [168]. Typically, many users watched the same movies. Therefore, like-minded users can be found for most users and recommendations can be given effectively. Similarly, most movies have been watched by at least some users and hence most movies can be recommended. This is different in the domain of research papers. There are typically few users but millions of papers, and only few users rated the same papers. Hence, finding like-minded users is often not possible. In addition, many papers are not rated by any users and therefore cannot be recommended.

### 3.2.3 Co-Occurrences

Six of the reviewed approaches are based on co-occurrences (9%). Three of those approaches analyze how often papers are *co-viewed* during a browsing session [61, 183, 291]. Whenever a user views a paper, those papers that had frequently been co-viewed with the browsed paper are recommended. Another approach uses proximity of co-citations to calculate document relatedness [142]: the higher the proximity of two references within a paper, the more related the cited papers are assumed to be. Pohl et al. compared the effectiveness of co-citations and co-downloads and found that co-downloads are only more effective than co-citations in the first two years after a paper is published [327].

Calculating co-occurrence recommendations is not always feasible. For instance, on arXiv.org, two thirds of all papers have no co-citations, and those that do usually have only one or two [327]. Despite all that, co-occurrence recommendations seem to perform quite well. Two popular research-paper recommender systems, bX and BibTip, both rely on co-occurrence recommendations and are delivering millions of recommendations every month [61, 291].

### 3.2.4 Graph Based

Eleven of the reviewed approaches utilize the inherent connections that exist in academia (16%). Based on these connections, the approaches build graph networks that typically show how papers are connected through citations [12, 219, 250]. Sometimes, graphs include authors [9, 235, 437], users/customers [182], venues [12, 235, 437], genes and proteins [9, 235], and the years the papers were published [235]. *Lao et al.* even included terms from the papers' titles in the graph, which makes their approach a mixture of graph and content based [235]. Depending on the entities in the graph, connections may be citations [12, 235, 250], purchases [182], "published in" relations, [12, 235, 437], authorship [9, 12, 437], relatedness between genes[19] [9], or occurrences of genes in papers [235]. Some authors connected entities based on non-inherent relations. For instance, *Huang et al.* and *Woodruff et al.* calculated text similarities between items and used the text similarity to connect papers in the graph [182, 413]. Other connections were based on attribute similarity[20], bibliographic coupling, co-citation strength [182, 413, 437], or demographic similarity [182]. Once a graph was built, graph metrics were used to find recommendation candidates. Typically there was one or several input papers, and from this input, random walks with restarts were conducted to find the most popular items in the graph [153, 219, 235].

[19] Relatedness between genes was retrieved from an external data source that maintained information about gene relatedness.

[20] Attribute similarity was calculated, e.g., based on the number of pages.

### 3.2.5 Global Relevance

In its simplest form, a recommender system adopts a one-fits-all approach and recommends items that have the highest global relevance. In this case, the relevance is not calculated user-specifically based on e.g. the similarity of user models and recommendation candidates. Instead, some global measures are utilized such as overall popularity. For instance, a movie-rental system could recommend those movies that were most often rented or that had the highest average rating over all users. In this case, the basic assumption would be that users like what most other users like.

From the reviewed approaches, many use global relevance as an additional ranking factor. For instance, five CBF approaches used global popularity metrics in their rankings [54, 164, 428]. They first determined a list of recommendation candidates with a user-specific CBF approach. Then, the recommendation candidates were re-ranked based on the global relevance metrics. Popular metrics were PageRank [54], HITS [164], Katz [164], citation counts [54, 164, 342], venues' citation counts [54, 342], citation counts of the authors' affiliations [342], authors' citation count [54, 342], h-index [54], recency of articles [54], title length [342], number of co-authors [342], number of affiliations [342], and venue type [342].

*Strohman et al.* report that the Katz measure strongly improved precision [369]. All variations that included Katz were about twice as good as those variations without. *Bethard and Jurafsky* report that a simple citation count was the most important factor, and age (recency) and h-index were even counterproductive [54]. They also report that considering these rather simple metrics doubled mean average precision compared to a standard content-based filtering approach.

### 3.2.6 Hybrid Recommendation Approaches

Approaches of the previously introduced recommendation classes may be combined in hybrid approaches. Many of the reviewed approaches have some hybrid aspects. For instance, several of the CBF approaches use global relevance attributes to rank the candidates, or graph methods are used to extend or restrict potential recommendation candidates. This type of hybrid recommendation technique is called "feature augmentation" [70]. It is only a weak kind of hybrid recommendation technique, as the primary technique is still dominant. In true hybrids, the combined concepts are more or less equally important [70, 71]. From the reviewed approaches, only those of the TechLens team, and to some extent from Papyres (Appendix F.5.2, p. 246), may be considered true hybrid approaches.

TechLens [110, 199, 211, 274, 275, 393] is certainly one of the most influential research-paper recommender systems, though it was not the first one as some claim (e.g. [326]). TechLens was developed by the GroupLens[21] team, but nowadays TechLens is not publicly available, although the GroupLens-team is still very active in the development and research of recommender systems in other fields. Between 2002 and 2010, Joseph A. Konstan, John Riedel, Sean M. McNee, Roberto Torres, and several others published six articles relating to research-paper recommender systems. Often, *McNee's et al.* article from 2002 is considered to be the original TechLens article [274]. However, the 2002 article 'only' introduced some algorithms for recommending citations, which severed as foundation for TechLens, which was introduced 2004 by *Torres et al.* [393]. Two articles about TechLens followed in 2005 and 2007 but added nothing new with respect to recommendations [199, 211]. In 2006, *McNee et al.* analyzed potential pitfalls of recommender systems [275]. In 2010, *Ekstrand et al.* published another article about the TechLens approaches, and enhanced them [110]. The most important TechLens articles are summarized in Appendix F.5.1, p. 243.

## 3.3 Survey of the Research Field and its Shortcomings

### 3.3.1 Neglect of User Modeling

Of the reviewed approaches, 79% require users to explicitly provide keywords [172, 295, 399], text snippet such as an abstract [54, 334, 428], or to provide a single paper as input to represent their interests [110, 301, 369]. This means that these approaches neglect the user modeling process, one of the most important parts of a recommender system. This makes the approaches very similar to classic search, or related document search [3, 159, 242], where users provide search terms or one input paper, and receive a list of search results or similar papers. Of course, neither classic search nor finding related documents are trivial tasks in themselves, but they neglect the user modeling process and we see little reason to label such systems as recommender systems.

---

[21] http://grouplens.org/

Only 21% of the reviewed approaches inferred information from the items the users interacted with. Most approaches that inferred information automatically used *all* papers that a user authored, downloaded, etc. [197, 286, 409]. This is not ideal. When inferring information automatically, a recommender system should determine those items that are currently relevant for the user-modeling process. For instance, papers being read ten years ago are probably not suitable to describe a user's current information needs. This aspect is called "concept drift" and it is important for creating meaningful user models. In the research-paper recommender systems community, concept drift is widely ignored: only three approaches considered concept drift in detail. *Middleton et al.* weight papers by the number of days since the user last accessed them [284]. *Watanabe et al.* use a similar approach [409]. *Sugiyama and Kan*, who utilize an user's authored papers, weight each paper based on the difference between a paper's publication year, and the year of the most recently authored paper [372]. In addition, they found that it makes sense to include only those papers that the user authored in the past three years [372].

Another important aspect about user modeling is the user-model size. While in search, user models (i.e. search queries) typically consist of a few words, user models in recommender systems may consist of hundreds or even thousands of words. Of the reviewed approaches, 91% did not report the user-model size, which leads us to the assumption that they simply used all features. Those few that reported on the user-model size usually stored fewer than 100 terms. For instance, Giles et al. utilized the top 20 words of the papers [139].

### 3.3.2 Focus on Accuracy

The research-paper recommender-system community focuses strongly on accuracy, and seems to assume that an accurate recommender system will lead to high user satisfaction. However, outside the research-paper recommender-system community it is widely known that many aspects beyond accuracy affect user satisfaction. For instance, users might become dissatisfied with accurate recommendations when they have no trust in the recommender system's operator [403], their privacy is not ensured [330], they need to wait too long for recommendations [330], or the user interfaces are not appealing to them [402]. Other factors that affect user satisfaction are confidence in a recommender system [336], data security [229], diversity [438], user tasks [275], item's lifespan [72] and novelty [433], risk of accepting recommendations [325], robustness against spam and fraud [93], transparency and explanations [167], time to first recommendation [168], and interoperability [76].

Among the reviewed articles, only a few authors considered aspects beyond accuracy, as shown in the following sections.

#### *3.3.2.1 Users' tasks*

*Torres et al.* from TechLens' considered a user's current task in the recommendation process, and distinguished between users who wanted to receive authoritative recommendations and novel recommendations [393]. *Torres et al.* showed that different recommendation approaches were differently effective for these tasks. The developers of *TheAdvisor* let users specify whether they are interested in classical or recent papers [217]. *Uchiyama et al.* found that students are typically not interested in finding papers that are "similar" to their input paper [395]. This finding is interesting because content-based filtering is based on the assumption that user want similar papers. However, the study from *Uchiyama et al.* was based on only 16 participants. As such, it remains uncertain how significant the results are.

#### *3.3.2.2 Diversity*

Diversity of recommendations was mentioned in a few articles, but really considered in depth only by *Vellino et al.* and *Küçüktunç et al. Vellino et al.* measured diversity as the number of different journals from which articles were recommended [399]. If recommendations were all from the same journals, diversity was zero. They compared diversity of a CF approach with the co-occurrence approach from bX and found that CF had a diversity of 60% while diversity of bX was 34%. *Küçüktunç et al.* from TheAdvisor published two articles about diversity in research-paper recommender systems [218, 220]. They provided a survey on diversification techniques in graphs, and proposed some new techniques to measure diversity.

#### *3.3.2.3 Layout*

*Farooq et al.* from CiteSeer analyzed which information users wanted to see when receiving recommendations in RSS feeds [115]. They found that the information to display varies on the type of recommendation. In one approach, *Farooq et al.* recommended papers that cited the user's papers. In this case, users preferred to see the citing paper's bibliographic data (title, author, etc.) and the context of the citation – the sentence in which the citation appeared. When papers were recommended that were co-cited with the users' papers, citation context was not that important. Rather, the users preferred to see the bibliographic data and abstract of the co-cited paper. When papers were recommended that had a similar content to the users' papers, users preferred to see bibliographic data and abstract. These findings are interesting because from the reviewed recommender systems the majority displays only the title and not the abstract.

As part of our work, we researched the impact of labeling and found that papers labeled as 'sponsored recommendation' performed worse than recommendations with a label that indicated that the recommendations were 'organic,' though the recommended papers were identical (cf. Appendix J, p. 271). It made also a difference whether paper recommendations were labeled as 'Sponsored' or 'Advertisement' although both labels indicate the same thing, namely that they are displayed for commercial reasons.

#### *3.3.2.4 User characteristics*

We also found that researchers who registered to a recommender system tended to have higher click-through rates than unregistered users (6.95% vs. 4.97%) (cf. Appendix H, p. 259). In addition, older users seem to have higher average click-through rates (40-44 years: 8.46%) than younger users (20-24 years: 2.73%) [52]. *Middleton et al.* also report differences for different user groups. Click-through rates in their recommender system Quickstep was around 9%, but only around 3.5% for Foxtrot, although both systems applied very similar approaches. However, Quickstep users were recruited from a computer science laboratory, while Foxtrot was a real-world system being offered to 260 faculty members and students (though only 14% of them used Foxtrot at least three times).

Click-through rates from the bX recommender are also interesting [390]. They varied between 3% and 10% depending on the university in which recommendations were shown (bX is providing more than 1,000 institutions with recommendations) [113]. This could have been caused by different layouts, and how the recommendations were presented, but it might also be caused by different backgrounds of the students.

#### *3.3.2.5 Time of usage*

*Middleton et al.* reported that the longer someone used the recommender system, the lower click-through rates became [286]. *Jack* reports the opposite, namely that precision increased over time (p=0.025 in the beginning, p=0.4 after six months) and depended on a user's library size (p=0.08 for 20 articles and p=0.40 for 140 articles) [187]. We showed that it might make sense to be "persistent" and show the same recommendations to the same users multiple times – even recommendations that users had clicked before were often clicked again (cf. Appendix I, p. 265).

#### *3.3.2.6 Recommendation medium*

User satisfaction also depends on the medium through which recommendations are made. *Middleton et al.* report that recommendations via email received only half the click-through rate as the same recommendations delivered via a website [286]. Of the reviewed recommender systems, only Docear [47] and Mendeley [187] provide recommendations through a desktop software; CiteSeer provided recommendations in a news feed [115]; and all others deliver their recommendations through websites. If and how click rates differ, when recommendations are delivered by desktop software or a website, remains unknown.

#### *3.3.2.7 Relevance and profile feedback*

Relevance feedback is a common technique to improve recommendations [336] but it is widely ignored in the research-paper recommender-system community. *Middleton et al.* showed that profile feedback is better than relevance feedback: allowing users to edit their user models is more effective than just learning from relevance feedback [286]. While *Bollacker et al.* from CiteSeer allowed their users to edit their profiles, they conducted no research on the effectiveness of that activity [236].

### 3.3.3 Lack of Transferring Research into Practice

Despite all the published articles and proposed approaches, we found only 24 research-paper recommender systems that could be used by real users (Table 3)[22]. Of these 24 recommender systems, eight (33%) never left the prototyping stage – and today only one of the prototypes is still publicly available. Of the remaining recommender systems, four are offline (25%), five are idling (31%)[23], and only seven are running and actively maintained (44%). However, from the seven active recommender systems, only four operators are involved with the recommender-system research community[22], publishing information about their systems.

[22] The recommender systems of Mendeley, CiteULike, and CiteSeer are counted twice because they offer or offered two independent recommender systems.

[23] We classified a recommender system as idling if no article was published or no changes were made at the system for a year.

Table 3: List of recommender systems

| Status | Name | Maturity | Research Oriented | Type | Presentation |
|---|---|---|---|---|---|
| Active | **BibTip** | Real System | No[1] | Stand-Alone | Webpage |
| Active | **bx** | Real System | No | Stand-Alone | Webpage |
| Active | **Docear** | Real System | Yes | On-Top | Software |
| Active | **Mendeley** | -- | -- | -- | -- |
| | Related Papers | Real System | Yes | On-Top | Software |
| | Suggest | Real System | Yes | On-Top | Software |
| Active | **RefSeer** | Real System | Yes | Stand-Alone | Webpage |
| Active | **Scholar Update** | Real System | No | On-Top | Webpage |
| Idle | **CiteULike** | -- | -- | -- | -- |
| | CF | Real System | No | On-Top | Webpage |
| | Item-Centric | Real System | No | On-Top | Webpage |
| Idle | **PubMed PRMA** | Real System | No | On-Top | Webpage |
| Idle | **ResearchGate** | Real System | No | On-Top | Webpage |
| Idle | **TheAdvisor** | Real System | Yes | Stand-Alone | Webpage |
| Idle | **Who Should I Cite?** | Prototype | Yes | Stand-Alone | Webpage |
| Offline | **CiteSeer** | -- | -- | -- | -- |
| | Alert | Real System | Yes | On-Top | Feed |
| | Related Documents | Real System | Yes | On-Top | Webpage |
| Offline | **Foxtrot** | Real System | Yes | Stand-Alone | Webpage, Email |
| Offline | **TechLens** | Real System | Yes | Stand-Alone | Webpage |
| Offline | **NSYSU-ETD** | Prototype | Yes | On-Top | Webpage |
| Offline | **OSUSUME** | Prototype | Yes | On-Top | ? |
| Offline | **Papits** | Prototype | Yes | On-Top | Webpage |
| Offline | **Papyres** | Prototype | Yes | On-Top | ? |
| Offline | **Pirates** | Prototype | Yes | Stand-Alone | ? |
| Offline | **Quickstep** | Prototype | Yes | Stand-Alone | Webpage |
| Offline | **Sarkanto & Synthese** | Prototype | Yes | Stand-Alone | Webpage |

Most of the real-world recommender systems apply simple recommendation approaches that are not based on recent research. For instance, PubMed seems still to use an approach introduced in 2007; ResearchGate is using a simple content-based filtering approach similar to classic search[24]; CiteULike apparently uses two approaches from 2008/2009; and BibTip and bX are using simple co-occurrence approaches. Whether RefSeer's is really applying all the results from their research remains also unclear. In other words, most of the reviewed research had no impact on real-world recommender systems.

[24] ResearchGate also applied other recommender systems, e.g. for persons or news, and it seems that these approaches are more sophisticated.

### 3.3.4 Lack of Persistence and Authorities

One reason the research seems not to be transferred into practice might be a lack of persistence and authorities in the field. From 327 authors who authored the 217 reviewed articles, 67% published only a single article (Figure 18). Only thirteen authors published more than five articles, but of these authors, several were co-authors publishing the same articles. This means, there is only a small number of groups that consistently do research in the field of research-paper recommender systems.

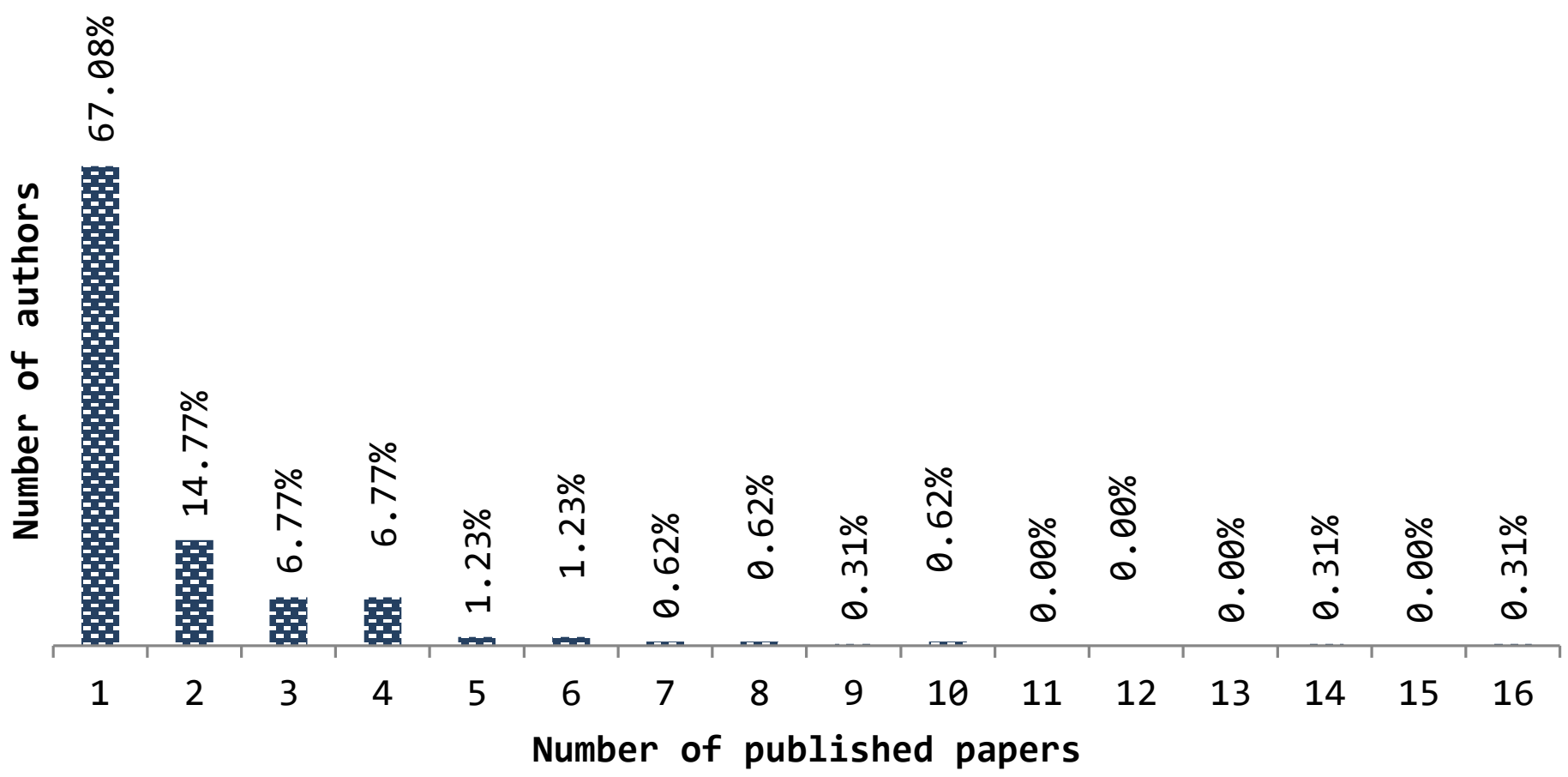

Figure 18: Papers per author

The most productive authors are C. Lee Giles and his co-author P. Mitra from CiteSeer/RefSeer (Table 4 and Table 5). No others have published as many articles (16) over as a long period of time (16 years) about as many different aspects of research-paper recommender systems. Other highly productive authors are A. Geyer-Schulz and his co-authors M. Hashler, and M. Jahn from BibTip. They published fourteen articles, but these were less often cited in the community. The articles are also narrower in scope than those of the CiteSeer authors. We authored ten papers between 2009 and 2013, including posters and short papers, and we concentrated on aspects beyond accuracy such as the impact of labeling recommendations and the impact of demographics on click-through rates. O. Küçüktunç and his co-authors E. Saule and K. Kaya from TheAdvisor published nine articles focusing on diversity and graph-based recommendations. J. A. Konstan, S. M. McNee, R. Torres, and J.T. Riedel, who are highly recognized authors in the field of recommender systems in general, developed TechLens and authored six articles relating to research-paper recommender systems during 2002 and 2010. Two of their articles influenced the work of several others and are among the most cited articles we have reviewed [274, 393]. W. W. Cohen, and his PhD student N. Lao, are also two productive authors. They authored six articles during 2008 and 2012 (some of which are unpublished). It stands out that the five most productive research groups all have access to real-world recommender systems.

Table 4: Most productive authors

| **Author** | **Paper Count** |
|---|---|
| C. Lee Giles | 16 |
| A. Geyer-Schulz | 14 |
| J. Beel | 10 |
| M. Hahsler | 10 |
| O. Küçüktunç | 9 |
| E. Saule | 8 |
| K. Kaya | 8 |
| S. Langer | 7 |
| M. Genzmehr | 7 |
| P. Mitra | 6 |
| J. A. Konstan | 6 |
| W. W. Cohen | 6 |
| B. Gipp | 6 |
| M. Jahn | 5 |
| N. Lao | 5 |

Table 5: Most productive author-groups

| **Author(s)** | **Max. Papers** |
|---|---|
| C. Lee Giles; P. Mitra (CiteSeer/RefSeer) | 16 |
| A. Geyer-Schulz; M. Hashler; M. Jahn (BibTip) | 14 |
| J. Beel; S. Langer, M. Genzmehr, B. Gipp (Docear) | 10 |
| O. Küçüktunç; E. Saule; K. Kaya (TheAdvisor) | 9 |
| J. A. Konstan; S.M. McNee; R. Torres, J.T. Riedl (TechLens) | 6 |
| W. W. Cohen; N. Lao | 6 |

### 3.3.5 Lack of Cooperation

Most articles were authored by multiple authors: the majority of papers had two (26.35%), three (26.35%) or four authors (20.36%) (Figure 19)[25]. Only 15% of the papers were authored by a single researcher. These numbers might indicate a high degree of collaboration, on first glance. However, it is noticeable that between the different co-author groups hardly any cooperation exists. The closest cooperation we could find was that *Giles* was part of a committee for a thesis that *Cohen* supervised [231]. No major authors of different groups ever co-authored any articles.

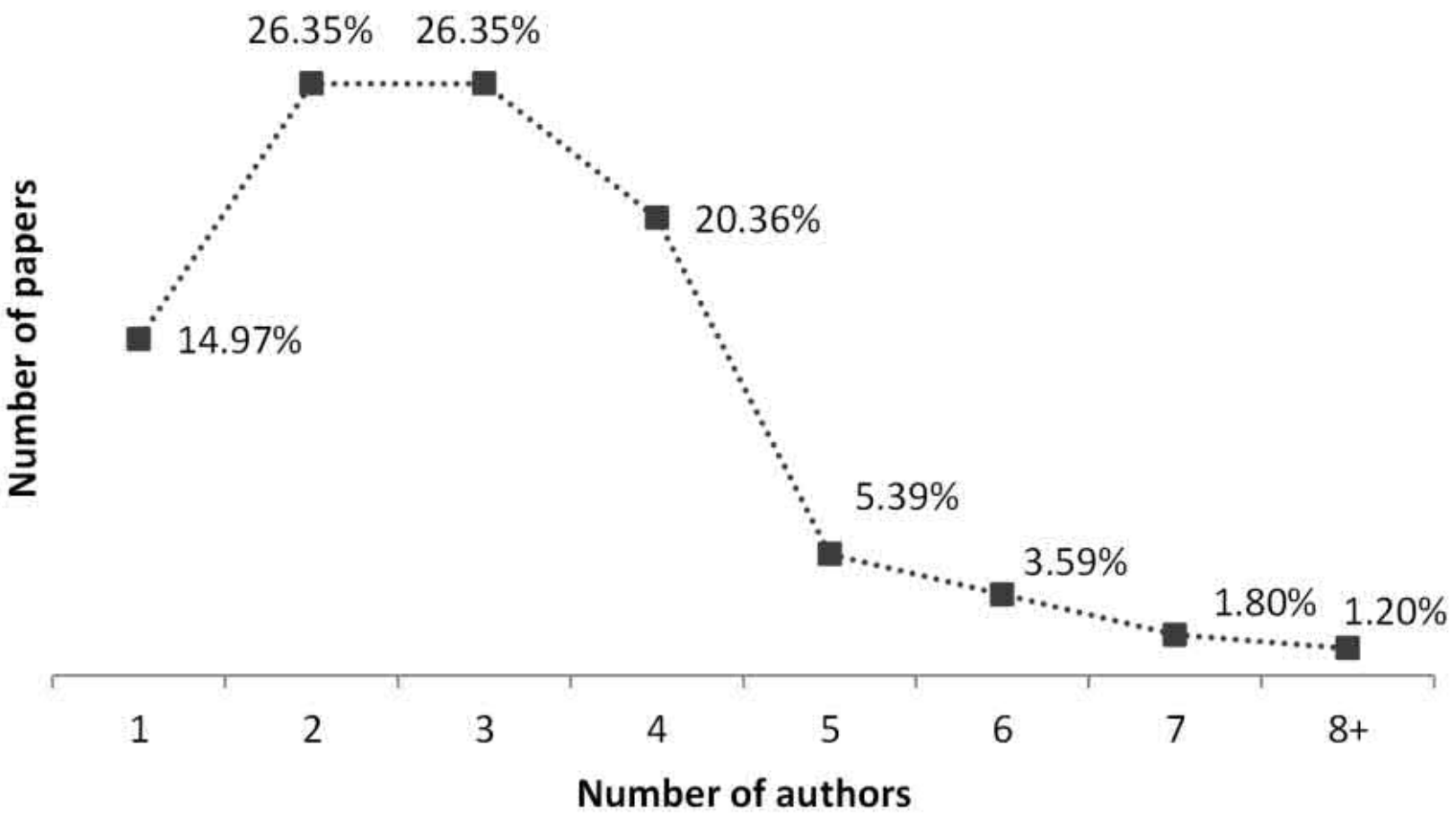

Figure 19: Number of authors of the reviewed papers

Many co-author groups seem to work alone and barely build their work based upon the work of peers, be it within or outside the research-paper recommender-system community. Among the reviewed articles, it barely happened that authors reported to have built their novel approach based upon an existing approach. This lack of cooperation also becomes apparent when looking at the citations. Although some of the reviewed articles gained many citations, these citations usually resulted from articles outside the research-paper recommender domain. For instance, the paper “Learning multiple graphs for document recommendations” attracted 63 citations since 2008 [437]. From these citations, only three were made by the reviewed articles. Another article, from the BibTiP developers, gained 24 citations since 2002 [130]. From the 24 citations, ten were self-citations and none was from the reviewed articles. Both examples are typical for most of the reviewed articles. One of the few articles that is constantly cited in the research-paper recommender community is an article about TechLens, which accumulated more than 100 citations [393]. However, most authors only cited the article for authoritative reasons. In the citing papers, TechLens is mentioned but, with few exceptions, its approaches are neither adopted nor used as baseline.

[25] Median author count was three, maximum count eleven.

### 3.3.6 Information Sparsity

Most authors provided sparse information about their approaches, which makes a re-implementation difficult, if not impossible. For instance, for 71% of the content-based filtering approaches, the authors did not report the weighting scheme they used (e.g. TF-IDF). The feature representation model (e.g. vector space model) was not reported for 59% of the approaches. For 69% of the approaches, authors did not report whether they removed stop words. For 67% of the CBF approaches no information was given on the fields the terms were extracted from (e.g. title or abstract). This means, when an evaluation reports promising results for an approach, other researchers would not know how to re-implement the approach in detail. If they tried, and guessed the specifics of an approach, the outcome would probably differ significantly from the original. This might cause problems in replicating evaluations, and reproducing research results – a serious shortcoming that is covered in more detail in the next section.

## 3.4 Survey of the Evaluations

Recommender-systems research is heavily dependent on thorough evaluations to assess the effectiveness of recommendation approaches and to decide which approaches to apply, either in practice or as a baseline for other evaluations. Among the key prerequisites for thorough evaluations are appropriate evaluation methods, a sufficient number of study participants, and a comparison of the novel approach against one or more state-of-the-art approaches [343]. In addition, the novel approach and its evaluation need to be thoroughly described. Only with such a description will readers be able to determine the soundness of the evaluation, re-implement the approach, and reproduce or replicate the results.

From the reviewed approaches, 21% haven not been evaluated by their authors [14, 59, 79, 85, 117, 126, 145, 297, 328, 341, 397, 404, 418], or were evaluated using convoluted methods that we could not follow [102, 172, 270, 409, 411, 419] [26]. In the remaining analysis, these approaches are ignored.

### 3.4.1 Evaluation Methods and their Adequacy

Of the evaluated approaches, 69% were evaluated with offline evaluations, 34% with lab-based user studies, 7% with an online evaluation, and 3% with qualitative user studies (Table 6)[27].

Table 6: Use of evaluation methods[27]

| Offline | User Study | Online | Qualitative |
|---|---|---|---|
| 69% | 34% | 7% | 3% |

Most user study participants rated only few recommendations and 17% of the studies were conducted with fewer than five participants [197, 387, 413]; 17% of the studies had five to ten participants [209, 250, 301]; 13% had 11-15 participants [61, 293, 434]; and 17% had 16-50 participants [142, 395, 408]. Only 25% were conducted with more than 50 participants [274, 295, 393]. The final 13% of the studies failed to mention the number of participants [179, 196, 288] (Table 7). Given these numbers, we conclude that most user studies were not large enough to arrive at meaningful conclusions.

Table 7: Number of participants in user studies

| Number of participants | n/a | <5 | 5-10 | 11-15 | 16-50 | >50 |
|---|---|---|---|---|---|---|
| Number of user studies | 13% | 17% | 17% | 13% | 17% | 25% |

Our review also indicates that the voiced criticism on offline evaluations (cf. 2.5.3.3, p. 24) applies to the offline evaluations in the field of research-paper recommender systems. Six of the approaches were evaluated using both an offline evaluation and a user study [110, 182, 250, 274, 387, 393]. Of these six evaluations, one did not compare its approach against any baseline [387]. In two evaluations, results from the offline evaluations were indeed similar to results of the user studies [110, 250]. However, the user studies had only five and 19 participants respectively, which led to statistically insignificant results. Three other studies reported contradicting results for offline evaluations and user studies [182, 274, 393] (two of these studies had more than 100 participants; the third study only had two participants). This means, offline evaluations could not reliably predict the effectiveness in the real world. Interestingly, the three studies with the most participants were all conducted by the authors of TechLens [110, 274, 393], who are also the only authors in the field of research-paper recommender systems discussing the potential shortcomings of offline evaluations [275]. It seems that other researchers in this field are not aware of – or chose not to address – problems associated with offline evaluations, although there has been quite a discussion outside the research-paper recommender-system community (cf. 2.5.3.3, p. 24).

[26] For the analysis, only 176 articles were reviewed that we found during a first round of literature search in 2012. Consequently, percentages relate to these 176 reviewed articles and not the 217 articles, that were reviewed for the rest of the survey.

[27] Some approaches were evaluated with several methods at the same time. Therefore, percentages do not add up to 100%.

### 3.4.2 The Operators' Perspective

Costs to build a recommender system, or implement an approach, were not reported in any reviewed article. Costs to run a recommender system were only reported by *Jack* from Mendeley [186]. He states that costs on Amazon's S3 were $66 a month plus $30 to update the recommender system that coped with 20 requests per second generated by 2 million users.

Important information relating to costs is runtime. Runtime information is crucial to estimate costs, and hence to estimate how feasible an approach will be to apply in practice. In one paper, runtimes of two approaches differed by a factor 600 [180]. For many operators, an approach requiring 600 times more CPU power than another would probably not be an option, particularly if differences in effectiveness are small. While this example is extreme, other runtime comparisons showed differences by a factor five or more, which also can affect the decisions on algorithm selection. However, information on runtime was only provided for 11% of the approaches.

Reporting on computational complexity is also important. For operators who want to offer their system for a large number of users, computational complexity is important for estimating the long-term suitability of an approach. An approach may perform well enough for a few users, but it might not scale well. Approaches with exponentially increasing complexity most likely will not be applicable in practice. However, computational complexity was reported for even fewer approaches than runtime.

### 3.4.3 Coverage

Coverage describes how many papers of those in the recommender's database potentially may be recommended [149, 166]. High coverage is important because it represents the number of recommendations a user may receive. As such, coverage is an important metric to judge the usefulness of a recommender system. For text-based approaches, coverage is usually 100%. For other approaches, coverage is typically lower. For instance, in collaborative filtering usually not all items are rated by users. Even though the unrated items might be relevant, they cannot be recommended with classic CF approaches. From the reviewed articles, only few consider coverage in their evaluations. He et al. judge the effectiveness of their approaches based on which approach provides the best tradeoff between accuracy and coverage [164]. The BibTip developers report that 80% of all documents have been co-viewed and can be used for generating recommendations [291]. *Pohl et al.* report that co-download coverage on arXiv is close to 100% while co-citation coverage is only around 30% [327]. The TechLens authors report that all of their hybrid and CBF approaches have 100% coverage, except pure CF which has a coverage of 93% [393].

### 3.4.4 Baselines

Another important factor in evaluating recommender systems is the baseline against which an algorithm is compared. For instance, knowing that a certain approach has a particular CTR is not useful if the CTRs of alternative approaches are unknown. Therefore, novel approaches should be compared against a baseline representative of the state-of-the-art approaches. Only then is it possible to quantify whether, and when, a novel approach is more effective than the state-of-the-art and by what margin.

Of the reviewed approaches, 19% were not compared against a baseline [111, 162, 177, 209, 258, 288, 293, 315, 366, 367, 387, 432, 434]. Another 71% of the approaches were compared against trivial baselines such as simple content-based filtering without any sophisticated adjustments. These trivial baselines do not represent the state-of-the-art and are not helpful for deciding whether a novel approach is promising. This is particularly troublesome since the reviewed approaches were not evaluated against the *same* trivial baselines. Even for a simple CBF baseline, there are many variables, such as whether stop words are filtered, which stemmer is applied, from which document field the text is extracted, etc. This means almost all reviewed approaches were compared against different baselines, and results cannot be compared with each other. Only 10% of the evaluated approaches were evaluated against approaches proposed by other researchers in the field. These evaluations allow drawing some conclusions on which approaches may be most effective.

It is interesting to note that in all evaluations, at least one of the novel approaches performed better than the baselines. No article reported on a non-effective approach. We can only speculate about the reasons: First, authors may intentionally select baselines such that their approaches appear favorable. Second, the simple baselines used in most evaluations achieve relatively poor results, so that any alternative easily performs better. Third, authors do not report their failures. Fourth, journals and conferences do not accept publications that report on failures. Whatever the reasons are, we advocate that reporting failures is desirable as it could prevent other researchers from doing the same experiments, and hence wasting time.

### 3.4.5 Offline Evaluation Metrics

In 69% of the offline evaluations, *precision* was used as evaluation metric (Table 8). Recall was used in 23%; F-measure and nDCG in 13%, and 15% were evaluated using other measures. Overall, results of the different measures highly correlated. That is algorithms, which performed well using precision also performed well using nDCG, for instance. However, there were exceptions. Zarrinkalam and Kahani tested the effectiveness of abstract and title against abstract, title, and citation context [428]. When *co-citation probability* was used as an evaluation metric, title and abstract were most effective. Based on recall, the most effective field combination was abstract, title, and citation context. With the nDCG measure, results varied depending on how the candidate set was generated and which ranking approach was used.

Table 8: Evaluation metrics

| Metric | Precision | Recall | F-Measure | nDCG | MRR | Other |
|---|---|---|---|---|---|---|
| Number of user studies | 69% | 23% | 13% | 13% | 8% | 15% |

### 3.4.6 Datasets and Architectures

Researchers and developers in the field of recommender systems can benefit from publicly available architectures and datasets[28]. *Architectures* help with the understanding and building of recommender systems, and are available in various recommendation domains such as e-commerce [314], marketing [243], and engineering [329]. *Datasets* empower the evaluation of recommender systems by enabling that researchers evaluate their systems with the same data. Datasets are available in several recommendation domains, including movies[29], music[30], and baby names[31].

Architectures of research-paper recommender systems have only been published by a few authors. The developers of *CiteSeer(x)* published an architecture that focused on crawling and searching academic PDFs [59, 324]. This architecture has some relevance for recommender systems since many task in academic search are related to recommender systems (e.g. crawling and indexing PDFs, and matching user models or search-queries with research papers). *Bollen and van de Sompel* published an architecture that later served as the foundation for the research-paper recommender system *bX* [61]. This architecture focuses on recording, processing, and exchanging scholarly usage data. The developers of *BibTiP* [132] also published an architecture that is similar to the architecture of bX (both bX and BibTip utilize usage data to generate recommendations).

Several academic services published datasets that eased the process of researching and developing research-paper recommender systems. *CiteULike*[32] and *Bibsonomy*[33] published datasets containing the social tags that their users added to research articles. The datasets were not originally intended for recommender-system research but are frequently used for this purpose [180, 197, 342]. *CiteSeer* made its corpus of research papers public[34], as well as the citation graph of the articles, data for author name disambiguation, and the co-author network [55]. CiteSeer's dataset has been frequently used by researchers for evaluating research-paper recommender systems [75, 107, 164, 180, 202, 320, 342, 393, 428]. *Jack et*

---

[28] Recommendation frameworks such as *LensKit or Mahout* may also be helpful for researchers and developers, but such frameworks are not the subject of this thesis.
[29] http://grouplens.org/datasets/movielens/
[30] http://labrosa.ee.columbia.edu/millionsong/
[31] http://www.kde.cs.uni-kassel.de/ws/dc13/
[32] http://www.citeulike.org/faq/data.adp
[33] https://www.kde.cs.uni-kassel.de/bibsonomy/dumps/
[34] http://csxstatic.ist.psu.edu/about/data

*al.* compiled a dataset based on the reference management software *Mendeley* [188]. The dataset includes 50,000 randomly selected personal libraries from 1.5 million users. These 50,000 libraries contain 4.4 million articles with 3.6 million of them being unique. For privacy reasons, *Jack et al.* only publish unique IDs of the articles and no title or author names. In addition, only those libraries having at least 20 articles were included in the dataset. *Sugiyama and Kan* released two small datasets[35], which they created for their academic recommender system [372]. The datasets include some research papers, and the interests of 50 researchers. The CORE project released a dataset[36] with enriched metadata and full-texts of academic articles, and that could be helpful in building a recommendation candidate corpus.

Of the reviewed approaches, 29% were evaluated using datasets from CiteSeer and 10% were evaluated using papers from ACM (Table 9). Other data sources included CiteULike (10%), DBLP (8%), and a variety of others, often not publicly available datasets (52%). Even when data originated from the same sources, it did not guarantee that the same datasets were used. For instance, no single CiteSeer dataset exists. Authors collected CiteSeer data at different times and pruned datasets differently. Some authors removed documents with fewer than two citations from the CiteSeer corpus [107], others with fewer than three citations [393], and others with fewer than four citations [102]. Other datasets were pruned even stronger: *Caragea et al.* removed papers having fewer than ten and more than 100 citations, as well as papers citing fewer than 15 and more than 50 papers [75]. From 1.3 million papers in the corpus, around 16,000 remained (1.2%). *Pennock et al.* removed documents from the corpus with fewer than 15 implicit ratings [320]: from originally 270,000 papers, 1,575 remained (0.58%). It is therefore safe to say that no two studies, performed by different authors, used the same dataset. This raises the question of the extent to which results based on different datasets are comparable.

Table 9: Source of datasets

| Dataset | CiteSeer | ACM | CiteULike | DBLP | Others |
|---|---|---|---|---|---|
| Number of user studies | 29% | 10% | 10% | 8% | 52% |

[35] http://www.comp.nus.edu.sg/~sugiyama/SchPaperRecData.html
[36] http://core.kmi.open.ac.uk/intro/data_dumps

It is commonly known that recommendation approaches perform differently on different datasets [64, 158, 201]. This is particularly true for the absolute effectiveness of recommendation approaches. For instance, an algorithm that achieved a recall of 4% on an IEEE dataset, achieved a recall of 12% on an ACM dataset [301]. The *relative* effectiveness of two approaches is also not necessarily the same with different datasets. For instance, because approach A is more effective than approach B on dataset I, does not mean that A is also more effective than B on dataset II. However, among the few reviewed approaches that were evaluated on different datasets, the effectiveness was surprisingly consistent.

Of the evaluated approaches, seven were evaluated on multiple offline datasets. Dataset combinations included CiteSeer and some blogs [298], CiteSeer and Web-kd [202], CiteSeer and CiteULike [180], CiteSeer and Eachmovie [320], and IEEE, ACM and ScienceDirect [301]. Only in one study did results differ notably among the different datasets. However, the absolute ranking of the approaches remained the same [180] (Table 10). In that article, the proposed approach (CTM) performed best on two datasets (CiteULike and CiteSeer), with a MRR of 0.529 and 0.467 respectively. Three of the four baselines performed similarly on the CiteSeer dataset (all with a MRR between 0.238 and 0.288). However, for the CiteULike dataset the TM approach performed four times as well as CRM. Consequently, if TM had been compared with CRM, rankings would have been similar on the CiteSeer dataset but different on the CiteULike dataset.

Table 10: MRR on different datasets

| | | Dataset | |
|---|---|---|---|
| **Rank** | **Approach** | **CiteSeer** | **CiteULike** |
| 1 | CTM | 0.529 | 0.467 |
| 2 | TM | 0.288 | 0.285 |
| 3 | cite-LDA | 0.285 | 0.143 |
| 4 | CRM | 0.238 | 0.072 |
| 5 | link-LDA | 0.028 | 0.013 |

Overall, a sample size of seven is small, but it gives at least some indication that the impact of the chosen dataset is rather low in the domain of research-paper recommender systems. This finding is interesting because in other fields it has been observed that different datasets lead to different results [64, 158]. Nevertheless, we doubt that pruning datasets should be considered good practice, in particular if only a fraction of the original data remains.

### 3.4.7 The Butterfly Effect: Unpredictable Results

The reproducibility of experimental results is the "fundamental assumption" in science [77], and the "cornerstone" for drawing meaningful conclusions about the generalizability of ideas [333]. Reproducibility describes the situation when (slightly) different ideas, scenarios, and evaluations lead to similar experimental results [77], whereas we define "similar results" as results that allow the same conclusions to be drawn. Conversely, if changes in the ideas, scenarios, or evaluations cause dissimilar results, i.e. results that do not allow the same conclusions to be drawn, we speak of non-reproducibility. Non-reproducibility is expected when significant changes are made to the ideas, scenarios, or evaluations. However, if minor changes are made but results are unexpectedly dissimilar, then we speak of the "butterfly effect".

*Reproducibility* should not be confused with *replicability*. Replicability is used to describe an exact copy of an experiment that uses the same tools, follows the same steps, and produces the same results [108]. Therefore, replicability is important when analyzing whether the original experiment was conducted thoroughly and whether the results can be trusted.

During the review, we found several examples of the butterfly effect, i.e. variations in experimental results that we considered unexpected. For instance, the developers of the recommender system *bx* report that the effectiveness of their recommender system varied by factor three at different institutions although the same recommendation approach was used [390]. *Lu et al.* reported that the *translation* model had twice the accuracy of the *language* model [258], but in another evaluation, accuracy was only 18% higher [162]. *Huang et al.* report that the *Context-aware Relevance Model* (CRM) and *cite-LDA* performed similarly, but in another evaluation by the same authors, CRM performed significantly worse than cite-LDA [180]. *Lu et al.* found that, sometimes, terms from the abstract performed better than terms from the body-text, while sometimes the opposite occurred [258]. *Zarrinkalam and Kahani* found that, sometimes, terms from the title *and* abstract were most effective, while sometimes terms from the title, abstract, *and* citation context were most effective [428]. *Bethard and Jurafsky* reported that citation counts *strongly* increased the effectiveness of their recommendation approach [54], while *He et al.* reported that citation counts *slightly* increased the effectiveness of their approach [164].

Probably most interesting, with respect to the butterfly effect, are some evaluations by the TechLens team (Table 11). The TechLens team evaluated several content-based (CBF) and collaborative filtering (CF) approaches for research-paper recommendations. In 2002, *McNee et al.* conducted an offline evaluation in which CF and CBF performed similarly [274]. However, their additional user study led to a different result – CBF outperformed CF. A user study by *Torres et al.* in 2004 report results similar to the user study by *McNee et al.* (CBF outperformed CF) [393]. However, the offline evaluation from *Torres et al.* contradicted the previous results – this time, CF outperformed CBF. In 2006, another user study by *McNee et al.* indicated that CF (slightly) outperforms CBF [275], which contradicts the previous user studies. In 2009, Dong et al., who are not affiliated with TechLens, evaluated the approaches of *Torres et al.* with an offline evaluation [107]. In this evaluation, CBF outperformed CF, contradicting the previous offline-results from *Torres et al*. In 2010, *Ekstrand et al.* found that CBF performed worse than CF in both an offline evaluation and user study, which again contradicts most of the previous findings [110].

Table 11: Results of different CBF and CF evaluations

| | McNee et al. 2002 | | Torres et al. 2004 | | McNee et al. 2006 | | Dong et al. 2009 | | Ekstrand et al. 2010 | |
|---|---|---|---|---|---|---|---|---|---|---|
| | Offline | User Std. | Offline | User Std. | Offline | User Std. | Offline | User Std. | Offline | User Std. |
| **CBF** | Draw | Win | Lose | Win | -- | Lose | Win | -- | Lose | Lose |
| **CF** | Draw | Lose | Win | Lose | -- | Win | Lose | -- | Win | Win |

The authors of the studies provide some potential reasons for the variations, such as different datasets, differences in user populations, and variations in the implementations. However, these reasons can only explain *some* of the variations, and overall we consider most of the contradictions to be unexpected.

We see the primary purpose of evaluations in aiding practitioners and researchers in identifying the most effective recommendation approaches (for a given scenario). Consequently, a practitioner who needed an effective recommendation approach, or a researcher who needed an appropriate baseline to compare a novel approach against, would not find much guidance in the existing evaluations. Similarly, the evaluations, as they currently are, do not help to conclude whether CF or CBF is more promising for research-paper recommender systems, or which of the approaches is most promising for mind maps.

## 3.5 Discussion and Summary

The literature survey was primarily conducted to find promising user-modeling approaches that could serve as basis for a mind-map-specific user-modeling approach. The survey should have further helped to find adequate evaluation methods and metrics. However, the survey revealed that most of the reviewed recommendation approaches neglected the user-modeling process. For instance, most approaches let users specify their interests, i.e. they do not perform user modeling at all. Those approaches that automatically inferred interests, mostly utilized all items of a user, i.e. they ignored issues like concept drift. Finally, for most approaches it was not reported how many features were stored in user models. Hence, most of the approaches are not adequate as a basis for mind-map-based user modeling. In addition, the evaluations of the approaches were mostly questionable. Several approaches were not evaluated at all, and of those who were, most were evaluated against trivial baselines, and with offline evaluations, which are subject to strong criticism. Most of the few user studies were also of little value since they were mostly conducted with a small number of participants. Consequently, it remains unclear, which of the reviewed approaches are most promising. Even if a few promising approaches could have been identified, there is little information about the specifics of the approaches. This would make a re-implementation difficult, if not impossible.

While the survey did not help in identifying specific recommendation approaches that might be promising, the survey did aid in finding adequate recommendation classes for user modeling based on mind maps.

Stereotype recommendations could be relevant for mind-mapping users: A user-modeling system could generalize about certain types of mind-mapping users and provide recommendations based on stereotypes. For instance, the mind-mapping software *Mindjet* is often used by lawyers. A user-modeling system could generalize that lawyers are interested in law books. Hence, Mindjet could implement a recommender system that recommends law books to its users. A similar approach could be applied with Docear.

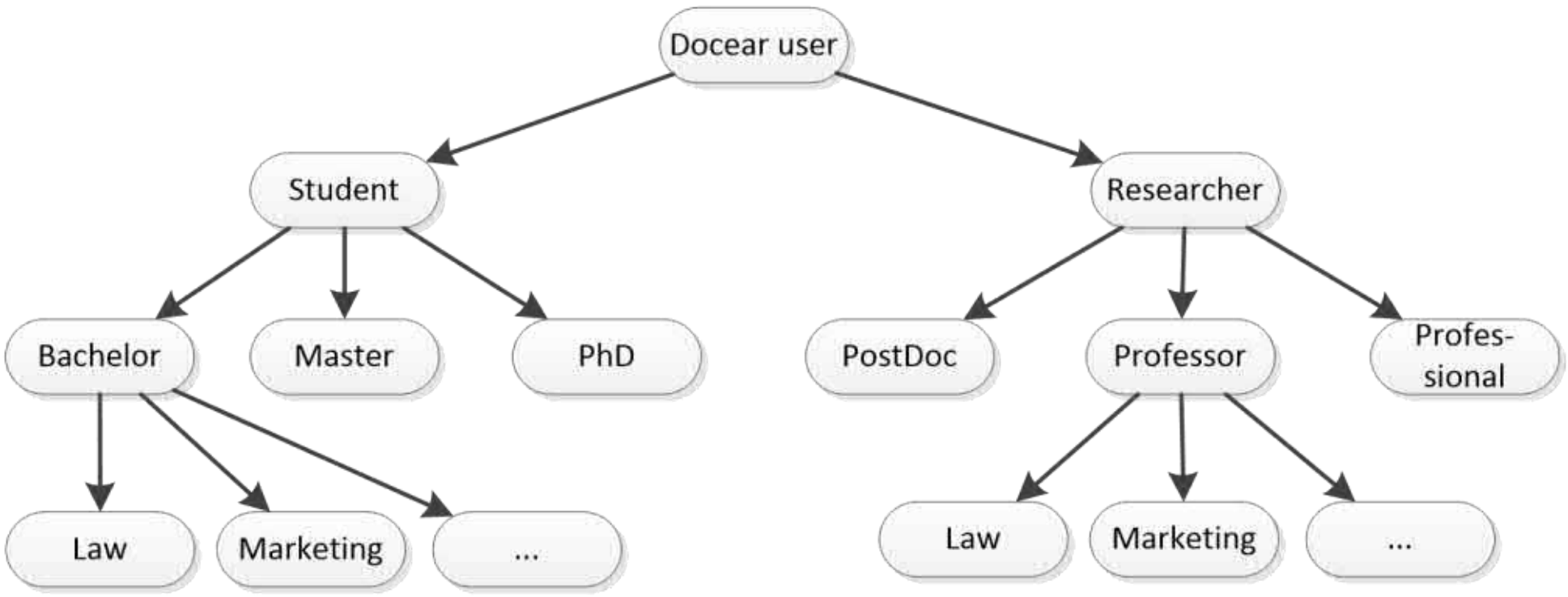

Figure 20: Illustration of a stereotype-tree of Docear users

Docear users are primarily students and researchers. In a simple scenario, Docear's user-modeling system could assign each user as a stereotype "student or researcher" and recommend academic literature that is presumably interesting to students and researchers – for instance, books about academic writing. In a more complex system, users could be classified as students and researchers, with researchers being further divided into Postdocs, professors, and professionals. Students could be further divided into Bachelor's, Master's, and PhD students (Figure 20). Each of these types could be further divided into fields of study/work, for instance law or marketing. Each stereotype level could be assigned a set of appropriate recommendations. For instance, a recommender system could recommend student loans to students in general, marketing textbooks to Bachelor's students in marketing, and so on.

Collaborative filtering seems to be of little relevance for mind-map-based user modeling. First, collaborative filtering is domain independent. As such, there is little need to consider special characteristics of mind maps. Second, a typical CF approach applied to mind maps would recommend mind maps. However, most users would probably not want to share their mind maps with other users. Theoretically, mind-map-based recommender systems could apply implicit collaborative filtering and infer ratings for e.g. research articles cited in mind maps. However, at least for Docear, this approach is not feasible due to high sparsity. We found that of 616,635 papers that are linked in the user's mind maps, only 224 papers (0.036%) were linked by two different users, three papers (0.00049%) were linked by three users, and no paper was linked by more than three users[37]. This means that 99.96% of the papers were only linked by a single user. In other words, barely any users have PDFs in common and if they do, they have at maximum three in common. This makes the application of (implicit) collaborative filtering infeasible. We also investigated the possibility of applying content-boosted CF [13, 280], but sparsity was still very high.

Co-occurrence recommendations could also be interesting for mind-mapping applications. In this case, relatedness of items linked in mind maps could be calculated. Items linked in the same mind map would assumed to be related, and the more often mind maps link them, the more related they would be. The concept of citation proximity analysis could also be applied to mind maps [142]; the closer two items were linked within a mind map, the higher their relatedness (Figure 21). We conducted an initial user study that showed promising results for such recommendations (cf. Appendix D, p. 205). However, since there are barely any users sharing the same PDF files (see previous paragraph), applying co-occurrence recommendations for Docear, seems unfeasible.

[37] In the analysis, we ignored papers that Docear automatically adds to user's mind-maps as part of demo-projects. We also ignored users and mind-maps that were identical.

Graph-based recommendations seem of little relevance for mind maps because mind maps are not inherently connected with each other. Hence, building a graph would be not possible. At most, a graph could be built if items linked in mind maps are additionally considered. However, again, the high sparsity makes this idea not feasible for Docear and probably most other mind-mapping applications as well. Global relevance metrics could be used to enhance recommendations in the field of mind maps. However, these metrics would not be mind-map specific, which is why we do neglect global relevance in the remainder, and focus on the characteristics of mind maps instead.

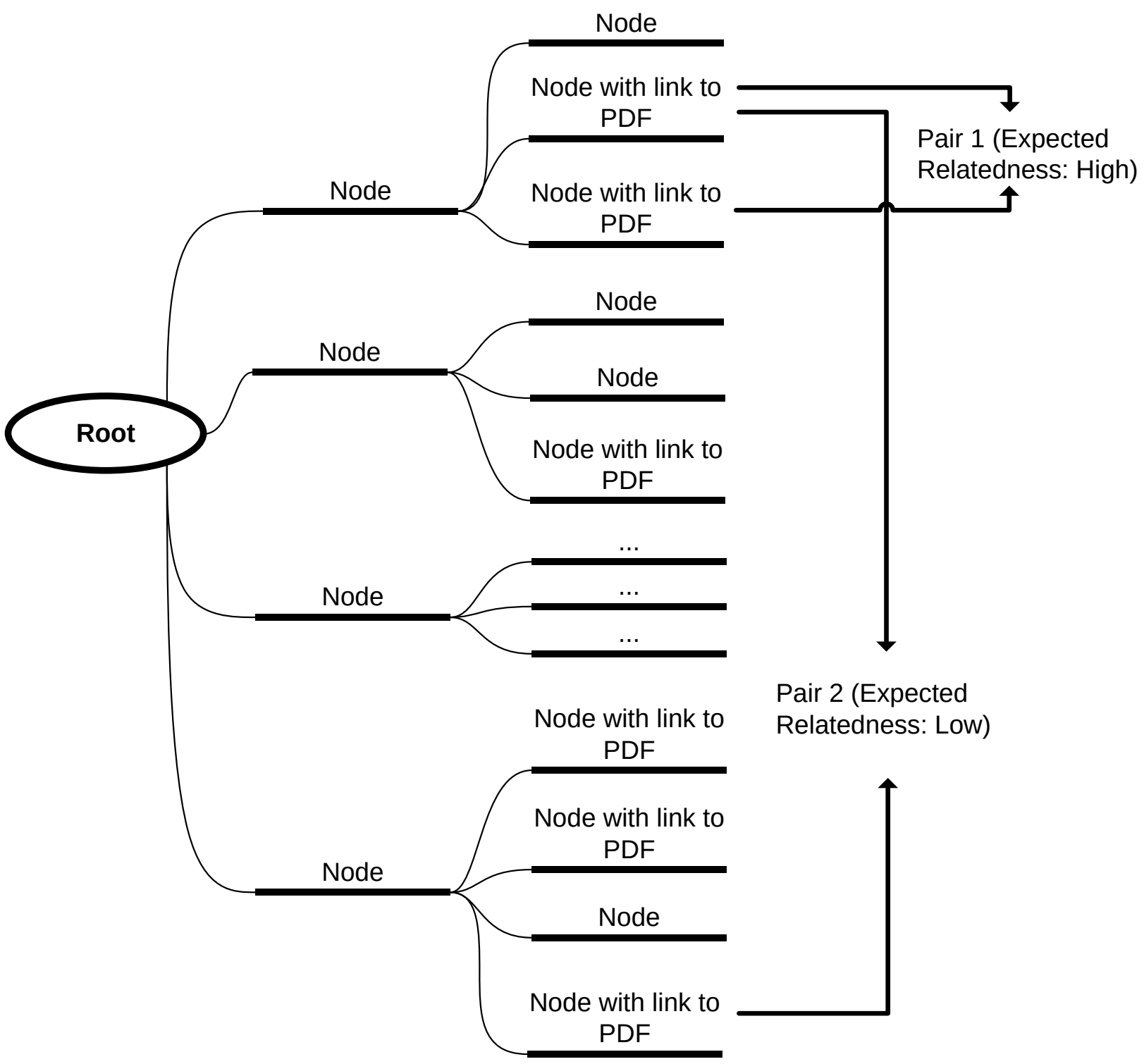

Figure 21: Item similarity based on proximity in mind maps

Eventually, content-based filtering seems to be the most promising user-modeling concept for mind maps. The few attempts that had been made in the field by MindMeister and Mindomo were content-based, and as our preliminary study showed, the effectiveness of some approaches was promising.

# 4. Methodology

This chapter presents how we built Docear's recommender system, compared different evaluation methods, found promising variables for mind-map-specific user modeling, and build a mind-map-specific user-modeling approach.

## 4.1 Development of Docear's Recommender System

Recommender-system researchers often conduct research on existing datasets (cf. Section 3.4.6, p. 55). However, the existing datasets and architectures are not helpful for developing and evaluating mind-map-specific user-modeling approaches (cf. 3.5, p. 59). In addition, we needed a real-world recommender system to compare results of different evaluation methods (Task 3). Therefore, as part of Task 2, we developed an architecture for a research-paper recommender system, implemented a recommender system based on this architecture, integrated it into Docear, and used this recommender system for our research.

Docear's recommender system applies primarily a content-based filtering approach, since collaborative filtering, co-occurrence recommendations, and graph-based recommendations are not feasible to apply with mind maps and Docear (cf. 3.5, p. 59). The basic idea is as follows (cf. Figure 22): For user $u$, a user model $um_u$ is created based on the mind map collection $MM_u$, whereas $MM_u$ consists of those mind maps $mm$ the user interacted with, i.e. $MM_u=\{mm_1, \ldots, mm_i\}$. In its simplest case $um_u$ is a bag of features $F$ that comprises all features $f$ contained in $MM_u$, i.e. $um_u = F(MM_u) = F(mm_1 \cup mm_2 \cup \ldots \cup mm_i)$. Similarly, the collection of research articles $A$, i.e. serves as recommendation candidates, consists of a number of articles, i.e. $A = \{a_1, \ldots, a_m\}$ that each has a certain number of features $f$. To give recommendations, those articles that have the most features in common with the user model are recommended.

The recommender system was mostly developed in JAVA, and displays recommendations to the users at the start-up of Docear, every five days (Figure 23). In addition, users may explicitly request recommendations at any time. Docear displays recommendations as a set of ten research papers, as long as ten papers are available to recommend, otherwise less recommendations are shown. A click on a recommendation opens the recommended PDF file in the user's web browser. For all users, the number of displayed and clicked recommendations was recorded, so we could calculate CTR for the evaluation. Users could also rate each recommendation set on a scale of one to five.

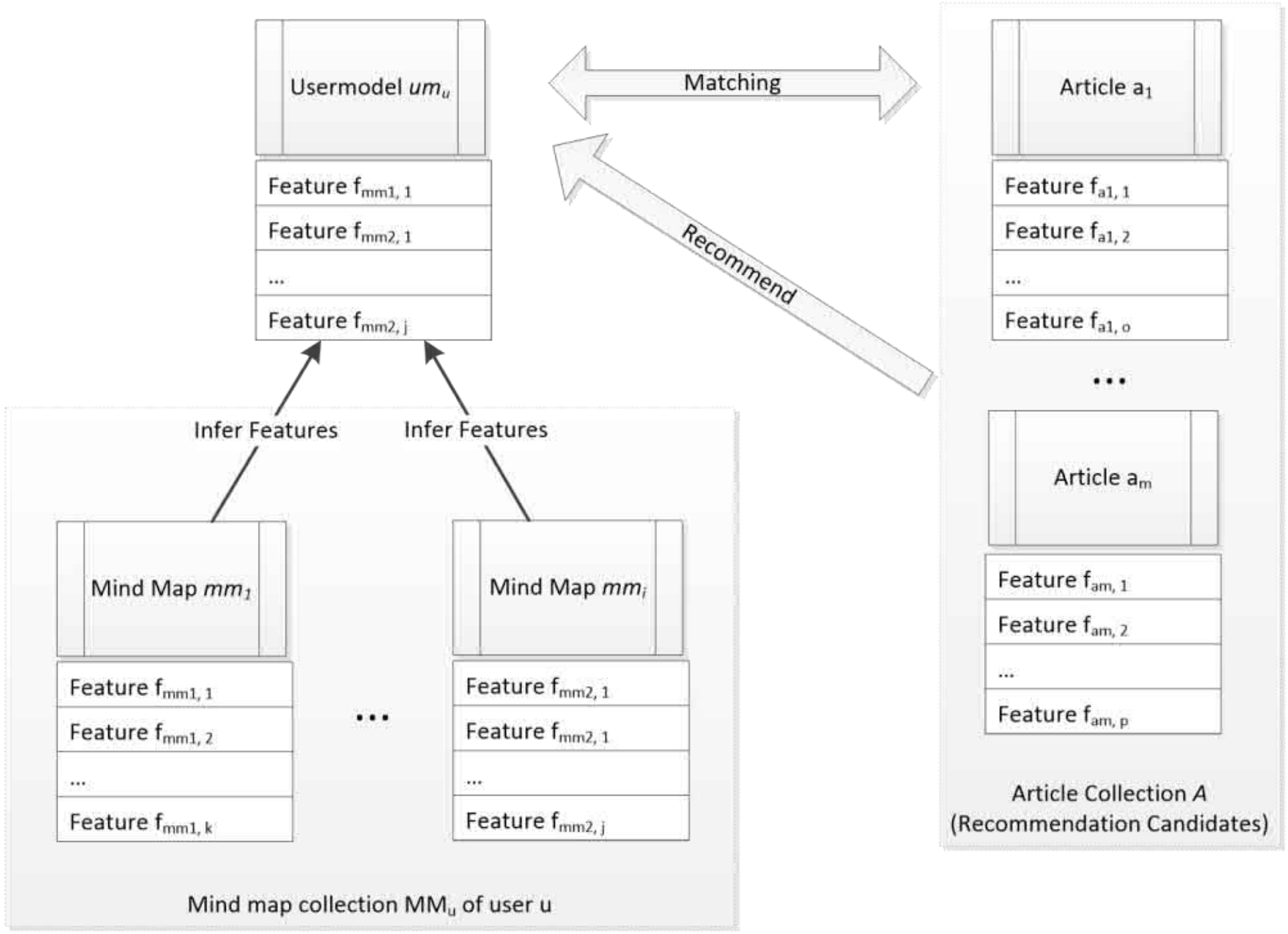

Figure 22: Content-based filtering with mind maps and research articles

Docear only displays the title of a recommendation. This might be not ideal because users would probably like to see further bibliographic information (cf. Section 3.3.2.3, p. 43). However, for most of the articles in Docear's corpus such information was not available, at least not in good quality. To not bias the recommendation process by sometimes displaying the abstract, and sometimes not displaying the abstract, we decided to only display the title. As shown later, this decision most likely has not negatively affected the results of the evaluation.

Most of the data that we collected with the recommender system is released as publicly available dataset (details follow in the subsequent chapter). By publishing the recommender system's architecture and datasets, we pursue three goals.

First, we want researchers to be able to understand, validate, and reproduce our research. Second, we want to support researchers when building their own research-paper recommender systems: Docear's architecture and datasets ease the process of designing one's own system, estimating the required development times, determining the required hardware resources to run the system, and crawling full-text papers to use as recommendation candidates. Third, we want to provide real-world data to researchers who have no access to such data. This is of particular importance, since the majority of researchers in the field of research-paper recommender systems have no access to real-world recommender systems.

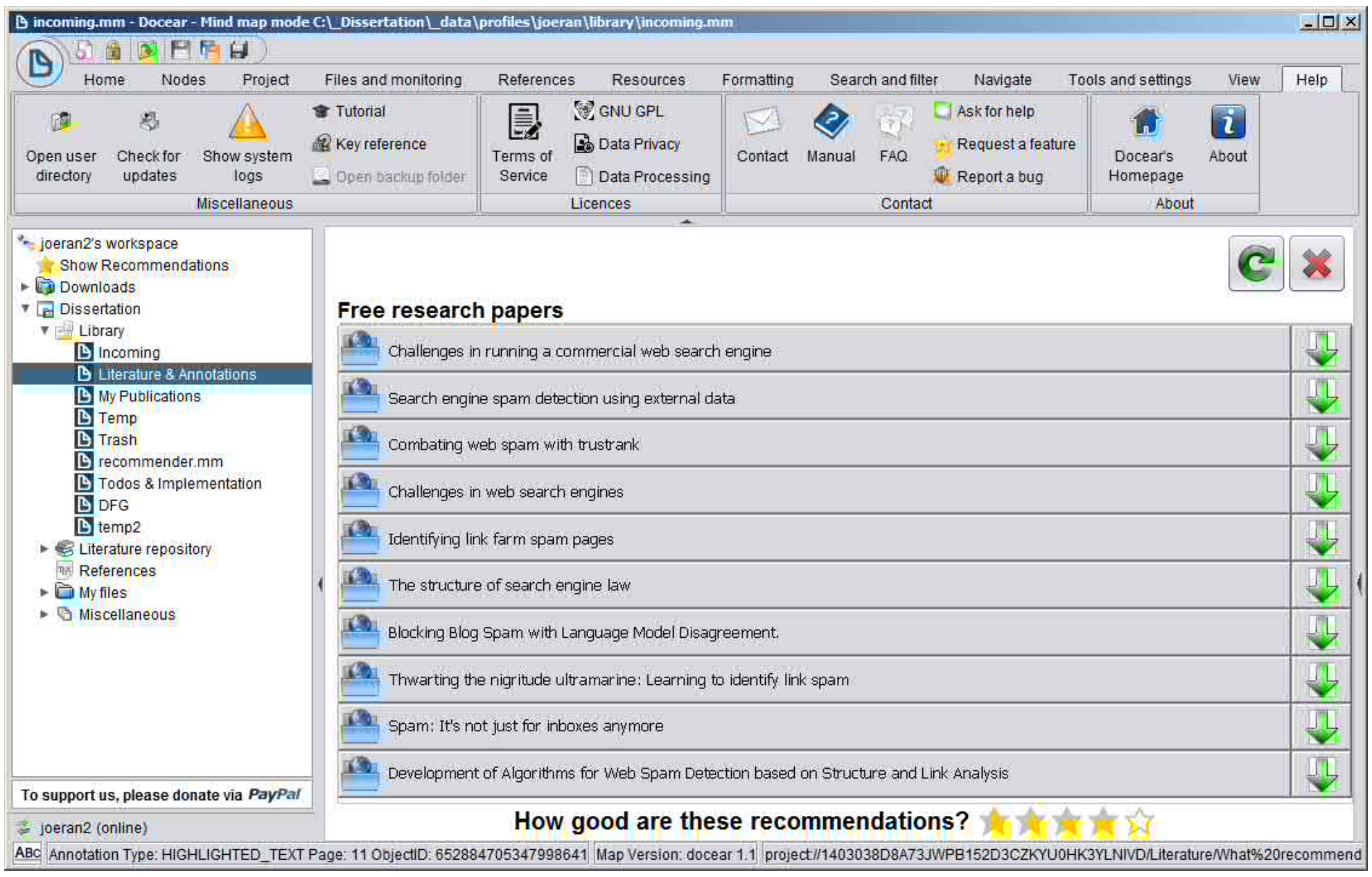

Figure 23: User-interface of Docear's recommender system

More details on the recommender system are presented in the subsequent chapter (5.1, p. 73).

## 4.2 Comparison of Evaluation Methods and Metrics

The literature survey did not help in identifying adequate evaluation methods and metrics. To research the adequacy of the different evaluation methods (Task 3), we measured the effectiveness of different user-modeling algorithms with an offline evaluation, user study, and online evaluation. We then analyzed the correlation of the results of the different evaluation methods and metrics. We expected that, ideally, all evaluation methods and metrics would lead to similar results. If concordance between the methods and metrics would be achieved, any of the methods and metrics could be used to evaluate our mind-map-specific user-modeling approach. If there were discord between the methods and metrics, a discussion would be necessary about which of the methods and metrics are adequate and under which circumstances.

We intended to conduct two user studies: one lab study and one real-world study. For the lab study, we wanted to recruit participants through our blog[38]. In our blog, we asked Docear's users to start Docear, request recommendations, click each of them, and read at least the abstract of each recommended paper. Users should then rate the relevance of the recommendations from 1 to 5 stars (Figure 23, p. 65), and if they wish, request new recommendations and continue this process for as long as they like. The study was intended to run from April to July 2014. We promoted the study in our newsletter (8,676 recipients), on Facebook (828 followers), on Twitter (551 followers), and on Docear's homepage (15,000 visitors per month). Despite 248 people reading the blog post, only a single user participated in the study. He rated three sets, each with ten recommendations. However, ratings of a single user are not suitable to receive meaningful results. Hence, we consider this user study as a failure, and focus on results of the real-world study. The real-world study was based on ratings that users provided during their normal work with Docear (Figure 23). Overall, 379 users rated 903 recommendation sets with 8,010 recommendations. The average rating was 2.82 (out of 5).

For the online evaluation, we measure acceptance rates of 45,208 recommendation sets displayed to 4,700 users from March 2013 to August 2014. Typically, each set of recommendations consists of ten recommendations, resulting in a total of 430,893 delivered recommendations. Acceptance is measured with the following metrics: Click-Through Rate (**CTR**) measures the ratio of clicked vs. delivered recommendations. Click-Through Rate over sets (**$\mathbf{CTR_{Set}}$**) is the mean of the recommendation sets' individual CTRs. For instance, if eight out of ten recommendations had been clicked in *set I*, and two out of five recommendations in *set II*, then for the two sets CTR would be $\frac{8+2}{10+5} = 66.67\%$ but $CTR_{Set}$ would be $\frac{8/10 + 2/5}{2} = 60\%$. We also calculated CTR over users (**$\mathbf{CTR_{User}}$**). $CTR_{User}$ levels the effect that a few power users might have. For instance, if *users A*, *B*, and *C* saw 100, 200, and 1,000 recommendations, and user *A* clicked seven, user *B* 16, and user *C* 300 recommendations, CTR would be $\frac{7+16+300}{100+200+1000} = 24.85\%$, but $CTR_{User}$ would be $\frac{7}{100} + \frac{16}{200} + \frac{300}{1000} = 12.36\%$, i.e. the impact of user C would be weaker. However, $CTR_{User}$ was only calculated for two analyses (the reason is discussed later). Link-Through Rate (**LTR**) describes the ratio of the displayed recommendations against those recommendations that actually had been clicked, downloaded, and linked in the user's mind map. Annotate-Through Rate (**ATR**) describes the ratio of recommendations that were annotated, i.e. a user opened a linked PDF in a PDF viewer, created at least one annotation (bookmark, comment, or highlighted text), and imported that annotation in Docear[39]. Cite-Through Rate (**CiTR**) describes the ratio of documents for which the user added some bibliographic data in the mind map, which strongly indicates that the user plans to cite that document in a future research paper, assignment, or other piece of academic work.

[38]http://www.docear.org/2014/04/10/wanted-participants-for-a-user-study-about-docears-recommender-system/

For the offline evaluation, we considered papers that users cited in their mind maps to be the *inferred ground-truth* (cf. 2.5.3.3, p. 24). For each Docear user, we created a copy of their mind maps, and removed the paper that was most recently added to the mind map. We then applied a randomly selected recommendation approach to the modified mind map. Overall, we calculated 118,291 recommendation sets. To measure the accuracy of the algorithm, we analyzed whether the removed paper was within the top10 (P@10) or top3 (P@3) of the recommendation candidates. We also calculated the Mean Reciprocal Rank (MRR), i.e. the inverse of the rank at which the removed paper was recommended. In addition, we calculated nDCG based on the 10 most recently added papers and 50 recommendation candidates. Our evaluation method is similar to other offline evaluations in the field of research-paper recommender systems, where the citations made in research papers are used as ground-truth. We display accuracy metrics as percentages in charts. Typically, such metrics are displayed as decimals between zero and one, but to display online and offline metrics in a single chart, we had to choose one unit. If not otherwise stated, all reported differences are statistically significant ($p<0.05$). Significance was calculated with a two-tailed *t*-test and $\chi^2$ test where appropriate.

[39] It should be noted that PDFs often contain annotations already when they are published. For instance, often PDFs contain the table of content as bookmarks. Consequently, results based on ATR should be considered with some skepticism.

## 4.3 Identification of Mind-Map-Specific User-Modeling Variables

To identify a number of variables that might affect user modeling based on mind maps (Task 4) we did a brainstorming session, leading to a number of potential variables. Due to time restrictions, we decided to implement and evaluate only a few variables that we considered most promising, and for which an evaluation with Docear was feasible. The variables we focused on included the number of mind maps to analyze, the number of nodes to utilize, the size of the user model, whether to use only visible nodes, and different weighting schemes including standard schemes like TF-IDF but also mind-map-specific weighting schemes based, for example, on the number of children a node has.

The variables are randomly arranged to assemble the final user-modeling algorithm, each time recommendations are generated. For instance, one algorithm might utilize *visible* nodes from the *2* most recently modified mind maps, weight the *terms* of these nodes with *TF-IDF*, and store the *25* highest weighted terms in the user model. Another algorithm might use the *250* most recently modified nodes (*visible and invisible)* among all mind maps, weight the *citations* of these nodes with *TF-only*, and store the *5* highest weighted citations in the user model.

A variable that we considered *not* feasible to analyze was the "position of a node". In most mind-mapping tools, users can arrange their nodes freely. We would assume that depending on the position of a node, its importance differs, and hence the weighting for user modeling should differ. For instance, a node far away from the root node could be weighted differently than a node in close proximity to the root node (illustrated in Figure 24). However, Docear positions nodes automatically, which is why we did not analyze the impact of a node's position.

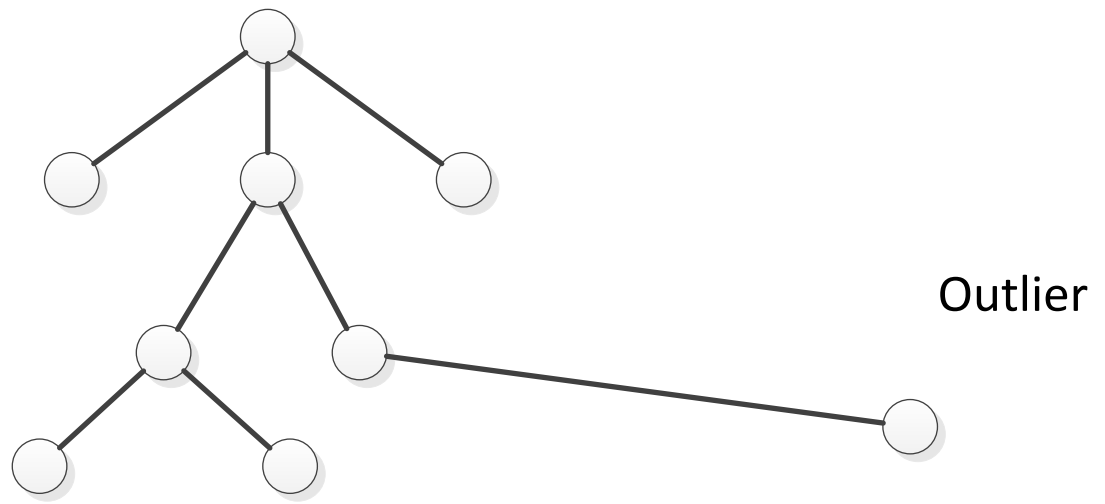

Figure 24: Mind map with an outlier node

We measured the effectiveness of the individual variables with CTR, since the comparison of the evaluation methods showed that CTR highly correlates with user satisfaction and reporting CTR has some inherent value. In addition, rather few users participated in the user study, and we did not receive enough ratings, to calculate statistically significant results for all variables based on ratings. To find the optimal values for each variable, we compared CTR for all approaches in which a variable had a certain value. For instance, to evaluate whether a user-model size of *10* or *100* terms was more effective, effectiveness of all algorithms with a user-model size of 10 was compared to the effectiveness of all algorithms with a user-model size of 100.

We analyzed the effects of the variables for both CBF based on citations and based on terms – we expected that the optimal values for the variables would differ for terms and citations. For term-based CBF variations, all reported differences are statistically significant ($p<0.05$), if not reported otherwise in the text. Significance was calculated with a two tailed *t*-test and $\chi^2$ test where appropriate. Results for citation based CBF are mostly not statistically significant, because the approach was implemented only a few months ago, and not all users have citations in their mind-maps. Therefore, an insufficient number of citation-based recommendations were delivered to produce significant results. Consequently, the focus of our research lies on the term-based CBF variations. We also report runtimes in the charts for informative reasons, but do not discuss the data. It should be noted that runtimes could significantly differ with different implementations, or on different hardware. Overall, runtimes are rather long. This is caused by recording many statistics and running some other services on the recommendation server.

Our methodology has a limitation since determining optimal values for each variable separately, ignores potential dependencies. For instance, only because a user-model size of 100 terms is most effective on average, and analyzing 500 nodes is most effective on average, does not mean that analyzing 500 nodes *and* a user-model size of 100 terms is the optimal combination. Ideally, we would have evaluated all possible variations to find the single best variation. However, for some variables, there are up to 1,000 possible values, and combining all these variables and values leads to millions of possible variations. Evaluating this many variations was not feasible for us. The second best option would have been a multivariate statistical analysis to identify the impact of the single variables. However, also for such an analysis we did not have enough data. Therefore, our methodology was the third best option. It will not lead to a single optimal combination of variables, but as our result will show, our methodology leads to a significantly better algorithm than the baselines, and the results help understanding the factors that affect effectiveness in mind-map-based user modeling.

## 4.4 Development of a Mind-Map-Specific User-Modeling Approach

To achieve our primary research objective, we identified the optimal value for each of the implemented variables, and combined the optimal values in a single algorithm. We them compared this algorithm against four baselines, to analyze whether this mind-map-specific user modeling performed better than the baselines.

One baseline was the stereotype approach that was chosen with a probability of 1%, whenever the recommendation process was triggered. To implement the stereotype approach, the recommender system generalizes over all users, and assumes that they are all researchers (which is not exactly true, because some users only use Docear for its mind-mapping functionality). The recommender system then recommends papers that are potentially interesting for researchers, i.e. books and research articles about academic writing that we manually added to the corpus. The stereotype does not consider any special characteristics of mind maps, and we implemented it as simple baseline and because stereotype recommendations have not been used before to recommend research papers.

The second, third and fourth baseline were those CBF variations that are rather obvious and that we already used in our initial study: **a)** the approach of MindMeister, in which only terms of the most recently modified node are analyzed for the user model ('modified' means 'created', 'edited' or 'moved'); **b)** all terms of the user's current mind map are used for the user model; **c)** all terms of all mind maps that the user ever created are utilized for the user model.

We did not compare our approach against any of the reviewed research-paper recommender approaches, for four reasons. First, it remains uncertain which of the 31 CBF approaches we should have used as baseline, due to the shortcomings in their evaluations (cf. Section 3.4, p. 50). Second, authors mostly provided sparse information, which would have made a re-implementation difficult, if not impossible. Third, the research-paper recommender approaches widely neglect the user modeling process, but the user modeling process was the focus of our research. Hence, the reviewed approaches are not adequate baselines for our purpose. Finally, and most importantly, the reviewed approaches are mostly not applicable to mind maps. This means, we could have implemented the reviewed approaches based on the PDFs of the users' or the bibliographic data entered in the mind maps. However, it would have been of little value to see that one of the reviewed approaches performs better on the users' PDFs than our novel approach on the user's mind maps. Besides, Docear's recommender system has no access to the users’ PDFs and hence could not apply general research-paper recommendation approaches (unless a function had been implemented to transfer the users’ PDFs to Docear’s server, which had been time consuming and had probably caused problems with copyright). Given these points, we conclude that comparing our novel approach against “standard” CBF baselines is the most sensible solution particularly since one of these baselines is the only approach that had been applied in practice in the domain of mind mapping.

# 5. Results & Discussion

In this chapter, we present and discuss the results of our work. Section 5.1 (p. 73) presents Docear's recommender system, i.e. its architecture and some datasets, that was developed a part of Task 2. The recommender system serves as framework for our research as it allows to run different user-modeling approaches and to evaluate them with different evaluation methods. Section 5.2 (p. 87) presents the analysis of the recommender-system evaluation methods (Task 3). We compared the outcomes of different methods when evaluating the same approaches, and we discuss the adequacy of the methods for our scenario and in general. Section 5.3 (p. 105) presents several variables that effect mind-map-based user modeling (Task 4). The effect of variables was evaluated with click-through rate, which showed to be the most sensible metric according to our previous analysis of evaluation methods. Finally, Section 5.4 (p. 121) presents Docear's mind-map-specific user-modeling approach that combines the optimal values of the variables in a single algorithm (Task 5). The approach is compared against several baselines, and proves to be about twice as effective as the most effective baseline.

## 5.1 Docear’s Recommender System[40]

### 5.1.1 Architecture

Docear itself is a JAVA desktop software with its source code hosted on *GitHub*[41]. Docear's recommender system is also primarily written in JAVA and runs on Docear’s web servers. To enable communication between the desktop software and the servers, we implemented a RESTful Web Service. Figure 25 illustrates the architecture and the particular components, which are explained in detail in the following sections, along with technical details.

---

[40] Parts of this chapter have been published as: Beel, Joeran, Stefan Langer, Bela Gipp, and Andreas Nürnberger. “The Architecture and Datasets of Docear’s Research Paper Recommender System.” In Proceedings of the 3rd International Workshop on Mining Scientific Publications (WOSP 2014) at the ACM/IEEE Joint Conference on Digital Libraries (JCDL 2014), 2014.

Please also note that all information in this chapter – including the datasets that we publish – is based on data that we collected before March 2014, while the following chapters are based on data that we collected until August 2014.

[41] https://github.com/Docear/

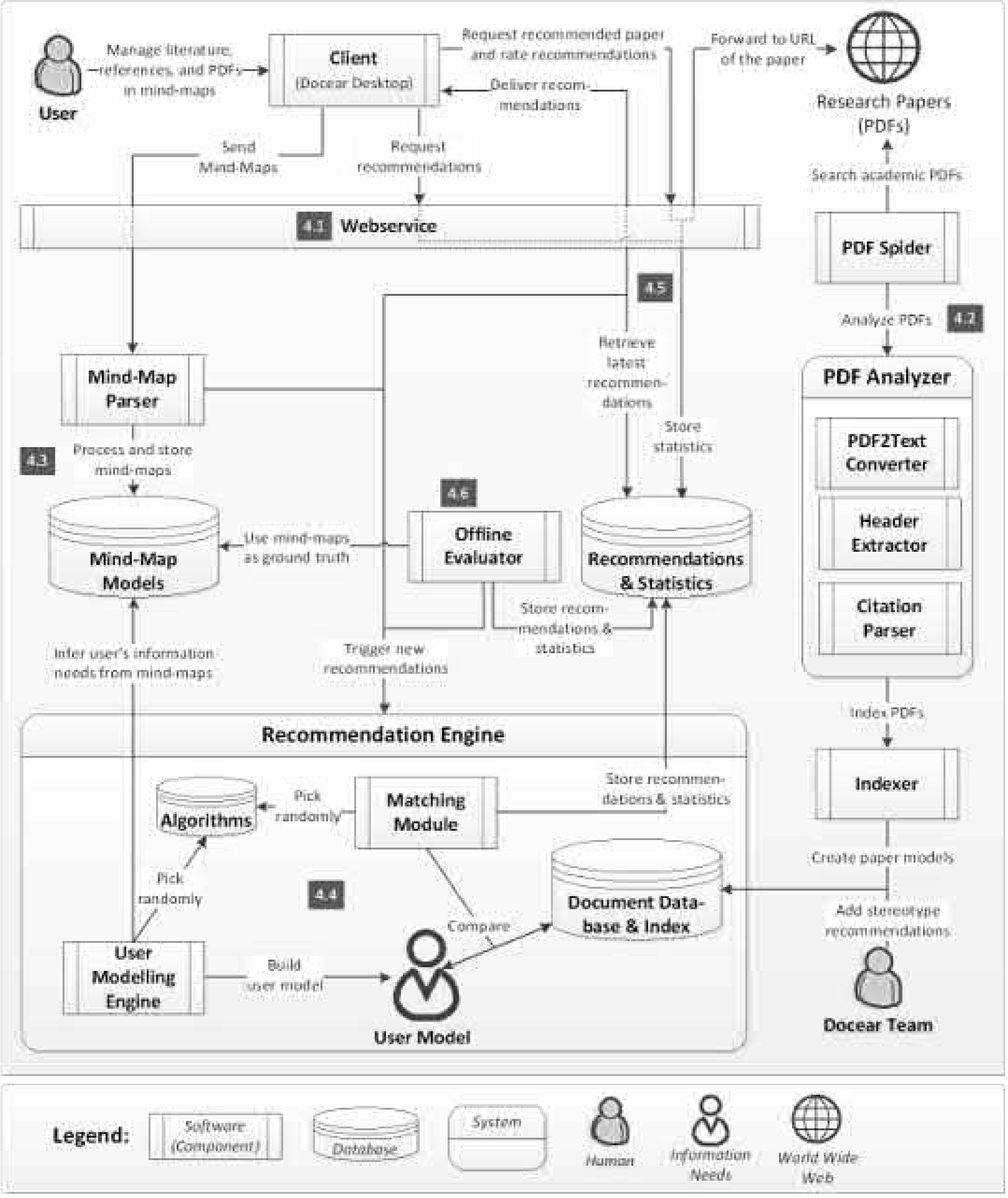

Figure 25: Architecture of Docear's recommender system

#### 5.1.1.1 Web Service / API

Docear's RESTful Web Service (based on *Jersey*)[42] is responsible for several tasks, including user registration and delivering recommendations. In Table 12, the most important methods relating to recommendations are listed. Third parties could use the Web Service, for instance, to request recommendations for a particular Docear user and to use the recommendations in their own application (if the third party knew the user's username and password). However, it should be noted that, for now, we developed the Web Service only for internal use, that there is no documentation available, and that the URLs might change without prior notification.

Table 12: POST and GET requests

| **Task** | **URL** | **Type** |
|---|---|---|
| **Upload a mind map** | *https://api.docear.org/user/{username}/mindmaps/* | POST |
| **Request recommendations** | *https://api.docear.org/user/{username}/recommendations/* | GET |
| **Confirm the receipt of recommendations** | *https://api.docear.org/user/{username}/recommendations/{recommendationsSetId}* | POST |
| **Download a recommended paper** | *https://api.docear.org/ user/{username}/ recommendations/fulltext/{hash}* | GET |
| **Send rating** | *https://api.docear.org/user/{username}/recommendations/{recommendationsSetId}* | POST |

#### 5.1.1.2 Building the corpus

The *Spider* crawls the Web for academic PDF files, which serve as recommendation candidates. Each PDF is converted into text, and the header information and citations are extracted. The text conversion is done with *jPod*[43], a PDF library we found to be more effective than the commonly used *PDFBox (cf. Appendix G.1, p. 249)*. The header extraction is done with *ParsCit*[44] and a tool that we developed and called *Docear's PDF Inspector (cf. Appendix G.2, p. 253)*. The citation extraction is also conducted with ParsCit, which we modified to identify the citation position within a text[45]. Once all information is extracted, it is indexed with *Apache Lucene*[46] and stored in Lucene's file-based data storage.

[42] http://jersey.java.net
[43] http://sourceforge.net/projects/jpodlib/
[44] http://aye.comp.nus.edu.sg/parsCit/

Instead of indexing the original citation placeholder with *[1]*, *[2]*, etc. the unique Docear ID of the cited document is indexed (e.g. *dcr_doc_id_54421*) (Figure 26). This allows to apply weighting schemes, such as TF-IDF to citations, i.e. CC-IDF [59], and searching with Lucene for documents that cite a certain paper. It also allows for the matching of user models and recommendation candidates based on terms *and* citations at the same time. For instance, a user model could consist of the terms and document-ID "*cancer, sun, dcr_doc_id_54421, skin"* and those papers would be recommended that contain the terms *cancer*, *sun* and *skin* and that cite the document *dcr_doc_id_54421*.

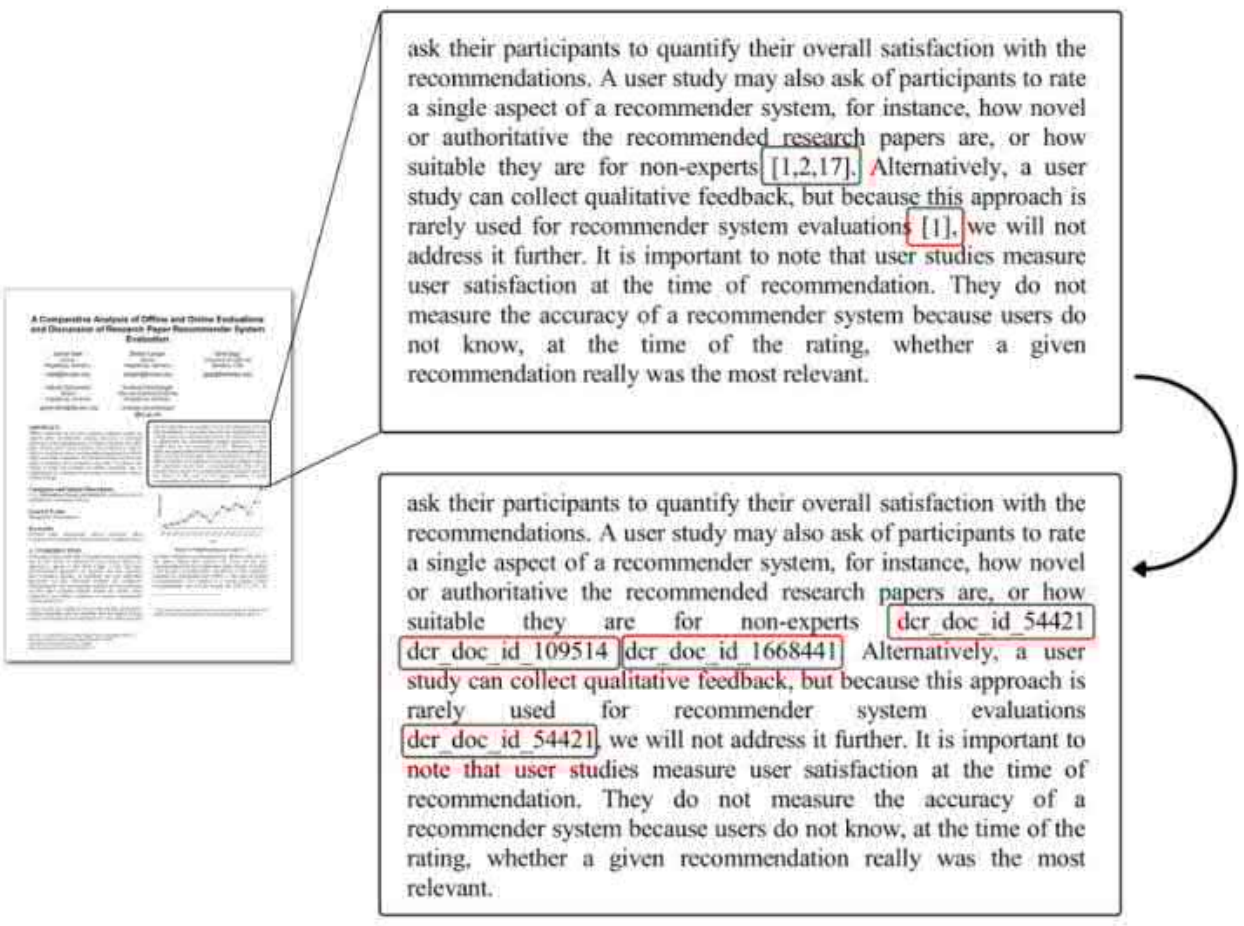

Figure 26: Converting in-text citations to Docear-IDs

In addition to the papers that were found by the Spider, we selected a few papers manually and added them to the corpus of recommendation candidates. These papers are about academic writing and search, i.e. topics we assume to be relevant for the majority of our users. These papers are recommended with the stereotype approach.

[45] Meanwhile, our modifications were integrated into ParsCit.
[46] http://lucene.apache.org

#### 5.1.1.3 Collecting information about users

Docear's recommender system needs access to the users' data, i.e. their mind maps, to be able to infer the users' information needs. To get access to the users' mind maps, Docear stores a copy of the mind maps in a temporary folder on the users' hard drive, whenever a mind map was modified and saved by the user. Every five minutes – or when Docear starts – Docear sends all mind maps located in the temporary folder to the Web Service. The Web Service forwards these mind maps, i.e. XML files, to the Mind-Map Parser (JAVA), which is based on *nanoXML*[47]. All nodes of the mind maps, including attributes (text, links to files, titles of linked PDFs, and bibliographic data) are extracted from the XML file and stored in a graph database (*neo4j*[48]). Citations in the mind maps are replaced with the corresponding Docear-IDs, similarly to the replace-process of citations in the research articles (cf. section 5.1.1.2 and Figure 26). A 'citation' in a mind map can either be a link to a PDF file, or the bibliographic data that is attached to a node. This means, if a node in a mind map links a PDF on the user's hard drive, the PDF is identified (via its title) and the link in the mind map is replaced with the Docear-ID of the cited article, linked PDF respectively. If the cited article is not already in Docear's database, the article is added and a new Docear-ID is created.

#### 5.1.1.4 Generating user models & recommendations

The recommendation engine is the central part of Docear's recommender system. It creates new user models and recommendations whenever new mind maps are uploaded to the server or after recommendations have been delivered to a user. Generating recommendations in advance has the disadvantage that a significant amount of computing time is wasted. Of all generated recommendations, only 21.3% were delivered to the users. In other words, 79.7% of the computing power could have been saved if recommendations were only created when they actually were needed. However, on average, it took 52 seconds to calculate one set of recommendations with a standard deviation of 118 seconds, and users would probably not want to wait so long for receiving recommendations. This rather long computing time is primarily caused by the many statistics that we calculate for each set of recommendations, along with a few algorithms that require intensive computing power. We also run several additional services on the recommendation servers that require a lot of computing power (e.g. PDF processing), and this slows down the recommendation process. If we would disable statistics, concentrate on a few algorithms, and use a dedicated server for the recommender system, it should be possible to generate recommendations in real-time. However, since we need the statistics, and want to evaluate different variations of the recommendation approaches, pre-generating recommendation seems the most feasible solution to us.

[47] http://nanoxml.sourceforge.net/orig/
[48] http://www.neo4j.org/

Docear's recommender system applies two recommendation approaches, namely stereotype recommendations and content-based filtering (CBF). Every time the recommendation process is triggered, one of these approaches is randomly chosen. The stereotype approach is chosen with a probability of 1%. The content-based filtering approach analyzes the users' mind maps and recommends research papers whose content is similar to the content of the mind maps. 'Similarity' is based on the number of terms or citations that user-models and research papers have in common. The user modeling process varies by a number of variables that are stored in the algorithms database (*MySQL* & *Hibernate*[49]). These variables are randomly arranged to assemble the final user-modeling algorithm, each time recommendations are generated. In the first step, the *feature type* to use from the mind maps is randomly chosen. The feature type may be terms, citations, or both. Then, a number of other variables are chosen such as the *number of mind maps* to analyze, the *number of features* the user model should contain, and the *weighting scheme* for the features. For instance, one randomly arranged algorithm might utilize the one hundred most recently created citations in the user's mind maps, weight the citations with CC-IDF, and store the five highest weighted citations as a user model. Another algorithm might utilize all the terms from the two most recently created mind maps, weight terms based on term frequency and store the 50 highest weighted terms as user model.

The *user model* is represented by a list of terms and/or citations that are supposed to describe the user's information needs. The user-modeling engine randomly chooses whether to store the user model as a weighted or un-weighted list in the database. An un-weighted list is a plain list of terms or citations such as *sun*, *skin*, *dcr_doc_id_54421*, *cancer* ordered by the terms' and citations' weight (the features are always sorted by weight, but the weight is discarded when storing the user model as un-weighted list). The weighted list is a vector in which the weights of the individual features are stored, in addition to the features themselves. Docear uses both weighted and un-weighted lists to research the differences in their effectiveness.

[49] http://hibernate.org/

The *matching module* is responsible for finding the appropriate recommendations for a given user model. To match user models and recommendation candidates, Apache Lucene is used, i.e. the user model is sent as a search query to Lucene. From Lucene's top 50 search results, a set of ten papers is randomly selected as recommendations. Choosing papers randomly from the top 50 results decreases the overall relevance of the delivered recommendations, yet increases the variety of recommendations, and allows for the analyzing of how relevant the search results of Lucene are at different ranks.

Matching user models with recommendation candidates is the same for both terms and citations. The user model, consisting of terms or citation IDs, is sent to Lucene. Lucene returned those research papers that are most relevant for the terms or citations (relevance is calculated with Lucene's default algorithm).

Once the recommendations are created, they are stored in the recommendation database (MySQL & Hibernate). The system stores for which user the recommendations were generated, by which algorithm, as well as some statistical information such as the time required to generate recommendations and the original Lucene ranking. The recommendations are not yet delivered to the user but only stored in the database.

##### *5.1.1.5 Delivering recommendations*

To display recommendations to a user, the Docear desktop software sends a request to the Web Service. The Web Service retrieves the latest created recommendations and returns them to Docear, which displays the recommendations to the user. The Web Service stores some statistics, such as when the recommendations where requested and from which Docear version. After recommendations are displayed to the user, a new set of recommendations is generated.

Each recommendation set receives a label that is displayed in Docear above the recommendations (Figure 23). Some labels such as "Free research papers" indicate that the recommendations are free and organic. Other labels such as "Research papers (Sponsored)" indicate that the recommendations are given for commercial reasons. For each user, the label is randomly chosen, when the user registers. The label has no effect on how the recommendations are actually generated. We randomly assign labels only to research the effect of different labels on user satisfaction (cf. Appendix J, p. 271).

When users click on a recommendation, a download request is sent to Docear's Web Service. The Web Service again stores some statistics, such as the time when the user clicked the recommendation. Then the user is forwarded to the original URL of the recommended paper. Forwarding has the disadvantage that papers occasionally are not available any more at the time of the recommendation since they were removed from the original web server. However, caching PDFs and offering them directly from Docear's servers might have led to problems with the papers' copyright holders.

#### *5.1.1.6 Offline evaluation*

The *Offline Evaluator* (JAVA) runs occasionally to evaluate the effectiveness of the different algorithms. The offline evaluator creates a copy of the users' mind maps and removes that citation that was most recently added to the mind maps. In addition, all nodes from the copy are removed that were created after the most recent citation was added. The offline evaluator then selects a random algorithm and creates recommendations for the users. The offline evaluator checks if the removed citation is contained in the list of recommendations and stores this information in the database. It is assumed that if an algorithm could recommend the removed citation, the algorithm was effective. The more often an algorithm could recommend a removed citation, the more effective it is.

#### *5.1.1.7 Technical details*

The recommender system runs on two servers. The first server is an Intel Core i7 PC with two 120GB SSDs, one 3 TB HDD, and 16 GB RAM. It runs the PDF Spider, PDF Analyzer, and the mind-map database, and its load is usually high, because web crawling and PDF processing require many resources. The second server is an Intel Core i7 PC with two 750 GB HDDs and 8 GB RAM. It runs all other services including the Web Service, mind-map parser, MySQL database, Lucene, and the offline evaluator. The server load is rather low on average, which is important, because the Web Service is not only needed for recommendations but also for other tasks such as user registration. While long response times, or even down times, for e.g. the PDF spider are acceptable, user registration should always be available.

### 5.1.2 Datasets

We publish four datasets relating to the research papers that Docear's spider found on the web (5.1.2.1), the mind maps of Docear's users (5.1.2.2), the users themselves (5.1.2.3), and the recommendations delivered to the users (5.1.2.4). The following sections provide only an overview of the most important data, particularly with regard to the randomly chosen variables. Please note that all variables are explained in detail in the *readme* files of the datasets, and the effects of most variables are presented in the following chapters. All datasets are available at http://labs.docear.org.

#### *5.1.2.1 Research papers*

The *research papers* dataset contains information about the research papers that Docear's PDF Spider crawled, and their citations.

The file *papers.csv* contains information about 9.4 million research articles. Each article has a unique *document id*, a *title*, a *cleantitle*, and for 1.8 million articles, a *URL* to the full-text is provided. The 1.8 million documents were found by Docears PDF Spider, and for each of these documents, titles were extracted with Docear's PDF Inspector or parsed from the web page that linked the PDF. The remaining 7.6 million documents in the dataset were extracted from the 1.8 million documents' bibliographies. In this case, no full-text URL is available and the document's title was extracted from the bibliography with ParsCit. Based on a small random sample of 100 documents, we estimate that 71% of the articles are written in English. Other languages include German, Italian, Russian, and Chinese. It also appears that the papers cover various disciplines, for instance, social sciences, computer science, and biomedical sciences. However, several of the indexed documents are of non-academic nature, and sometimes, entire proceedings were indexed but only the first paper was recognized.

Document disambiguation is only based on the documents' "cleantitle". To generate a cleantitle, all characters are transformed to lowercase, and only ASCII letters from *a* to *z* are kept. If the resulting cleantitle is less than half the size of the original title, the original title is used as cleantitle – this prevents e.g. Chinese titles to be shortened to a string of length zero. If two documents have the same cleantitle, the documents are assumed identical. Comparing documents only based on such a simplified title is certainly not very sophisticated but it proved to be sufficiently effective for our needs.

The file *citations.csv* contains a list of 572,895 papers with 7.95 million citations. These numbers mean that of the 1.8 million PDFs, 572,895 PDFs could be downloaded and citations could be extracted, and on average, each of the PDFs contained around 14 references. The dataset also contains information where citations occur in the full-texts. For each *citing->cited* document pair, the position of a citation is provided in terms of character count, starting from the beginning of the document. This leads to 19.3 million entries in *citations.csv*, indicating that, on average, each cited paper is cited around three times in a citing document. The dataset allows building citation networks and hence calculating document similarities, or the document impact. Since the position of the citations is provided, document similarity based on citation proximity analysis could be calculated, which we developed during the past years [142] and which is an extension of co-citation analysis.

Due to copyright reasons, full-texts of the articles are not included in the dataset. Downloading the full-text is easily possible, since the URLs to the PDFs are included (as long as the PDFs are still available on the Web).

#### *5.1.2.2 Mind maps / user libraries*

Every month, 3,000 to 4,000 newly created and modified mind maps are uploaded to Docear's server. Some mind maps are uploaded for backup purposes, but most mind maps are uploaded as part of the recommendation process.

The file *mindmaps.csv* contains information on 52,202 mind maps created by 12,038 users who agreed that we could publish their information. Docear does not only store the latest version of a mind map but keeps each revision. Information about 390,613 revisions of the 52,202 mind maps is also included in *mindmaps.csv*. This means, on average there are around seven to eight revisions per mind map. All mind maps and revisions in the dataset were created between March 2012 and March 2014. There are three different types of mind maps. First, there are mind maps in which users manage academic PDFs, annotations, and references (Figure 5). These mind maps represent data similar to the data included in the Mendeley dataset (cf. Chapter 3.4.6, p. 55). While Mendeley uses the term "personal libraries" to describe a collection of PDFs and references, Docear's mind maps represent also collections of PDFs and references but with a different structure than the ones of Me ndeley. Second, there are mind maps to draft assignments, research papers, theses, or books (Figure 23). These mind maps differ from the first type as they typically contain only few PDFs and references, but they include additional data such as images, LaTeX formulas, and more text.

The third type of mind maps, are “normal” mind maps that users create to brainstorm, manage tasks, or organize other information. Due to privacy concerns, this dataset does not contain the mind maps themselves but only metadata. This includes a list of all the mind maps and revisions, their file sizes, the date they were created, and to which user they belong. The data may help to analyze how often researchers are using reference management software, for how long they are using it, and how many papers they manage in their mind maps, personal collections respectively.

The file *mindmaps-papers.csv* contains a list 473,538 papers that are linked eight million times in 12,994 mind maps. This means, of the 52,202 mind maps, 24.8% contain at least one link to a PDF, and PDFs are linked 17 times in a mind map on average. The paper-IDs in *mindmaps-papers.csv* are anonymized and do not correlate with paper-IDs from the research paper dataset, nor does *mindmaps-papers.csv* contain titles of the linked papers. It should also be noted that the 473,538 papers are not necessarily contained in *papers.csv* as *papers.csv* contains only information of the publicly available PDFs and their citations. These limitations were made to ensure the privacy of our users.

#### *5.1.2.3 Users*

There are three types of users in Docear, namely local users, registered users, and anonymous users. *Local users* chose not to register when they install Docear. Consequently, they cannot use Docear’s online services such as recommendations or online backup, and we do not have any information about these users, nor do we know how many local users there are. *Registered users* sign-up with a username, a password, and an email address and they can use Docear’s online services. During the registration process, these users may provide information about their age and gender. Between March 2012 and March 2014, around 1,000 users registered every month, resulting in 21,439 registered users. *Anonymous users* decline to register but still want to use some of Docear’s online services. In this case, Docear automatically creates a user account with a randomly selected user name that is tied to a users’ computer. Anonymous users cannot login on Docear’s website, but they can receive recommendations as their mind maps are transferred to Docear’s servers, if they wish to receive recommendations. Due to spam issues, no new anonymous users were allows since late 2013. Until then, around 9,500 anonymous user accounts were created by non-spammers.

The file *users.csv* contains anonymized information about 8,059 of the 21,439 registered users, namely about those who activated recommendations and agreed to have their data analyzed and published. Among others, the file includes information about the users' date of registration, gender, age (if provided during registration), usage intensity of Docear, when Docear was last started, when recommendations were last received, the number of created mind maps, number of papers in the user's mind maps, how recommendations were labeled, the number of received recommendations, and click-through rates.

The file *users_papers.csv* contains a list of 6,726 users and 616,651 papers that the users have in their collections, i.e. mind maps. This means, on average, each user has linked or cited 92 documents in his or her mind maps. The paper IDs in *users_papers.csv* do not correlate with the IDs from the research paper dataset, to ensure the users' privacy.

The users-dataset may help to identify how differences between users affect users' satisfaction with recommendations. For instance, we found that older users are more likely to click on recommendations than younger users (cf. Appendix H, p. 259), and that the labelling of recommendations has an effect on user satisfaction (cf. Appendix J, p. 271). The dataset also allows analyses about the use of reference managers, for instance, how intensive researchers are using Docear.

#### *5.1.2.4 Recommendations*

Between March 2013 and March 2014, Docear delivered 31,935 recommendation sets with 308,146 recommendations to 3,470 users[50]. Of the delivered sets, 38.7% were explicitly requested by the users. The remaining 62.2% were delivered automatically when the Docear Desktop software was started. Among the 308,146 recommendations, there were 147,135 unique documents. In other words, from Docear's 1.8 million documents, 9% were actually recommended. The recommendation dataset splits into two files.

The file *recommendation_sets.csv* contains information about the 31,935 delivered recommendation sets. This includes the number of recommendations per set (usually ten), how many recommendations were clicked, the date of creation and delivery, the time required to generate the set and corresponding user models, and information on the algorithm that generated the set. There is a large variety in the algorithms. We stored whether stop words were removed, which weighting scheme was applied, whether terms and/or citations were used for the user modelling process, and several other variables were applied that are described in more detail in the dataset's readme file.

[50] The analyses in the following chapters are based on this data and additional data that we collected until August 2014.

The file *recommendations.csv* contains information about the 308,146 recommendations that Docear delivered. This information includes all details contained in r*ecommendation_sets.csv* and additional information, such as at which position a recommendation was shown, and which document was recommended (again, we anonymized the paper IDs).

## 5.2 Adequacy of Evaluation Methods and Metrics

Task 3 was to identify adequate evaluation methods. Hence, we measured the effectiveness of different recommendation approaches, and their variations, with a user study, an online evaluation, and an offline evaluation. This section first presents the results of the evaluations (p. 87). It follows a discussion about the adequacy of online-evaluation metrics (p. 96), adequacy of online evaluations and user studies (p. 98), and adequacy of offline evaluations (p. 99).

### 5.2.1 Results of the Evaluations

#### *5.2.1.1 Effectiveness of recommendation approaches*

We evaluated the effectiveness of stereotype recommendations, CBF based on terms, and CBF based on citations with an online evaluation, offline evaluation, and user study. The user study and online evaluation both led to the same ranking of the approaches[51]: Term-based CBF performed best, i.e. CTR, $CTR_{Set}$, DTR, LTR, CiTR, and ratings were highest; citation-based CBF performed second best; and the stereotype approach performed worst, but still reasonable (Figure 27).

On average, LTR was around one third of CTR. For instance, LTR for the stereotype approach was 1.46% while CTR was 4.11%. This means that one third of the recommendations that had been clicked were actually downloaded and linked in the mind maps. ATR was around half of LTR for the CBF approaches. This means that users annotated about half of the recommendations that they downloaded[52]. However, for the stereotype approach, ATR was only 0.18%, i.e. $\frac{1}{8}$ of LTR. Similarly, CiTR for the stereotype approach was only $\frac{1}{75}$ of LTR, while CiTR for term- and citation-based CBF was around $\frac{1}{4}$ of LTR. Apparently, stereotype recommendations were rarely annotated or cited, yet users cited every fourth content-based recommendation that they downloaded[53].

---

[51] To compare term- and citation based recommendations, we only compared recommendations when the corresponding set contained ten recommendations, original ranks were 10 or lower, and the users' mind-maps contained at least one citation.

[52] Many PDFs contain already annotations before they are downloaded. For instance, some publishers create bookmarks for the chapters of a paper. Hence, results for ATR should be considered with skepticism.

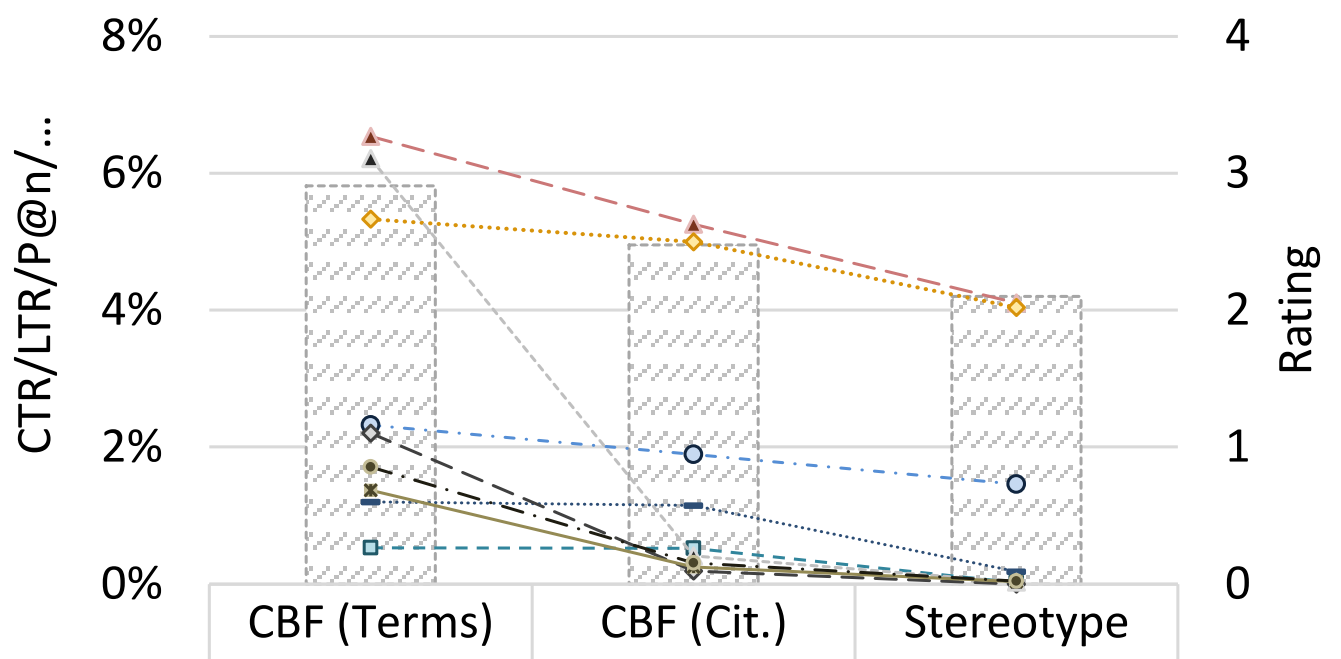

| | CBF (Terms) | CBF (Cit.) | Stereotype |
|---|---|---|---|
| Ratings | 2.91 | 2.48 | 2.10 |
| CTR | 6.53% | 5.25% | 4.11% |
| CTR (Set) | 5.33% | 5.00% | 4.04% |
| LTR | 2.32% | 1.89% | 1.46% |
| ATR | 1.20% | 1.15% | 0.18% |
| CiTR | 0.53% | 0.52% | 0.02% |
| P@3 | 2.20% | 0.19% | 0.00% |
| P@10 | 6.21% | 0.41% | 0.03% |
| nDCG | 1.37% | 0.25% | 0.03% |
| MRR | 1.71% | 0.31% | 0.04% |

Approach

Figure 27: Effectiveness of recommendation approaches[54]

The offline evaluation led to the same overall ranking than the online evaluation and user study. However, all four offline metrics attest that term-based CBF has significantly better effectiveness than citation based CBF (around four to ten times as effective), while user study and online evaluation only attest a slightly higher effectiveness. In addition, the effectiveness of the stereotype approach in the offline evaluation is close to zero, while user study and online evaluation show a reasonable effectiveness.

[53] Please note that users did not really cite the papers in their own publications but retrieved metadata for these papers, which we interpret as a citation.

[54] The comparison of term and citation-based recommendations was based on a "fair" comparison, i.e. only recommendations delivered to users who had made at least 1 citation were considered and recommendation sets including ten recommendations.

#### *5.2.1.2 Effect of user-model size*

We researched not only the effectiveness of distinct recommendation approaches, but variables such as the extent of the user-model size. User-model size describes how many terms (or citations) are stored to represent the users' information needs. Whenever recommendations are requested, Docear randomly selected a user-model size between 1 and 1000 terms. For term-based CBF, the highest ratings (3.26) were given for recommendations that were based on user models containing 26 to 100 terms (Figure 28). All online metrics, except CiTR[55], confirmed the results of the user study. The offline metrics led to slightly different results and showed the highest effectiveness for user models containing 101 to 250 terms.

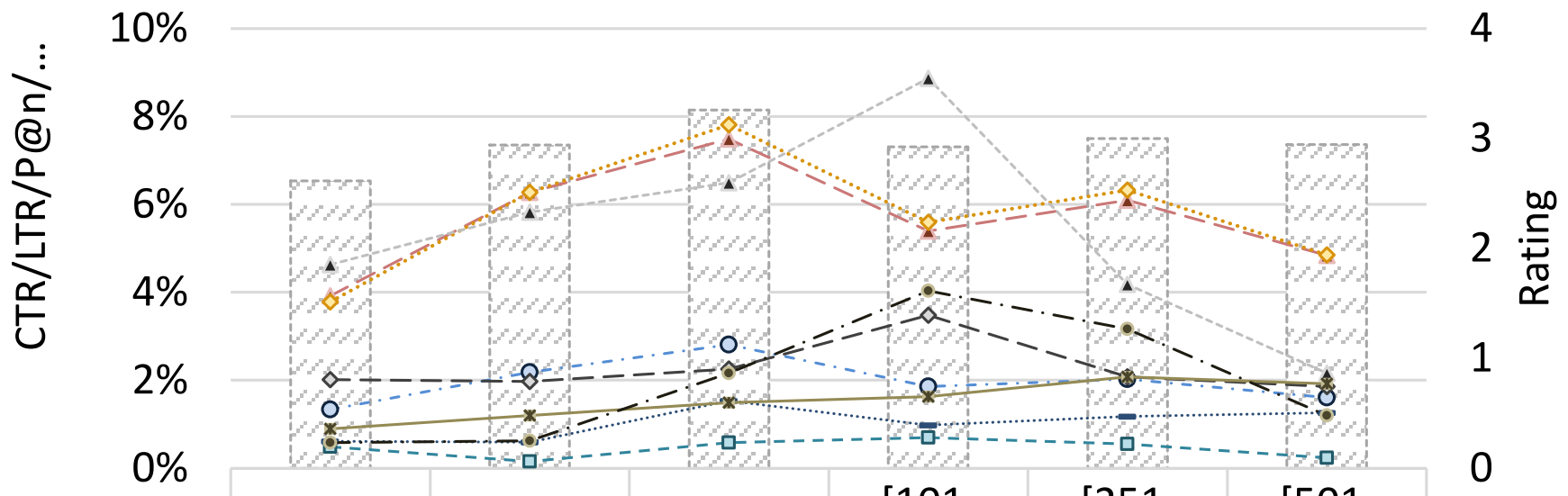

| | [1;10] | [11;25] | [26;100] | [101; 250] | [251; 500] | [501; 1,000] |
|---|---|---|---|---|---|---|
| Ratings | 2.62 | 2.94 | 3.26 | 2.92 | 3.00 | 2.94 |
| CTR | 3.92% | 6.27% | 7.48% | 5.40% | 6.09% | 4.84% |
| CTR (Set) | 3.78% | 6.27% | 7.81% | 5.60% | 6.32% | 4.85% |
| LTR | 1.33% | 2.18% | 2.81% | 1.86% | 2.03% | 1.60% |
| ATR | 0.61% | 0.59% | 1.53% | 0.98% | 1.18% | 1.26% |
| CiTR | 0.49% | 0.15% | 0.58% | 0.69% | 0.55% | 0.24% |
| P@3 | 2.01% | 1.97% | 2.25% | 3.48% | 2.07% | 1.85% |
| P@10 | 4.63% | 5.82% | 6.49% | 8.87% | 4.18% | 2.16% |
| nDCG | 0.90% | 1.20% | 1.49% | 1.63% | 2.08% | 1.92% |
| MRR | 0.58% | 0.62% | 2.16% | 4.04% | 3.17% | 1.20% |

Number of Terms

Figure 28: Effectiveness based on user-model size

[55] The differences for CiTR were statistically not significant.

#### 5.2.1.3 *Effect of number of nodes*

Docear's mind maps often contain thousands of nodes. We assumed that analyzing too many nodes might introduce noise into the user models. Therefore, Docear randomly selected how many of the *x* most recently modified nodes, should be utilized for extracting terms. Based on user ratings, analyzing between 50 and 99 nodes is most effective (Figure 29)[56]. As more nodes were analyzed, the average ratings decreased. CTR, $CTR_{Set}$, LTR, and CiTR also showed an optimal effectiveness for analyzing 50 to 99 nodes. Based on ATR, the optimal number of nodes is larger, but results were statistically not significant. The offline metrics indicate that analyzing a larger number of nodes might be sensible, namely 100 to 499 nodes.

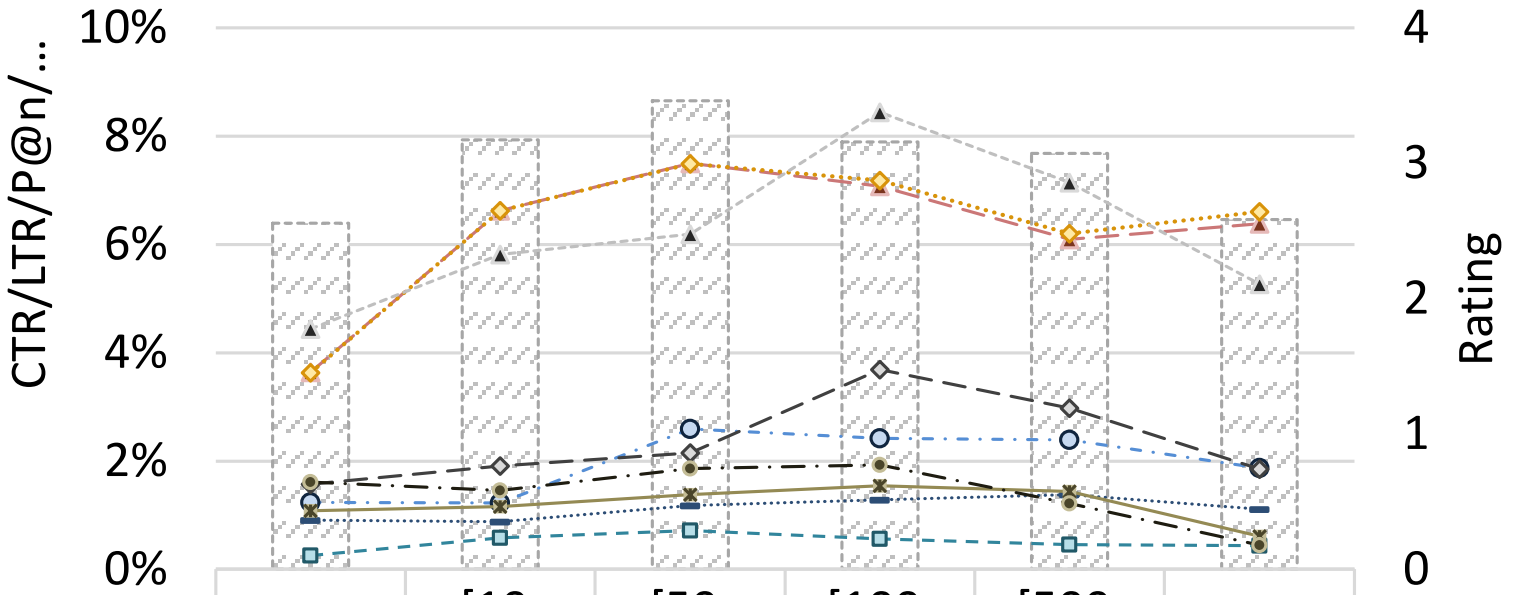

| | [1;9] | [10; 49] | [50; 99] | [100; 499] | [500; 999] | 1,000+ |
|---|---|---|---|---|---|---|
| Ratings | 2.56 | 3.17 | 3.46 | 3.16 | 3.07 | 2.58 |
| CTR | 3.64% | 6.62% | 7.50% | 7.08% | 6.09% | 6.38% |
| CTR (Set) | 3.63% | 6.62% | 7.49% | 7.18% | 6.20% | 6.60% |
| LTR | 1.24% | 1.23% | 2.60% | 2.42% | 2.39% | 1.87% |
| ATR | 0.91% | 0.88% | 1.18% | 1.28% | 1.38% | 1.11% |
| CiTR | 0.25% | 0.58% | 0.72% | 0.56% | 0.45% | 0.44% |
| P@3 | 1.57% | 1.91% | 2.15% | 3.69% | 2.97% | 1.85% |
| P@10 | 4.44% | 5.81% | 6.19% | 8.44% | 7.14% | 5.27% |
| nDCG | 1.09% | 1.16% | 1.38% | 1.54% | 1.44% | 0.61% |
| MRR | 1.61% | 1.45% | 1.86% | 1.92% | 1.22% | 0.45% |

Number of Nodes

Figure 29: Effectiveness based on the number of nodes to analyze

[56] All results are based on recommendations to users who had created 1,000 nodes and more.

##### 5.2.1.4 *Effect of node-selection method*

Another variable we tested was the node modification type (Figure 30). The recommender system chose randomly, whether to utilize only nodes that were newly *created*, nodes that were *moved*, nodes that were *edited*, or nodes with any type of *modification* (created, edited, or moved). Utilizing moved nodes only, resulted in the highest ratings on average (3.31). The online metrics CTR, $CTR_{Set}$, and LTR as well as the offline metric MRR also have the highest effectiveness when utilizing moved nodes. Results for ATR and CiTR differ, but are statistically not significant. Based on P@N, utilizing all modified nodes is most effective, based on nDCG utilizing newly created nodes is most effective.

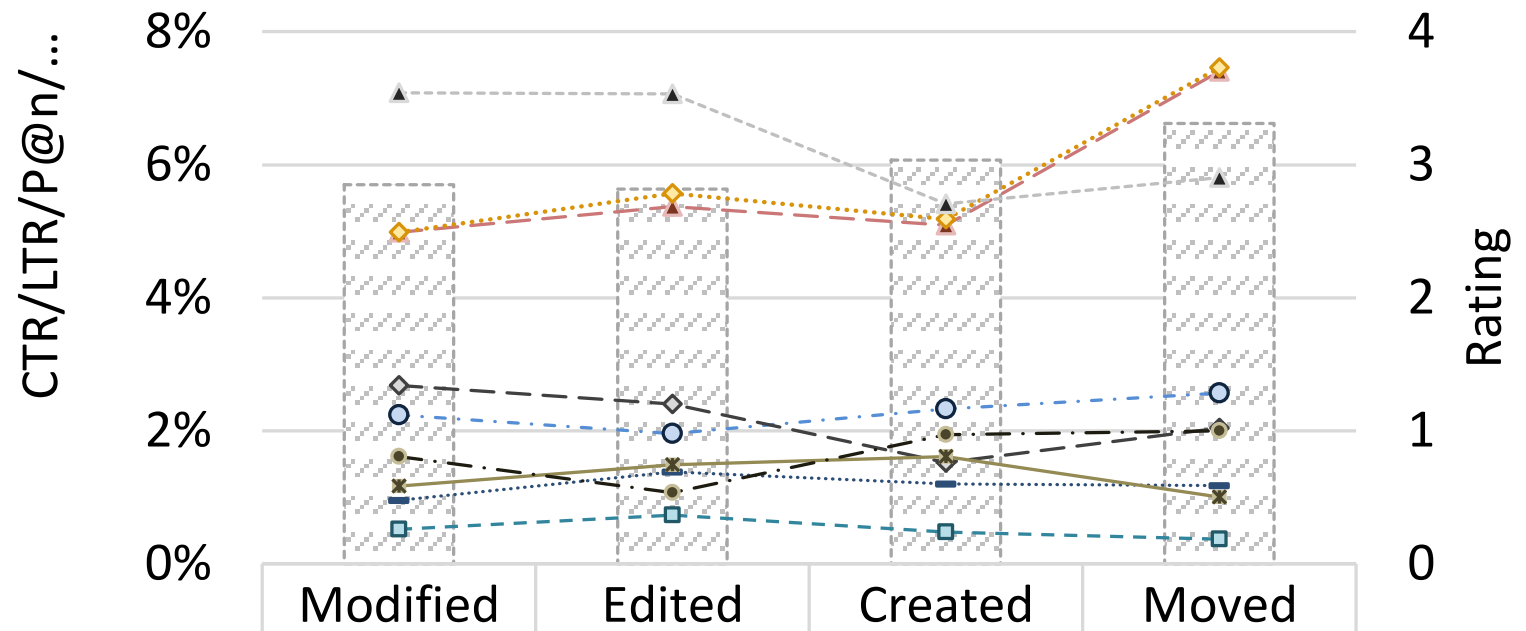

| | Modified Nodes | Edited Nodes | Created Nodes | Moved Nodes |
|---|---|---|---|---|
| Ratings | 2.85 | 2.82 | 3.04 | 3.31 |
| CTR | 4.99% | 5.38% | 5.09% | 7.40% |
| CTR (Set) | 4.98% | 5.57% | 5.18% | 7.46% |
| LTR | 2.24% | 1.96% | 2.33% | 2.57% |
| ATR | 0.95% | 1.38% | 1.20% | 1.18% |
| CiTR | 0.52% | 0.73% | 0.48% | 0.37% |
| P@3 | 2.68% | 2.40% | 1.52% | 2.04% |
| P@10 | 7.08% | 7.07% | 5.41% | 5.81% |
| nDCG | 1.17% | 1.49% | 1.61% | 1.00% |
| MRR | 1.61% | 1.07% | 1.94% | 2.00% |

Node Selection Method

Figure 30: Effectiveness based on the node modification type

##### 5.2.1.5 *Effect of stop-word removal*

When the recommender system removed stop-words, the average rating was 3.16 compared to 2.88 when no stop-words were removed (Figure 31). All other metrics, except ATR, also showed a higher effectiveness when stop-words were removed, but, again, results for ATR were statistically insignificant.

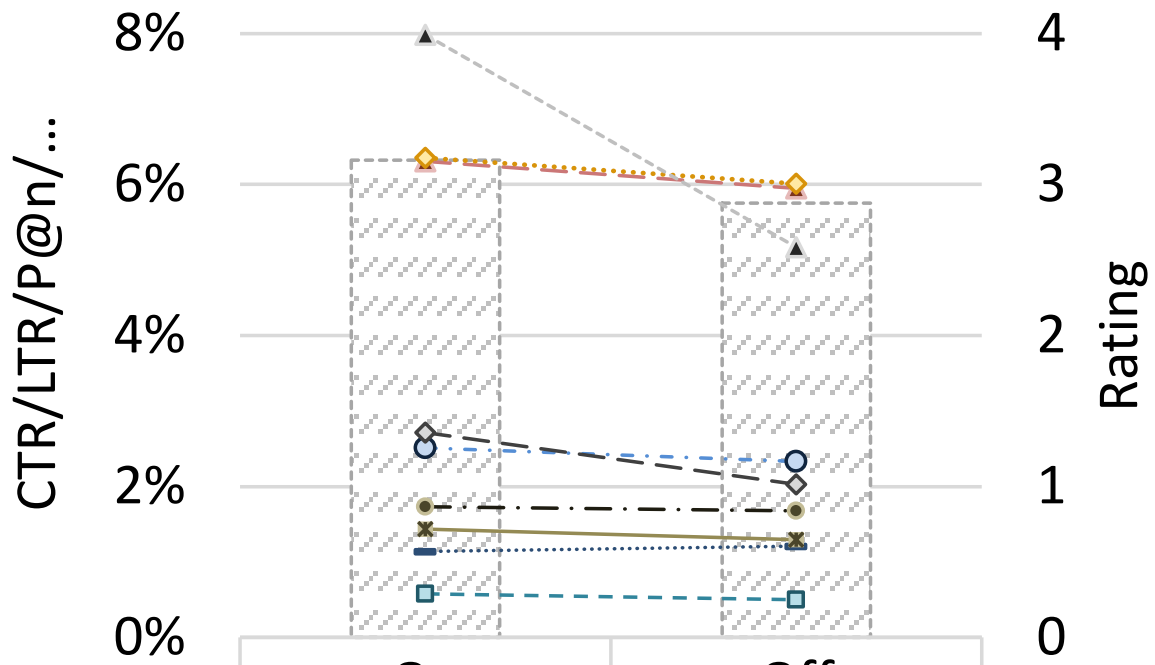

| | On | Off |
|---|---|---|
| Ratings | 3.16 | 2.88 |
| CTR | 6.31% | 5.94% |
| CTR (Set) | 6.35% | 6.01% |
| LTR | 2.51% | 2.33% |
| ATR | 1.14% | 1.21% |
| CiTR | 0.58% | 0.50% |
| P@3 | 2.71% | 2.03% |
| P@10 | 7.98% | 5.17% |
| nDCG | 1.44% | 1.29% |
| MRR | 1.73% | 1.68% |

Stop-Word Removal

Figure 31: Effectiveness of stop-word removal

### 5.2.1.6 *Effect of user types*

Docear's recommender system is open to both registered and unregistered/anonymous users (cf. Chapter 5.1.2.3, p. 83), and we were interested whether there would be differences in the two users groups with respect to recommendation effectiveness. CTR and $CTR_{Set}$ show a clear difference between the two user types (Figure 32). Registered users had an average CTR of 5.32% while unregistered users had an average CTR of 3.86%. $CTR_{User}$ is also higher for registered users (4.00%) than for anonymous users (3.77%), but the difference is not that strong. LTR and ATR also show a (slightly) higher effectiveness for registered users. The offline evaluation contradicts the findings of the online evaluation: P@3, P@10, and MRR indicate that recommendations for registered users were about half as effective as for anonymous users, and nDCG showed no statistically significant difference between the user groups.

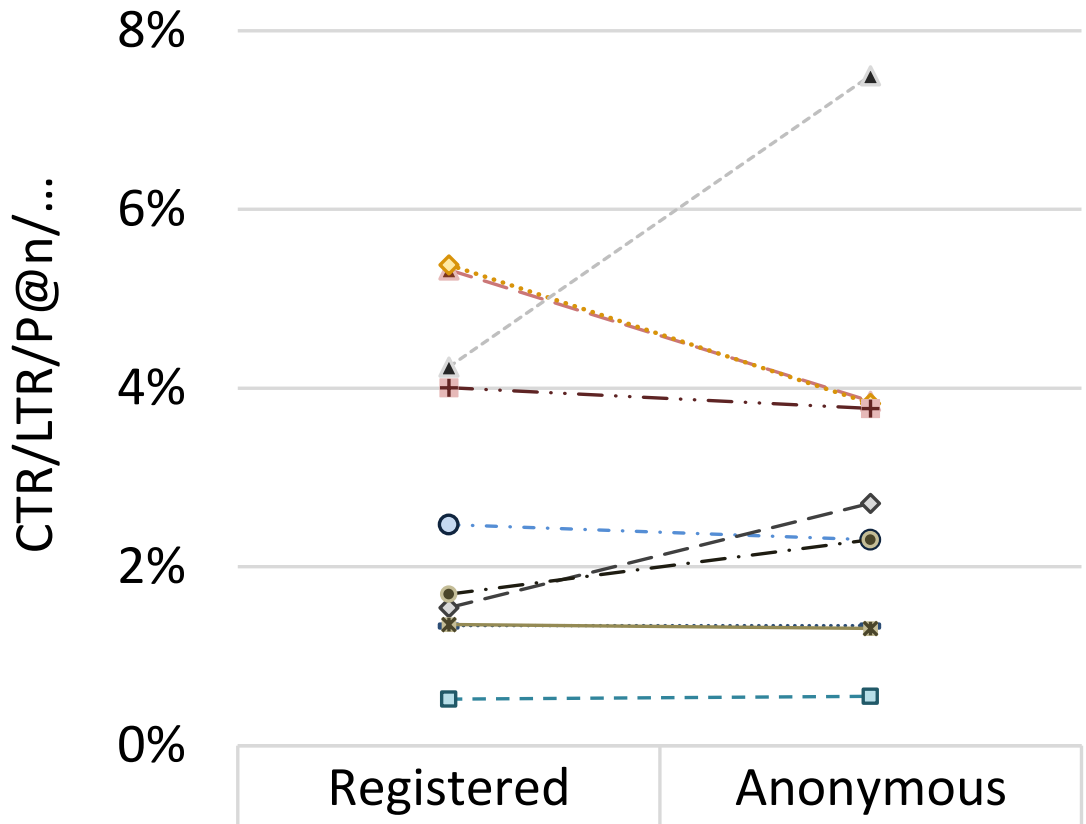

| | Registered | Anonymous |
|---|---|---|
| CTR | 5.32% | 3.86% |
| CTR (Set) | 5.38% | 3.83% |
| CTR (User) | 4.00% | 3.77% |
| LTR | 2.47% | 2.30% |
| ATR | 1.34% | 1.34% |
| CiTR | 0.52% | 0.55% |
| P@3 | 1.54% | 2.71% |
| P@10 | 4.24% | 7.50% |
| nDCG | 1.35% | 1.31% |
| MRR | 1.69% | 2.30% |

User Type

Figure 32: Effectiveness by user type (registered and anonymous)[57]

### *5.2.1.7 Effect of labels*

For each user, Docear randomly determined whether to display an organic label (e.g. “Free Research Papers”), a commercial label (e.g. “Research Papers [Sponsored]”), or to display no label at all (cf. Appendix J, p. 271). For each user a fix label was randomly selected once, i.e. a particular user always saw the same label. The label had no impact on how recommendations were generated. This means, if recommendation effectiveness would differ for a particular label, then only because users would value different labels differently.

[57] User ratings were only introduced in early 2014. Since no new anonymous users have been allowed since late 2013, there are no ratings made by anonymous users. Hence, we could only compare results of online and offline evaluations for the two types of users.

In the user study, there were no significant differences for the three types of labels in terms of effectiveness: the ratings were around 2.9 on average for all labels (Figure 33). Based on CTR, $CTR_{Set}$, and LTR, displaying no label was most effective. In addition, commercial labels were slightly, but statistical significantly, more effective than organic labels. Based on $CTR_{User}$, commercial recommendations were least effective, organic labels were most effective, and 'no label' was second most effective. ATR and CiTR led to statistically not significant results, and offline metrics could not be calculated for this kind of analysis.

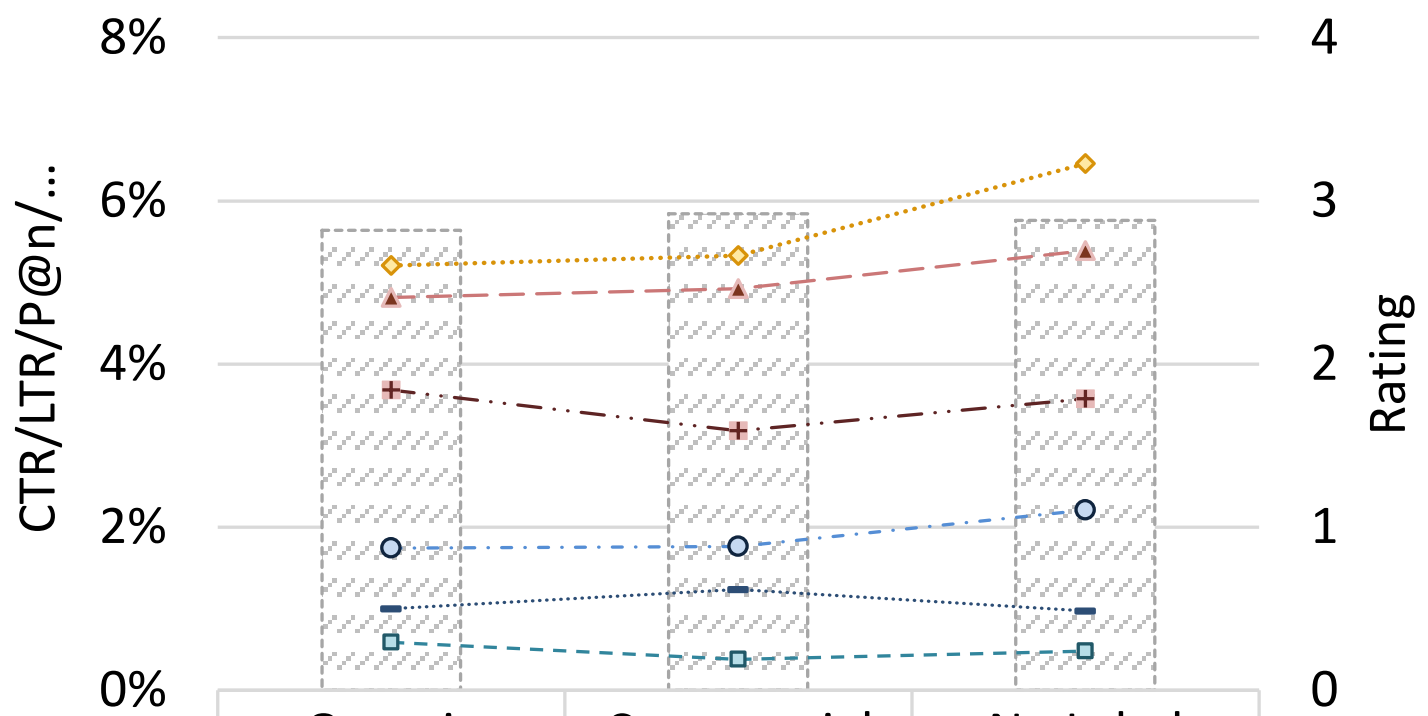

| | Organic | Commercial | No Label |
|---|---|---|---|
| Ratings | 2.82 | 2.92 | 2.88 |
| CTR | 4.82% | 4.92% | 5.39% |
| CTR (Set) | 5.21% | 5.33% | 6.46% |
| CTR (User) | 3.68% | 3.18% | 3.57% |
| LTR | 1.75% | 1.76% | 2.21% |
| ATR | 1.00% | 1.23% | 0.97% |
| CiTR | 0.59% | 0.38% | 0.48% |

Labels

Figure 33: Effectiveness of labels

### 5.2.1.8 *Effect of trigger*

Two triggers in Docear lead to displaying recommendations. First, Docear displays recommendations automatically every five days when Docear starts. Second, users may explicitly request recommendations at any time. The user study shows a similar effectiveness for both types of trigger with an average rating between 2.8 and 2.9 (Figure 34)[58]. Interestingly, the online evaluation shows a significantly higher effectiveness for requested recommendations than for automatically displayed recommendations. For instance, CTR for requested recommendations is 2.5 times higher than for automatically displayed recommendations (9.14% vs. 3.67%). Conducting an offline evaluation was not possible for this type of analysis.

[58] The small difference in the average rating is statistically insignificant.

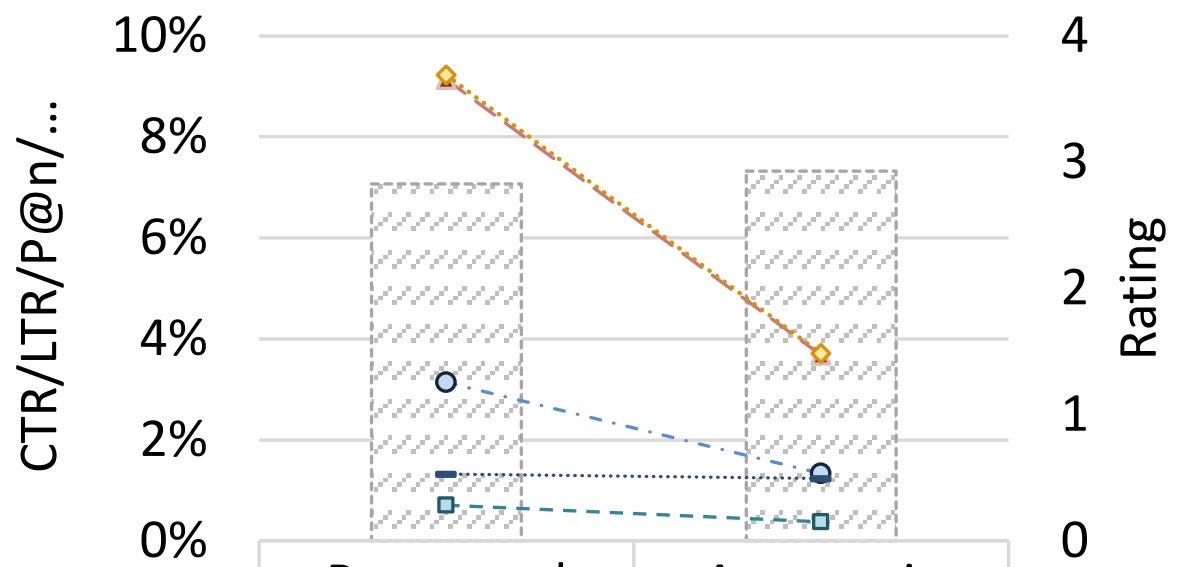

| | Requested | Automatic |
|---|---|---|
| Ratings | 2.83 | 2.93 |
| CTR | 9.14% | 3.67% |
| CTR (Set) | 9.23% | 3.71% |
| LTR | 3.14% | 1.33% |
| ATR | 1.32% | 1.24% |
| CiTR | 0.71% | 0.38% |

Trigger

Figure 34: Effectiveness by trigger

### *5.2.1.9 Correlation of the evaluation metrics*

We calculated the Pearson correlation coefficient for the different evaluation metrics (Table 13). Both CTR and $CTR_{Set}$ show a strong positive correlation with ratings ($r$=0.78). Correlation of all other metrics, both offline and online, with ratings is between 0.52 (CiTR) and 0.67 (nDCG). This means that CTR and $CTR_{Set}$ are most suitable to approximate ratings in our scenario. If the goal is to approximate CTR, then ratings, obviously, is the most adequate metric ($r$=0.78), followed by LTR ($r$=0.73). The other metrics have rather low correlation coefficients with CTR; the worst are nDCG ($r$=0.28) and MRR ($r$=0.30). Among the offline metrics, P@3 and P@10 correlate well ($r$=0.92), which is to expect. MRR and nDCG also show a reasonable strong correlation ($r$=0.71), while correlation of P@10 and MRR ($r$=0.56) and P@10 and nDCG ($r$=0.55) is rather weak.

Table 13: Correlation of the different metrics

| Correlation of Ratings and … | |
|---|---|
| Ratings | -- |
| CTR | 0.78 |
| CTR (Set) | 0.78 |
| DTR | 0.65 |
| ATR | 0.61 |
| CiTR | 0.52 |
| P@3 | 0.62 |
| P@10 | 0.65 |
| MRR | 0.55 |
| nDCG | 0.67 |

| Correlation of CTR and … | |
|---|---|
| Ratings | 0.78 |
| CTR | -- |
| CTR (Set) | 0.97 |
| DTR | 0.73 |
| ATR | 0.53 |
| CiTR | 0.42 |
| P@3 | 0.41 |
| P@10 | 0.48 |
| MRR | 0.30 |
| nDCG | 0.28 |

| Correlation of … | |
|---|---|
| P@3 and P@10 | 0.92 |
| P@10 and MRR | 0.56 |
| P@10 and nDCG | 0.55 |
| nDCG and MRR | 0.71 |

### 5.2.2 Adequacy of Online-Evaluation Metrics

Among the online metrics, CTR and $CTR_{Set}$ seem to be the most adequate metrics, at least for Docear’s scenario. CTR and $CTR_{Set}$ had the highest correlation with users’ ratings, are easiest to calculate, provided more often statistically significant results than the other metrics, and CTR is commonly used in other fields such as e-commerce and search engines. CTR and $CTR_{Set}$ (and LTR) also provided the more plausible results for the stereotype recommendations. Based on CTR, the stereotype approach was reasonably effective, while the approach was ineffective based on ATR and CiTR. The result based on CTR seems more plausible since the recommendations were about academic writing and most of Docear’s users should be interested in improving their writing skills. However, there is little reason for someone who is doing research in a particular research field, to annotate or even cite an article about academic writing even if the article was useful. Hence, judging stereotype recommendations based on ATR or CiTR seems inadequate to us.

However, ATR and CiTR might be more sensible measures than CTR in other scenarios. For instance, imagine two algorithms called “A” and “B”. Both are the same content-based filtering approaches but B additionally boosts papers published in reputable journals.[59] In the online evaluation, users would probably see no difference between the titles of the recommendations created with the two approaches, assuming that authors publishing in reputable journals do not formulate titles that are significantly different from titles in other journals. Consequently, recommendations of the two algorithms would appear to be similarly relevant and received similar CTR. However, most people would probably agree that algorithm B would be preferable to algorithm A in practice. Therefore, in this example, ATR and CiTR might be more appropriate than CTR.

[59] For this example we ignore the question how reputability is measured

Measuring CTR, while displaying only the title of recommendations, was criticized by some reviewers of our previous publications. The reviewers argued that titles alone would not allow thorough assessment of recommendations and CTR could therefore be misleading. In some scenarios, such as the example above with the two algorithms, one being boosted by journal reputation, this criticism could indeed apply. However, in the scenario of Docear, the results do not indicate that displaying only the title led to any problems or bias in the results since CTR correlates well with those metrics that are based on a more thorough assessment of the recommendations (e.g. user ratings or ATR).

Compared to CTR, $CTR_{user}$ smoothed the effect of variables that strongly affected a few users. For instance, $CTR_{user}$ was highest for organic labels, lowest for commercial labels, and mediocre for no labels – a result that one would probably expect. In contrast, CTR was highest for no label, second highest for commercial recommendations, and lowest for organic recommendations – a result that one would probably not expect. After looking at the data in detail, we found that a few users who received many recommendations (with no label) “spoiled” the results. Hence, if the objective of an evaluation was to measure overall user satisfaction, $CTR_{user}$ was probably preferable to CTR because a few power users will not spoil the results. However, applying $CTR_{user}$ requires more users than applying CTR, since $CTR_{user}$ requires that users receive recommendations based on the same parameters of the variables and not per recommendation set. For instance, to calculate $CTR_{user}$, each user must always see the same label, each user model must always be the same size for a particular user, and recommendations must always be based on terms *or* citations for a particular user. In contrast, to calculate CTR, users may receive recommendations based on terms in one occasion, and recommendations based on citations in another occasion, or user models could differ in size, different weighting schemes could be used etc. Consequently, to receive statistically significant results, $CTR_{user}$ requires more users than CTR. At least for Docear, calculating $CTR_{user}$ for variables such as user-model size, number of nodes to analyze, features to utilize (terms or citations), and weighting schemes is not feasible since we would need many more users than Docear currently has.

Considering the strong correlation of CTR and ratings, the more plausible result for stereotype recommendations, and the rather low number of users being required, we conclude that CTR is the most appropriate online metric for our scenario. This is not to mean that in other scenarios other online metrics such as $CTR_{User}$ or ATR might not be more sensible.

### *5.2.3* Adequacy of Online Evaluations & User Studies

Ratings in the user study correlated strongly with CTR ($r$=0.78). This indicates that explicit user satisfaction (ratings) is a good approximation of the acceptance rate of recommendations (CTR), and vice versa. Only in two cases, CTR and ratings contradicted each other, namely for the impact of labels (cf. Section 5.2.1.7) and the trigger (cf. Section 5.2.1.8). Both of these analyses relate to human factors. For the analyses relating to the algorithms and their variations in user-model size, number of nodes to analyze, etc., CTR and ratings always led to the same conclusions. These differences indicate that when the accuracy of recommendation algorithms is to be evaluated, both CTR and ratings are equally well suitable. However, which of the two metrics to use when it comes to evaluating human factors?

We argue that none of the metrics is generally more authoritative than another metric. Ultimately, the authority of user studies and online evaluations depends on the objective of the evaluator, and operator of the recommender system respectively. If, for instance, the operator receives a commission per click on a recommendation, CTR was to prefer over ratings. If the operator is interested in user satisfaction, ratings were to prefer over CTR. Ideally, both CTR and ratings, should be considered when making a decision about which algorithm to apply in practice or to choose as baseline, since they both have some inherent value. Even if the operator's objective was revenue, and CTR was high, low user satisfaction would not be in the interest of the operator. Otherwise, users would probably ignore recommendations in the end, and hence stop clicking them. Similarly, if the objective was user satisfaction, and ratings were high, a low CTR would not be in the interest of the operator: a low CTR means that many irrelevant recommendations are given, and if these could be filtered, user satisfaction would probably further increase. Therefore, ideally, researchers should evaluate their approaches with both online evaluation and user study when it comes to evaluating human factors. However, if researchers do not have the resources to conduct both types of evaluation, or if the analysis focuses on recommendation algorithms with low impact of human factors, we suggest that conducting either a user study or an online evaluation should still be considered "good practice".

### 5.2.4 Adequacy of Offline Evaluations

Our research shows only a mediocre correlation of offline evaluations with user studies and online evaluations. Sometimes, the offline evaluation could predict the effectiveness of algorithms in the user study or online evaluation quite precisely. For instance, the offline evaluation was capable of predicting whether removing stop-words would increase the effectiveness. The optimal user-model size and number of nodes to analyze were also predicted rather accurately (though not perfectly). However, the offline evaluation remarkably failed to predict the effectiveness of citation-based and stereotype recommendations. If one had trusted the offline evaluation, one had never considered stereotype and citation-based recommendations to be a worthwhile option.

The uncertain predictive power of offline evaluations, questions the often proclaimed purpose of offline evaluations, namely to identify a set of promising recommendation approaches for further analysis. However, this does not necessarily mean that conducting offline evaluation is meaningless. To assess the adequacy of offline evaluations, we propose that the following three questions need to be answered:

1. Can we identify scenarios where offline evaluations will have predictive power for how recommendation approaches will perform in online evaluations and user studies? If we can, we should use offline evaluations only in these scenarios.
2. Do results of offline evaluations have inherent value? If they do, it would not matter if they contradicted results of online evaluations of user studies. Instead, results of offline evaluations would have inherent value and would be worth to be reported in publications, similar to CTR and ratings.
3. Are offline evaluations inherently flawed? If they are, we should abandon them entirely.

In the following sections, we attempt to answer these questions. We do not claim that our answers are definitive but we hope to stimulate a discussion that will eventually lead to widely accepted answers.

#### 5.2.4.1 *Finding scenarios for which offline evaluations have predictive power*

A common criticism on offline evaluations is the ignorance of human factors (cf. Section 3.3.2, p. 42). At least for some of our analyses, human factors *might* have caused the non-predictive power of offline evaluations.

For instance, on first glance we expected that Docear's recommendation approaches create equally relevant recommendations for both anonymous and registered users. However, the offline evaluation showed higher effectiveness for anonymous users than for registered users while we saw the opposite in the online evaluation. Although we find these results surprising, the influence of human factors *might* explain the difference: It could be that anonymous users are more concerned about privacy than registered users[60]. Users concerned about their privacy, might worry that when they click a recommendation, some unknown, and potentially malicious website, opens. This could be the reason that anonymous users, who tend to be concerned about their privacy, click recommendations not as often as registered users, and CTR is lower on average. Nevertheless, the higher accuracy for anonymous users in the offline evaluation might still be plausible. If anonymous users tended to use Docear more intensively than registered users, the mind maps of the anonymous users would be more comprehensive and hence more suitable for user modeling and generating recommendations, which would lead to the higher accuracy in offline evaluations. This means that although mind maps of anonymous users might be more suitable for user modeling, the human factor "privacy concerns" causes the low effectiveness in online evaluations.

If human factors have an impact on recommendation effectiveness, we must question whether one can determine scenarios in which human factors have *no* impact. Only in these scenarios, offline evaluations would be an appropriate tool to approximate the effectiveness of recommendation approaches in online evaluations or user studies. In scenarios like our analysis of registered vs. anonymous users, it is apparent that human factors may play a role, and that offline evaluations might be not appropriate. For some of our other experiments, such as whether to utilize terms or citations, we could see no plausible influence of human factors, yet offline evaluations could not predict the performance in the user study and online evaluation. Therefore, and assuming that results of offline evaluations have no inherent value, we would propose abandoning offline evaluations, as they cannot reliably fulfil their purpose. However, could the results of offline evaluations have some inherent value?

---

[60] If users register, they have to reveal private information such as name and email address. If users are concerned about revealing this information, they probably tend to use Docear as anonymous user.

#### *5.2.4.2 The inherent value of offline evaluations*

Offline evaluations, online evaluations, and user studies typically measure different types of effectiveness (cf. Section 3.4.1, p. 51). One might therefore argue that comparing the results of the three methods is like comparing apples, peaches, and oranges, and that the results of each method have some inherent value. For online evaluations and user studies, such an inherent value doubtlessly exists (see previous section).

An inherent value for offline evaluations would exist if those persons who compiled the ground-truth, better knew which items were relevant than current users who decide to click, download, or rate an item. This situation is comparable with a teacher-student situation. Teachers know which books their students should read, and although students might not like the books, or had not chosen the books themselves, the books might be the best possible choice to learn about a certain subject. Such a teacher-student situation might apply to offline evaluations.

In case of *expert-datasets*, one might argue that topical experts, who compile the dataset, can better judge relevance of certain items than average users who use the recommender system. For instance, if experts were asked to compile an introductory reading list on recommender systems for undergraduate students, they could probably better select the most relevant documents than the students themselves could. Therefore, results from offline evaluations based on expert-datasets might be more authoritative than results obtained from online evaluations or user studies based e.g. on undergraduate students. However, an expert-created list for undergraduate students would not be suitable for PhD students who wanted to investigate the topic of recommender systems in more depth. Thus, another expert list would be needed for PhD students, another for senior researchers, and another for foreign language students, etc. Overall, there would be an almost infinite number of lists required to cater to all user backgrounds and information needs. Such a comprehensive dataset does not exist and probably never will. In addition, today's expert-datasets, such as TREC and MeSH, focus on specific use-cases and the datasets were not created for recommender-system evaluation. For instance, MeSH allows the determination of the similarity of documents. Recommending similar documents might be one use case for a recommender system, but there are many more. Considering the mentioned limitations, we conclude that offline evaluations based on *expert-datasets* might theoretically have some inherent value, and provide even more authoritative results than online evaluations and user studies, but in practice, appropriate datasets will probably never be available, except perhaps for some niche recommender systems.

*Inferred ground-truths* do not suffer the problem of overspecialization and should typically represent a large variety of use-cases. Therefore, in principle, evaluations based on inferred ground-truths (e.g. from citations) could be more authoritative than online evaluations or user studies. For instance, before a researcher decides to cite a document – which would add the document to the ground-truth – the document was ideally carefully inspected and its relevance was judged according to many factors such as the publication venue, the article's citation count, or the soundness of its methodology. These characteristics usually cannot be evaluated in an online evaluation or user study. Thus, one might argue that results based on personal-collection datasets might be more authoritative than results from online evaluations and user studies.

There is also a plausible example in which results based on an inferred ground-truth may be more authoritative than e.g. CTR. Recapitulate the previous example with two content-based filtering approaches, called "A" and "B," where B boosts papers that were published in reputable journals (cf. Section 5.2.2, p. 96). In the online evaluation, both approaches would probably receive similar CTR. In contrast, an offline evaluation based on an inferred ground-truth might predict a better performance for approach B, because articles from reputable journals are probably more often cited than articles from non-reputable journals. Hence, if citations were used as ground-truth, articles from reputable journals were more often contained in the ground-truth, and would more often be considered a good recommendation than articles from less reputable journals. As a result, algorithm B would show better accuracy than algorithm A. In this scenario, the offline evaluation would have identified the best algorithm while CTR did not.

Assuming that offline evaluations could be more authoritative than user studies and online evaluations, the following question arises: How useful are recommendations that might objectively be most relevant to users when users do not click, read, or buy the recommended item, or when they rate the item negatively? In contrast to teachers telling their students to read a particular book, a recommender system cannot force a user to accept a recommendation. We argue that an algorithm that is not liked by users, or that achieves low CTR, can never be considered useful. Only if two algorithms performed similarly or if both approaches had at least a mediocre performance in an online evaluation or user study, an additional offline evaluation might be used to decide which of the two algorithms is more effective. However, this means that offline evaluations had to be conducted *in addition* to user studies or online evaluations, and not beforehand or as only evaluation method. Consequently, a change in the current practice of recommender-systems evaluation was required.

#### *5.2.4.3 The fundamental flaw of inferred ground-truths*

While inferred ground-truths look promising on first glance, we see a fundamental problem: inferred ground-truths are supposed to contain *all* items that are relevant for recommendation (cf. 2.5.3.3, p. 24). To compile such a ground-truth, comprehensive knowledge of the domain is required. It should be apparent that most users do not have comprehensive knowledge of their domain (which is why they need a recommender system). Consequently, ground-truths are incomplete and contain only a fraction of relevant items, and perhaps even irrelevant items. If the ground-truth is inferred from citations, the problem becomes even more apparent. Many conferences and journals have space restrictions that limit the number of citations in a paper. This means that even if authors were aware of all relevant literature – which they are not – they would only cite a limited amount of relevant articles.

Citation bias further enforces the imperfection of citation-based ground-truths. Authors cite papers for various reasons and these do not always relate to the paper’s relevance to that author [259]. Some researchers prefer citing the most recent papers to show they are “up-to-date” in their field. Other authors tend to cite authoritative papers because they believe this makes their own paper more authoritative or because it is the popular thing to do. In other situations, researchers already know what they wish to write but require a reference to back up their claim. In this case, they tend to cite the first appropriate paper they find that supports the claim, although there may have been more fitting papers to cite. Citations may also indicate a “negative” quality assessment. For instance, in Chapter 3, we cited several papers that we considered of little significance and excluded from the in-depth review. These papers certainly would not be good recommendations. This means that even if authors were aware of all relevant literature, they will not always select the most relevant literature to cite.

When incomplete or even biased datasets are used as ground-truth, recommender systems are evaluated based on how well they can calculate such an imperfect ground-truth. Recommender systems that recommend papers that are not contained in the imperfect dataset, but that might be equally relevant, would receive a poor rating. A recommender system might even recommend papers of higher relevance than those in the offline dataset, but the evaluation would give the algorithm a poor rating. In other words, if the incomplete status quo – that is, a document collection compiled by researchers who are not aware of all literature, who are restricted by space and time constraints, and who typically do biased citing – is used as ground-truth, a recommender system can never perform better than the imperfect status quo.

We consider the imperfection to be a fundamental problem. To us, it seems plausible that the imperfection is also reason why offline metrics could not predict the effectiveness of citation-based and stereotype recommendations in the online evaluations and user study. As long as one cannot identify the situations in which the imperfection will affect the results, we propose that inferred ground-truths should not be used to evaluate research-paper recommender systems.

## 5.3 Mind-Map-Specific User-Modeling Variables

In this section, the results of Task 4 are presented. The task was to identify variables that affect user modeling based on mind maps, and measure the impact of the variables. To accomplish this task, content-based filtering algorithms were randomly assembled, and the impact on the user modeling effectiveness was measured with click-through rate.

### 5.3.1 Mind-Map & Node Selection

#### *5.3.1.1 Mind-map selection*

When utilizing mind maps for user modeling, one central question is *which mind maps* to analyze, and *which parts* of the mind maps to analyze. We experimented with a few variables to answer this question.

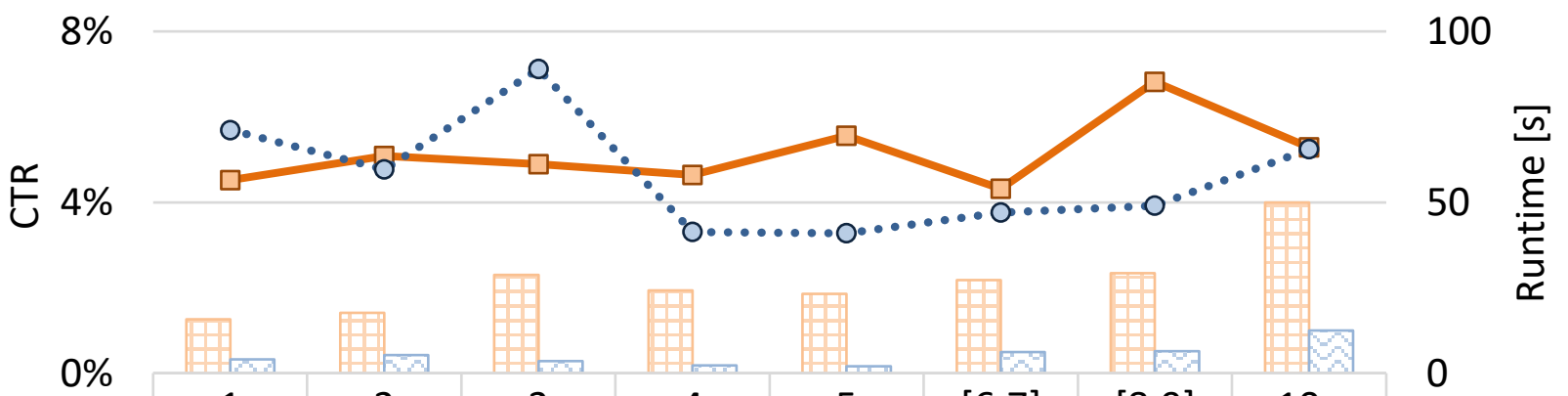

| | 1 | 2 | 3 | 4 | 5 | [6;7] | [8;9] | 10+ |
|---|---|---|---|---|---|---|---|---|
| Rntm (Terms) | 16 | 18 | 29 | 24 | 23 | 27 | 29 | 50 |
| Rntm (Citat.) | 4 | 5 | 4 | 2 | 2 | 6 | 7 | 13 |
| CTR (Terms) | 4.52% | 5.09% | 4.90% | 4.64% | 5.56% | 4.32% | 6.82% | 5.28% |
| CTR (Citat.) | 5.69% | 4.77% | 7.12% | 3.31% | 3.28% | 3.76% | 3.92% | 5.25% |

Number of mind-maps

Figure 35: CTR by the number of mind maps to analyze (all users)

We hypothesized that analyzing all mind maps of a user is not the most effective method: If too many, or too old mind maps are analyzed, this could introduce noise in the user model. To test this hypothesis, Docear's recommender system randomly used the *x* most recently modified mind maps, regardless of *when* they were modified. An initial analysis shows a slight tendency that CTR increases, the more mind maps are analyzed (Figure 35). When only a user's most recent mind map is utilized, CTR is 4.52% on average. Utilizing eight or nine mind maps resulted in the highest CTR (6.82%). However, these results might be misleading since the analysis is based on recommendations for all users including those who created only few mind maps: for these users it would not be possible to analyze the eight or nine most recently created mind maps. Therefore, we did the same analysis for users who created at least eight mind maps (Figure 36). In this analysis, no statistically significant difference could be found for the number of utilized mind maps. Judging by these numbers, it seems that the number of the most recently modified mind maps is not an effective variable to optimize the user modeling process.

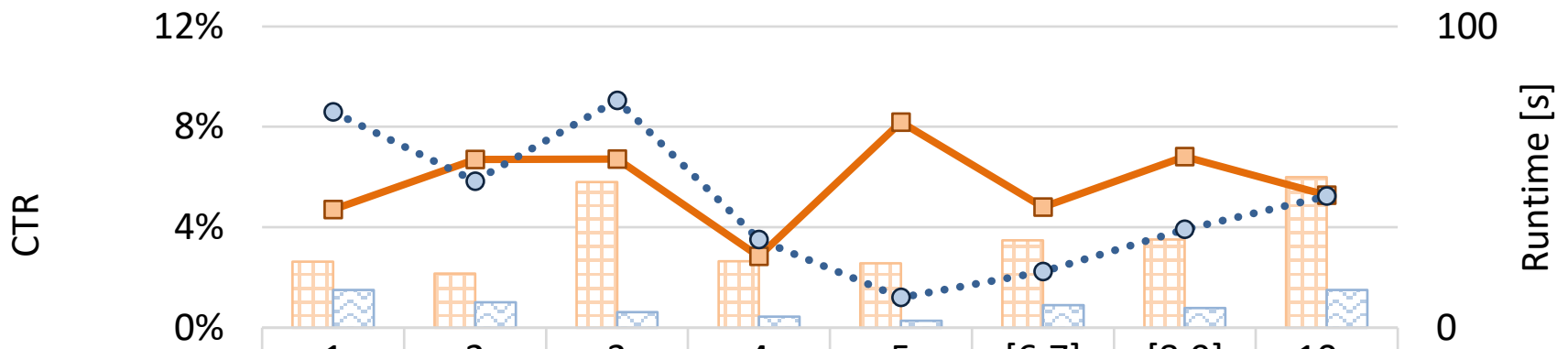

| | 1 | 2 | 3 | 4 | 5 | [6;7] | [8;9] | 10+ |
|---|---|---|---|---|---|---|---|---|
| Rntm (Terms) | 22 | 18 | 48 | 22 | 21 | 29 | 29 | 50 |
| Rntm (Citat.) | 12 | 9 | 5 | 4 | 2 | 7 | 7 | 13 |
| CTR (Terms) | 4.71% | 6.71% | 6.72% | 2.85% | 8.18% | 4.80% | 6.82% | 5.28% |
| CTR (Citat.) | 8.59% | 5.84% | 9.06% | 3.51% | 1.20% | 2.24% | 3.92% | 5.25% |

Number of mind-maps

Figure 36: CTR by number of mind maps to analyze (8+ mind maps)

### 5.3.1.2 *Node selection*

As an alternative to using the *x* most recently modified *mind maps*, Docear analyzed the *x* most recently modified *nodes*. For example, if *x=50*, the terms (or citations) contained in the 50 most recently modified nodes are used. The intention is that users might be working in different sections of several mind maps, and only those actively edited sections are relevant for user modeling. The analysis shows that the more nodes are used, the higher CTR becomes (Figure 37). While CTR is 3.66% on average when one to nine nodes are used, CTR increases to 6.60% when 1,000 and more nodes are used. Interestingly, it is the opposite for citations: the more nodes with citations are used, the lower CTR becomes.

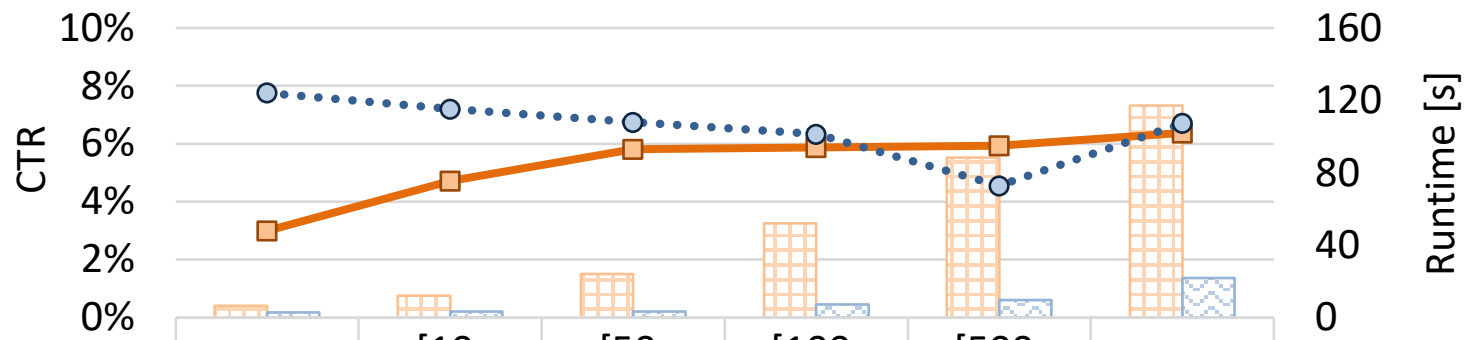

| | [1;9] | [10; 49] | [50; 99] | [100; 499] | [500; 999] | 1,000+ |
|---|---|---|---|---|---|---|
| Rntm (Terms) | 7 | 12 | 24 | 52 | 88 | 117 |
| Rntm (Citat.) | 3 | 3 | 3 | 7 | 10 | 22 |
| CTR (Terms) | 3.00% | 4.72% | 5.82% | 5.87% | 5.93% | 6.38% |
| CTR (Citat.) | 7.76% | 7.21% | 6.75% | 6.34% | 4.56% | 6.71% |

Number of nodes

Figure 37: CTR by the number of nodes to analyze (all users)

However, these results, again, might be misleading since not all users have created e.g. thousand nodes. Consequently, the CTR for e.g. one to nine nodes includes recommendations for all users, but the CTR for analyzing 1,000 or more nodes only considers recommendations to users who created at least 1,000 nodes. Therefore, we performed the previous analysis again, but for users who have created at least 1,000 nodes (Figure 38). This time, a saturation appears. When the 50 to 99 most recently created nodes are used, CTR is highest (7.50%). When more nodes are analyzed, CTR decreases. We did the same analysis for other user groups, and results were always the same – using only the 50 to 99 most recently modified nodes led to the highest CTRs on average. With regard to citations, the results slightly change. For users with 1,000 or more nodes, using the most recent 10 to 49 citations is most effective (8.36% vs. 7.32% for using 1 to 9 citations).

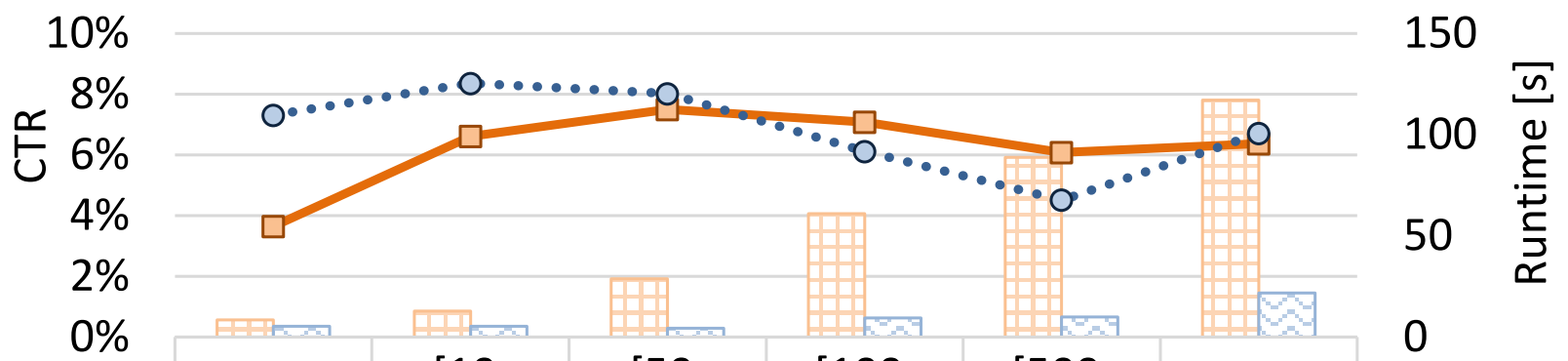

| | [1;9] | [10; 49] | [50; 99] | [100; 499] | [500; 999] | 1,000+ |
|---|---|---|---|---|---|---|
| Rntm (Terms) | 8 | 13 | 29 | 61 | 89 | 117 |
| Rntm (Citat.) | 5 | 5 | 4 | 10 | 10 | 22 |
| CTR (Terms) | 3.64% | 6.62% | 7.50% | 7.08% | 6.09% | 6.38% |
| CTR (Citat.) | 7.32% | 8.36% | 8.02% | 6.12% | 4.52% | 6.71% |

Number of nodes

Figure 38: CTR by the number of nodes to analyze (1,000+ nodes available)

Selecting a fix number of nodes might not be the most effective criteria. The most recent, for example, 75 nodes could include nodes that were modified some years ago. Such nodes would probably not represent a user's current information needs any more. Therefore, Docear's recommender system randomly used all nodes that were modified within the past *x days* (Figure 39). When the recommender system utilized only those nodes that were modified on the current day, CTR was 3.81% on average[61]. When nodes from the last two or three days were utilized, CTR increased to 5.52%. CTR was highest, when nodes modified during the past 61 to 120 days were used (7.72%), and remained high when nodes of the past 121 to 180 days were used. When nodes were used that were modified more than 180 days

[61] The analysis was done only for users being registered since at least 360 days

ago, CTR began to decrease. Apparently, the interests of Docear's users change after a few months.

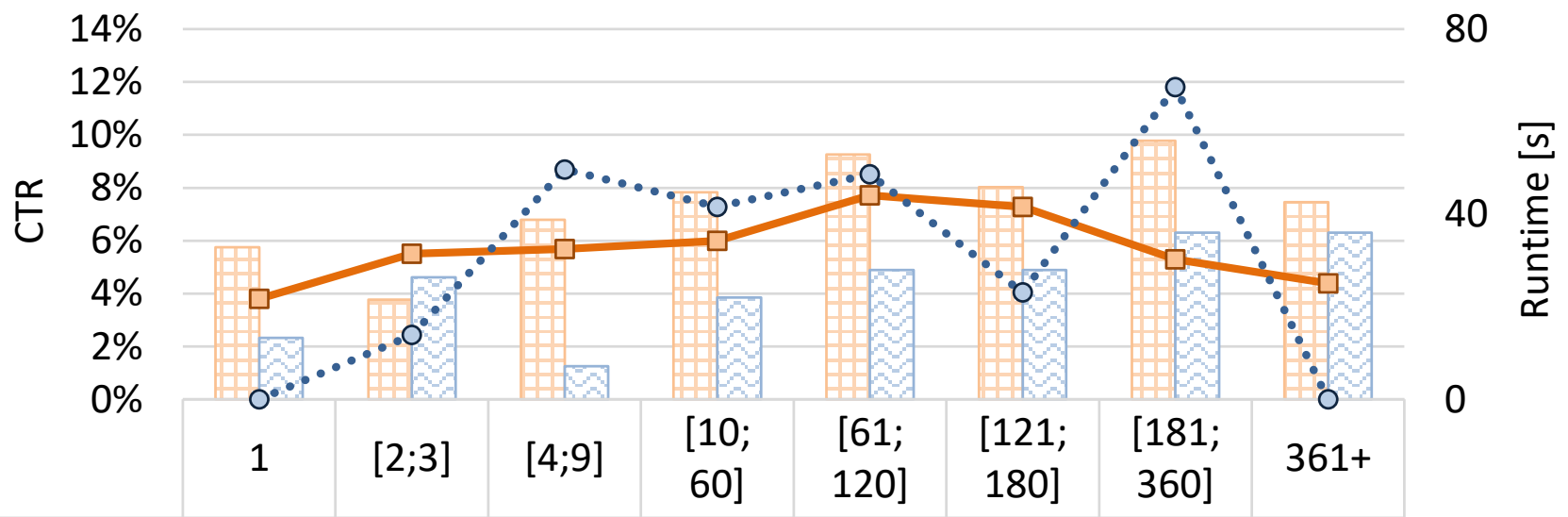

| | 1 | [2;3] | [4;9] | [10; 60] | [61; 120] | [121; 180] | [181; 360] | 361+ |
|---|---|---|---|---|---|---|---|---|
| Rntm (Terms) | 33 | 22 | 39 | 45 | 53 | 46 | 56 | 43 |
| Rntm (Citat.) | 13 | 26 | 7 | 22 | 28 | 28 | 36 | 36 |
| CTR (Terms) | 3.81% | 5.52% | 5.68% | 6.00% | 7.72% | 7.29% | 5.29% | 4.39% |
| CTR (Citat.) | 0.00% | 2.44% | 8.70% | 7.28% | 8.50% | 4.05% | 11.81% | 0.00% |

Number of days

Figure 39: CTR for nodes analyzed in the past *x* days[62]

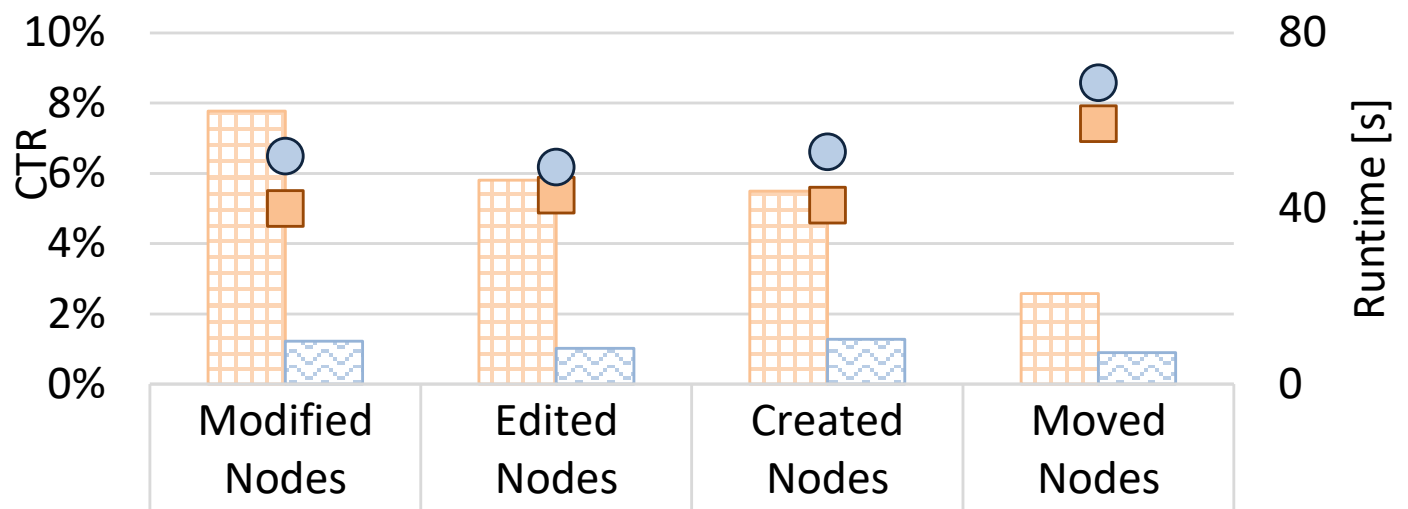

| | Modified Nodes | Edited Nodes | Created Nodes | Moved Nodes |
|---|---|---|---|---|
| Rntm (Terms) | 62 | 46 | 44 | 21 |
| Rntm (Citat.) | 10 | 8 | 10 | 7 |
| CTR (Terms) | 4.99% | 5.38% | 5.09% | 7.40% |
| CTR (Citat.) | 6.48% | 6.18% | 6.61% | 8.57% |

Modification type

Figure 40: Node modification type

Another variable we tested was the node modification type (Figure 40). The recommender system chose randomly, whether to utilize only nodes that were newly *created*, nodes that were *moved*, nodes that were *edited*, or nodes with any type of *modification* (created, edited, *or* moved). Utilizing moved nodes only, resulted in the highest CTR (7.40%) on average, while the other modification types achieved CTRs around 5%. We find this interesting, because this result indicates that the evolution of a mind map might be important for user modeling, and certain actions (e.g. moving nodes) indicate a high significance of certain nodes.

[62] For users being registered since more than 360 days

Most mind-mapping tools allow folding a node, i.e. to hide its children. In Docear, this is indicated by a circle at the end of node (Figure 5, p. 12). We hypothesized that nodes that are hidden, are currently not relevant for describing the user's information needs. Therefore, Docear's recommender system randomly chose whether to use only *visible* nodes, *invisible* nodes, or *all* nodes. When using visible nodes only, CTR increased from 6.00% (analyzing all nodes) to 7.61% (Figure 41). Using only invisible nodes led to a CTR of 4.89% on average. This indicates once more that by selecting a few meaningful nodes, a better effectiveness can be achieved than by examining simply all nodes.

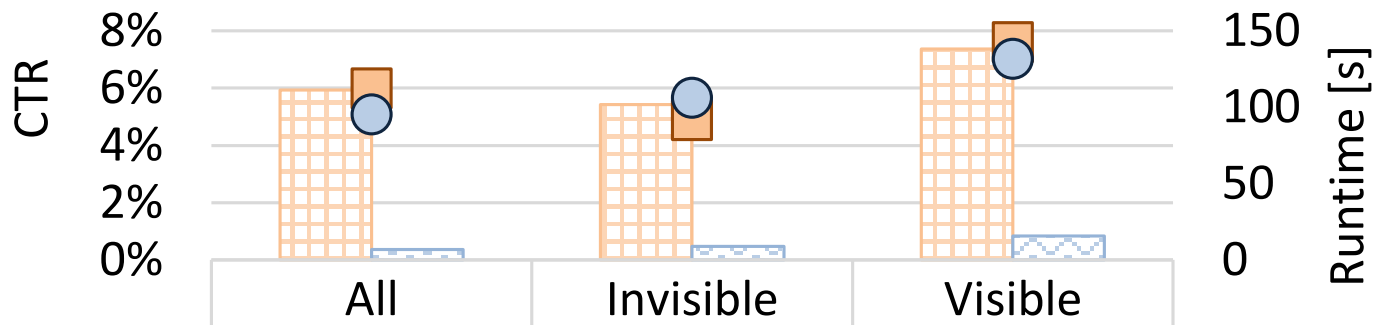

| | All | Invisible | Visible |
|---|---|---|---|
| Rntm (Terms) | 111 | 102 | 138 |
| Rntm (Citat.) | 7 | 9 | 16 |
| CTR (Terms) | 6.00% | 4.89% | 7.61% |
| CTR (Citat.) | 5.08% | 5.67% | 7.03% |

Node visibility

Figure 41: Node visibility as selection criteria (at least 100 nodes analyzed)

### *5.3.1.3 Node extension*

We hypothesized that the relation among nodes is important. This means the terms of a node might be more meaningful when the node's context is known with regard to the neighboring nodes. The most common neighbors are *parents*, *children*, and *siblings*. For instance, in Figure 5 (p. 12), the author's information needs seem rather vague when looking only at one node, “Google Scholar indexed invisible text”. In combination with the (grand) parent “(Academic) Search Engine Spam”, the author's interests become clearer. Therefore, we experimented with extending the original selection of nodes. After the system chose the relevant nodes to examine with one of the previously introduced methods, the recommender system randomly chose whether to add siblings, parents, or children to the original selection. This process is illustrated in Figure 42 where two nodes were originally selected and the direct parent, children, and siblings were selected as extension to the original nodes.

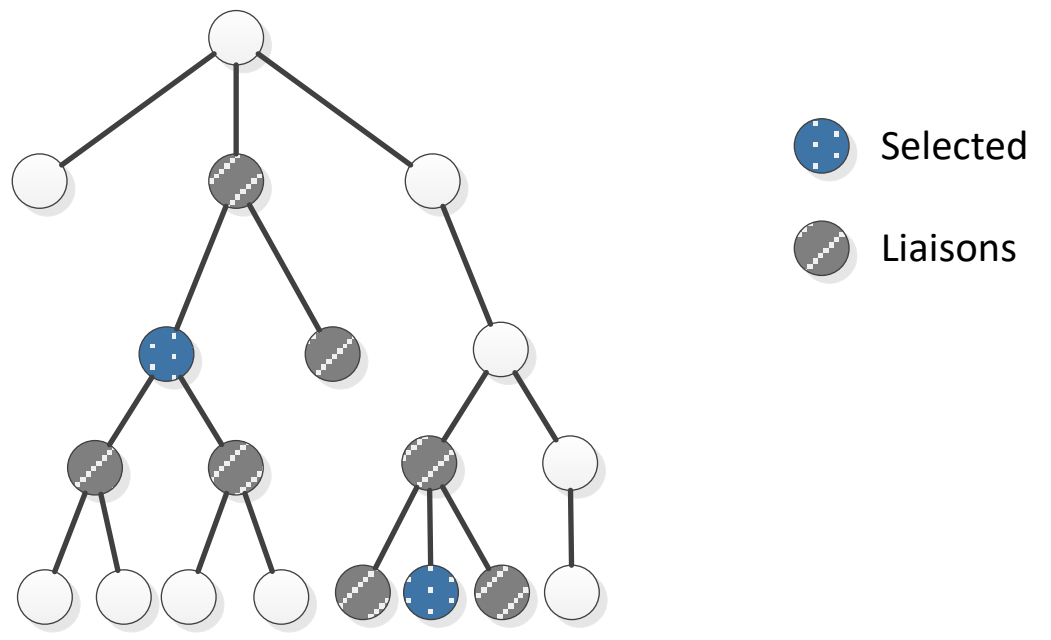

Figure 42: Extending the original node selection

Adding siblings resulted in a CTR of 5.73% compared to 5.10% for not adding siblings (Figure 43). Adding parent-nodes decreased CTR to 5.36% compared to 5.46% for not adding them. Adding children increased CTR from 5.22% to 5.61%. Differences are small but significant. In addition, when the recommender system combined all factors, i.e. adding siblings and children but ignoring parents, CTR was 6.18% on average, which is a significant improvement, compared to not extending nodes (4.84%). One might suspect that extending the original node selection was only more effective because the extension caused more nodes to be used, and the more nodes are used, the higher CTR tends to become. However, this suspicion is not correct. For instance, when 100 to 499 nodes were selected, and no children or siblings were added, CTR was 5.15% on average. When, 10 to 50 nodes were selected and after adding children and siblings 100 to 499 nodes were used, CTR was 5.45%. This indicates that selecting a few recently modified nodes, and extending them with their siblings and children, is more effective than selecting simply more nodes based only on the modification date.

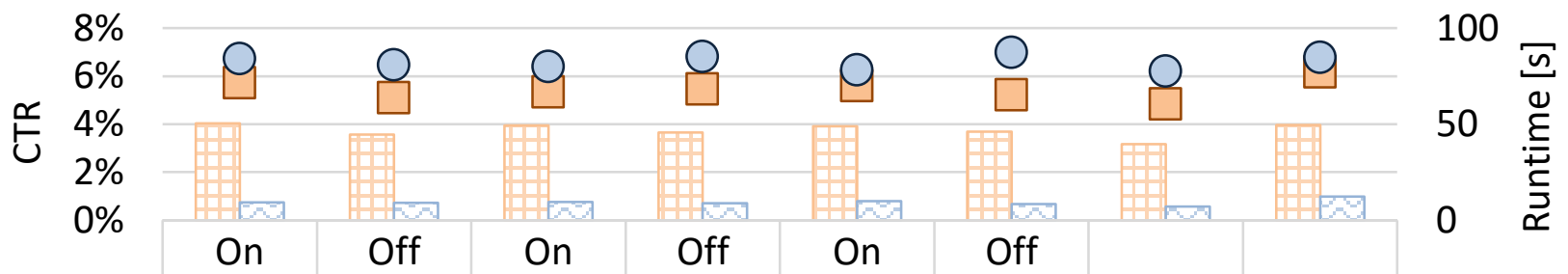

| | On | Off | On | Off | On | Off | | |
|---|---|---|---|---|---|---|---|---|
| | Include Siblings | | Include Parents | | Include Children | | All Off | Bst. Fctrs. Cmbnd. |
| Rntm (Terms) | 51 | 45 | 49 | 46 | 49 | 46 | 40 | 50 |
| Rntm (Citat.) | 9 | 9 | 9 | 9 | 10 | 9 | 7 | 12 |
| CTR (Terms) | 5.73% | 5.10% | 5.36% | 5.46% | 5.61% | 5.22% | 4.84% | 6.18% |
| CTR (Citat.) | 6.74% | 6.47% | 6.41% | 6.82% | 6.27% | 6.99% | 6.21% | 6.80% |

Extension method

Figure 43: Extending the original node selection

### 5.3.2 Node Weighting

Often, user-modeling applications weight features (e.g. terms) that occur in a certain document field (e.g. in title) stronger than features occurring in other document fields (e.g. the body text). Mind maps have no fields for title, abstract, headings, or body text. Instead, mind maps have nodes, which have a certain *depth*, i.e. their distance from the root node. We hypothesized that the depth of a node might indicate the importance of the node. For instance, in Figure 5 (p. 12), the node “Scopus” has a depth of 2, and we would assume that the term “Scopus” describes the user’s interests with a different accuracy than the node “Academic Search Engines” that has a depth of 1.

To test the hypothesis, Docear’s recommender system randomly chose whether to weight terms of a node stronger or weaker, depending on its depth. If the nodes were to be weighted stronger the deeper they were, the weight of a node (1 by default) was multiplied with **a)** the absolute depth of the node; **b)** the natural logarithm of the depth; **c)** the logarithm to base 10 of the depth; or **d)** the square root of the depth. If the resulting weight was lower than 1, e.g., for ln (2), then the weight was set to 1. If nodes were to receive less weight the deeper they were, then the original weight of 1 was multiplied with the reciprocal of the metrics a) – d). If the resulting weight was larger than 1, e.g., for ln(2), the weight was set to 1. In the following charts, we provide CTR for the mentioned metrics. However, the differences among the metrics are not statistically significant. Hence, we concentrate on comparing the overall CTR, i.e. the CTR of weighting nodes stronger or weaker the deeper they are regardless of the particular metric.

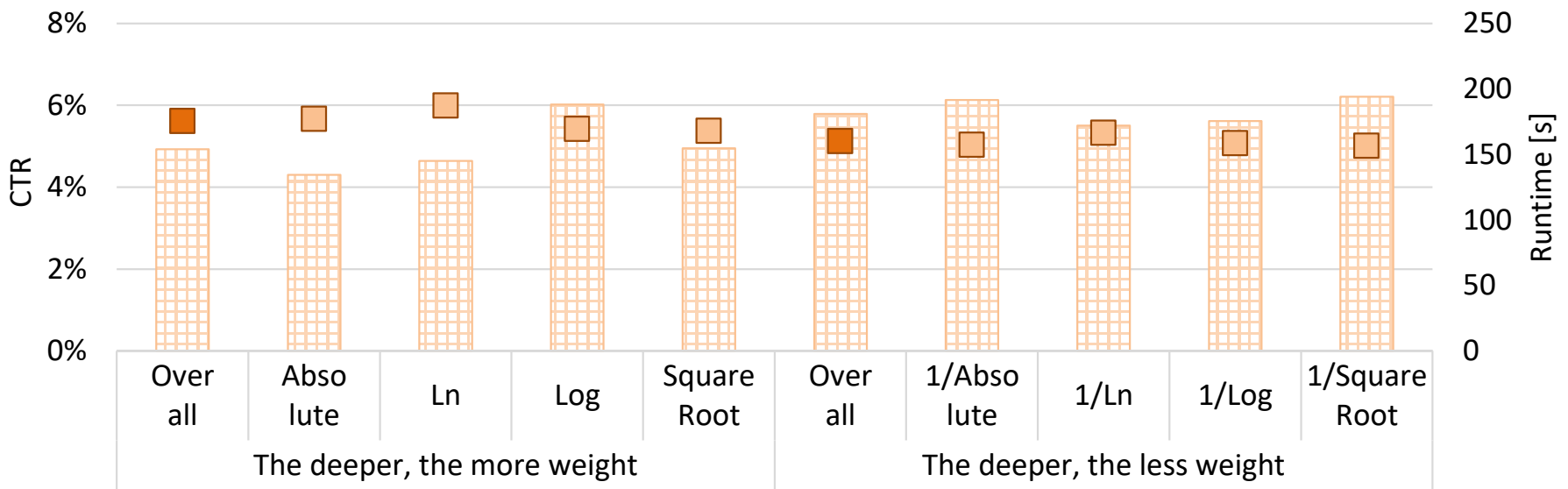

| | The deeper, the more weight | | | | | The deeper, the less weight | | | | |
|---|---|---|---|---|---|---|---|---|---|---|
| | Overall | Absolute | Ln | Log | Square Root | Overall | 1/Absolute | 1/Ln | 1/Log | 1/Square Root |
| Rntm (Terms) | 154 | 134 | 145 | 188 | 155 | 181 | 192 | 172 | 176 | 194 |
| CTR (Terms) | 5.61% | 5.68% | 6.00% | 5.43% | 5.38% | 5.12% | 5.04% | 5.33% | 5.07% | 5.02% |

Node depth metric

Figure 44: Weighting based on node depth

Results show that when nodes are weighted stronger the deeper they are in a mind map, CTR increases (Figure 44). Weighting them stronger, led to a CTR of 5.61% on average, while weighting them weaker led to a CTR of 5.12% on average.

We also experimented with other metrics that are based on the number of children, the number of siblings, and the number of words contained in a node. Figure 42 illustrates this. Node A is a leaf node because it has no children. In contrast, node B has two children, which in turn have two children each, too. We hypothesized that the number of children indicates how important a node is to infer a user's interests. Therefore, Node A would be weighted differently than Node A. Similarly, Node C also is a leaf but has more siblings than Node A. Therefore, Node C might be weighted differently than Node A.

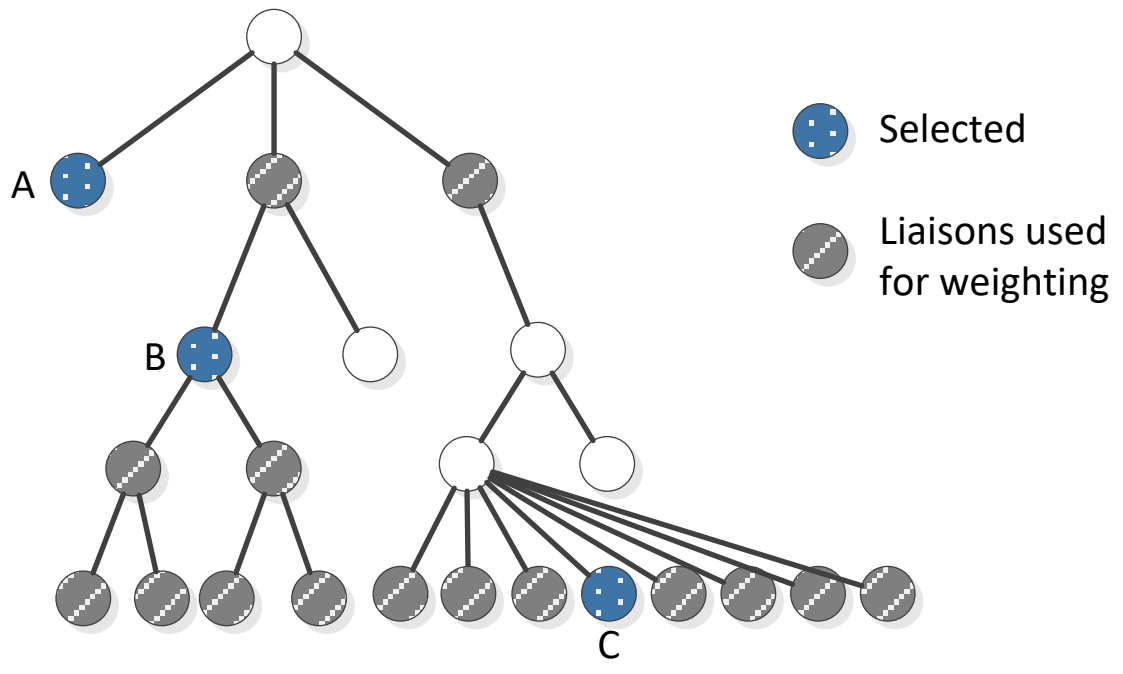

Figure 45: Weighting based on children and siblings

Based on our experiments, CTR increases when nodes are weighted stronger the more children a node has (Figure 46). Weighting them stronger led to a CTR of 5.17% on average, while weighting them weaker led to a CTR of 4.97%. However, the difference was statistically not significant. Weighting based on the number of siblings had a significant effect (Figure 47). Weighting nodes stronger the more siblings they have led to a CTR of 5.40%, compared to 5.01% for weighting them weaker. Weighting nodes based on the number of terms they contained led to no significant differences (Figure 48).

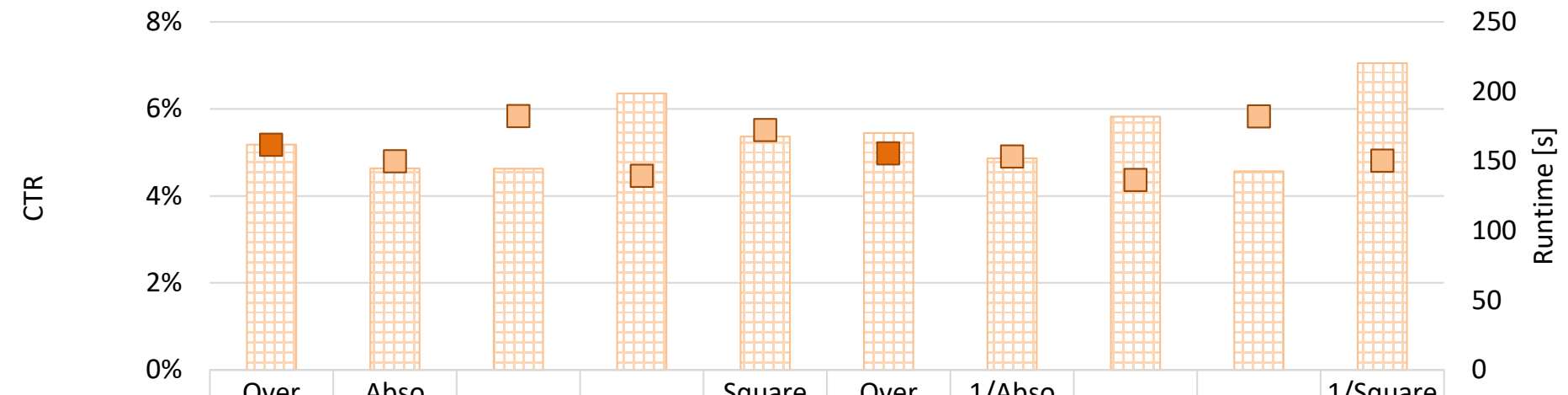

| | The more children, the more weight | | | | | The more children, the less weight | | | | |
|---|---|---|---|---|---|---|---|---|---|---|
| | Over all | Abso lute | Ln | Log | Square Root | Over all | 1/Abso lute | 1/Ln | 1/Log | 1/Square Root |
| Rntm (Terms) | 162 | 145 | 145 | 199 | 168 | 170 | 152 | 182 | 143 | 220 |
| CTR (Terms) | 5.17% | 4.80% | 5.84% | 4.46% | 5.51% | 4.97% | 4.91% | 4.37% | 5.83% | 4.81% |

Children metric

Figure 46: Weighting based on the number of children

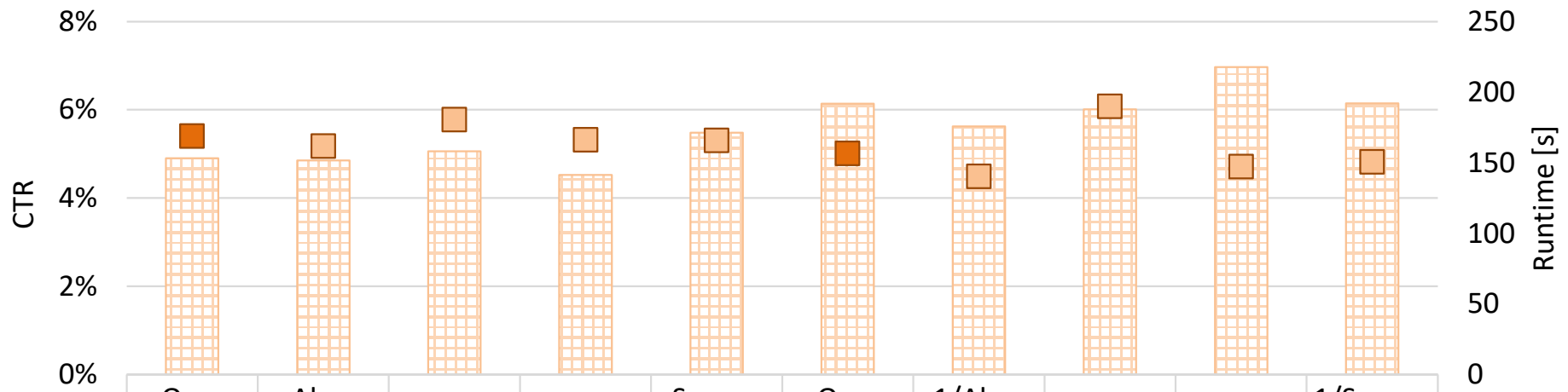

| Sibling metric | Overall | Absolute | Ln | Log | Square Root | Overall | 1/Absolute | 1/Ln | 1/Log | 1/Square Root |
|---|---|---|---|---|---|---|---|---|---|---|
| | The more siblings, the more weight | | | | | The more siblings, the less weight | | | | |
| Rntm (Terms) | 153 | 152 | 158 | 141 | 171 | 192 | 176 | 188 | 218 | 192 |
| CTR (Terms) | 5.40% | 5.17% | 5.77% | 5.31% | 5.31% | 5.01% | 4.49% | 6.07% | 4.71% | 4.81% |

Sibling metric

Figure 47: Weighting based on the number of siblings

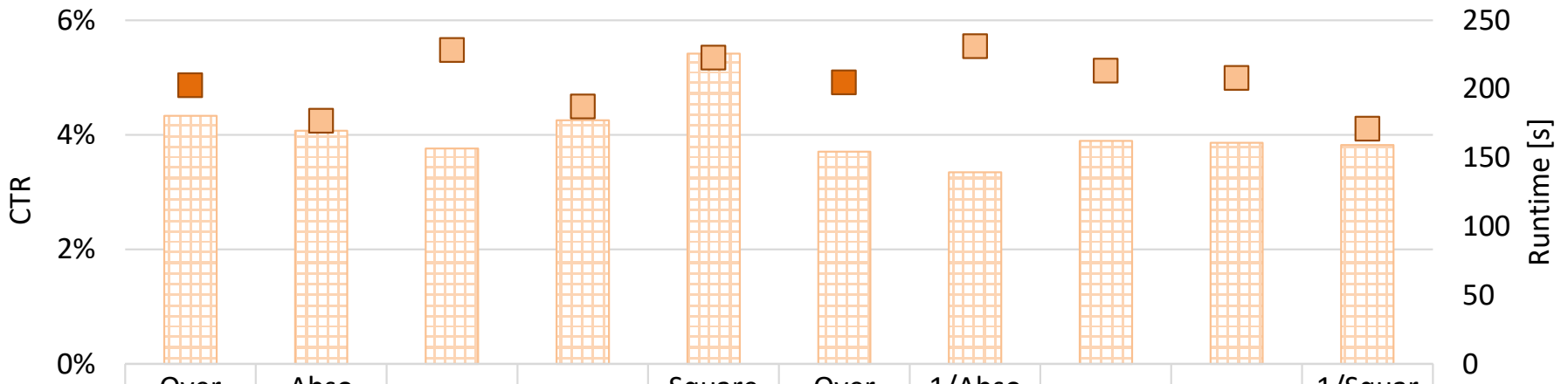

| Word count metric | Overall | Absolute | Ln | Log | Square Root | Overall | 1/Absolute | 1/Ln | 1/Log | 1/Square Root |
|---|---|---|---|---|---|---|---|---|---|---|
| | The more the better | | | | | The more the worse | | | | |
| Rntm (Terms) | 180 | 170 | 157 | 177 | 226 | 154 | 139 | 162 | 161 | 159 |
| CTR (Terms) | 4.86% | 4.24% | 5.48% | 4.49% | 5.35% | 4.91% | 5.54% | 5.12% | 4.99% | 4.10% |

Word count metric

Figure 48: Weighting based on the number of terms contained in a node

After the individual weights are calculated, the weights need to be combined into a single node weighting score. We experimented with four different schemes to combine the scores. The most effective scheme was using the *sum* of all individual scores (CTR = 6.38%). Using only the maximum score (*max*), multiplying the scores (*product*) or using the average score (*avg*) led to CTRs slightly above 5% (Figure 49).

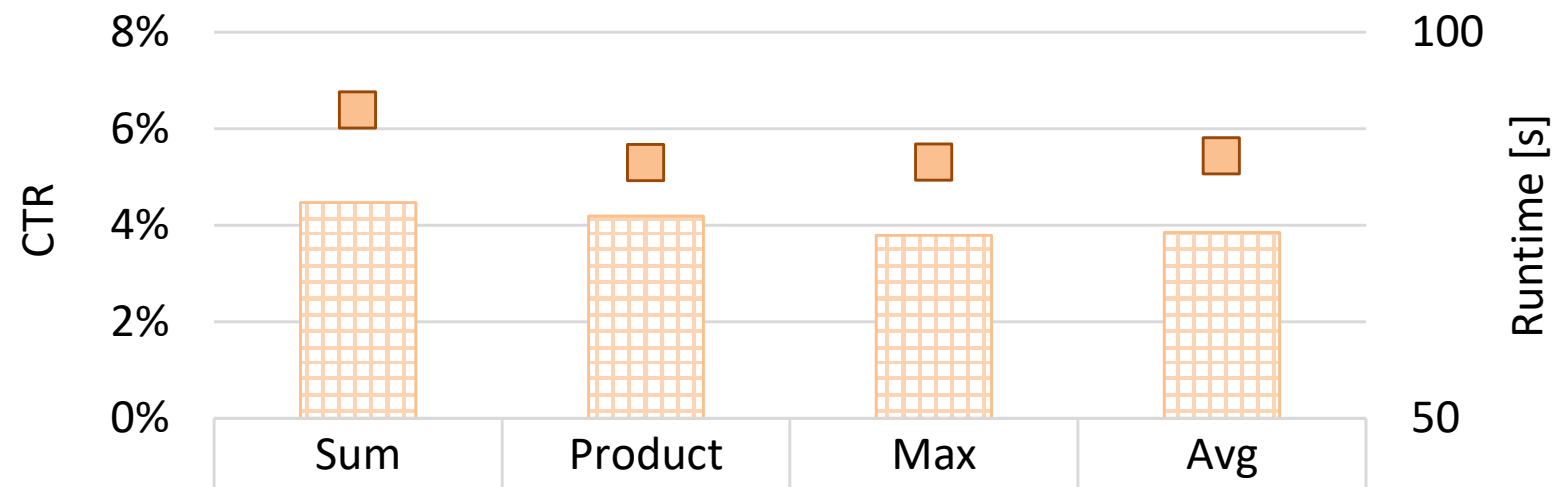

| | Sum | Product | Max | Avg |
|---|---|---|---|---|
| Rntm (Terms) | 78 | 76 | 74 | 74 |
| CTR (Terms) | 6.38% | 5.29% | 5.30% | 5.43% |

Combination scheme

Figure 49: Combining the node weights

### 5.3.3 Feature Weighting

After *nodes* are weighted, the nodes' *features* inherit the weight of the nodes, and they are additionally weighted with a randomly chosen weighting scheme. This means, if a node has a weight of eight, then all terms (or citations) of that node receive a weight of eight, and this weight was multiplied with one of the following weighting schemes: plain term or citation frequency *(TF-Only)*, *TF-IDF*, and a novel metric that we call TF-ID*u*F[63]. TF-ID*u*F is similar to TF-IDF but based on the inverse document frequency in the user's document corpus, instead of the standard document corpus. Hence, TF-ID*u*F, weights a term stronger the more often it occurs in the user's mind maps (or nodes) that are currently selected for user modeling, but the less mind maps of the user contain this term. The rationale is that when users use a term frequently that they did not use frequently before, this term is of particular importance. In addition, if users are using terms for a longer time, they probably have already received recommendations for that term.

TF-IDF is commonly found to be more effective than TF-only [263]. Our analysis confirms this well-known finding – TF-IDF outperformed TF-only for terms with a CTR of 5.10% vs. 4.13% (Figure 50). However, to the best of our knowledge, it has not been empirically shown that TD-IDF is superior to TF-only when applied to citations, i.e. CC-IDF (cf. Section 3.2.1, p. 36). In Docear's recommender system, CC-IDF led to lower CTRs (5.75%) than TF-only, CC respectively, when applied to citations (6.07%).

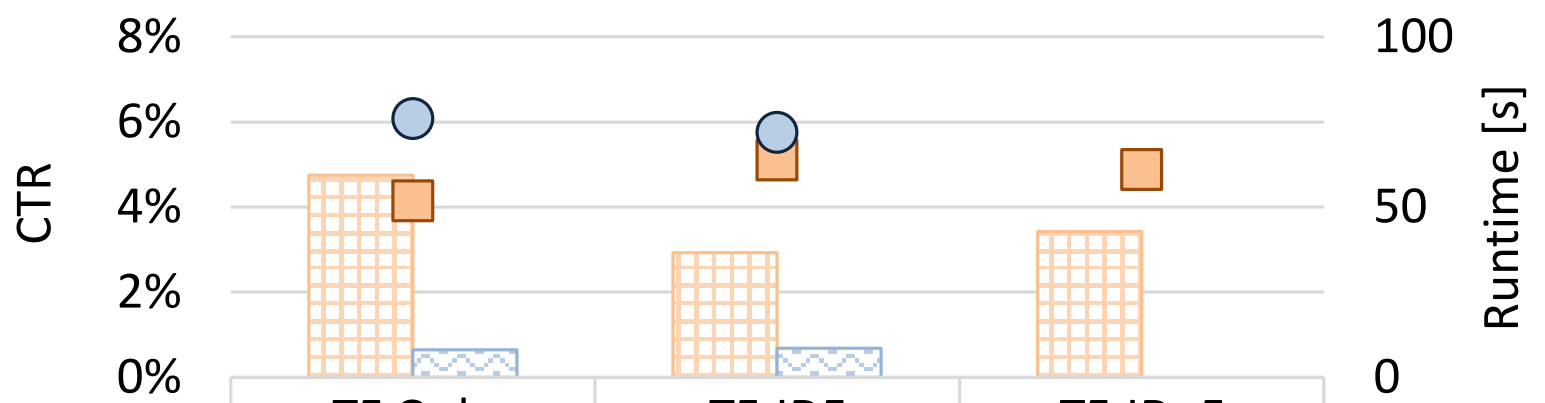

| | TF Only | TF-IDF | TF-IDuF |
|---|---|---|---|
| Rntm (Terms) | 59 | 37 | 43 |
| Rntm (Citat.) | 8 | 8 | |
| CTR (Terms) | 4.13% | 5.10% | 4.88% |
| CTR (Citat.) | 6.07% | 5.75% | |

Weighting scheme

Figure 50: CTR of different weighting schemes[63]

[63] TF-ID*u*F was only calculated for terms, not yet for citations.

We find this result surprising and can only speculate about the reason. One explanation might be the following: The rationale behind IDF for weighting terms is that terms occurring in many documents of the corpus (e.g. *the*, *and*, *he*, *she*, *etc.*), do not describe the content of a documents well. This rational seems plausible to us. However, the rationale does not necessarily apply for citations. Citations occurring in many documents of the corpus might still describe the citing document well, maybe even better than little cited papers. For instance, this thesis cites, among others, reference [59] and [345]. Reference [59] is about research-paper recommender systems and received more than 300 citations according to Google Scholar. This means, many papers in the corpus contain a citation to [59]. Reference [345] is about news recommendations and received only four citations. [59] is certainly more relevant to describe my thesis than [345] and hence, CC-IDF would have led to suboptimal results when weighting the two papers. Of course, this is only one example, and there might be other examples in which CC-IDF were to prefer over TF-only. Further research is necessary to explore this issue.

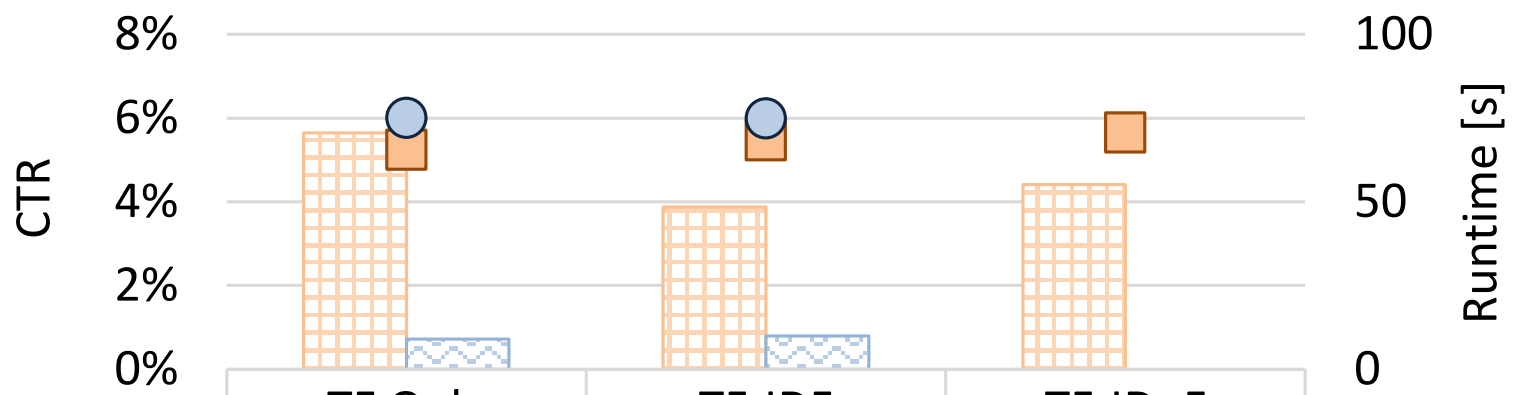

| | TF Only | TF-IDF | TF-IDuF |
|---|---|---|---|
| Rntm (Terms) | 71 | 48 | 55 |
| Rntm (Citat.) | 9 | 10 | - |
| CTR (Terms) | 5.24% | 5.45% | 5.65% |
| CTR (Citat.) | 6.00% | 5.99% | |

Weighting scheme

Figure 51: CTR of different weighting schemes (500+ nodes)[63]

Our novel metric TF-ID*u*F (CTR = 4.88%) was slightly less effective than TF-IDF for terms (CTR = 5.10%) but more effective than TF-only (Figure 50). When we repeated the analysis for those user-modeling processes that analyzed at least 500 nodes, TF-ID*u*F became slightly more effective than TF-IDF (Figure 51). This shows that the use of terms within a user's "personal corpus" may be an important measure about a term's relevancy, in particular when the personal corpus is large. Further research is necessary to explore the potential of TF-ID*u*F. Probably, TF-ID*u*F is particularly interesting when there is no access to the global corpus, and hence TF-IDF cannot be calculated. A combination of TF-IDF and TF-ID*u*F might also be interesting.

### 5.3.4 User-Model Size

Just because utilizing e.g. the 50 most recently moved nodes is most effective, does not mean that necessarily all features of these nodes need to be stored as user model. Therefore, Docear's recommender system randomly selected to store only the *x* highest weighted features as user model. For user modeling based on at least 50 nodes, CTR is highest (8.81%) when user models contain the 26 to 50 highest weighted terms (Figure 52). User models containing less, or more, terms achieve significant lower CTRs. For instance, user models with one to ten terms have a CTR of 3.92% on average. User models containing more than 500 terms have a CTR of 4.84% on average. Interestingly, CTR for citations continuously decreases the more citations a user model contains[64]. Consequently, a user-model size between 26 and 50 seems most sensible for terms, and a user-model size of ten or less for citations.

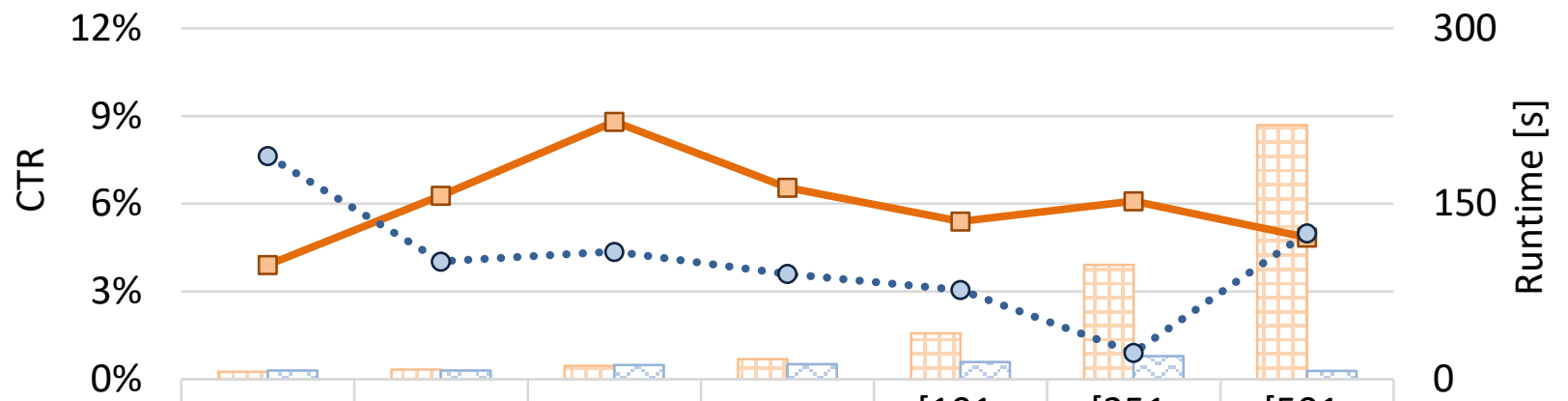

| | [1;10] | [11;25] | [26;50] | [51;100] | [101; 250] | [251; 500] | [501; 1,000] |
|---|---|---|---|---|---|---|---|
| Rntm (Terms) | 7 | 8 | 11 | 17 | 39 | 98 | 217 |
| Rntm (Citat.) | 7 | 7 | 12 | 13 | 15 | 20 | 7 |
| CTR (Terms) | 3.92% | 6.27% | 8.81% | 6.56% | 5.40% | 6.09% | 4.84% |
| CTR (Citat.) | 7.63% | 4.03% | 4.36% | 3.61% | 3.06% | 0.91% | 5.00% |

Number of terms/citations

Figure 52: CTR by user-model size (feature weight not stored)

The previous analysis was based on un-weighted lists of terms and citations, i.e. the user model contained only a list of the features without any weight information. If the weights were stored in the user model, and used for the matching process, the picture changes (Figure 53). In this case, CTR has a peak for user models containing between 251 and 500 terms (8.13%). Interestingly, this CTR is similar to the maximum CTR for the optimal non-weighted user-model size (8.81% for 26 to 50 terms). We find this surprising because we expected weighted lists to be more effective than un-weighted lists. The results for weighted citations are also surprising – CTR varies and shows no clear trend. We have no explanation for the results and hence see a need for further research[65].

[64] Analysis for 500 nodes or more being analyzed. The high CTR for user models with 501 and more citations is statistically not significant.

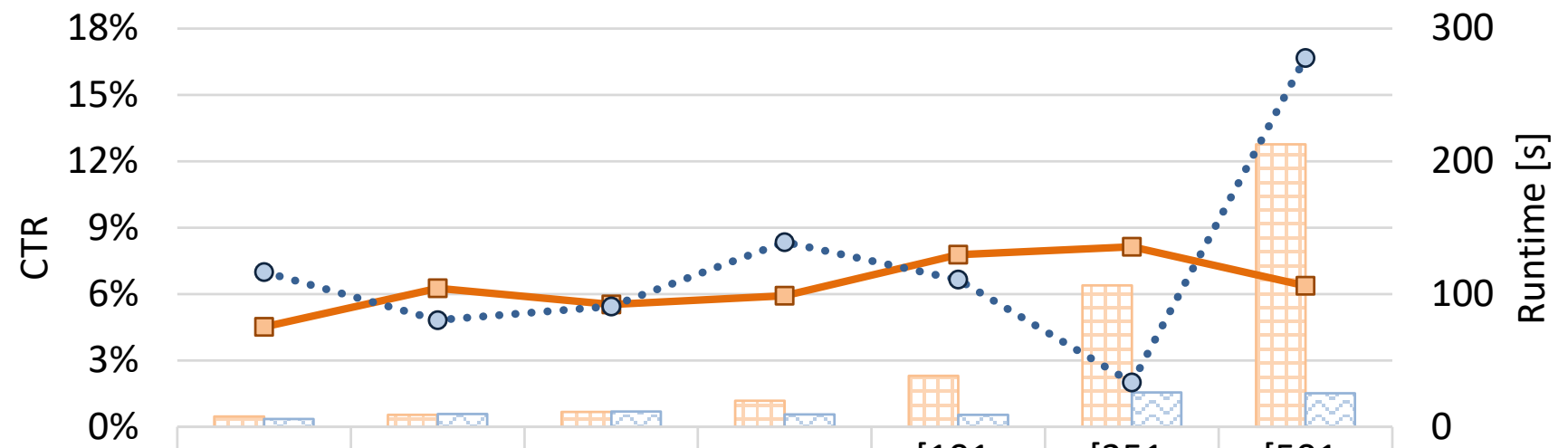

| | [1;10] | [11;25] | [26;50] | [51;100] | [101; 250] | [251; 500] | [501; 1,000] |
|---|---|---|---|---|---|---|---|
| Rntm (Terms) | 8 | 9 | 11 | 20 | 38 | 106 | 213 |
| Rntm (Citat.) | 6 | 10 | 11 | 9 | 9 | 26 | 25 |
| CTR (Terms) | 4.51% | 6.27% | 5.53% | 5.92% | 7.77% | 8.13% | 6.37% |
| CTR (Citat.) | 6.99% | 4.82% | 5.44% | 8.34% | 6.65% | 2.00% | 16.67% |

Number of terms/citations

Figure 53: CTR by user-model size (feature weight stored)

### 5.3.5 Citations vs. Terms

Based on all delivered recommendations, citation-based recommendations have an average CTR of 6.10%, while term-based recommendations have an average CTR of 5.07% (Figure 54). Hence, on first glance, it appears that citation-based recommendations are more effective than term-based recommendations. On second glance, one realizes that citation-based recommendations are only possible if users have at least one citation in their mind maps. Consequently, novel users, without citations in their mind maps receive only term-based recommendations. These users tend to have lower CTRs than users with more comprehensive mind maps. In addition, citation-based user modeling often returns less than ten recommendation candidates, which might affect CTR as well. Therefore, we made a "fair" comparison and compared citation- and term-based recommendations for which at least 10 recommendation candidates were returned; the original rank of the recommendation candidates was between one and ten; and the users had made at least one citation in their mind map. In this comparison, term-based recommendations outperform citation-based recommendations with a CTR of 6.53% vs. 5.25%.

[65] We double-checked all data and the source code, and are quite confident that there are no flaws in the user modeling process.

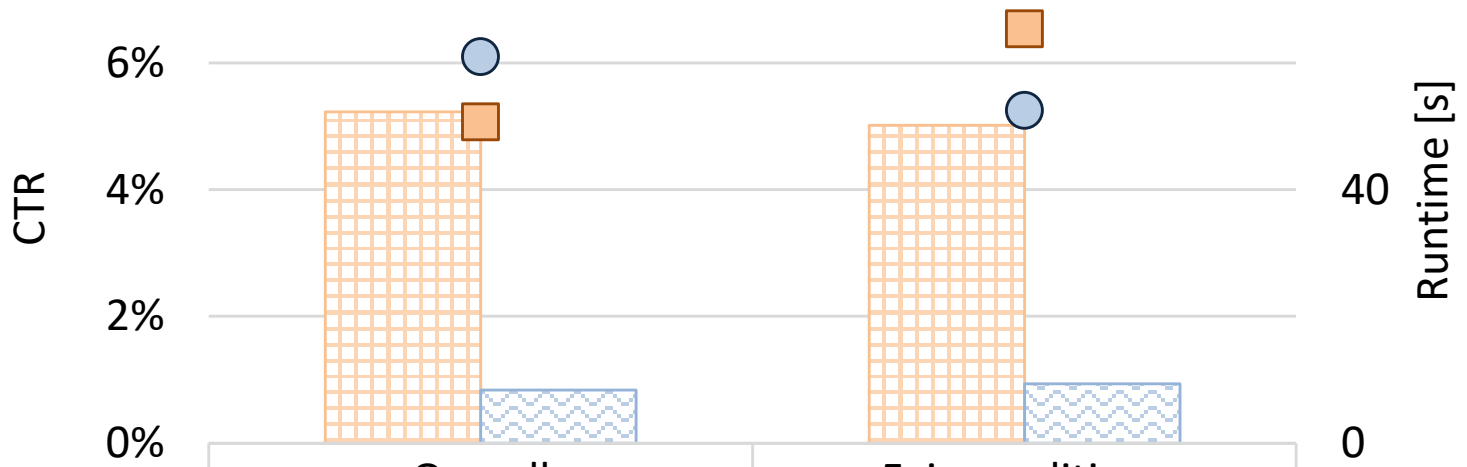

| | Overall | Fair conditions |
|---|---|---|
| Rntm (Terms) | 52 | 50 |
| Rntm (Citat.) | 8 | 9 |
| CTR (Terms) | 5.07% | 6.53% |
| CTR (Citat.) | 6.10% | 5.25% |

Figure 54: Citation vs. term-based citations, overall and under “fair” conditions

### 5.3.6 Mind maps vs. Other Items

Our main goal was to build a recommender system that considers the unique characteristics of mind maps. However, we were also curious to explore how user modeling based on mind maps compares to user modeling based on other items, research papers in particular. Therefore, we evaluated seven approaches that utilized **(a)** terms of *nodes* in the users’ mind maps **(b)** terms from the titles of the users’ PDF files **(c)** terms from the titles of the user’s citations **(d)** terms from the titles of the PDFs *and* citations, **(e)** terms from the nodes of mind maps, *and* titles of citations, **f)** terms from the nodes of mind maps, *and* titles of PDFs files **g)** terms from all mentioned sources.

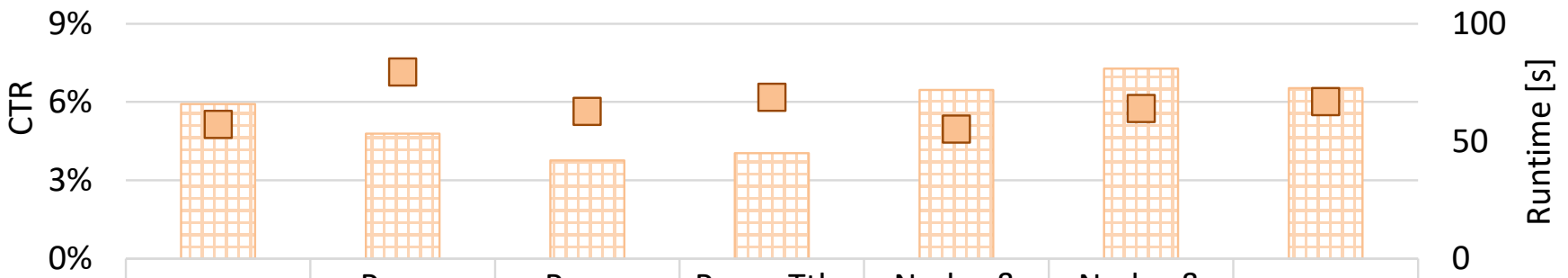

| | Nodes | Paper Titles (Cited) | Paper Titles (PDF) | Paper Ttls. (Cited + PDF) | Nodes & Titles (Cited) | Nodes & Titles (PDF) | All |
|---|---|---|---|---|---|---|---|
| Rntm (Terms) | 66 | 53 | 42 | 45 | 72 | 81 | 73 |
| CTR (Terms) | 5.13% | 7.14% | 5.63% | 6.17% | 4.96% | 5.74% | 6.00% |

Source of the terms

Figure 55: CTR based on the source from which terms were extracted

The highest CTR (7.14%) was achieved for utilizing terms from the cited paper titles (Figure 55). Utilizing terms from the titles of all PDFs that users had linked in their mind maps led to a CTR of 5.63% on average. Utilizing terms from the mind maps’ nodes led to a CTR of 5.13% on average. The other approaches also achieved CTRs around 5% and 6%. We would not conclude that titles of citations are generally more effective than nodes in mind maps. The approaches were all rather simple. With appropriate enhancements, all approaches probably could perform more effectively (in the next section we show that a mind-map-specific user-modeling approach is twice as effective as the simple node approach). In addition, cited papers are not commonly available in mind-mapping applications, but only in Docear. However, based on the numbers we conclude that mind maps are in the same league for user modeling, as are research papers, since the CTRs are comparable. The numbers support our hypothesis that developers of mind-mapping tools should integrate recommender systems in their tools, and that these recommender systems will achieve similar performances as recommender systems in other domains.

### 5.3.7 Additional Observations

During our research, we made a few observations that do not necessarily relate to mind-map-specific user modeling, but that might be interesting for the general recommender-system community.

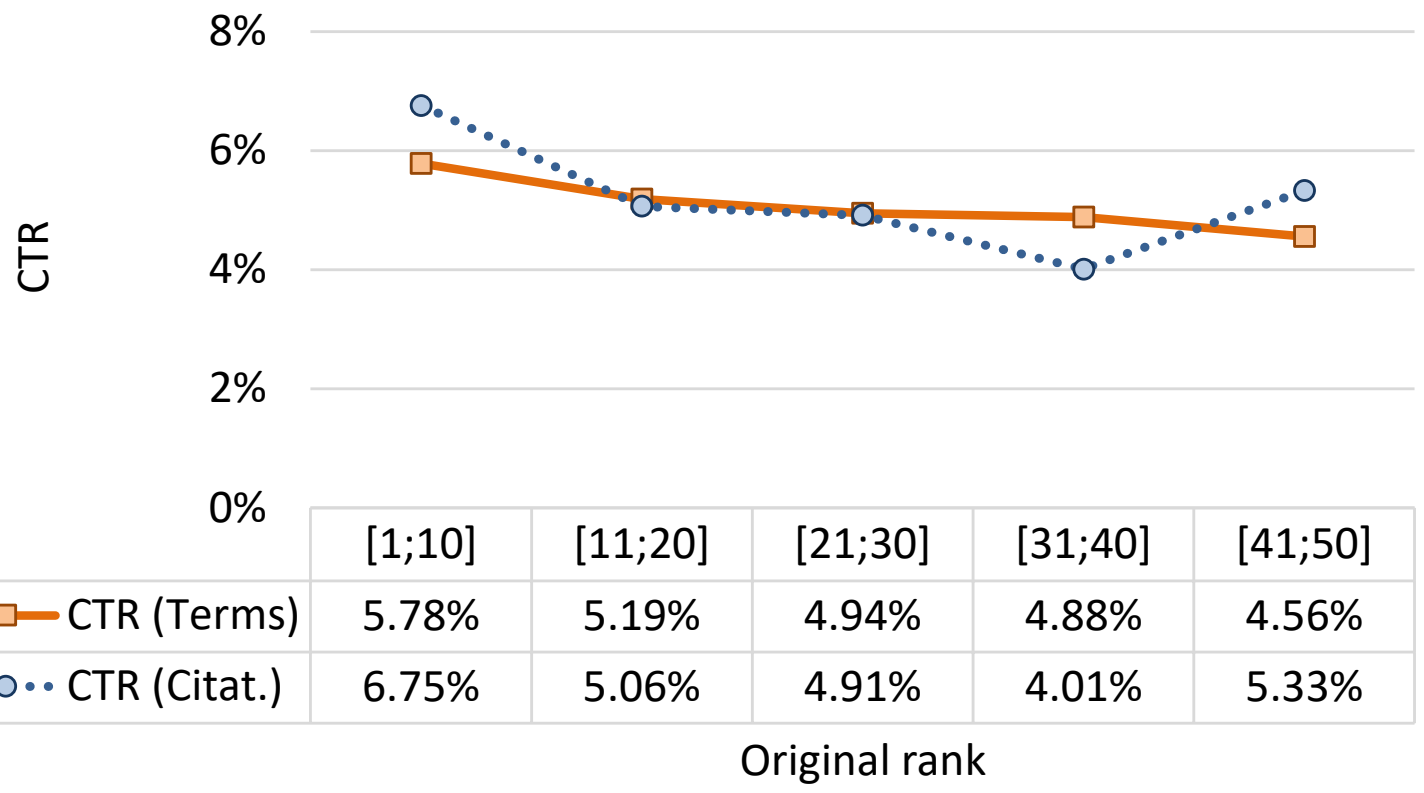

| | [1;10] | [11;20] | [21;30] | [31;40] | [41;50] |
|---|---|---|---|---|---|
| CTR (Terms) | 5.78% | 5.19% | 4.94% | 4.88% | 4.56% |
| CTR (Citat.) | 6.75% | 5.06% | 4.91% | 4.01% | 5.33% |

Figure 56: CTR based on the recommendation's original rank

To provide greater variety to users, Docear's recommender system randomly chose the ten final recommendations from the top 50 recommendation candidates. This method increases variety but decreases CTR on average (Figure 56). Those recommendations that were originally among the top 10 candidates, achieved a CTR of 5.78% on average. For lower ranked recommendations, CTR continuously decreased down to 4.56% for recommendations that were among the top 41 to 50 candidates. For citation-based recommendations, the trend was similar.

Before Docear displays the finally chosen recommendations, the recommendations are shuffled. This allowed us to analyze the effect that the display-rank of a recommendation has on CTR (Figure 57). Those recommendations displayed at position 1 had the highest CTR on average (6.73%), while recommendations shown in the middle of the list had the lowest CTR (4.37% on position 5). For the last positions, CTR again increased a little bit (5.31% for position 10). This means, just the recommendation rank made a difference in CTR of up to 50% (6.73% vs. 4.37%). The discovery that the recommendation rank affects CTR is not new, but, to the best of our knowledge, it has not been empirically quantified in the domain of research-paper recommender systems.

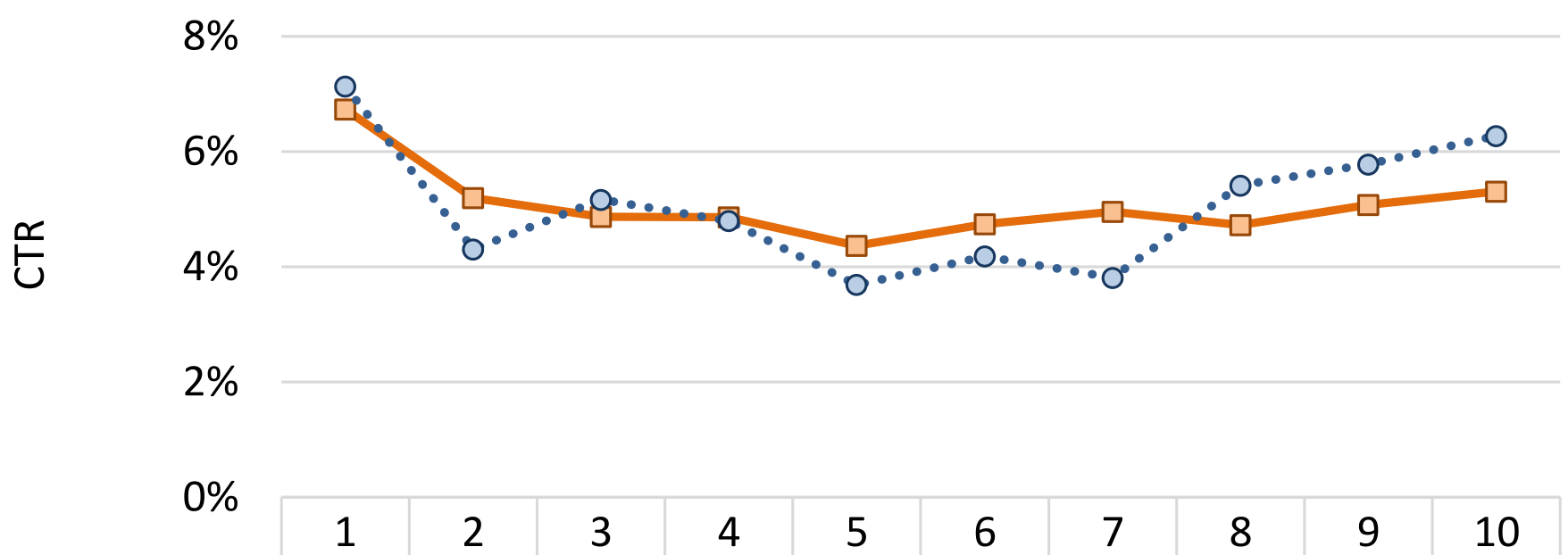

| | 1 | 2 | 3 | 4 | 5 | 6 | 7 | 8 | 9 | 10 |
|---|---|---|---|---|---|---|---|---|---|---|
| CTR (Terms) | 6.73% | 5.19% | 4.87% | 4.86% | 4.37% | 4.74% | 4.96% | 4.73% | 5.08% | 5.31% |
| CTR (Citat.) | 7.13% | 4.31% | 5.17% | 4.80% | 3.69% | 4.18% | 3.81% | 5.41% | 5.78% | 6.27% |

Presentation rank

Figure 57: CTR based on the rank at which a recommendation was displayed

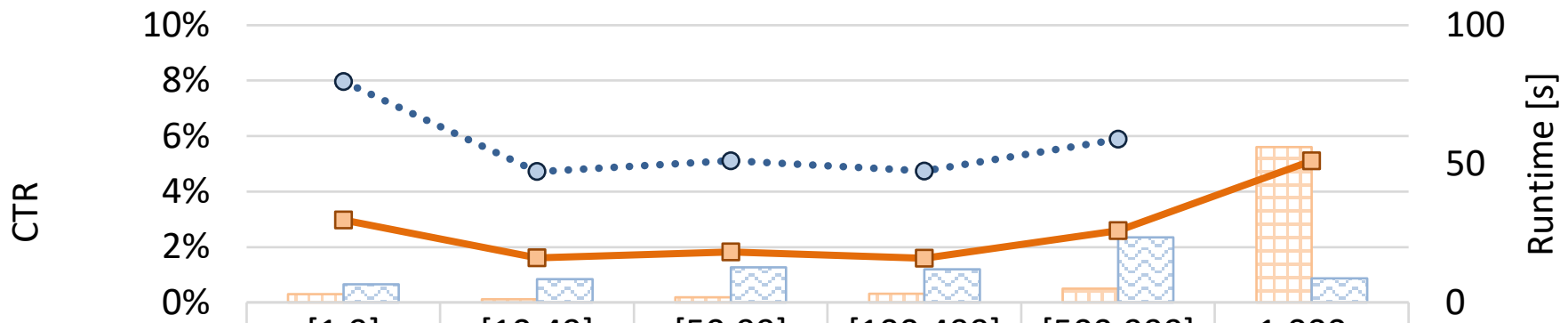

| | [1;9] | [10;49] | [50;99] | [100;499] | [500;999] | 1,000+ |
|---|---|---|---|---|---|---|
| Rntm (Terms) | 3 | 1 | 2 | 3 | 5 | 56 |
| Rntm (Citat.) | 7 | 8 | 13 | 12 | 23 | 9 |
| CTR (Terms) | 2.97% | 1.61% | 1.82% | 1.60% | 2.60% | 5.10% |
| CTR (Citat.) | 7.96% | 4.72% | 5.11% | 4.74% | 5.88% | |

Number of recommendation candidates

Figure 58: CTR based on available recommendation candidates

We also observed that when Lucene returned less than 1,000 recommendation candidates for a particular user model, the CTR tended to be lower as if 1,000 or more candidates were returned, in particular for term-based recommendations (Figure 58). For 1,000 or more recommendation candidates, CTR was 5.10% on average. In contrast, when 50 to 99 candidates were returned, CTR was 1.82% on average. This means, the overall CTR could be increased when user models that lead to less than 1,000 recommendation candidates are discarded, and a new user model is created instead. We also observed that CTR was comparatively high, when Lucene returned less than ten recommendation candidates. While this might seem surprising on first glance, there is a plausible explanation. If less than ten candidates are returned, then less than ten recommendations can be displayed. However, the less recommendations are shown in general, the higher the CTR tends to be. For citation-based recommendations, the maximum number of recommendation candidates never was above 1,000.

## 5.4 Docear's Mind-Map-Specific User-Modeling Approach

The so-far results make clear that numerous variables influence the effectiveness of mind-map-based user modeling. We combined the optimal values of these variables in a single algorithm as follows: The algorithm used the 75 most recently moved nodes from the past 90 days that were visible. If less than 75 moved and visible nodes were available, then *up to* 75 most recently modified nodes from the past 90 days were used instead. The nodes were extended by their children and siblings. Nodes were weighted based on depth and number of siblings (we used the *ln* weighting and summed the individual scores). The terms of these nodes were additionally weighted with TF-IDuF (stop-words were removed). The 35 highest weighted terms were stored in the user model as un-weighed list. This user model was used for the matching process with the recommendation candidates.

Among the baselines, using all terms from all mind maps and from a single mind map performed alike in terms of CTR (Figure 59) and ratings (Figure 60). Using terms from only one node – the approach that MindMeister applied – resulted in the lowest CTR (1.16%) and ratings (1.63). Stereotype recommendations performed comparably reasonable with a CTR of 3.08% and a rating of 1.97 on average. Overall, CTR of the baselines tends to be lower than in our initial study (cf. Appendix B, p. 179). However, since our previous evaluation, we added several new variables, and some might have decreased CTR on average. In addition, Docear's Web Spider was not running in the past couple of months. This means, no new recommendation candidates were added to the corpus. Hence, long-time users probably often received recommendations they had received previously, which decreases average CTR (Appendix H, p. 259). The reasonable effectiveness of stereotype recommendations might seem surprising, considering how rarely used this approach is in the recommendation community. Nevertheless, the result is plausible. Most of Docear's users are researchers and therefore they should be interested in books about academic writing, and hence click the corresponding recommendations. Even though the ratings are not very high, further research about stereotype recommendations might be promising.

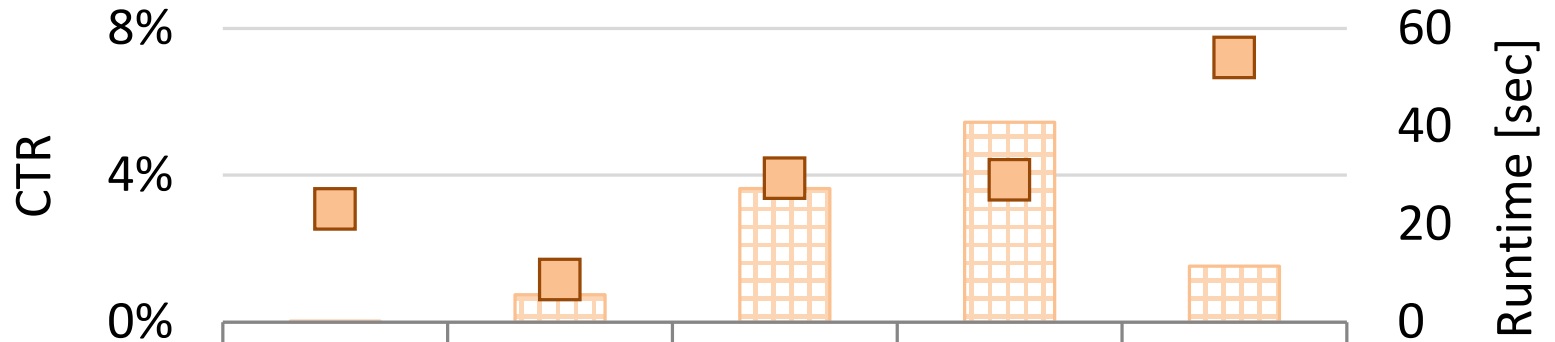

| | Stereo type | Single Node | Single Mind-Map | All Mind-Maps | Docear |
|---|---|---|---|---|---|
| Rntm [s] | 0.20 | 5.57 | 27.24 | 40.86 | 11.37 |
| CTR | 3.08% | 1.16% | 3.92% | 3.87% | 7.20% |

Figure 59: Docear’s mind-map-specific approach vs. baselines (I)

Docear’s mind-map-specific user modeling algorithm significantly outperformed all baselines and achieved a CTR of 7.20% on average (Figure 59). This is nearly twice as high as the best performing baseline and six times as high as MindMeister’s approach, the only approach that had been applied in practice thus far. User ratings also show a significantly higher effectiveness for Docear’s approach (3.23) than for the best performing baseline (2.53)[66]. Because we experimented only with a few variables, and the experiments were of relative basic nature, we are convinced that more research could further increase the effectiveness.

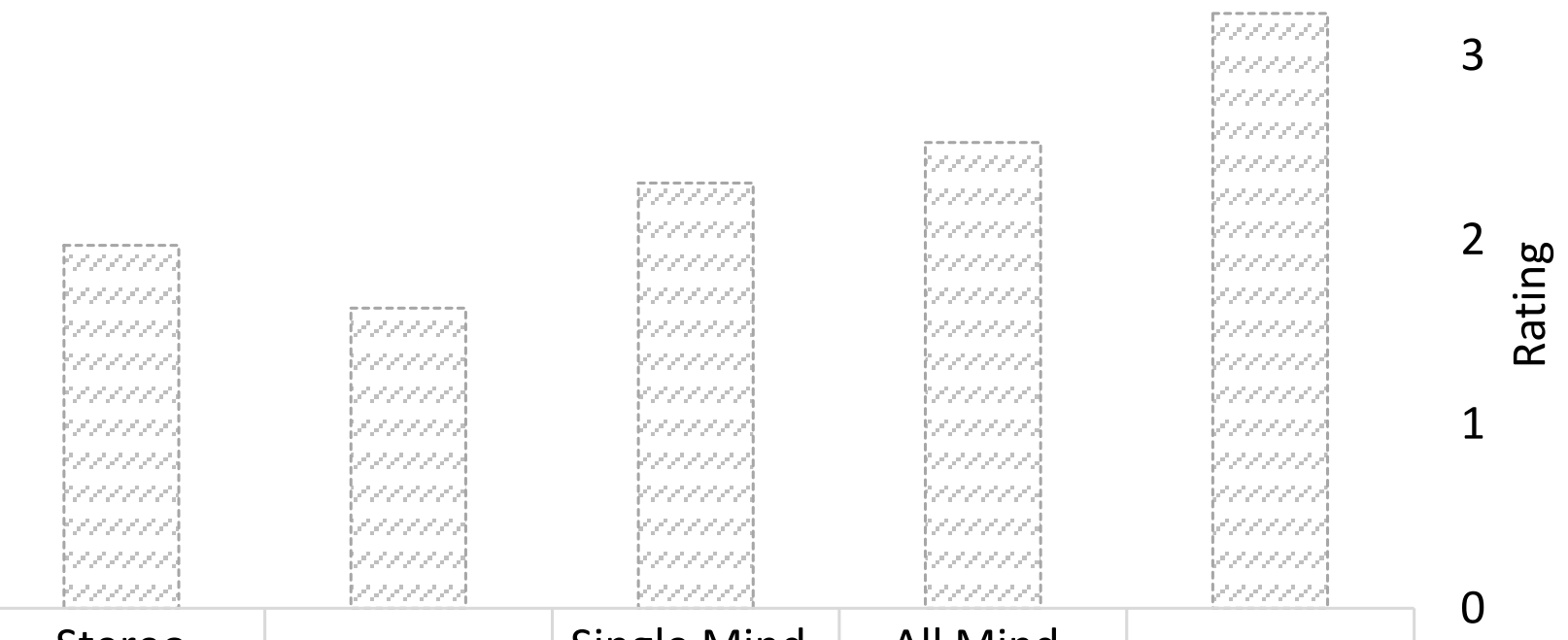

| | Stereo type | Single Node | Single Mind-Map | All Mind-Maps | Docear |
|---|---|---|---|---|---|
| Ratings | 1.97 | 1.63 | 2.31 | 2.53 | 3.23 |

Figure 60: Docear’s mind-map-specific approach vs. baselines (II)

[66] Differences are statistically not significant due to a small number of users the ratings are based on.

# 6. Summary and Contributions

## 6.1 Overview

Mind maps are widely neglected by the user modeling and recommender-systems community, and applications such as recommender systems are rarely found in mind-mapping tools. Given the popularity of recommender systems in other domains, we assumed that recommender systems could also benefit the more than two million mind-mapping users. Hence, we defined the following research objective:

**Develop an effective user-modeling approach based on mind maps**

By pursuing this objective, we expected to obtain a user-modeling approach that is more effective than standard user-modeling approaches applied to mind maps. Such an approach would enable developers of mind-mapping applications to integrate, for instance, recommender systems in their applications for providing additional value to their users.

The research objective let to three research questions and the following answers:

i. Which existing user-modeling approaches could serve as a basis for mind-map-based user modeling?

Based on a literature survey and research that we conducted, we concluded that content-based filtering (CBF) is the most promising recommendation *class* for the application of mind maps. However, from the CBF *approaches* that we reviewed, none seemed particularly promising for the application of mind maps. This was primarily due to the neglect of user modeling in most recommendation approaches, inadequate evaluations, and sparse information in publications. Hence, standard user-modeling approaches, such as utilizing all terms from a mind map, seemed to be the most plausible starting position for mind-map-based user modeling. As our initial study showed, such approaches achieved a decent effectiveness in terms of click-through rate.

ii. Could the effectiveness of the existing approaches be increased by adjusting them to the special characteristics of mind maps?

We identified a number of mind-map-specific variables that potentially would affect user modeling. We evaluated these variables and combined them into a single mind-map-specific user-modeling approach that was around twice as effective as the best baseline. Hence, the answer to the question is that the effectiveness of user modeling can increase significantly when the user-modeling process is adjusted to the special characteristics of mind maps.

iii. How should the effectiveness of user-modeling approaches be measured?

A key prerequisite for our research was the use of adequate evaluation methods to measure the effectiveness of user-modeling and recommendation approaches. Our research shows that both user studies and online evaluations generally are adequate methods, at least in our scenario. In online evaluations, click-through rate seem most adequate since CTR correlates best with user rating, and CTR is directly relevant for operators that run pay-per-click schemes. Offline evaluations provided less meaningful results, and at least in the field of research-paper recommender systems, they suffer from some inherent flaws, which is why we concluded not to use them.

The answers to the three research questions were achieved by pursuing five tasks that are summarized in the subsequent sections, along with the contributions that were made while pursuing the tasks.

## 6.2 Task 1: Survey Related Work

To find promising recommendation approaches (question *i*) and adequate evaluation methods (question *iii*), we reviewed 217 research articles in the field of research-paper recommender systems and articles about recommender-systems evaluation. We used all 217 articles for quantitative analyses regarding citation counts and other bibliographic metrics. For an in-depth analysis, 90 articles were excluded because we considered them of little significance (due to serious English errors, or non-original ideas and inadequate evaluations), or because they were published too late to be included in the in-depth review.

The survey leads to three contributions.

### 6.2.1 Contribution 1: Overview of Research-Paper Recommender Systems

The first contribution is an overview of the existing approaches in the field of research-paper recommender systems (cf. Chapter 3, p. 29 and Appendix F, p.

219). This overview aids researchers and practitioners in gaining an understanding of the research field and the recommendation approaches being applied.

We found that content-based filtering (CBF) is the predominant recommendation class in the field of research-paper recommender systems. Of 70 reviewed approaches, 34 used CBF (49%). From these approaches, the majority utilized plain terms contained in the documents. Some used n-grams, or topics based on LDA. A few approaches also utilized non-textual features such as citations or authors. The most popular model to store item representations was the Vector Space Model.

Only nine approaches (13%) applied collaborative filtering, and none of them used explicit ratings. Implicit ratings were inferred from the number of pages the users read, users' interaction with the papers (downloads, edits, views, etc.), and citations. The main problem of collaborative filtering for research papers seems to be sparsity. André Vellino compared the available (implicit) ratings on Mendeley (research papers) and Netflix (movies), and found that the number of ratings on Netflix differs from the number on Mendeley by a magnitude of three.

Six of the reviewed approaches (9%) were co-occurrence recommendations. Three of them analyzed how often papers were *co-viewed* during a browsing session. In that approach, whenever users browsed a paper, the system recommended those papers that had frequently been co-viewed with the browsed paper in the past. Another approach used co-citations to calculate document relatedness. The higher the proximity of two references within a paper, the more related they are assumed to be. *Pohl et al.* compared the effectiveness of co-citations and co-downloads and found that co-downloads are only more effective than co-citations in the first two years after a paper's publication.

Eleven recommendation approaches built graphs to generate recommendations (16%). Such graphs typically included papers that were connected via citations. Some graphs included authors, users/customers, venues, genes and proteins, and publishing years of the papers. *Lao et al.* even included terms from the papers' titles in the graph. Depending on the entities in the graph, connections included citations, purchases, "published in" relationships, authorship, relatedness between genes, and occurrences of genes in papers. Some authors connected the entities based on non-inherent relations. For instance, *Huang et al.* and *Woodruff et al.* calculated text similarities between items and used the text similarity to connect papers in the graph. Other connections were based on attribute similarity, bibliographic coupling, co-citation strength, and demographic similarity.

### 6.2.2 Contribution 2: Identification of Several Shortcomings in Research-Paper Recommender-Systems Research

The second contribution of the survey is an analysis of the research-paper recommender field in general and its major shortcomings.

One shortcoming relates to the evaluations. Of the reviewed approaches, 21% were not evaluated at all. Of the remaining approaches, 19% were not evaluated against a baseline, and most of the other approaches were compared only to simple baselines. The evaluations were mostly conducted with offline evaluations based on inferred ground-truths, which are subject to various criticisms. In addition, the offline evaluations were often based on datasets pruned in ways that we would consider inadequate[67]. The majority of the user studies (58%) had fewer than 16 participants, which also raises doubts of the significance of these evaluations. Only 7% of the approaches were evaluated with online evaluations in real recommender systems with real users.

Another shortcoming was related to too little information provided by the authors. The sparsity of information makes a re-implementation of the approaches difficult, if not impossible. For instance, most authors did not report on the text fields they utilized, and which weighting schemes were used.

In addition, much of the research was done in “the ivory tower”: research results often are neither transferred into practice, nor considered by peers. Despite the large number of research articles, there are only a handful of active recommender systems, and most of them apply simple recommendation approaches that are not based on any recent research results. As such, the extensive research that has been conducted in the past 16 years has apparently had only a minor impact on the practice of research-paper recommender systems in the real world. Additionally, several of the active recommender systems do not engage in the research community and seem reluctant to publish information on their systems.

Many researchers also seem to be unaware of developments in related research domains such as scientometrics or the reviewer-assignment problem, and the major co-author groups in the domain of research-paper recommender systems seem not to cooperate with each other.

[67] For instance, Pennock et al. removed all documents with fewer than 15 implicit ratings from the corpus. Therefore, 1,575 papers remained from the original 270,000 (0.58%). Results based on such datasets do not allow drawing reliable conclusions how the approaches might perform in real-world recommender systems

The majority of authors took no note of the fact that user satisfaction depends not only on accuracy, but also on factors such as privacy, data security, diversity, serendipity, labeling, and presentation. The operator perspective was also widely neglected. Information about runtime was provided for only 11% of the approaches. Complexity was covered by very few authors, and the actual costs of running a recommender system were only reported in a single article. We also noted that too many authors neglected the user-modeling process: 79% of the approaches let their users provide keywords, text snippets, or a single input paper to represent their information needs. Only a few approaches automatically inferred information from the users' authored, tagged, or otherwise-connected papers.

The analysis made clear that one cannot currently identify promising research-paper recommendation approaches, neither in general nor for the purpose mind-map-based user modeling. This means that we could not completely answer research question *i*. Hence, we decided to use standard content-based filtering approaches as the basis for our research instead of any of the reviewed approaches.

### 6.2.3 Contribution 3: Showing the Need for More Research on Recommender-Systems Evaluation

The survey revealed that there is uncertainty in the community about how to evaluate recommender systems. The most common applied evaluation method is offline evaluations, but this method is subject to criticism. However, user studies and online evaluations are also not without criticism. Hence, the survey showed the need for a more thorough analysis of the adequacy of evaluation methods and metrics for both Docear's particular scenario and in general.

## 6.3 Task 2: Develop a Recommender System for Docear

Since existing recommendation datasets and architectures are not suitable for researching mind-map-based user modeling, we decided to develop a research-paper recommender system for Docear. This recommender system was the foundation of our research and answering the second and third research questions. In addition, the development of the recommender system led to two further contributions, namely the architecture and the datasets that we published.

### 6.3.1 Contribution 4: Docear's Recommender-System Architecture

Docear's recommender-system architecture is unique in the domain of research-paper recommendations and mind-map-based user modeling. Most of the previously published architectures are rather brief, and architectures such as bX and BibTip focus on co-occurrence based recommendations. These approaches are primarily relevant for recommender systems with many users. Docear's architecture is comprehensive, explaining the individual components, the required hardware, and the integrated software libraries. Hence, the architecture should provide a good introduction for new researchers and developers on how to build a research-paper recommender system (based on mind maps). Due to the focus on content-based filtering, the architecture is also relevant for building recommender systems for applications with rather few users.

### 6.3.2 Contribution 5: Docear's Datasets

The datasets are also unique. While the research-paper dataset is rather small, and the metadata is probably of rather low quality, the dataset contains 1.8 million URLs to freely accessible full-text articles from various research fields and languages, and the dataset contains information where citations in a paper occur. The mind-map dataset is smaller than the dataset, e.g. of Mendeley, but it was not pruned, and hence allows for analyses for users with less than 20 papers in their collections. The dataset also contains information on how often a paper appears in a mind map. This information could be used to infer implicit ratings that are not only binary (linked/not linked) but also to weight the implicit rating. The datasets about Docear's users and recommendations contain extensive information, including user demographics, the number of received and clicked recommendations, and specifics about the algorithms with which recommendations were created. This data allows for analyses that go beyond those that we already performed, and should provide a rich source of information for researchers who are interested in recommender systems or for the use of reference managers.

## 6.4 Task 3: Identify Adequate Evaluation Methods and Metrics

Most recommender systems are evaluated with offline evaluations, although offline evaluations are subject to strong criticism. We did not want to take the risk of measuring the effectiveness of our research with inappropriate methods. Therefore, we analyzed and discussed the adequacy of offline evaluations and its two alternatives, namely user studies and online evaluations. The discussion was based on the comparison of results from three different evaluation methods. As part of an online evaluation, Docear displayed 45,208 recommendation sets with 430,893 recommendations to 4,700 users from March 2013 to August 2014. In a user study, 379 users rated 903 recommendation sets with 8,010 recommendations, and in an offline evaluation, 118,291 recommendation sets were generated and analyzed for their accuracy. To the best of our knowledge, we are first in the field of recommender systems to compare the three evaluation methods and various metrics, and to provide a detailed discussion on the appropriateness of the evaluations methods, metrics, and ground-truths. The research led to two contributions.

### 6.4.1 Contribution 6: Showing the Inadequacy of Offline Evaluations

One contribution was the confirmation that offline evaluations based on inferred ground-truths only sometimes predict recommender effectiveness in online evaluations and user studies. For instance, based on our offline evaluation, the stereotype approach would have never been considered a worthwhile option for further evaluation. In practice however, stereotype recommendations received reasonable ratings, click-through rates and link-through rates that were not much lower than for the other approaches. Similarly, the offline evaluation indicated that *term*-based CBF was – depending on the metric – around five to fifteen times more effective than *citation*-based CBF. Based on these numbers, citation-based CBF would have never been considered a promising approach. In the online evaluation and user study, however, citation-based CBF was only slightly less effective than term-based CBF, and certainly an interesting approach.

To assess the adequacy of offline evaluations, we concluded that three questions needed to be answered:

1. Can we identify scenarios where offline evaluations will have predictive power?

If the community could determine which factors affect the predictive power of offline evaluations, offline evaluations could be applied only in scenarios where the factors are not present. However, we assume that it will not be possible to determine such scenarios. Hence, we conclude that offline evaluations, based on inferred ground truths, should be abandoned unless their results are shown to have inherent value.

2. Do results of offline evaluations have some inherent value?

Results of offline evaluations might have inherent value that could make offline evaluations a worthwhile evaluation method, even if results do not correlate with results from online evaluations and user studies. Such inherent value might exist in a student-teacher scenario, when those compiling a ground-truth know what is relevant better than the users receiving recommendations. A student-teacher scenario might occur, in particular, for offline evaluations based on expert-ground-truths, e.g. datasets compiled by experts in their field. However, expert ground-truths suffer from the problem of overspecialization and we doubt that there will ever be an appropriate expert-dataset to comprehensively evaluate (research-paper) recommender systems for different research fields and groups of users such as undergraduates, postgraduates, doctoral students, professors, and foreign students. Inferred ground-truths, i.e. datasets, inferred e.g. from citations or users' personal document collections, do not suffer from overspecialization. Hence, theoretically, inferred ground-truths could have inherent value but they suffer from a fundamental problem.

3. Are offline evaluations generally flawed?

We argue that inferred ground-truths are generally flawed, at least in the domain of research-paper recommender systems. Since researchers do not have perfect knowledge of their domains, the datasets are incomplete. If datasets are based on citations, the datasets additionally suffer from citation bias that makes the datasets biased. Consequently, evaluations based on inferred ground-truths only assess how accurately a recommendation approach recommends the imperfect ground-truths. Such an assessment is not useful. We conclude that offline evaluations based on inferred ground-truths should probably not be used for evaluating (research-paper) recommender systems.

### 6.4.2 Contribution 7: Showing the Adequacy of Online Evaluations and User Studies

Another contribution was to show that ratings in user studies strongly correlated with online-evaluation metrics, particularly with CTR. However, in some situations ratings and CTR led to different results. For instance, CTR and ratings led to different results when comparing the effect of labels and the trigger to generate recommendations. When analyzing things such as user-model size, the number of nodes to utilize, and stop-word removal, CTR and ratings strongly correlated. Apparently, a discrepancy between CTR and ratings is more likely for measuring the effect of factors that do not directly relate to the recommendation algorithm but to human factors. Therefore, we conclude that both online evaluations and user studies are equally well suited for evaluating recommender systems. Ideally, both methods should be used, but also applying only one of the two methods should be considered good practice.

Regarding metrics in online evaluations, there are some noteworthy differences. Annotation-through rate (ATR) and citation-through rate (CiTR) have the advantage of being based on thorough assessments of the recommendations. However, they require more users and delivered recommendations to receive statistically significant results compared to click-through rate (CTR) or link-through rate (LTR). Consequently, applying ATR and CiTR is only feasible in large-scale recommender systems. In addition, ATR and CiTR predicted user satisfaction for stereotype recommendations incorrectly. As such, at least for the scenario of Docear, CTR seems most appropriate. For other scenarios, a thorough assessment of the appropriateness of metrics is needed for each online evaluation. Ideally, multiple metrics would be used.

## 6.5 Task 4: Identify Mind-Map-Specific User-Modeling Variables

We experimented in Docear's recommender system with several variables that were randomly assembled to create user models. For instance, the recommender system randomly chose whether to store, for example, the 10, 50, 100, or 1000 highest weighted terms or citations as a user model. Experimenting with the variables led to the following contribution.

### 6.5.1 Contribution 8: Identification and Evaluation of Mind-Map-Specific Variables

We showed that several variables affect the effectiveness of user modeling based on mind maps. Based on our research, the following variables have an effect: **a)** the number of analyzed nodes. It seems that the terms of the most recently modified 50 to 99 nodes are sufficient to describe the users' information needs. Using more, or fewer, nodes decreased the average CTR. **b)** Time restrictions were important. It seems that utilizing nodes that were created more than four months ago decreased CTR. **c)** CTR increased when only nodes were used that were recently *moved* by a user, instead of using nodes that were created or edited. **d)** Using only nodes that were visible in the mind map also increased effectiveness compared to using both visible and invisible nodes. **e)** Extending the originally selected nodes by adding siblings and children increased the average CTR slightly, but statistically significantly. This indicates that the full meaning of nodes becomes only clear when their neighbor nodes are considered. **f)** We also found that weighting nodes and their terms based on node depth and the number of siblings, increased CTR. The deeper a node, and the more siblings it has, the more relevant its terms to describe the users' information needs. The separate weights should be combined by their sum. **g)** The final user model should contain the highest weighted 26 to 50 terms if the user model is stored as un-weighted list. If weights are stored, it seems that larger user models are sensible. However, more research is needed to clarify this.

## 6.6 Task 5: Develop a Mind-Map-Specific User-Modeling Approach

Our research goal was to develop a mind-map-specific user-modeling approach. Therefore, we combined the identified variables in a single algorithm and compared this algorithm against several standard user-modeling approaches. The experiments led to three contributions.

### 6.6.1 Contribution 9: Evaluation of Standard User-Modeling Approaches Applied to Mind Maps

We showed that standard user-modeling approaches could be reasonably effective when applied to mind maps. However, the effectiveness varied depending on which standard approach was used. When user models were based on all terms of users' mind maps, the click-through rate (CTR) was around 4%[68]. When only terms from the most recently modified node were used, CTR was 1.16%. These results led us to conclude that user modeling based on mind maps is not trivial, and minor differences in the approaches lead to significant differences in effectiveness.

[68] CTR of the standard approaches was lower in the final evaluation than in our initial study (6%). See section 5.4 for details and an explanation.

### 6.6.2 Contribution 10: A Mind-Map-Specific User-Modeling Approach

When the variables were combined in their apparently favorable way, this mind-map-specific user-modeling approach outperformed standard user-modeling approaches applied to mind maps by a factor of nearly two (CTR of 7.20% vs. 3.92%). Compared to the approach that was applied in practice by MindMeister (using only the last modified node), our approach increased effectiveness by a factor of six (CTR of 7.20% vs. 1.16%). The user study confirmed the results of the online evaluation: the mind-map-specific user-modeling approach was significantly more effective than the baselines (rating of 3.23 vs. 2.53).

### 6.6.3 Contribution 11: Demonstrating the Potential of Mind Maps as Source for User Modeling

We compared the effectiveness of user modeling based on mind maps with the effectiveness of user modeling based on the user's PDFs and citations. We found that user modeling based on the citations' titles was most effective (CTR = 7.14%), while user modeling based on the mind maps had an average CTR of 5.13%. On one hand, this shows that for the particular scenario of Docear, it might be more sensible to utilize users' citations instead of mind maps, or maybe to combine them. However, most mind-mapping applications do not have access to users' PDFs or citations. Therefore, even though standard user modeling based on mind maps might be slightly less effective than standard user modeling based on other items, the results show that the effectiveness is in the same league. Hence, we see no reason why developers of mind-mapping applications should not integrate recommender systems in their applications – particularly because a mind-map-specific user modeling approach can further increase effectiveness. Consequently, we would expect that recommender systems in mind-mapping applications would lead to benefits similar to the benefits of recommender systems in other domains.

## 6.7 Further Contributions

As part of our research, we made the following further contributions.

We introduced TF-ID*u*F, a weighting scheme that is equally effective as TF-IDF, and that might be combined with TF-IDF (cf. Section 5.3.3, p. 114). In addition, we were first who empirically compared CC-IDF, i.e. TF-IDF applied to citations, with plain citation frequency (cf. Section 5.3.3, p. 114). The results indicate that CC-IDF might be less effective than a simple citation-count measure. However, more research is needed to clarify this. In the domain of research-paper recommender systems, the finding that a user-model size should be between 26 and 50 terms is also novel (cf. Section 5.3.4, p. 116). The finding that researchers' interests shift after about four months might also prove useful for other research-paper recommender systems (cf. Section 5.3.1.2, p. 106). To the best of our knowledge, it has also not been shown that the recommendation rank can affect CTR by up to 50% in the field of research-paper recommendations (cf. Section 5.3.5, p. 117).

As part of recommender-system development, we also developed *SciPlore Xtract* (cf. Appendix G.1, p. 249), and its successor *Docear's PDF Inspector* (cf. Appendix G.2, p. 253). Both tools extract titles from academic PDF files by applying a simple heuristic: the largest text on the first page of a PDF is assumed the title. This simple heuristic achieves accuracy of around 70% and outperforms machine-learning-based tools like ParsCit in both run-time and accuracy. Docear's PDF Inspector was released under the free open source license GPL 2+ at http://www.docear.org, written in JAVA, and runs on any major operating system. The dataset for its evaluation is also publicly available at http://labs.docear.org.

We conducted an exploratory study of 19,379 mind maps created by 11,179 users from the mind mapping applications Docear and MindMeister (cf. Appendix C, p. 193). The objective was to find out how mind maps are structured, what information they contain, and to identify potential information-retrieval applications that could utilize mind maps. The results include the discovery that a typical mind map is rather small, with 31 nodes on average (median), whereas each node usually contains between one to three words. The number of hyperlinks tends to be rather low but depends upon the mind mapping application. Most mind maps are edited only over one (60.76%) or two days (18.41%). A typical user creates around 2.7 mind maps (mean) a year. However, there are exceptions, which create a long tail. One user created 243 mind maps, the largest mind map contained 52,182 nodes, one node contained 7,497 words, and one mind map was edited on 142 days.

The analysis of the mind maps led to the preliminary study that created eight ideas about how mind maps could be utilized by information retrieval applications (cf. Appendix B, p. 179). We evaluated the feasibility of the eight ideas, based on estimates of the number of available mind maps, an analysis of the content of mind maps, and an evaluation of the users' acceptance of the ideas. We concluded that user modelling is the most promising application with respect to mind maps, which eventually led to the development of our mind-map-specific user-modeling approach.

Finally, to enhance Docear's recommender system, and to show that non-accuracy factors have a significant impact on recommender effectiveness (cf. 3.3.2, p. 42), we evaluated the impact of labeling, demographics, and persistence on the effectiveness of recommender systems.

With respect to labels, we showed that organic recommendations are preferable to commercial recommendations, even when they point to the same freely downloadable research papers (cf. Appendix J, p. 271). Simply the fact that users perceive recommendations as commercial decreased their willingness to click them. We further showed that the exact labeling of recommendations matters. For instance, recommendations labeled as "advertisements" performed worse than those labeled as "sponsored" did. Similarly, recommendations labeled as "*Free* Research Papers" performed better than those labeled as "Research Papers" did.

We also analyzed how click-through rates vary between research-paper recommendations previously shown to the same users and recommendations shown for the very first time. Our research indicates that recommendations should only be given once. Click-through rates for "fresh," i.e. previously unknown recommendations, are twice as high as for known recommendations. However, results also show that some users are "oblivious." Users frequently clicked on recommendations they already knew. In one case, the same recommendation was shown six times to the same user and the user clicked it each time. Overall, around 50% of clicks on re-shown recommendations were such "oblivious-clicks."

We further showed the importance of considering demographics and other user characteristics when evaluating (research-paper) recommender systems (cf. Appendix H, p. 259). We analyzed 37,572 recommendations delivered to 1,028 users and found that older users clicked more often on recommendations than younger ones. For instance, 20-24 years old users achieved click-through rates (CTR) of 2.73% on average while CTR for users between 50 and 54 years was 9.26%. Gender only had a marginal impact (CTR males 6.88%; females 6.67%) but other user characteristics such as whether a user was registered (CTR: 6.95%) or not (4.97%) had a strong impact. Due to the results, we argue that research articles on recommender systems should report detailed data on their users to make results better comparable (to learn about the demographics of Docear's users, please see Appendix H, p. 259).

Overall, 45 publications resulted from the work done as part of this dissertation [16–19, 21–52, 141–147, 230, 252].

# 7. Outlook

Our literature review revealed several shortcomings in current research-paper recommender-systems research. To eliminate these shortcomings, we consider it crucial that the community discusses and develops frameworks and best-practice guidelines for research-paper recommender-systems evaluation. This should include an analysis and discussion of how suitable offline evaluations are, to what extent datasets should be pruned, the minimum number of participants in user studies, and which factors influence the outcome of evaluations (e.g. user demographics). Ideally, a set of reference approaches that could be used as baselines would be implemented. In addition, more details on implementation are needed, based on a discussion of the information needed in research articles. It is also crucial to discover why apparently minor differences in algorithms, datasets, evaluations, etc. lead to major variations in evaluation results. Until the reasons for these variations are found, scholars cannot rely on existing research results because it is unclear whether the results can be generalized to any new recommendation scenario.

A step towards using the full potential behind research-paper recommender systems could be to establish a platform for researchers to publish and communicate, such as appropriate conferences or workshops focusing solely on research-paper recommender systems. An open-source recommender framework containing the most promising approaches could help bring the research results into practice. Such a framework would also help new researchers in the field access a number of baselines with which they could compare their own approaches. A framework could either be built from scratch or based on existing frameworks such as MyMediaLite[69], LensKit[70], Mahout[71], Duine[72], RecLab Core[73], easyrec[74], or Recommender101[75]. The community would probably also benefit from considering research results from related disciplines. In particular, research about user modeling and scientometrics seems highly promising to us, as well as research from the general recommender-systems community about non-accuracy aspects.

[69] http://www.mymedialite.net/
[70] http://lenskit.grouplens.org/
[71] http://mahout.apache.org/
[72] http://www.duineframework.org/
[73] http://code.richrelevance.com/reclab-core/
[74] http://easyrec.org/
[75] http://ls13-www.cs.uni-dortmund.de/homepage/recommender101/index.shtml

Our comparison of the different evaluation methods showed offline evaluations to be probably inadequate for evaluating research-paper recommender systems. However, the offline dataset by Docear might not be considered an optimal dataset due to the large number of novice users. A repetition of our analysis on other datasets, with more advanced users might lead to more favorable results for offline evaluations (nevertheless, our criticism about the imperfection of inferred ground-truths remains). Future research should also analyze the extent to which the limitations of offline datasets for research-paper recommender systems apply to other domains. We also conclude that the differences of CTR, LTR, ATR, and CiTR need more research to find out when which metrics are most appropriate. Finally, offline evaluations based on explicit ground-truths have been neglected in our analysis since they are not used in the domain of research-paper recommender systems, and we are not familiar with their use. We know that such ground-truths are widely used in other recommender domains (e.g. movies), and propose that a thorough analysis and discussion of explicit ground-truths is highly needed.

Our research on mind-map-specific user modeling showed that several variables affect the user-modeling effectiveness. So far, the values for the variables are only rough suggestions. For instance, our finding that the optimal user-model size is between 26 and 50 terms is still rather vague. Hence, more research is required to specify the optimal values of the variables. There are also more potential variables that we have not yet analyzed but that might be promising. For instance, the evolution of mind maps over time might enhance the effectiveness of mind-map-specific user modeling. We could imagine that weighting nodes by the intensity of use (e.g. how often a node was edited, opened, or moved) might provide valuable information. We also advocate research on the differences of content and the structure of mind maps created for different purposes, such as brainstorming or literature management. This might provide valuable insights on the characteristics of mind maps. More research is also needed to explore dependencies among the variables. This requires more advanced statistical analysis of the variables. This, however, requires research in large-scale recommender systems with significantly more users than Docear has. It should also be noted that our research was based only on Docear, which is a unique mind-mapping software tool, because it focuses on researchers. Additional research with other mind-mapping tools seems desirable. This is particularly true because most mind-mapping tools focus on certain groups of users, and it would be interesting to explore whether there is *one* mind-map-specific user-modeling approach that suits all mind-mapping applications, or whether each application needs to apply a different approach. Finally, most of our results regarding citations were statistically not significant. It would also be interesting to research in more detail how citations, or hyperlinks, could be exploited to enhance user modeling, or realize some of our other ideas how mind maps could benefit information-retrieval applications.

Overall, the results of our research reinforced our astonishment that mind maps are being disregarded by the user-modeling and recommender-system community. Our research showed the potential of mind-map-specific user modeling, and we hope that the results initiate a discussion that encourages other researchers to do research in this field. Our results should also help practitioners to implement a decently effective user-modeling approach. We hope this encourages developers of mind-mapping tools to integrate recommender systems in their software, which would create additional value for the millions of mind-mapping users.

# Appendix

# A List of Publications

The following research articles were published during pursuing my PhD. Some parts of the doctoral thesis were published in these articles.

<u>2014</u>

<u>2013</u>

2011

# B Preliminary Study[76]

## B.1 Introduction

Information retrieval (IR) applications utilize many items beyond the items' original purpose. For instance, emails are intended as a means of communication, but Google utilizes them for generating user profiles and displaying personalized advertisement [150]; social tags can help to organize private webpage collections, but search engines utilize them for indexing websites [441]; research articles are meant to publish research results, but they, or more precisely their references, are utilized to analyze the impact of researchers and institutions [189].

We propose that mind-maps are an equally valuable source for information retrieval as are social tags, emails, research articles, etc. Consequently, our research objective was to identify, how mind-maps could be used to empower IR applications. To achieve our objective, we 1) analyzed the extent to which mind-mapping is used, to decide if mind-map based IR is a field worth researching, 2) brainstormed how mind-maps might be utilized by IR applications, 3) analyzed the feasibility of the ideas, and 4) implemented a prototype of the most promising idea, which – to anticipate the result – is a recommender system that creates user models based on mind-maps. All estimates are based on data collected from our own mind-mapping software *Docear* [33, 47], *Google Trends* and the mind-mapping tools' websites.

We hope to stimulate a discussion that encourages IR and user modelling researchers to further analyze the potential of mind-maps. We believe that researchers will find this new research field rewarding, and the results will enable developers of mind-mapping tools to devise novel services for their millions of users.

[76] This chapter has been published as: Beel, Joeran, Stefan Langer, Marcel Genzmehr, and Bela Gip. "Utilizing Mind-Maps for Information Retrieval and User Modelling." In *Proceedings of the 22nd Conference on User Modelling, Adaption, and Personalization (UMAP)*, edited by Vania Dimitrova, Tsvi Kuflik, David Chin, Francesco Ricci, Peter Dolog, and Geert-Jan Houben, 8538:301–313. Lecture Notes in Computer Science. Springer, 2014.

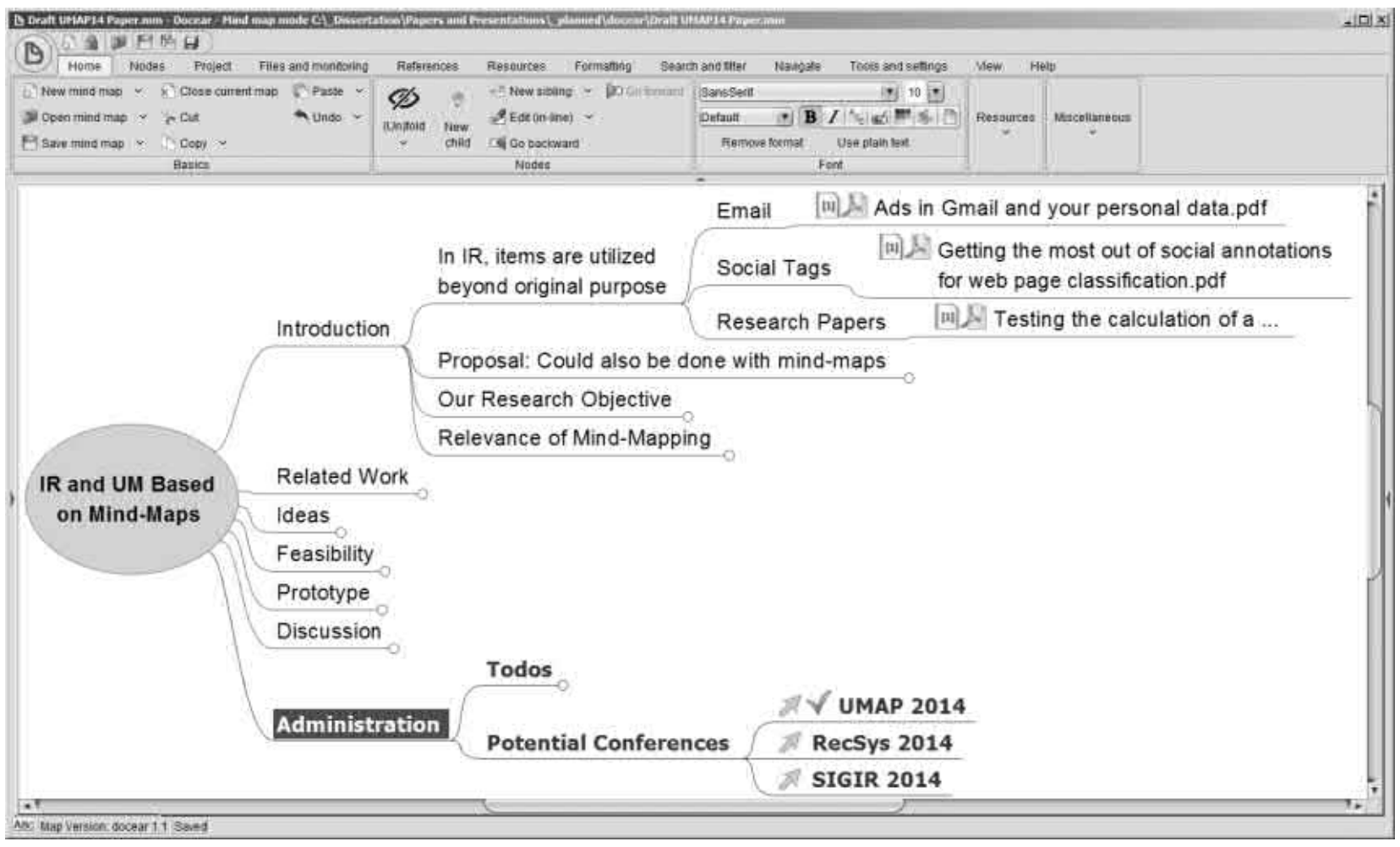

Figure 61: Mind-map example (draft of this chapter)

## B.2 Related Work

Mind-maps are typically used to develop ideas and organize information. As such, they are often used for tasks including brainstorming, project management, and document drafting. Figure 61 shows an example of a mind-map, created with our mind-mapping software *Docear (http://docear.org)* [33]. We created the mind-map to represent a draft of this chapter. The root node represents the title. From the root node, child nodes branch to represent each chapter, additional child nodes branch off for each paragraph, sentence and reference. We also added a list of relevant conferences, to which we planned to submit the paper. Red arrows indicate a link to a website. A PDF icon indicates a link to a PDF file on the hard drive. A “circle” on a node indicates that the node has child nodes that are currently hidden.

There has been plenty of research showing the effectiveness of mind-mapping as a learning tool [302]; creating mind-maps automatically from full-text streams [67]; and evaluating whether paper-based or electronic mind-mapping is more effective [261]. To the best of our knowledge, mind-maps have not been researched with regard to information retrieval or user modelling. However, there are two types of information retrieval applications, which utilized mind-maps in practice.

The first type of application is a search engine for mind-maps. Several mind-mapping tools, for instance *XMind* and *MindMeister*, allow their users to publish their mind-maps in so called “mind-map galleries”. These galleries are similar to photo galleries. They show thumbnails of mind-maps that users uploaded to the gallery. Visitors of the galleries may search for mind-maps containing certain keywords, and download the corresponding mind-maps. According to *MindMeister*, around 10% of mind-maps being created by their users are published in the galleries[77]. The other mind-maps remain private.

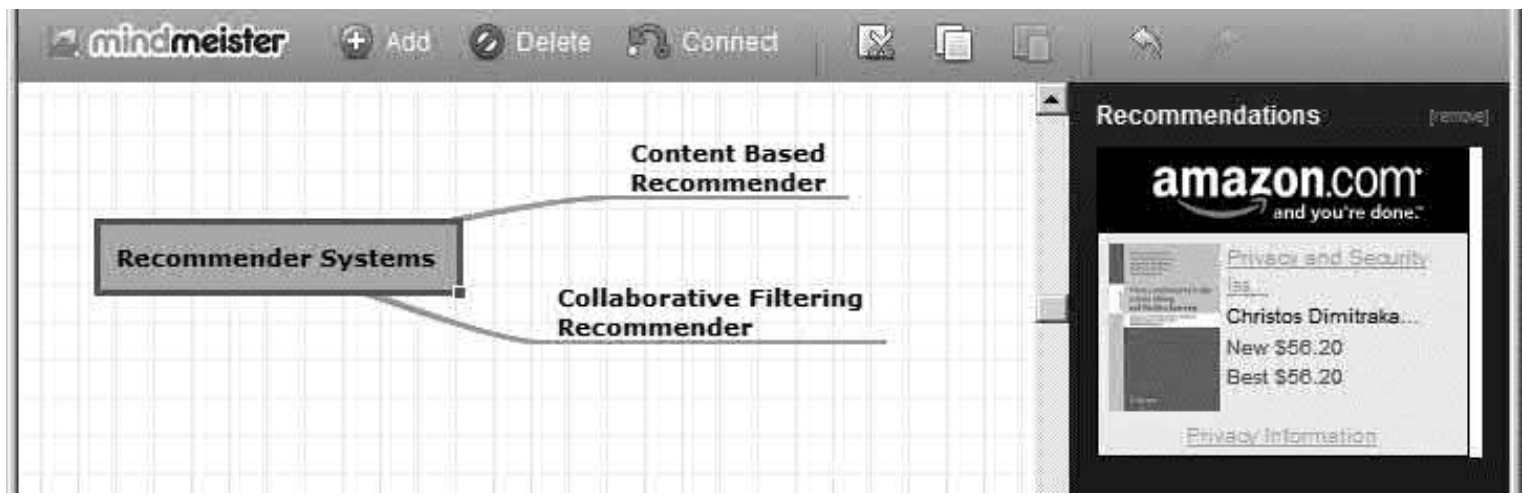

Figure 62: Personalized advertisement in MindMeister

The second type of application is a user modelling system. Only two companies – *MindMeister* and *Mindomo* – implemented such a system to generate user models and display personalized advertisement. *MindMeister* extracted the terms of the node that a user last edited or created – typically, a node contains two or three terms [40]. These terms were sent to Amazon's Web Service as search query. Amazon returned book recommendations matching the search query, which *MindMeister* displayed in a window besides the mind-map (Figure 62). *Mindomo* had a similar concept, only that Google AdSense instead of Amazon was used. Meanwhile, both companies abandoned personalized advertisement, though they still offer and actively maintain their mind-mapping tools. In an email, *Mindomo* said that "people were not really interested" in the advertisement[1].

## B.3 Popularity of Mind-Mapping

Some reviewers of previous papers were skeptical whether there is enough interest in mind-mapping to justify the effort for researching the potential of IR applications utilizing mind-maps. We believe this skepticism to be unfounded, because, as shown in the next paragraphs, there is a significant number of mind-mapping tools and users who could benefit from the research.

The popularity of *mind-mapping,* based on search volume, is similar to the popularity of e.g. *note taking*, *file management*, or *crowdsourcing*, and significantly higher than for *reference management*, *user modelling*, *recommender systems*, or *information retrieval* (Figure 63). The website *Mind-Mapping.org* lists 142 mind-mapping tools being actively maintained, although some tools offer mind-mapping only as secondary feature in addition to other visualization techniques, such as concept maps or Gantt charts. When discontinued tools are included in the count, there are 207 tools. Of the 'pure' mind-mapping tools, i.e. those that focus on mind-mapping functionality, *XMind* is the most popular tool, based on search volume (25%) (Figure 64)[78]. Other popular tools are *FreeMind* (23%), *MindManager* (13%), and *MindMeister* (8%). The search volume for *XMind* is in the same league as search volume for the *Dropbox* alternative o*wnCloud*, the reference manager *Zotero*, or the Blog *TechCrunch*, and the volume is significantly higher than for academic conferences such as *UMAP*, *SIGIR*, or *RecSys* (Figure 65).

[77] Email from MindMeister's CEO Michael Hollauf, June 28, 2011. Permission for publication was granted.

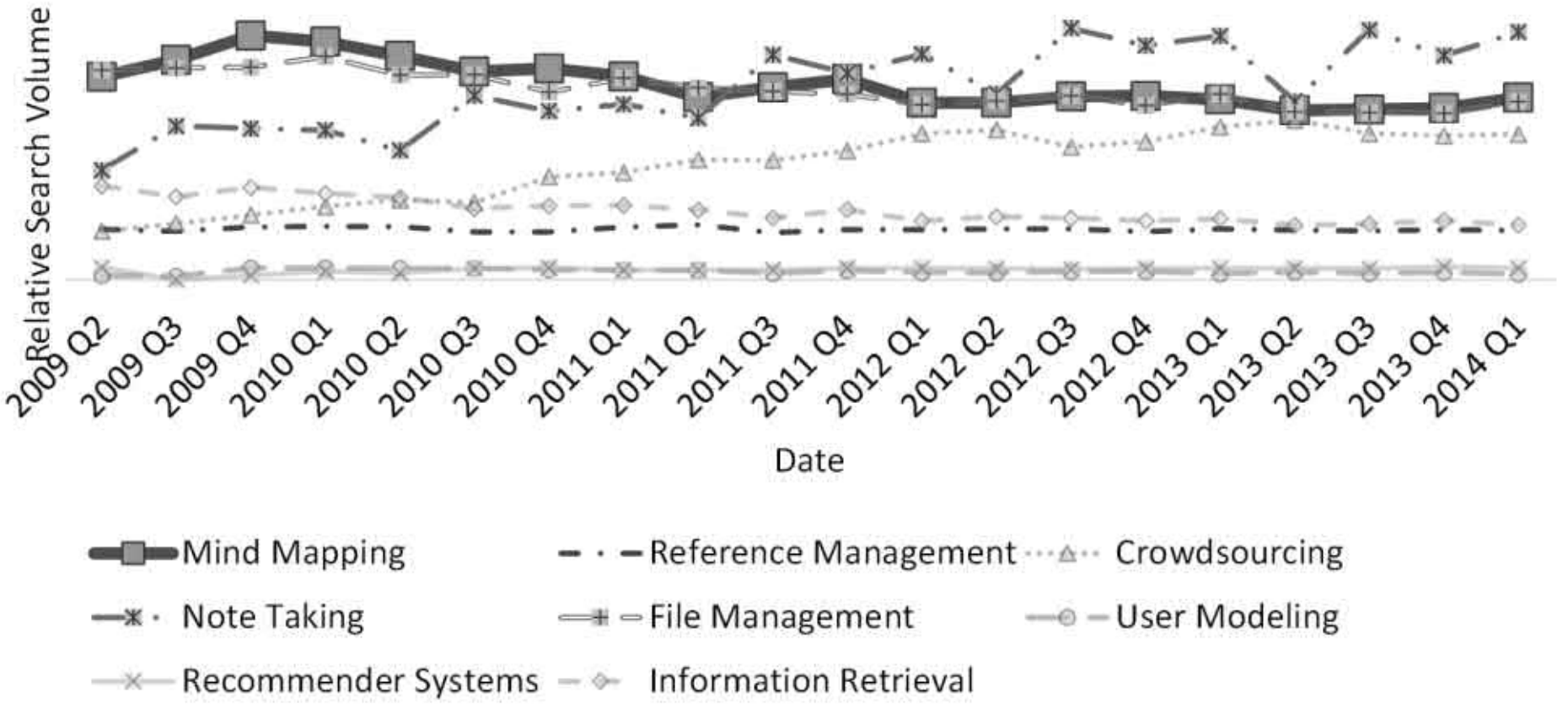

Figure 63: Search volume for selected search terms

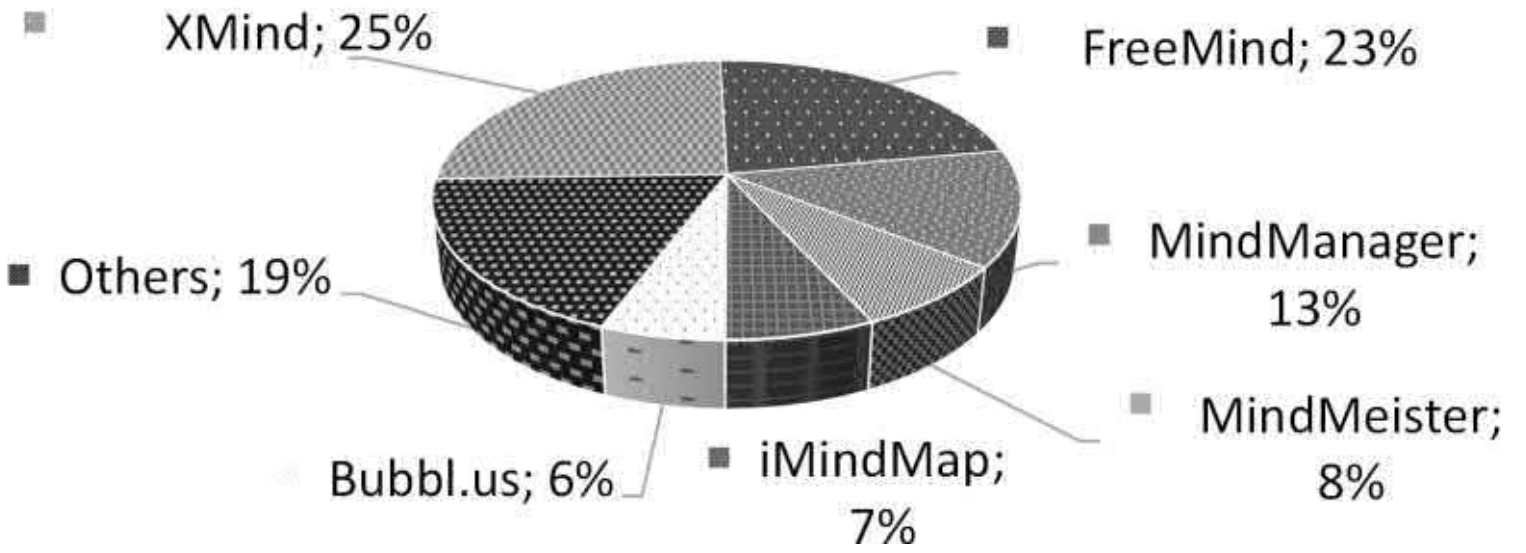

Figure 64: Search volume for mind-mapping tools

According to the tools' websites, *XMind* has more than 1 million users, *Bubbl.us* more than 1.5 million, *MindManager* more than 2 million, and *MindMeister* more than 2.5 million users. In sum, this makes 7 Million users for four tools that accumulate 52% of the search volume (Figure 64). Interpolating from the search volume, we can estimate that the remaining tools (48% of the search volume) must have around 6.5 million users. This results in a total of around 13.5 million mind-map users. To us, it seems likely that these numbers also include inactive users. For our own mind-mapping software *Docear*, 10 to 20% of the users who registered in the past years, are active, i.e. they started Docear in the past month. Based on this information, we may estimate the numbers of active mind-map users to be between 1.35 and 2.7 million.

[78] All numbers relating to search volume are based on Google Trends http://www.google.com/trends/. Search volume is calculated relatively by Google, as such there are no numbers to display on the y-axis.

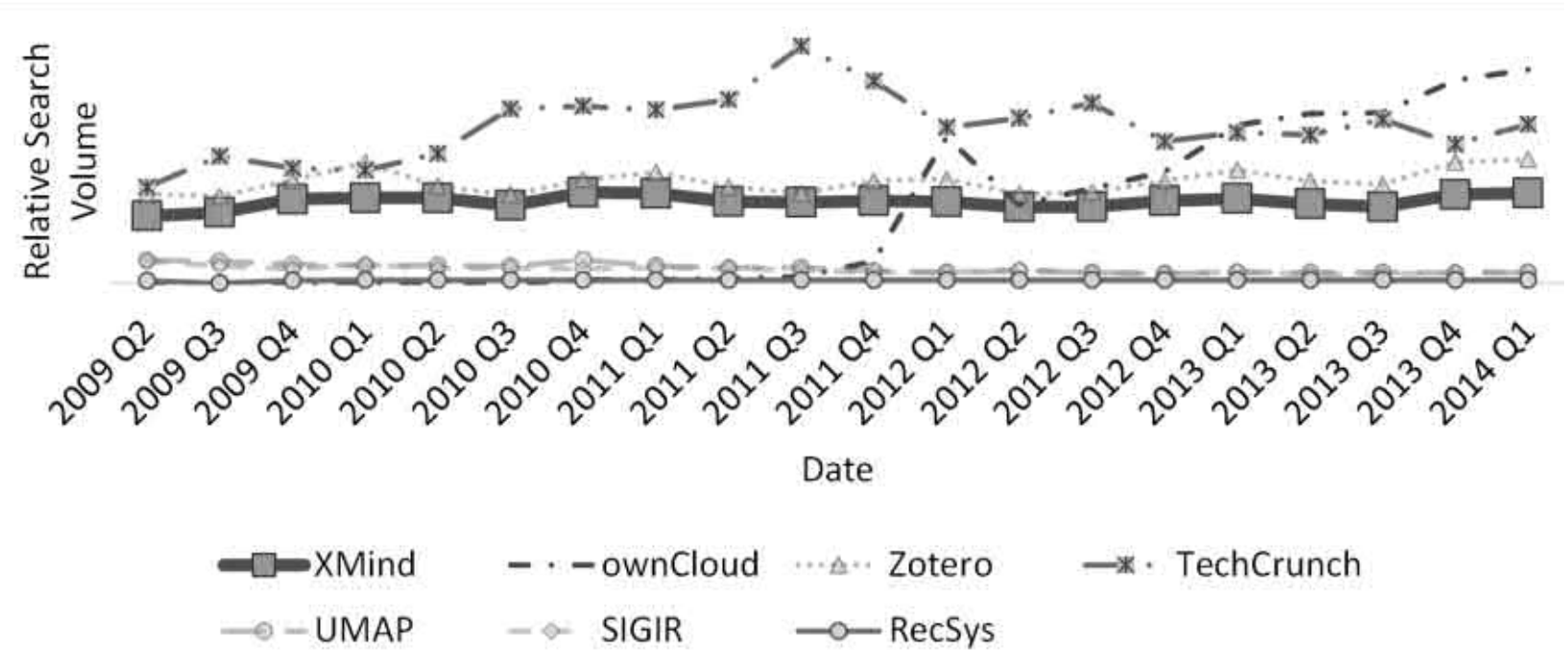

Figure 65: Search volume for "XMind" and other selected search terms

The claimed user counts do not always correlate with the search volume. For instance, MindMeister accumulates less than 8% of the search volume, and claims 2.5 million users. In contrast, XMind accumulates 25% of the search volume, but reports only around 1 million users. We assume that these differences result from different registration and usage concepts. MindMeister is a web-based tool that requires everyone to register. XMind is a desktop software that can also be used without registration. As such, our estimate remains a rough guess. However, another estimate leads to a similar result. The open source mind-mapping software *FreeMind* was downloaded 1.4 million times in the past 12 month (we considered only downloads of the latest stable release)[79]. Assuming, that the number of active users is around 1/3 of users who downloaded the software in the past year, leads to the estimate that FreeMind has around 450,000 active users. Interpolating from the search volume (22.58%), leads to an estimate of 2 million active mind-map users.

[79] http://sourceforge.net/projects/freemind/files/stats/timeline

We believe that these numbers indicate a substantial interest in the topic of mind mapping, and the active user base justifies the effort to research the potential of utilizing mind-maps for IR applications.

## B.4 Mind-Map based IR Applications

We developed eight ideas, how mind-maps could be utilized beyond their original purpose. These ideas are briefly described in the following paragraphs, and were originally published in [43]. For more details refer to [19, 37].

*Search Engines for Mind-Maps*: Mind-maps contain information that probably is not only relevant for the given authors of a mind-map, but also for others. Therefore, a search engine for mind-maps might be an interesting application.

*User Modelling*: Analog to analyzing users' authored research papers, emails, etc., user modelling systems could analyze mind-maps to identify users' information needs and expertise. User models could be used, for instance, for personalized advertisements, or by recommender systems, or expert search systems. For instance, when employees create mind-maps, we would assume that the mind-maps would be suitable to infer the employees' expertise. This information could be used by an expert search system. As described previously, *Mindomo* and *MindMeister* implemented user modelling systems, but Mindomo reported that users were not interested in the results. Hence, they removed the system from their mind-mapping application. Apparently, user modelling based on mind-maps is not trivial and does not always lead to satisfying results.

*Document Indexing / Anchor Text Analysis*: Mind-maps could be seen as neighboring documents to those documents being linked in the mind-maps, and anchor text analysis could be applied to index the linked documents with the terms occurring in the mind-maps. Such information could be valuable, e.g., for classic search engines.

*Document Relatedness*: When mind-maps contain links to web pages or other documents, these links could be used to determine relatedness of the linked web pages or documents. For instance, with citation proximity analysis [142], documents would be assumed to be related that are linked in close proximity, e.g. in the same sentence. Such calculations could be relevant for search engines and recommender systems.

*Document Summarization*: Mind-maps could be utilized to complement document summarization. If a mind-map contains a link to a webpage, the node's text, and maybe the text of parent nodes, could be interpreted as a summary for the linked web page. Such summaries could be displayed by search engines on their result pages.

*Impact Analysis*: Mind-maps could be utilized to analyze the impact of the documents linked within the mind-map, similar to PageRank or citation based similarity metrics. This information could be used by search engines to rank, e.g., web pages, or by institutions to evaluate the impact of researchers and journals.

*Trend Analysis*: Trend analysis is important for marketing and customer relationship management, but also in other disciplines [82]. Such analyses could be done based on mind-maps. For instance, analyzing mind-maps that stand for drafts of academic papers would allow estimating citation counts for the referenced papers. It would also predict in which field new papers can be expected.

*Semantic Analysis*: A mind-map is a tree and nodes are in hierarchical order. As such, the nodes and their terms are in direct relationship to each other. These relationships could be used, for instance, by search engines to identify synonyms, or by recommender systems to recommend alternative search terms or social tags.

## B.5 Feasibility

We evaluated the ideas' feasibility in three steps. First, we estimated whether there are enough mind-maps and mind-map users available to realize the ideas. Second, we analyzed whether the content of mind-maps is suitable for realizing the ideas. Finally, we gauged whether users are accepting the ideas.

### B.5.1 Mind-Map Users and (Public) Mind-Maps

Most of the ideas hinge on the availability of a large number of mind-maps. It is also important to distinguish between public and private mind-maps. If many mind-maps were available publicly, the ideas could be realized by anyone. If mind-maps were private, i.e. only available to the developers of the mind-mapping tools, only these developers could realize the ideas.

There are more than 300,000 mind-maps in public galleries, 50% of them in the gallery of MindMeister, 20% in the gallery of Mindomo, and 16% in the gallery of XMind (Figure 66)[80]. Over the years, the number of public mind-maps increased from 67,167 in 2010 to 303,084 in 2014. Given, that MindMeister's users published around 62,000 mind-maps between 2013 and 2014, we estimate that MindMeister's users created approximately 620,000 mind-maps during that period, since around 10% of mind-maps being created are also published[77]. Interpolating these numbers with the search volume (Figure 64), we can estimate that overall 4.6 million mind-maps were created between 2013 and 2014. Another estimate confirms this number: Mind-map users create between 2 and 3 mind-maps per year on average [40]. A calculation with 2.5 mind-maps per year, and 2 million mind-map users, leads to an estimate of 5 million mind-maps created per year. Considering that mind-mapping tools have been used for many years, a few dozens of millions mind-maps must exist on the computers of mind-map users.

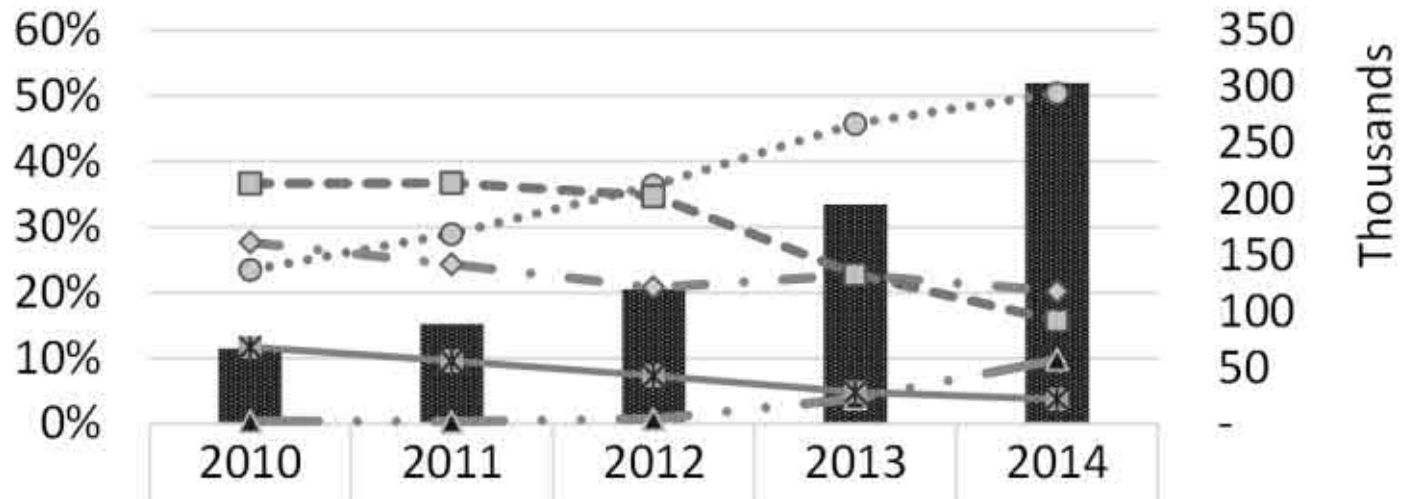

| | 2010 | 2011 | 2012 | 2013 | 2014 |
|---|---|---|---|---|---|
| Public Mind-Maps | 67,167 | 88,624 | 119,778 | 195,087 | 303,084 |
| MindMeister | 23% | 29% | 36% | 46% | 50% |
| Mindomo | 28% | 24% | 21% | 23% | 20% |
| XMind | 37% | 37% | 35% | 23% | 16% |
| Mindjet | 0% | 0% | 1% | 4% | 10% |
| Others | 12% | 10% | 7% | 5% | 4% |

Figure 66: Public mind-maps

### B.5.2 Content of Mind-Maps

We recently analyzed the content of 19,379 mind-maps, created by 11,179 *MindMeister* and *Docear* users [40]. On average, mind-maps contained a few dozens of nodes, each with two to three words on average. Some mind-maps even contained a few thousand nodes, with some nodes containing more than a hundred words. This amount of nodes, and words, is comparable to the number of words in emails or web pages. Since emails and web pages are successfully utilized by information retrieval applications, the content of mind-maps might be suitable for those ideas that depend on the existence of terms. However, the number of links in mind-maps is low. Almost two thirds of the mind-maps did not contain any links to files, such as academic articles or other documents (63.88%), and most of the mind-maps that did contain links, contained only few of them. Links to webpages were not available in 92.37% of Docear's mind-maps and 75.27% of MindMeister's mind-maps. Consequently, those ideas based on link-analysis seem less attractive.

[80] Over the past four years, we retrieved the numbers of mind-maps each year directly from the webpages of the galleries.

### B.5.3 User Acceptance

We evaluated the user acceptance of the eight ideas with our mind-mapping software *SciPlore MindMapping* [35]. 4,332 users were shown at first start a settings dialog. In this dialog, users could (un)select four options relating to the different ideas we proposed (Figure 67). It was randomly chosen whether options were pre-selected.

When all options were pre-selected, 61% of the users accepted user modelling to receive recommendations based on their mind-maps (Figure 67). 38% of the users accepted that the content of their mind-maps could be utilized e.g. for anchor text analysis. 32% of users agreed that SPLMM could also analyze the content of the documents they linked in their mind-maps. Usage mining, i.e. the general analysis of how users are making use of a software, was accepted by 48% of the users.

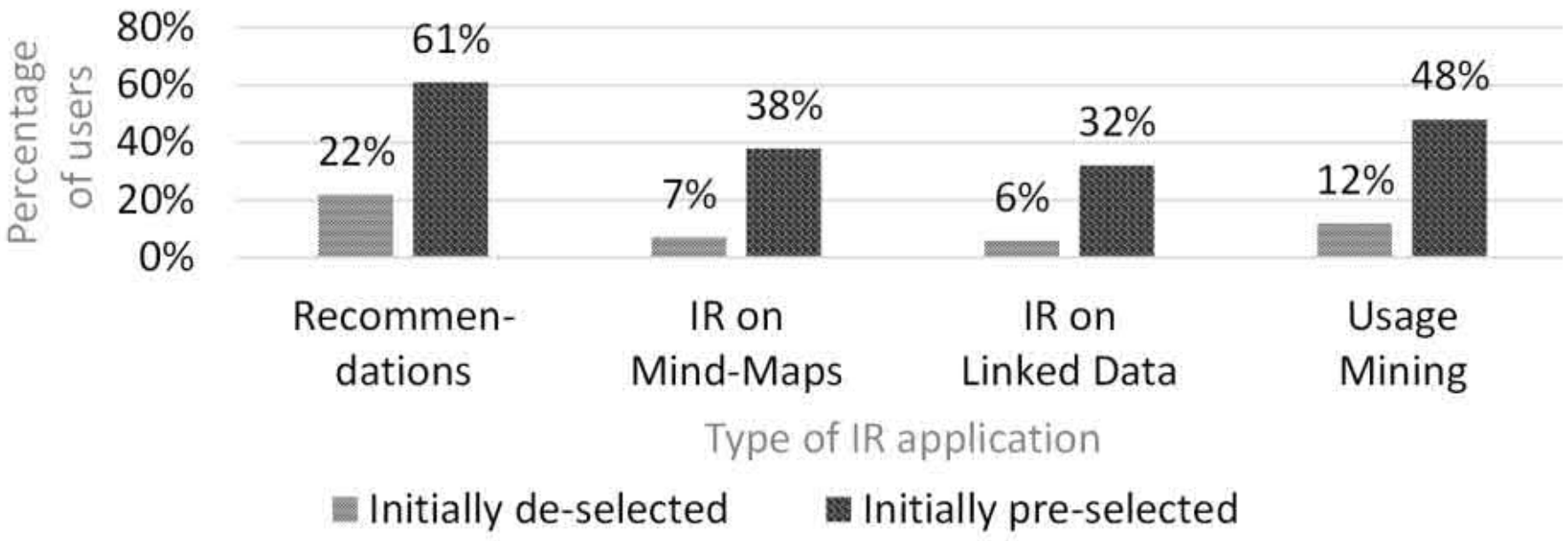

Figure 67: User acceptance of IR on their mind-maps

If options were not pre-selected, fewer users allowed the analysis of their data. 22% activated recommendations, 7% activated information retrieval on mind-maps, 6% activated IR on the linked documents, and 12% activated usage mining.

### B.5.4 Discussion of the Feasibility

Due to the generally few links available in mind-maps, anchor text analysis, calculating document relatedness, document summarization, and impact analysis seem less feasible for the majority of mind-mapping tools (Table 14). However, there might be exceptions, for instance in the case of Docear. Docear's mind-maps contain comparatively many links to PDF files, because most users are researchers who manage their academic papers with Docear. Assuming that Docear's users create enough mind-maps, the link-based ideas might be interesting to pursue.

Those ideas that depend on the availability of terms seem more feasible, considering the content of mind-maps. However, only a small number of mind-maps are publicly available (around 300,000). This makes the ideas less interesting for third parties who do not offer their own mind-mapping software. The same is true for developers of mind-mapping software with only a few users. A mind-map search engine or trend analysis using for example only 50,000 mind-maps, cannot attract many people. For the major players, such as XMind, FreeMind, or MindMeister, this might be different. They potentially have access to millions of mind-maps, which should be sufficient to achieve reasonable results. One idea is also relevant for the less popular mind-mapping tools, namely user modelling. User modelling, more precisely recommender system, personalized advertisement, or expert search, should be well applicable even with few users. User modelling has also the highest acceptance rate among the users. User Modelling for a recommender system was accepted by 61% or the users. User acceptance of the other ideas was lower. Around 10% of mind-maps are published, and around 30-40% of users accept IR to enhance external applications.

Table 14: Feasibility of the ideas

| | Mind Map Availability | | | | |
|---|---|---|---|---|---|
| | For 3rd parties | For MM tool developers | Content Suitability | Users' Acceptance | Overall |
| Search Engine | Low | Depends | Good | Low | Low |
| Document Indexing | Low | Depends | Low | Medium | Low |
| Document Relatedness | Low | Depends | Low | Medium | Low |
| Document Summarization | Low | Depends | Low | Medium | Low |
| Impact Analysis | Low | Depends | Low | Medium | Low |
| Trend Analysis | Low | Depends | Medium | Medium | Medium |
| Semantic Analysis | Low | Depends | Good | Medium | Medium |
| User Modeling | --- | Good | Good | Good | Good |

Overall, user modelling seems to be the most promising idea: The content of mind-maps is suitable, user acceptance is rather high, and user modelling is relevant for all developers of mind-mapping software, and companies whose employees use mind-maps. In addition, user modelling *directly* benefits the mind-mapping tools and may be fundamentally important for a company. For instance, Google is generating almost its entire profit from personalized advertisements [151], and Amazon is also making a significant amount of revenue through its recommender system [205]. In contrast, applications such as semantic analysis are usually not fundamental to a company's business.

However, user modelling based on mind-maps already had been implemented, but results indicate that it is not as promising as our analysis suggests. MindMeister and Mindomo created user models for displaying personalized advertisement but both abandoned this after a while. This leads to the question, whether mind-maps actually can successfully be utilized by user modelling systems.

## B.6 Prototype

To analyze whether user modelling based on mind-maps can be done effectively, we integrated a recommender system into our mind-mapping tools *SciPlore MindMapping* (SPLMM) [35], and its successor *Docear* [33]. Both tools are primarily used by researchers. Therefore, the recommender system recommends research papers. We implemented different recommendation approaches that we evaluated using click-through rate (CTR), i.e. the ratio of clicked recommendations against the number of displayed recommendations. Please note that due to space restriction we may only provide superficial information on the recommender system and its evaluation. We are about to publish a paper that will present the architecture of Docear's recommender system in more detail, as well as a discussion on the suitability of CTR as an evaluation metric for recommender systems. These papers will be available soon at http://www.docear.org/publications/.

For SPLMM, we implemented an approach similar to MindMeister's approach. Each time, a user modified, i.e. edited or created, a node, the terms of that node were send as search query to Google Scholar. Google Scholar's Top 3 results were shown in a separate window above the currently opened mind-map. Between July and December 2011, 78,698 recommendations were displayed, of which 221 were clicked, i.e. an overall CTR of 0.28% was achieved (Figure 68). A CTR of 0.28% is low. If MindMeister and Mindomo should have achieved similarly CTRs, it is no surprise that they abandoned the personalized advertisement.

In Docear, we integrated a new recommender system [47]. The new system showed recommendations only when users explicitly requested them, or automatically every five days on start-up of Docear. Recommendations were based on Docear's own document corpus, consisting of around 1.8 million full-text articles. The recommender system used four different approaches and displayed 21,445 recommendations between July 2012 and February 2013. The first approach made use of the terms of the last modified node, similar to the approach of SPLMM. This led to a CTR of 1.17% (Figure 68). The reasons why CTR was around four times higher than CTR in SPLMM, may be manifold. Maybe, the lower frequency of displaying recommendations (every five days instead of continuously) or the source (Docear's corpus vs Google Scholar), influenced CTR. However, 1.17% is still a rather low CTR. The second approach utilized the most frequent words of the user's current mind-map. This increased CTR to 6.12%. When the most frequent words of *all* mind-maps were utilized, CTR was also above 6%. For the fourth approach, we manually compiled a list of ten research articles relating to academic writing. Most of Docear's users are researchers and therefore we assumed that these articles would be relevant to most of Docear's users. When recommendations were given based on this approach – the stereotype approach [337] – CTR was 4.99%.

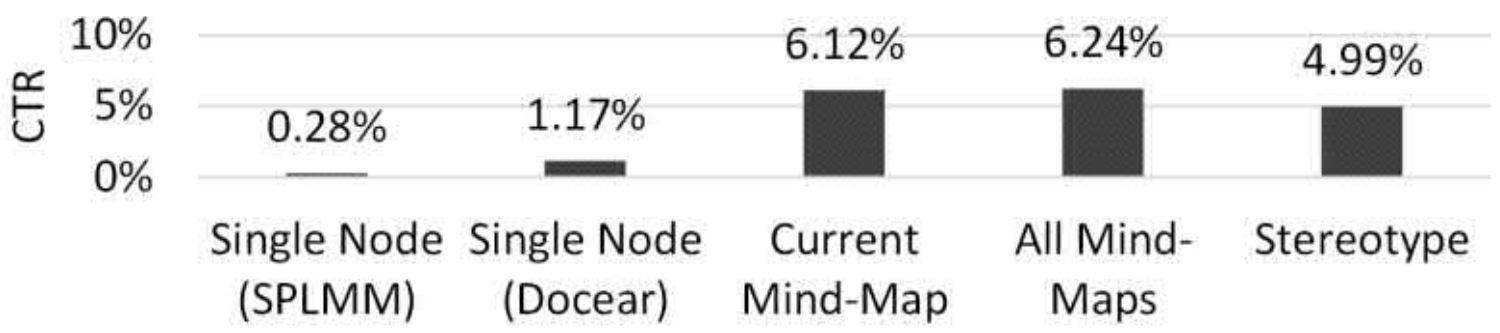

Figure 68: CTR of different approaches

The results show that a single node, typically containing two to three words, does not express users' information needs thoroughly. Instead, entire mind-maps are needed for analysis. To analyze this in more detail, we modified the recommender system, so it randomly chose the number of nodes to analyze. The results show that there is a strong correlation between the number of nodes analyzed and the CTR (Figure 69). When the recommender system utilized only the last 1 – 9 modified nodes, CTR was 3.16% on average. When 10 to 49 nodes were utilized, CTR increased to 4% on average. Utilizing between 500 and 999 nodes resulted in the highest CTR (7.47%). When more than 1,000 nodes were utilized, CTR began to decrease (though, the difference is not statistically significant).

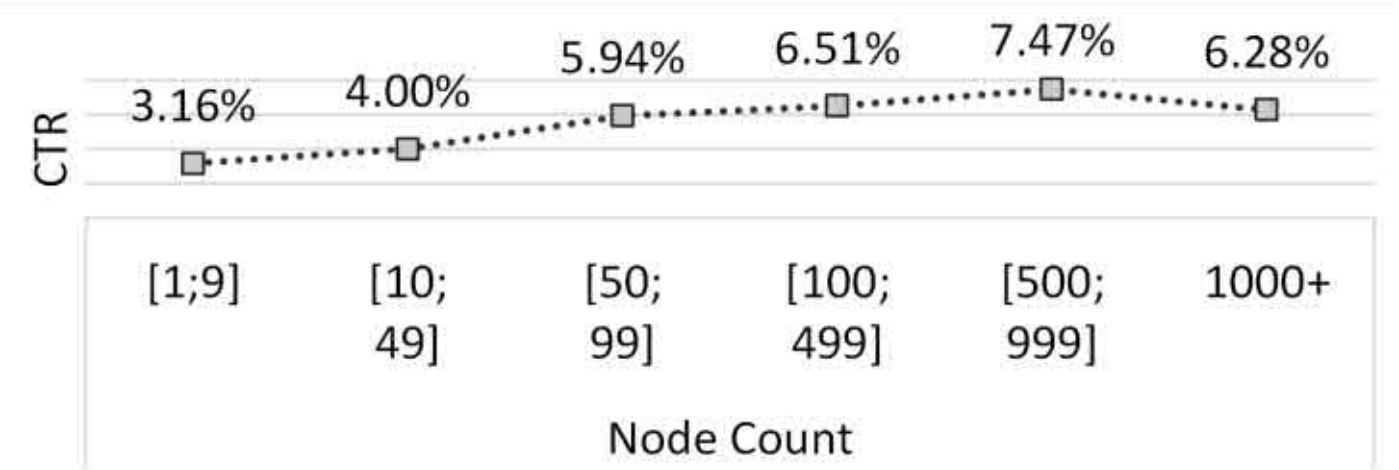

Figure 69: CTR by number of analyzed nodes

## B.7 Summary

Mind-maps have thus far been widely neglected by the information retrieval community. We found that there are more than 100 mind-mapping tools and that, based on search volume, the popularity of mind-mapping is comparable to the popularity of note taking, file management, or crowdsourcing. Popular mind-mapping tools, such as XMind, are as popular as popular reference management software (e.g. Zotero), or Tech Blogs (e.g. TechCrunch). Overall, we estimated, there are around 2 million people who actively create mind-maps using a mind-mapping software. Based on these numbers, we conclude that it is worth to research whether the developers of mind-mapping tools, and their users, might benefit from new applications, which utilize mind-maps.

We presented eight ideas of how mind-maps could be utilized to enhance information retrieval applications: search engines for mind-maps could help to find interesting information; user modelling based on mind-maps could enable the implementation of recommender systems, personalized advertisement, and expert search; anchor text analysis applied to mind-maps could enhance the indexing of webpages and other documents; similarly, anchor-text analysis could enhance the summarization of webpages and documents being linked in mind-maps; citation and link analysis could help to calculate document relatedness, which might be useful to enhance search engines or recommender systems; similarly, citation and link analysis in mind-maps could be used for impact and trend analysis; finally, semantic analyses could be applied to mind-maps to identify synonyms and other relationships of words,

Not all ideas are equally feasible. We analyzed the content of mind-maps and learned that mind-maps often do not contain any citations or links. In addition, there are only around 300,000 mind-maps publicly available, although around 5 million mind-maps are created each year. The user's acceptance to utilize their mind-maps was mediocre. 38% of the users allowed the use of their mind-maps for e.g. anchor text analysis, 61% accepted recommendations based on their mind-maps. We concluded that, out of the eight ideas, user modelling is the most feasible use case. The content of mind-maps is suitable for user modelling, the users' acceptance seems reasonably high, and user modelling is relevant for all developers of mind-mapping software, not only the major players.

We implemented a prototype of a user modelling system, namely a research paper recommender system, and, overall, results are promising. While the most simple user modelling approach – utilizing terms of the currently edited or created node – performed poorly (CTRs around 1% and lower), utilizing terms of users' entire mind-maps achieved click-through rates above 6%. This shows that user modelling based on mind-maps is not trivial, and strongly depends on the applied approaches. Further research is required to identify the unique characteristics of mind-maps, and to use these characteristics successfully in user modelling systems such as expert search, and recommender systems.

# C Exploratory Analysis of Mind-Maps[81]

We conducted an exploratory study of mind-maps, which was originally published in 2011 [40]. The objective was to find out how mind maps are structured and which information they contain. Results include: A typical mind map is rather small, with 31 nodes on average (median), whereas each node usually contains between one to three words. In 66.12% of cases, there are few notes, if any, and the number of hyperlinks tends to be rather low, too, but depends upon the mind mapping application. Most mind maps are edited only on one (60.76%) or two days (18.41%). A typical user creates around 2.7 mind maps (mean) a year. However, there are exceptions, which create a long tail. One user created 243 mind maps, the largest mind map contained 52,182 nodes, one node contained 7,497 words, and one mind map was edited on 142 days.

## C.1 Introduction

Millions of people are using mind maps for brainstorming, note taking, document drafting, project planning, and other tasks that require hierarchical structuring of information. Figure 70 shows a mind map which was created as draft for this chapter. As all mind maps, it has a central node (the root) which represents the main topic the mind map is about. From this root node, child-nodes branch out, in order to describe sub-topics. Each node may contain an arbitrary number of words. This way, a mind map is comparable to an outline but with stronger focus on the graphical representation. Mind maps created on a computer may also contain links to files, hyperlinks to websites (in Figure 70 indicated by red arrows), pictures, and notes (indicated by yellow note icons).

In this chapter, we present the initial results of an exploratory study of 19,379 mind maps. The overall research objective was to find out how mind maps are structured and what information they contain. To our knowledge, this is the first study of its kind. We therefore aimed at a broad overview to determine further areas of interesting research.

[81] This chapter has been published as: Beel, Joeran, and Stefan Langer. "An Exploratory Analysis of Mind Maps." In Proceedings of the 11th ACM Symposium on Document Engineering (DocEng'11), 81–84. ACM, 2011.

## C.2 Related Work

There is lots of research on content and structure of other documents: Web pages, emails, academic articles, etc. have all been analyzed thoroughly in the past (e.g. [1-3]). With respect to mind maps, there is mostly research about the effectiveness as learning tool (e.g. [4]).

The lack of analyses of mind maps is not surprising. Emails, web pages, etc. had to be thoroughly researched to make information retrieval tasks, for instance, indexing, and spam detection, effectively possible. Such information retrieval tasks have never been applied to mind maps, and therefore the need for knowledge about mind map content and structure was low.

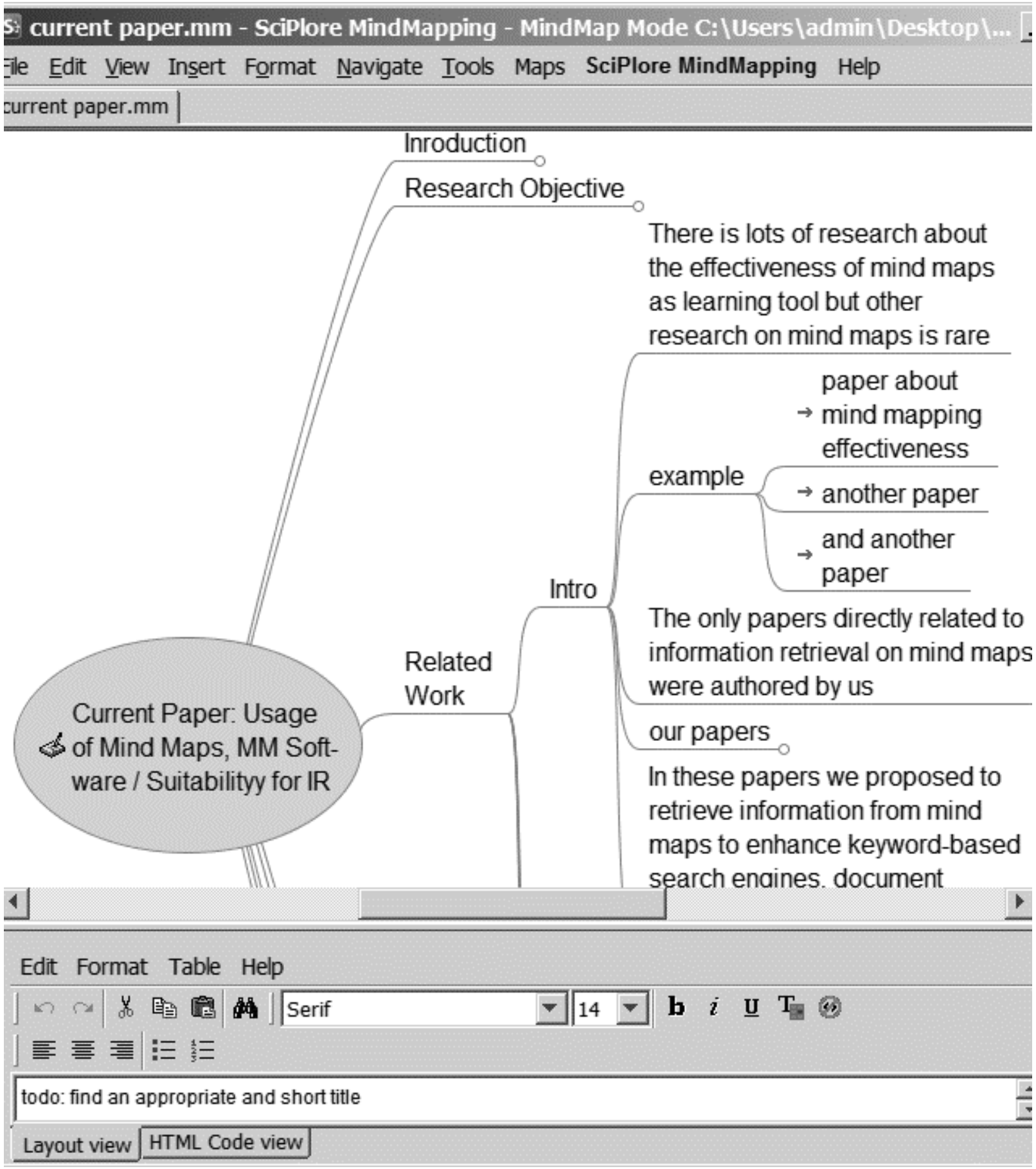

Figure 70: Screenshot of a mind mapping software

However, recently we proposed to apply information retrieval tasks to mind maps to enhance keyword-based search engines, document recommender systems, and user profile generation [5]. To do this effectively, knowledge about the content and structure of mind maps is required.

There was only one paper we found that is somewhat related: a survey from the *Mind Mapping Software Blog* [6]. For this survey 334 participants answered questions about their use of mind mapping software. However, the survey was based on 334 self-selected participants from a single source (readers of the *Mind Mapping Software Blog*). Accordingly, it seems likely that predominantly very active mind mapping users participated in the survey and results are not representative. In addition, the survey focused on the usage of mind mapping software rather than the content and structure of mind maps.

## C.3 Methodology

We conducted an exploratory study on 19,379 mind maps created by 11,179 users from the two mind mapping applications *Docear*[82] and *MindMeister*[83] (the latter one is abbreviated as 'MM' in figures and tables).

Docear is a mind mapping application for Windows, Linux and Mac, focusing on academic literature management, and developed by ourselves [7]. 2,779 users agreed to have their mind maps analyzed. They created 7,506 mind maps between April 1, 2010 and March 31, 2011.

MindMeister is a web-based mind mapping application. 8,400 users published 11,873 mind maps in MindMeister's public mind map gallery[84] between February 2007 and October 2010. For our study these public mind maps were downloaded in XML format via MindMeister's API[85], parsed, and analyzed.

Numbers include only mind maps containing six or more nodes[86], and that were not being edited between April 1, 2011 and the day of the analysis (June 2, 2011). This way it is ensured that mind maps in the beginning of their life-cycle do not spoil the results but only "mature" mind maps were analyzed.

[82] http://docear.org
[83] http://mindmeister.com
[84] http://mindmeister.com/maps/public
[85] http://mindmeister.com/services/api
[86] A random sample of 50 mind maps showed that the vast majority of mind maps with five or fewer nodes were created for testing purposes and did not contain valuable content.

We were particularly interested in finding out whether differences existed for different types of mind maps and between the two mind mapping applications. Therefore, mind maps were grouped based on their size, measured by the number of nodes. Mind maps with 6 to 35 nodes were considered as ‘tiny’, with 36 to 100 nodes as ‘small’, with 101 to 350 nodes as ‘medium’, with 351 to 1000 nodes as ‘large’ and with more than 1000 nodes as ‘very large’. In the data set, the majority of mind maps were tiny (52.47%) or small (31.40%) as shown in Figure 71.

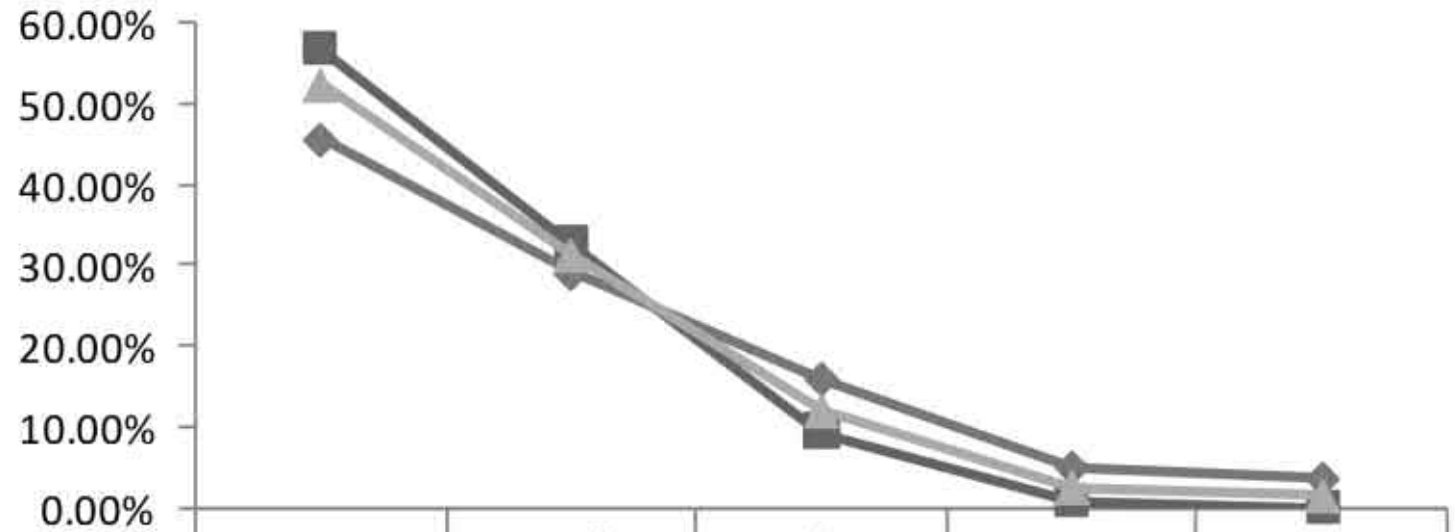

| | Tiny: 6-35 nodes | Small: 36-100 nodes | Medium: 101-350 nodes | Large: 351-1k nodes | Very Large: >1k nodes |
|---|---|---|---|---|---|
| Docear | 45.68% | 29.16% | 16.15% | 5.20% | 3.81% |
| MM | 56.77% | 32.81% | 9.48% | 0.83% | 0.10% |
| Total | 52.47% | 31.40% | 12.06% | 2.54% | 1.52% |

Figure 71: Distribution of mind maps based on size (number of nodes)

## C.4 Results & Interpretation

### C.4.1 Mind Maps per User

Figure 72 shows the number of mind maps users created. The majority of MindMeister users created, or we should say published, exactly one mind map (81.26%). Only 2.32% of MindMeister users published five or more mind maps. In contrast, 56.75% of Docear users created one mind map and 11.36% created five or more mind maps. On average (mean), users created 2.7 mind maps (Docear) during the 12 month period of data collection, respectively 1.4 (MindMeister) during ~3.5 years. The highest number of mind maps created by one user was 243 for Docear and 73 for MindMeister. It has to be noted that numbers of MindMeister and Docear are only limitedly comparable, as we did only analyze MindMeister mind maps that were published by their users. It can be assumed that most users who published mind maps on the Web, created further private mind maps that were not publicly available.

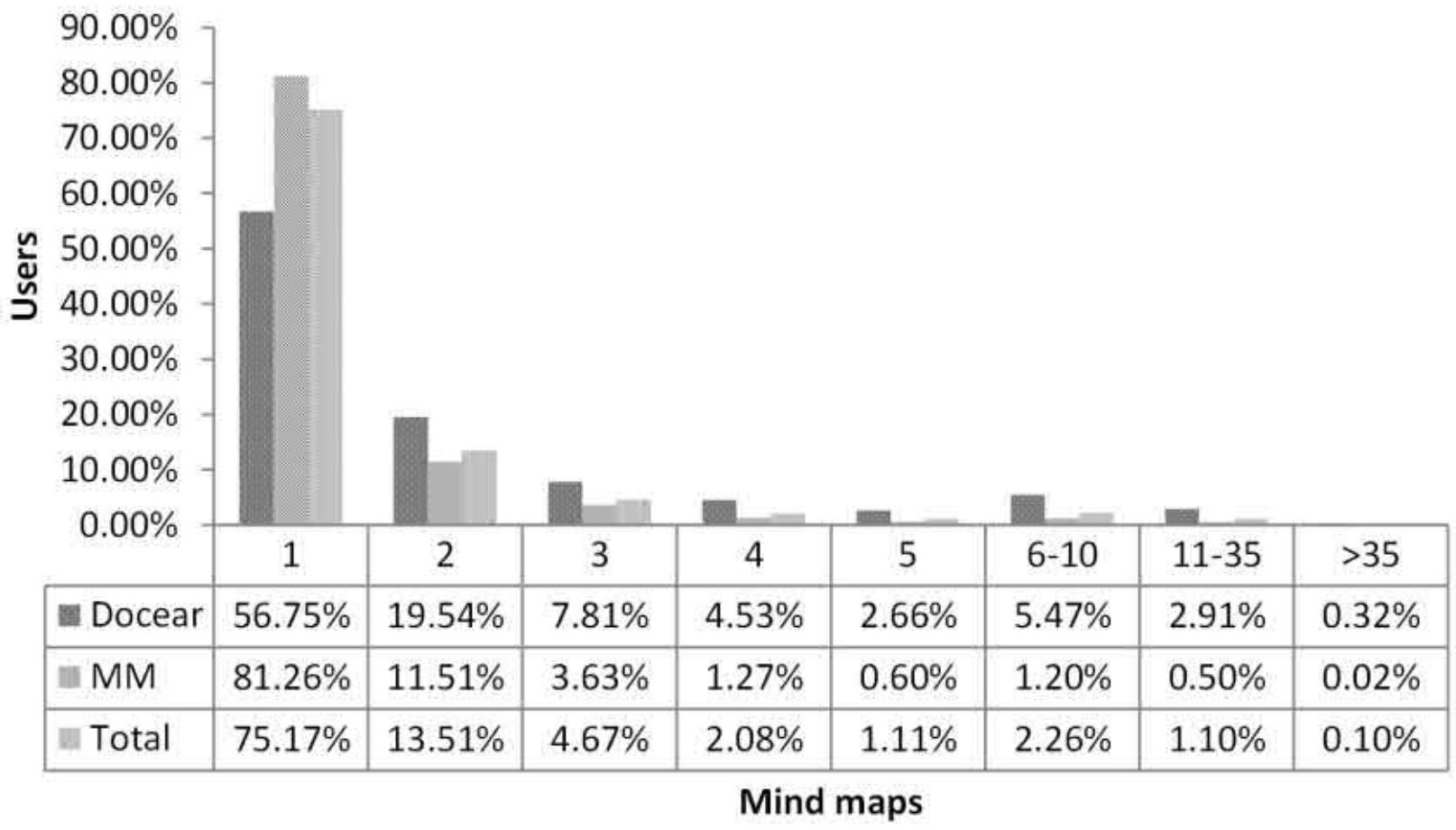

| | 1 | 2 | 3 | 4 | 5 | 6-10 | 11-35 | >35 |
|---|---|---|---|---|---|---|---|---|
| Docear | 56.75% | 19.54% | 7.81% | 4.53% | 2.66% | 5.47% | 2.91% | 0.32% |
| MM | 81.26% | 11.51% | 3.63% | 1.27% | 0.60% | 1.20% | 0.50% | 0.02% |
| Total | 75.17% | 13.51% | 4.67% | 2.08% | 1.11% | 2.26% | 1.10% | 0.10% |

Figure 72: Number of created mind maps per user

### C.4.2 Nodes per mind map

As mentioned in the methodology and shown in Figure 71, most mind maps were rather small. On average, Docear mind maps contained 232 nodes (mean), respectively 41 nodes (median). MindMeister mind maps contained 51 nodes (mean), respectively 31 (median). Docear mind maps tended to be larger than MindMeister mind maps. For instance, while only 0.10% of MindMeister mind maps were 'very large', 3.81% of Docear mind maps were. The largest Docear mind map contained 52,182 nodes (and there are several more mind maps containing 10,000+ nodes); the largest MindMeister mind map contained 2,318 nodes.

### C.4.3 File Links

In a mind map, users may link to files on their hard drive. Figure 73 shows the distribution of mind maps containing a certain number of links (for Docear mind maps only since MindMeister does not provide this feature). Well over half of mind maps do not contain any links to files (63.88%).

Table 15: File types linked in mind maps

| PDFs | Images | Documents | HTML | Excel/CSV | PowerPoint | MP3s | Other |
|---|---|---|---|---|---|---|---|
| 89.58% | 1.26% | 0.53% | 0.47% | 0.42% | 0.34% | 0.27% | 7.14% |

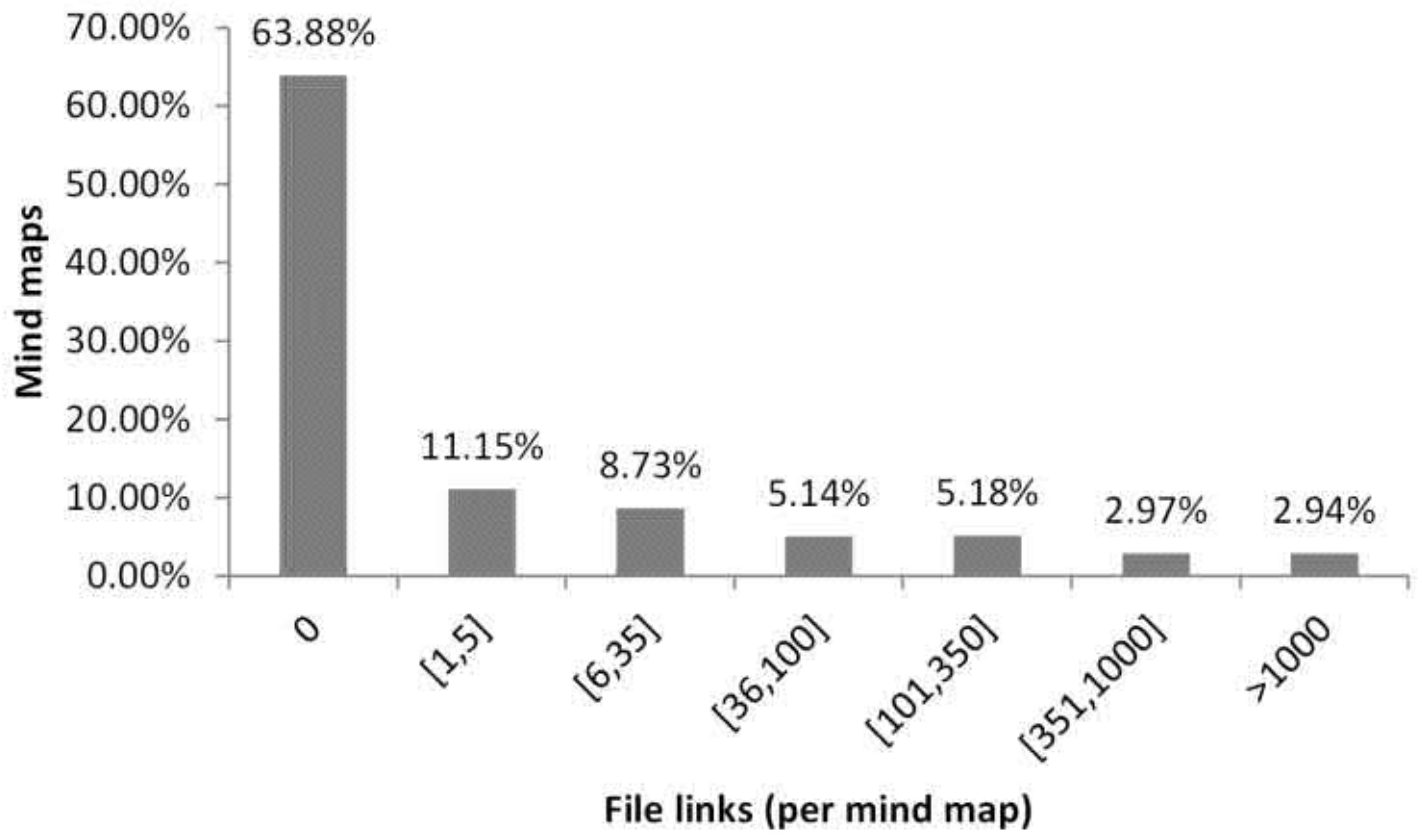

Figure 73: Number of file-links in Docear mind maps

However, some users make heavy use of the feature. 2.94% of mind maps contained more than 1,000 links to files and 2.97% of mind maps contained between 351 and 1,000 links. The highest number of links in a mind map was 52,138 and all 7,506 Docear mind maps together contained 1,184,547 links to files on the users' hard drives. This does not mean that 1,184,547 different files were linked. Most users linked the same file multiple times in a mind map.

From all links, 89.58% pointed to PDF files (Table 15). Other files being linked included images (.gif, .png, .jpeg, .tiff), MP3s and text documents (.doc, .docx, .odt, .rtf, .txt), but with much smaller frequency.

### C.4.4 Hyperlinks

Looking at all mind maps, 81.57% do not contain a single hyperlink to a website (Figure 74). However, there are differences between Docear and MindMeister. While 92.37% of Docear mind maps do not contain hyperlinks at all, only 75.27% of MindMeister mind maps do not contain any hyperlinks. In other words: 7.63% of Docear mind maps and 24.73% of MindMeister mind maps contain at least one hyperlink.

Larger mind maps more often contain hyperlinks when compared to smaller mind maps. For instance, around 20% of Docear's (very) large mind maps but only 3.94% of tiny mind maps contain hyperlinks. Similarly, around 40% of MindMeister's (very) large mind maps but only 22% of tiny mind maps contain hyperlinks.

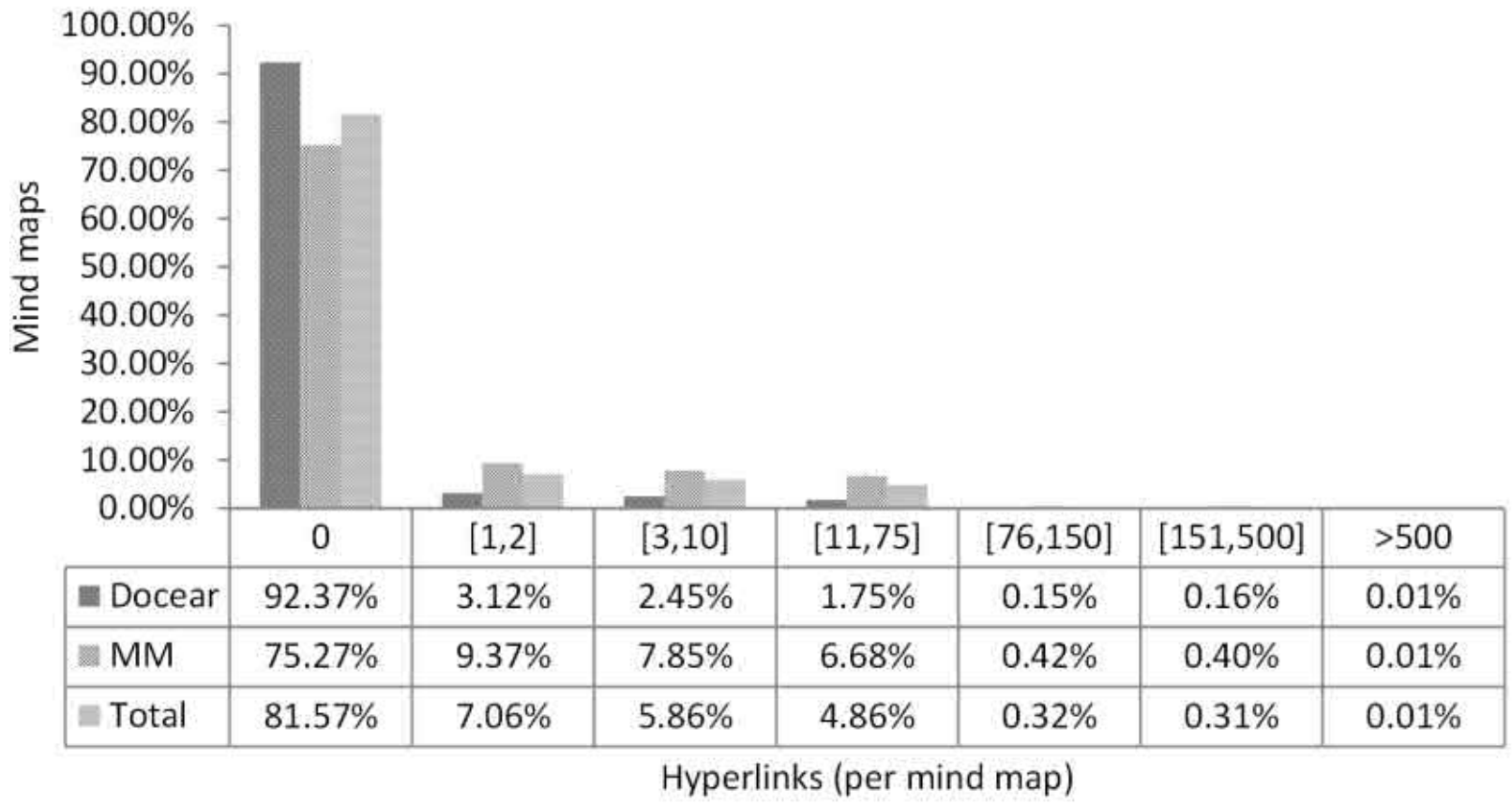

Figure 74: Number of hyperlinks in mind maps

### C.4.5 Notes

Most mind mapping software tools (such as Docear and MindMeister) allow users to add notes to a node. Many users do not use this feature – 66.12% of mind maps do not contain any notes (Table 16). Results are similar for both, MindMeister and Docear mind maps.

Table 16: Number of notes in mind maps

| | | Amount of Notes | | | | | | |
|---|---|---|---|---|---|---|---|---|
| | | **0** | **[1,2]** | **[3,10]** | **[11,75]** | **[76,150]** | **[151,500]** | **>500** |
| Mind Maps Size | **Tiny** | 68.66% | 19.53% | 8.00% | 3.81% | 0.00% | 0.00% | 0.00% |
| | **Small** | 65.72% | 15.02% | 9.58% | 9.53% | 0.15% | 0.00% | 0.00% |
| | **Medium** | 59.58% | 13.73% | 9.92% | 14.37% | 1.97% | 0.43% | 0.00% |
| | **Large** | 52.15% | 11.86% | 11.66% | 17.59% | 4.29% | 2.25% | 0.20% |
| | **Very large** | 61.74% | 6.04% | 7.38% | 16.44% | 3.69% | 3.36% | 1.34% |
| | **Total** | 66.12% | 17.01% | 8.81% | 7.42% | 0.45% | 0.16% | 0.03% |

### C.4.6 Words per node

Figure 75 shows the distribution of words per node (everything separated by whitespace characters was assumed to be a word). Nodes in mind maps generally contain few words. Nearly 1/3 of all 2,352,584 nodes contained a single word (29.91%). Only 8.25% of nodes contained more than ten words.

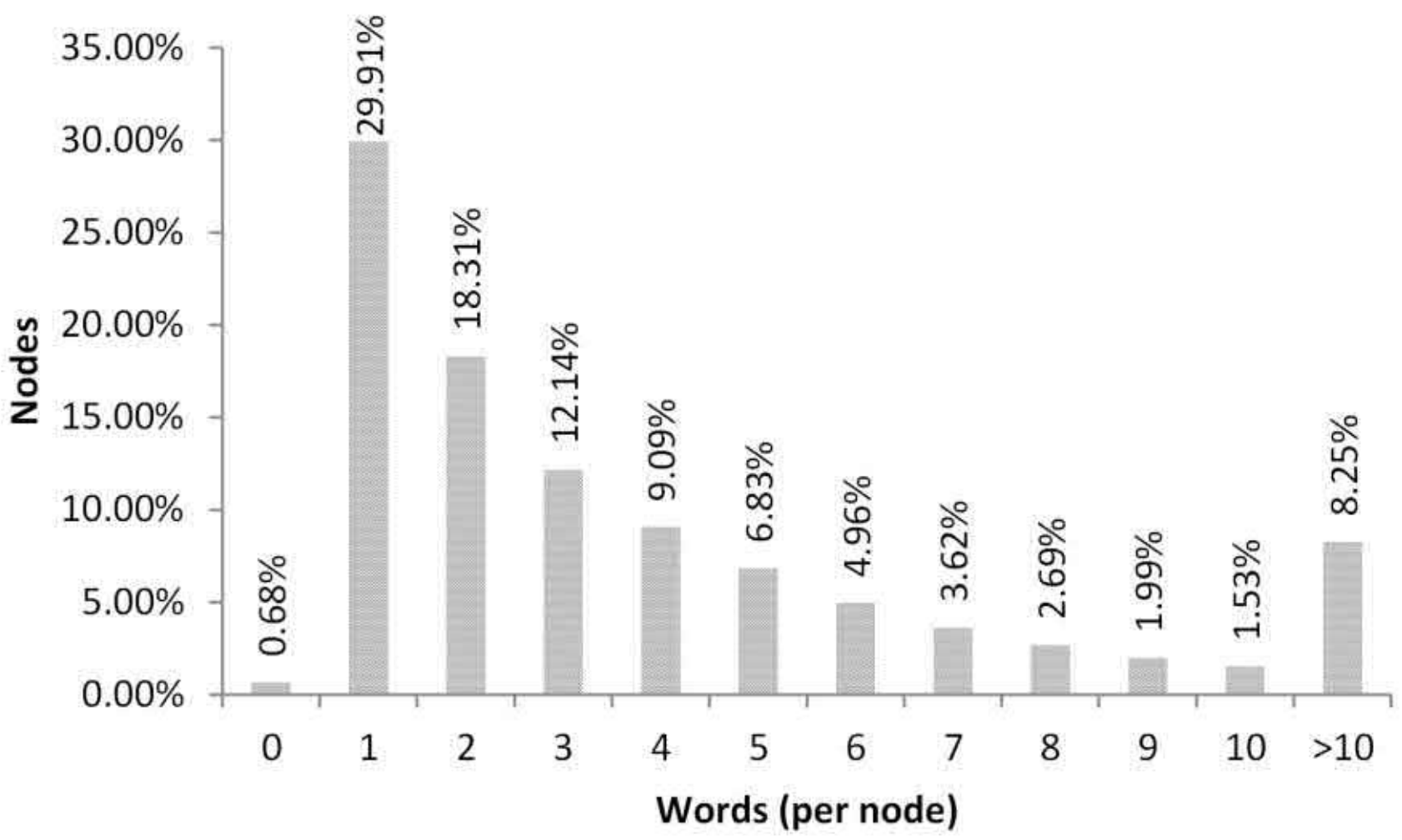

Figure 75: Number of words per node

However, there is a long tail in the distribution – the maximum word count for a node was 7,497 for Docear and 1,184 for MindMeister. Although the most frequent word count per node is one, mean is 4.80 words per node and median is 3. There is a slight tendency that the larger mind maps are, the more words their nodes contain. Details are provided in Table 17.

Table 17: Number of words per node by mind map size

| | Word count per node | | | |
|---|---|---|---|---|
| | **Mean** | **Median** | **Modal** | **Max** |
| **Tiny maps** | 4.67 | 2 | 1 | 1,874 |
| **Small maps** | 4.45 | 2 | 1 | 687 |
| **Medium maps** | 5.07 | 2 | 1 | 1,463 |
| **Large maps** | 5.76 | 3 | 1 | 2,723 |
| **Very large maps** | 4.60 | 3 | 1 | 7,497 |

Also, the deeper a node is in a mind map (further out on the branch), the more words it tends to contain. While root nodes (level 0) contain 3.03 words on average (mean), respectively 2 (median), nodes in level 5 contain 5.11 words on average (mean), or 3 (median) respectively (also Figure 76).

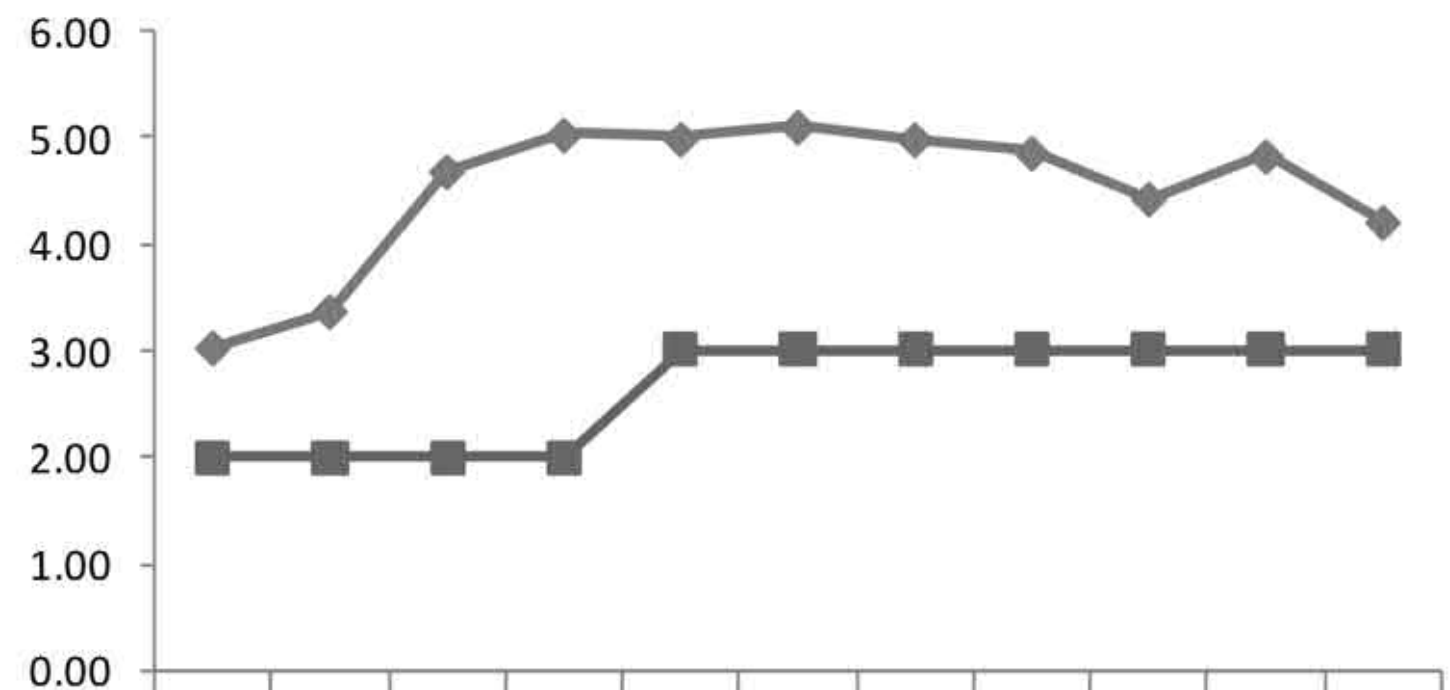

| | Level 0 | Level 1 | Level 2 | Level 3 | Level 4 | Level 5 | Level 6 | Level 7 | Level 8 | Level 9 | Level 10 |
|---|---|---|---|---|---|---|---|---|---|---|---|
| mean | 3.03 | 3.37 | 4.69 | 5.04 | 5.00 | 5.11 | 4.99 | 4.87 | 4.43 | 4.83 | 4.21 |
| median | 2 | 2 | 2 | 2 | 3 | 3 | 3 | 3 | 3 | 3 | 3 |

Figure 76: Number of words per node based on node level

Results are similar for both, Docear and MindMeister mind maps. Except, the median word count for Docear is three, and for MindMeister two.

### C.4.7 Days Edited

The majority of mind maps seem to be used for rather short term activities such as brainstorming or maybe taking meeting-minutes.

Figure 77 shows on how many days mind maps were edited[87]. 60.76% of mind maps were edited only during a single day[88]. However, also a large proportion of mind maps were edited on several days, and a small fraction (0.55%) even on more than 25 days. On average, mind maps were edited on one day (median), respectively 2.36 days (mean). The maximum was 142 days.

## C.5 Interpretation & Summary

For some features, there appear to be significant differences between mind maps created with Docear and those created with MindMeister. However, most of the differences can be attributed to the special functionality of the corresponding software. For instance, Docear offers special features for literature management such as automatically importing PDF bookmarks as new nodes to a mind map.

[87] Data was available for Docear mind maps only.

[88] Creation of a mind map was counted as one edit. All edits made during one day were combined.

Accordingly, it was expected that Docear mind maps would be larger, in terms of number of nodes. Concerning this case, probably MindMeister numbers are more representative for other mind maps than Docear's are.

On the other hand, when estimating the number of mind maps per user, Docear's numbers are probably more suitable for generalizations, as we could only analyze public mind maps of MindMeister users.

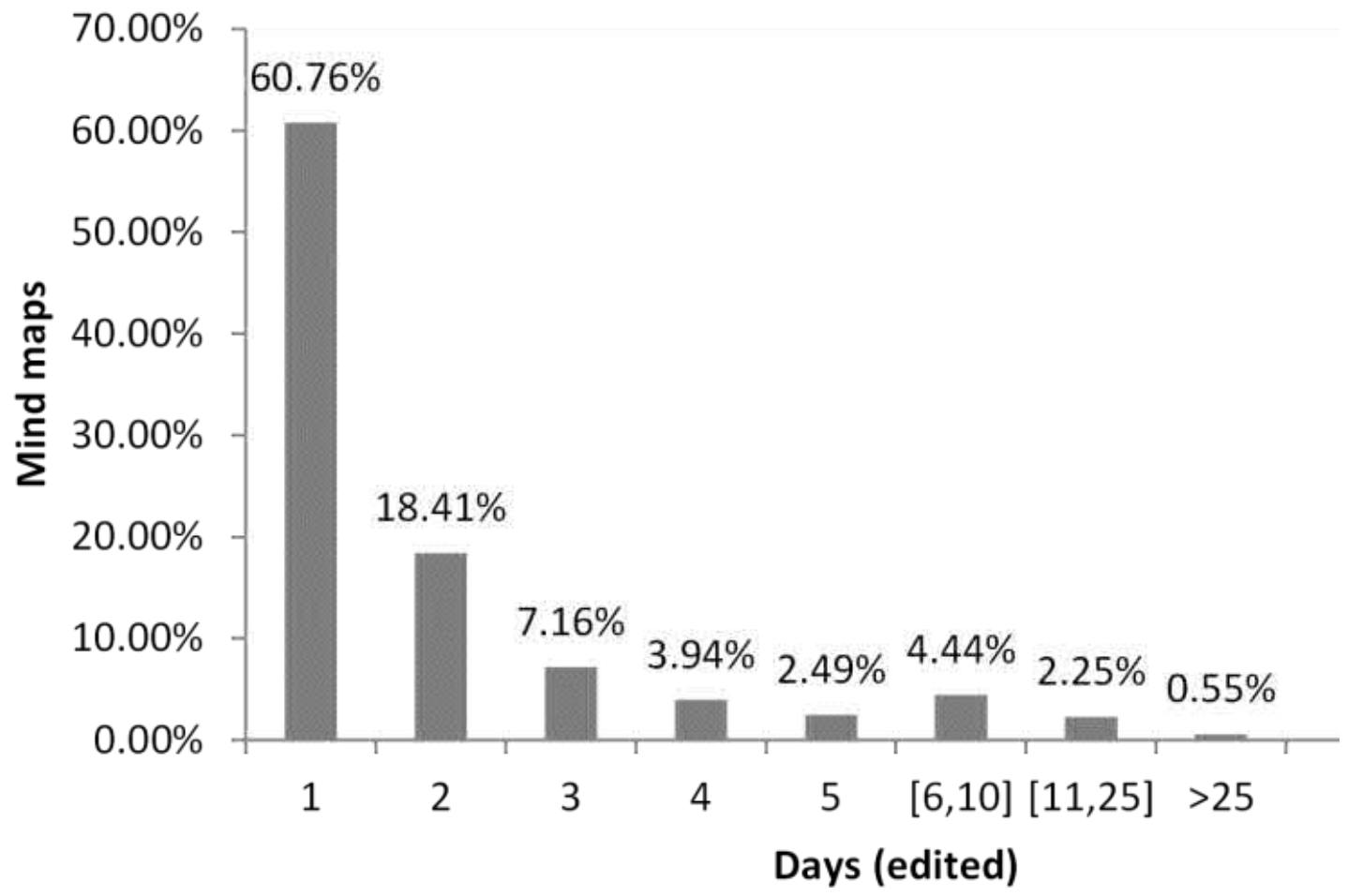

Figure 77: Number of days mind maps were edited

The study showed that a 'typical' (average) mind map is rather small, with a few dozen nodes (31 was the median for MindMeister mind maps), whereas each node contains probably between one to three words (more for large mind maps or nodes deeper in a mind map). The mind map probably contains few if any notes (66.12%). The number of hyperlinks depends on the mind mapping application and tends to be rather low, too. Probably the mind map was edited only on one (60.76%) or two days (18.41%) and it is expected that a typical user creates around 2.7 mind maps a year (mean, Docear).

However, these are only averages. Most results followed a power-law distribution with a long tail. There was one user who created 243 mind maps (and several users more created 10+ mind maps). The largest mind map in the data set contained 52,182 nodes (and several more with 10,000+ nodes existed), there was one node containing 7,497 words (and several more nodes with 100+ words existed), one mind map was edited on 142 days (and several more were edited a few dozen times) and several mind maps contained a few hundred notes.

## C.6 Outlook

For future research, analysis of the evolvement of mind maps could be interesting. Maybe there are different patterns how mind maps evolve and are used by users. Also, differences between user types should be analyzed. In addition, the content of mind maps has only been analyzed superficially, yet. It would be interesting to know what exactly the content is and what mind maps are used for exactly (brainstorming, literature management, etc.). A more detailed analysis should also look at the extremes and outliers (e.g. the node with 7,497 words).

Most importantly, mind maps need to be compared to other types of documents and consequences for information retrieval needs to be drawn. What does it mean when nodes usually contain one to three words? Are they comparable to search queries which usually consist of a similar number of terms? If so, can approaches for search query recommender easily be adopted to create a 'node recommender'? Are mind maps with a few dozen nodes comparable to a user's collection of social tags which usually also consist of a few dozen tags each with one or two words? If so, can approaches for user modeling based on social tags easily be applied to model the interests of mind map users? And are mind maps, which contain a few thousands nodes or words, comparable to web pages, academic articles, or emails? If so, what does this mean for the ability to apply information retrieval on mind maps? All these questions need to be answered in further research.

# D Link Analysis in Mind Maps[89]

## D.1 Introduction

Mind mapping is a common method to structure and visualize ideas, manage electronic literature and to draft documents. Some users do link in their mind map to external documents such as PDFs or websites. Some even cite scholarly literature, for instance by adding BibTeX keys to a mind map's node (Figure 78 for an example). In a recent paper we proposed to analyze these links and references to determine the relatedness of those documents that are linked in the mind map [37][90].

The basic idea is that two documents are related if they are both linked by a mind map. In addition, it was assumed that the closer the links occur in the mind map, the higher related the linked documents are. If the assumption proves to be right, *Link Analysis in Mind Maps* (LAMM) could be used to enhance search engines and document recommender systems since these systems often present related documents to their users.

We conducted a brief experiment to test the proposed idea and present the results in this chapter. The focus of this chapter lies on calculating the relatedness of scholarly literature and on enhancing research paper recommender systems as we plan to integrate LAMM into our academic search engine and research paper recommender system *SciPlore*[91]. However, it's highly probable that the results would be similar for other kind of documents linked by a mind map such as websites.

In the next section, related work about research paper recommender systems and citation analysis is presented. It is then followed by a section showing the methodology which has been used to evaluate LAMM. Finally, the results, a discussion, and an outlook towards future work conclude.

---

[89] This chapter has been published as: Beel, Joeran, and Bela Gipp. "Link analysis in mind maps: a new approach to determining document relatedness." In Proceedings of the 4th International Conference on Ubiquitous Information Management and Communication, 38. ACM, 2010.

[90] We do not distinguish between linking files and referencing scholarly literature, for instance with a BibTeX key. Citations, links to files on the user's hard drive and hyperlinks to websites are all considered as 'link'.

[91] http://www.sciplore.org

## D.2 Related Work

Several attempts have been made to establish research paper recommender systems [2, 57, 145, 153, 381, 393]. Some of them use citation analysis to determine the degree of relatedness between two papers. An overview of different citation analysis approaches for determining the relatedness of research papers is given in [257]. At this time, our research focuses on *co-citation analysis* [267] and its extension *citation proximity analysis* [142].

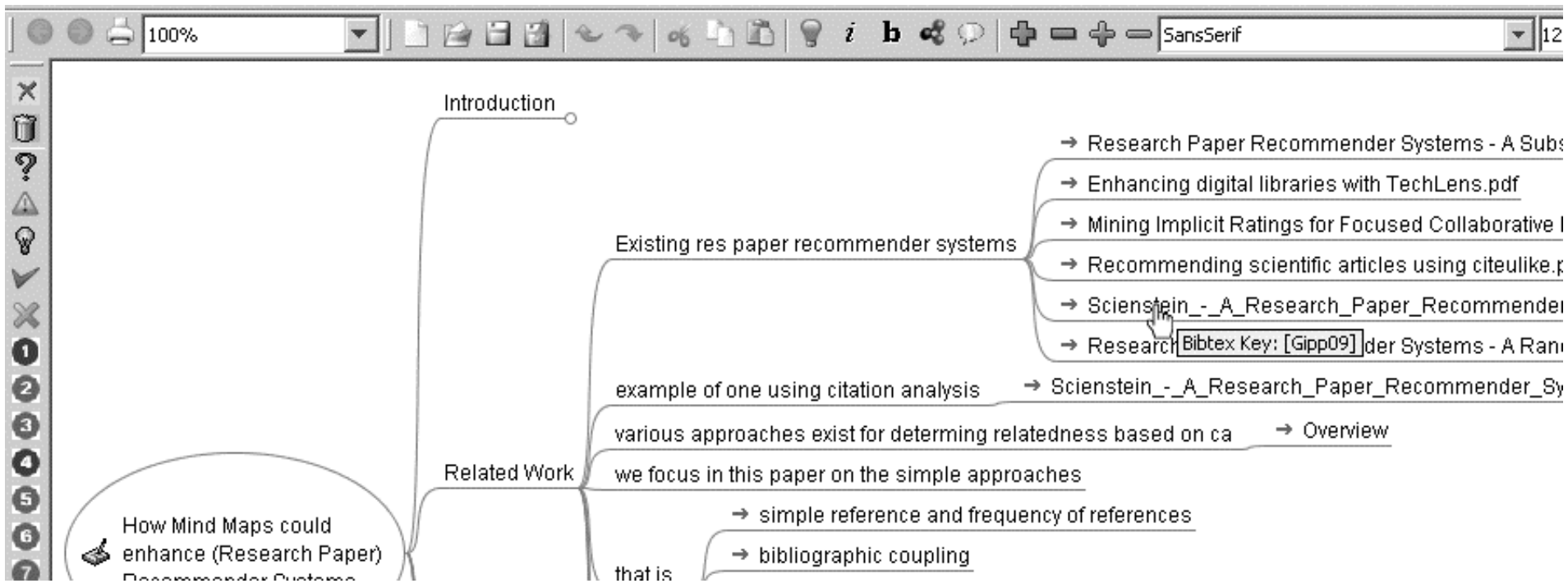

Figure 78: Mind map draft of a paper

According to co-citation analysis, two papers A and B are related if a third paper C references both. If more than one paper reference paper A and B together, their relatedness is supposed to be even higher. Citation proximity analysis additionally considers the location of citations in the full text: Two papers A and B are supposed to be more highly related when they are closely referenced by a third paper C in the text. For instance, if paper C references paper A and B in the same sentence, A and B are likely to be highly related. If paper C references paper A in the beginning of a 100-page document and paper B at the end, their relatedness is probably not nearly as high.

Co-citation analysis and citation proximity analysis can be used by research paper recommender systems to make item-based recommendations: If paper A and B are related, paper B may be recommended to those users interested in paper A (but not knowing paper B yet).

However, co-citation analysis and citation proximity analysis have to cope with some drawbacks.

1. **Availability of Data:** Co-citation analysis and citation proximity analysis cannot be applied to all research papers due to a lack of (correct) data [240, 259]: many research papers are not cited at all; citation databases such as ISI Web of Knowledge do not cover all available publications; and due to technical difficulties, citations are not always recognized correctly, which in turn leads to incorrect data in citation databases.

2. **Robustness of Data:** Citations are often considered as biased because authors do cite papers they should not cite and do not cite papers they should cite [259]. Accordingly, citation based recommender systems might provide irrelevant recommendations.
3. **Timeliness of Data:** Publishing scientific articles is a slow process and it takes months or even years before they are published and citations are received. Accordingly, documents recommended based on citation analysis are, at the very least, several months old.
4. **Metrics:** There exist metrics for measuring the relatedness of research papers based on citation analysis (for instance, *coupling strength* [359] or the *citation proximity index* [142]). However, to our knowledge, each metric focuses solely on one citation analysis approach and no combining metric exists yet. Consequently, relatedness of research papers based on citations cannot be measured and expressed thoroughly.

Summarized, citation analysis applied to scholarly literature can do a good job in identifying related articles, but there is room for improvement.

## D.3 Methodology

Our intention was to conduct an experiment to obtain first indications if *Link Analysis in Mind Maps* (LAMM) might be suitable for determining research paper relatedness. Two assumptions were researched:

1. Two research papers A and B are related if at least one mind map links them both
2. Two research papers A and B are more highly related the more closely they are linked within a mind map

As part of the experiment, five mind maps were analyzed which were originally created for drafting research papers, respectively Masters Theses[92]. That means each of the mind maps links at least to a few PDF files representing academic articles. From each mind map, links (respectively citations) to three articles were extracted and pairs were built (Figure 21 for illustration). The first pair was built from the first and second link in a mind map. Since the distance between them was low, we expected this pair to be 'highly related'. The second pair was built from the first and last link in the mind maps. Here, the distance between the links was high. Accordingly, we expected the corresponding articles to be less closely related.

[92] Two mind maps represented drafts of our own papers and three mind maps were created by some of our students for their Masters' theses.

To test our assumptions, titles and abstracts of the linked PDFs were extracted. Since five mind maps were analyzed, five pairs with low distance (expected relatedness = (very) high) and five pairs with high distance (expected relatedness =low) existed. In addition, five 'control pairs' of papers were created. We created these pairs in a way that they should appear as not being related to each other at all[93].

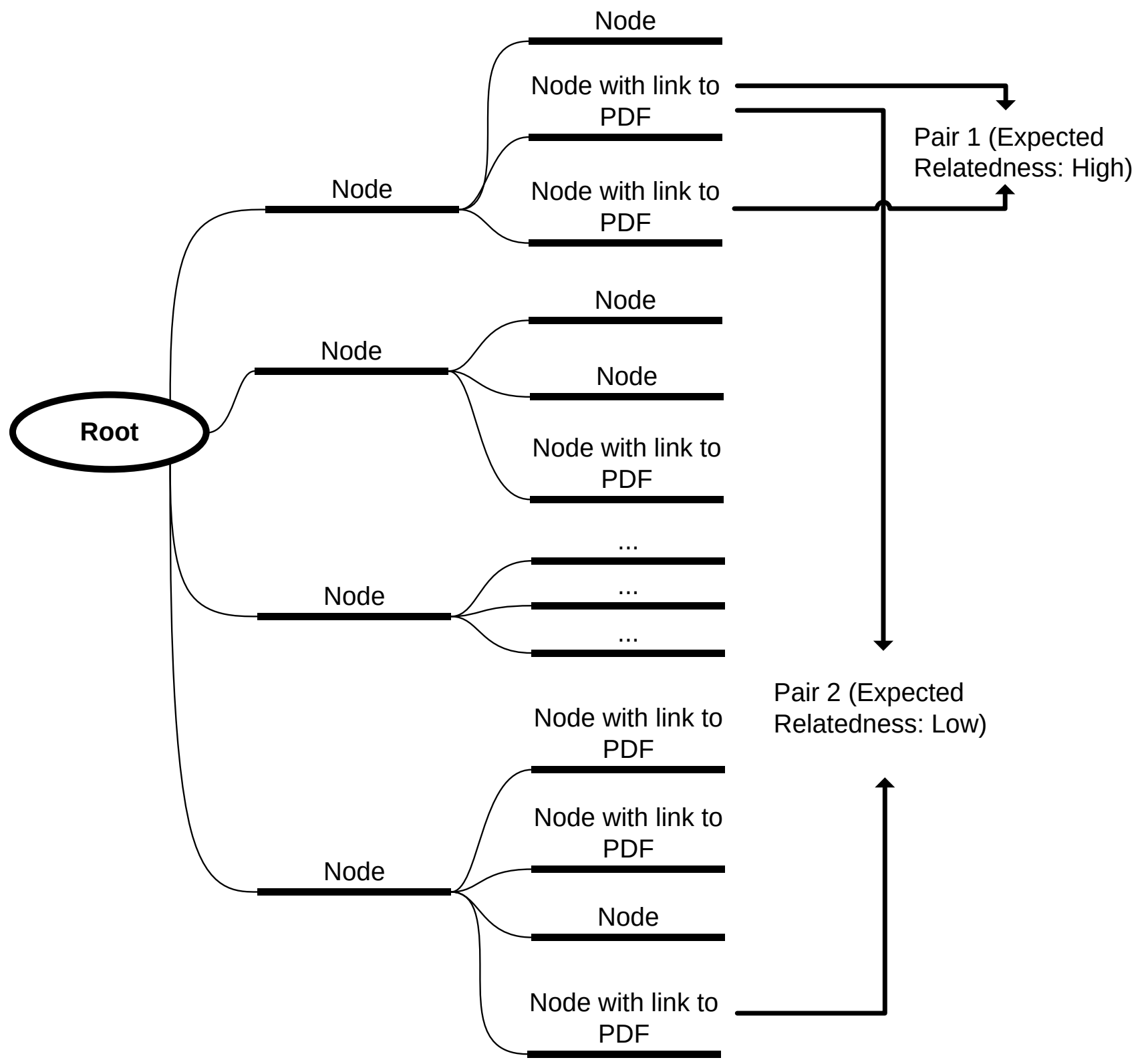

Figure 79: Link extraction from the mind maps (illustration)

[93] The papers were taken from the SciPlore database, were not linked by any of the mind maps and did not cite each other.

All pairs were shown to five participants[94] and the participants had to rate the relatedness of the pairs on a scale from 1 to 5 (1 = not related, 5 = highly related). For evaluation, ratings were painted in a scatter plot for each participant as well as the overall rating (mean and median). A more detailed statistical analysis was not considered necessary, since the graphs showed quite clear results and the amount of data was too little for extensive statistic analyses.

## D.4 Results

Figure 80 shows the results. On average (mean), those pairs linked closely together in the mind maps were considered significantly more often (highly) related than those pairs not linked closely together. The control pairs, which were not linked by any mind map, were all rated as not related, on average.

Some outliers exist: On average, pair 2 in mind map 2 was considered higher related than pair 1 in mind map 2. In addition, pair 2 of mind map 3 and pair 1 of mind map 5 were rated as almost not related. However, this is not surprising since mind maps are usually used for drafting a paper and therefore variances are to be expected.

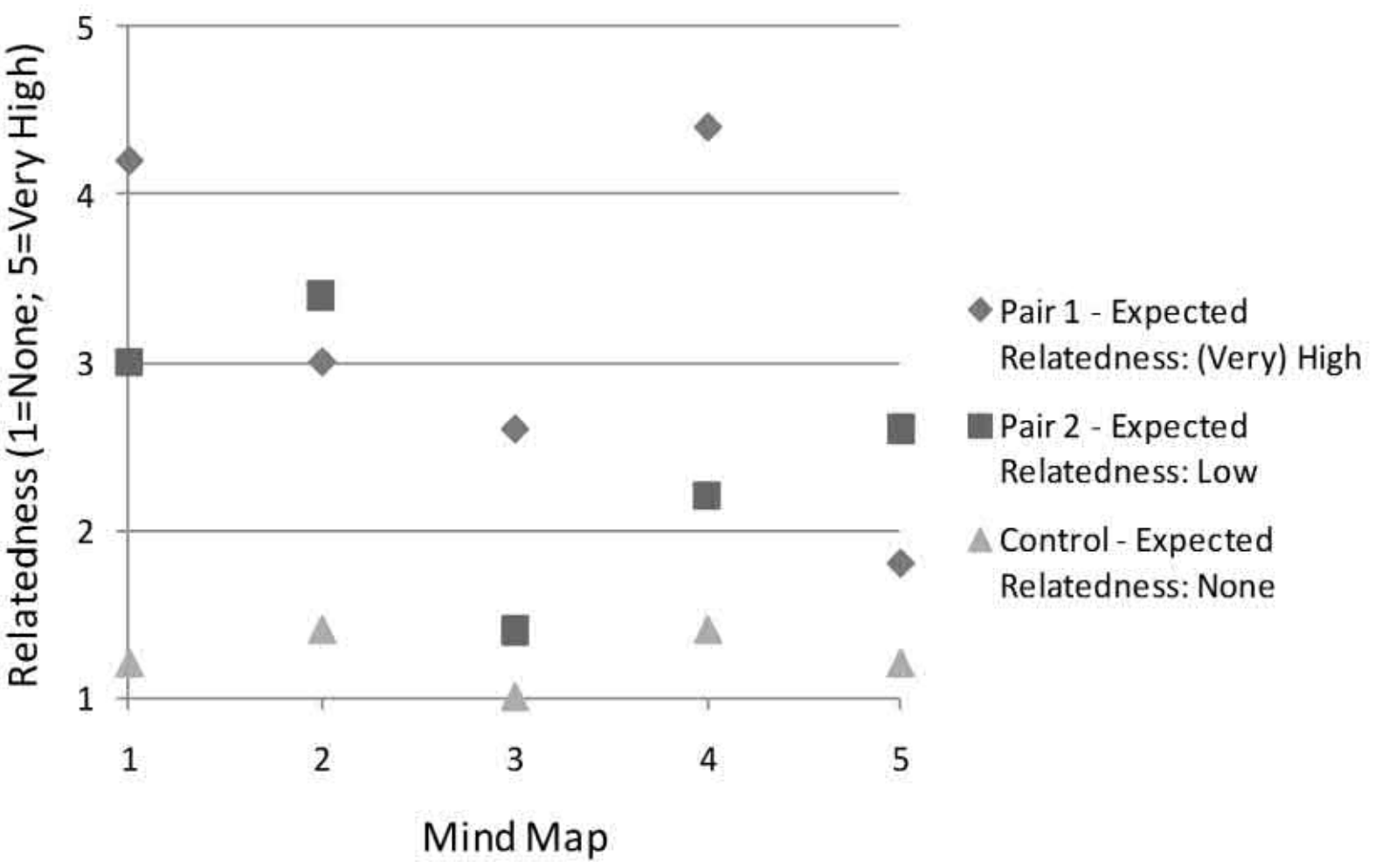

Figure 80: Relatedness of pairs in mind maps (mean)

---

[94] None of the participants were involved in creating the mind maps. The pairs of papers were distributed to the participants without their knowledge of the pairs being linked by a mind map or not. Each participant was shown all 15 pairs at once.

## D.5 Discussion

Overall, the results are a first indication that mind maps can be used to calculate research papers' relatedness. However, it needs to be emphasized that all five mind maps were created by our students and ourselves and hence came from the same 'school of thought'. It's very possible that other researchers use mind maps in a different way, which would then lead to variations in the results.

In addition, similar problems as for classic ciation analysis are to be expected for *Link Analysis in Mind Maps*. These problems are related to data availability, robustness, timeliness, and metrics and are discussed in the following sections.

### D.5.1 Availability of Data

Data availability seems to be the main challenge LAMM will have to face. It is unknown how many researchers use mind maps and how many are willing to share their data. It could be that the number is rather low. Nevertheless, mind mapping is a popular application. For instance, the mind mapping tool *FreeMind* is downloaded over a 150,000 times a month [363], more than 1.5 million people use *MindManager* [287] and there exist dozens of tools more [105]. Even platforms for sharing mind maps exist already[95]. On our website sciplore.org we also offer a special mind mapping software for researchers which will enable us to collect mind maps [35].

Overall, we are confident, that sufficient data can be collected that makes LAMM worth researching. Certainly, it will never replace citation analysis in scholarly literature or hyperlink analysis on websites but LAMM could serve as a complement for both.

Technical problems (in terms of identifying references) should be equal or even less for LAMM than for classic citation analysis. If users link to a unique identifier such as a BibTeX key, the corresponding metadata should be easily extractable from the user's bibliographic database. If the user links a PDF file, at least the title should be easily identifiable from the PDF, in most cases[96].

---

[95] For instance, http://www.mappio.com, http://share.xmind.net, and http://www.mindmeister.com/maps/public/

[96] We developed a tool for extracting titles from PDFs. First tests are promising.

### D.5.2 Robustness of Data

All social media platforms do have to cope with spam and fraud as soon as they become successful. There is no reason to assume this would be different if mind maps were used for calculating relatedness of documents. However, most social media platforms also find a way to cope with fraud and spam. If only mind maps of 'trusted' users were used, serious spam and fraud could probably be prevented successfully. Trustworthiness of users probably could be determined in cooperation with social networks, other community websites or by usage data of mind mapping software.

### D.5.3 Timeliness of Data

With LAMM, timeliness has a clear advantage over classic citation analysis. Mind maps do not need to be published in journals or at conferences. They could be analyzed the moment they are created. This would enable research paper recommender systems to recommend new publications faster than with classic citation based approaches.

### D.5.4 Appropriate Metrics

LAMM could use the same metrics that are used for citation analysis. Perhaps slight modifications would have to be made, but overall, metrics should be very similar (and so the advantages and disadvantages of citation based metrics).

## D.6 Summary & Future Research

We presented L*ink Analysis in Mind Maps* (LAMM). LAMM is an approach for determining the relatedness of documents by applying methods from hyperlink and citation analysis to mind maps. The basic idea is: If two documents A and B are linked or referenced by a mind map, these articles are likely to be related. Consequently, a recommender system could recommend document B to those users liking document A. In addition, we proposed that two documents are higher related when their proximity in the mind map is higher. In a small study (five mind maps and five participants) we obtained first indications that our assumptions could be true. The participants rated research articles that were linked in high proximity in the mind map, as more highly related than those articles linked within low proximity. Advantages and problems of LAMM in comparison to classic citation analysis were also discussed. Especially in respect to timeliness, MMCA seems likely to outperform classic citation analysis. On the other hand, data availability is likely to be a much larger problem than it is for citation analysis.

Overall, LAMM might prove to be a promising field of research having the chance to complement classic citation analysis and enhance research paper recommender systems in the long run. However, there is a need for more research since many questions remain unanswered:

- How many researchers are using mind maps?
- How many are willing to share them?
- How can spam and fraud be prevented?
- Which metrics should be used to measure relatedness?
- How should these metrics be combined with existing ones based on citations and other techniques (for instance, based on text mining and collaborative filtering)?

While we focused on determining relatedness of scholarly literature, LAMM could be applied equally well to other document types such as web pages.

# E Docear4Word

## E.1 Introduction

Reference management probably is the most tiring task for students and researchers. They have to re-type and format bibliographic information over and over, for each paper, assignment or thesis. This is particularly frustrating if they need to change citation styles in a document. This might become necessary, for instance, because a supervisor changes his mind on his favorite citation style, or a paper is submitted to another journal, which requires a different citation style than the previous journal.

In the past decades, many software tools evolved to facilitate this workflow. Commercial tools such as Endnote and Citavi enable researchers to maintain a database with all the bibliographic data of their references. These so called ‘reference managers’ usually offer add-ons for Microsoft Word allowing users to insert and format references and bibliographies in a convenient way. Also some open source tools offer such add-ons, for instance Zotero. However, all these tools use proprietary data formats. Accordingly, a Microsoft Word add-on from one of these tools (e.g. Endnote) works only with corresponding data format of that particular tool.

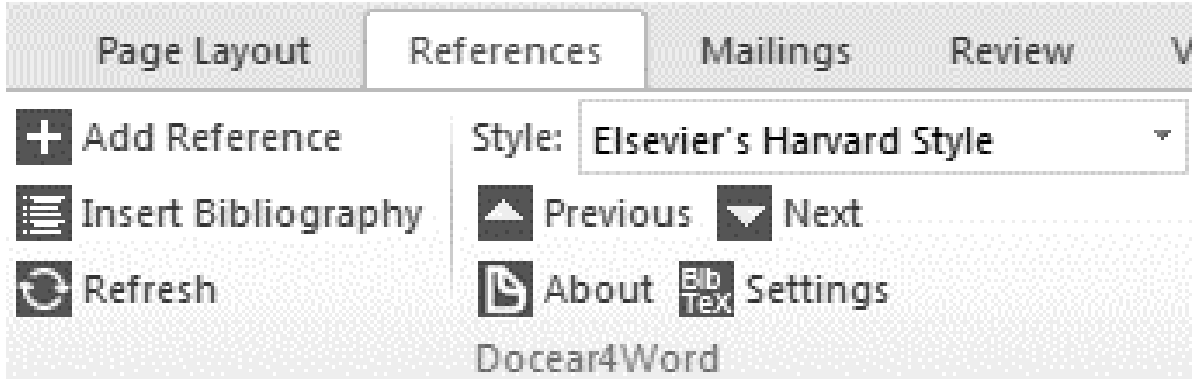

Figure 81: Docear4Word ribbon in Microsoft Word 2010

There is one alternative to the proprietary formats, namely BibTeX. BibTeX was created by Oren Patashnik in 1988 [317] and is the de-facto standard to store references. There are many reference mangers directly supporting BibTeX, for instance JabRef, BibDesk and our own reference manager Docear [33]. Even proprietary tools such as Endnote usually allow exporting their database to BibTeX. There is a Microsoft Word add-on for BibTeX-based databases named BibTeX4Word[97]. However, BibTeX4Word requires the installation of additional tools and is difficult to setup and use. For instance, in the Blog MedicalNerds more than 250 comments were made on BibTeX4Word, mostly questioning about the usage – especially installing new citation styles is complicated [276].

[97] http://www.ee.ic.ac.uk/hp/staff/dmb/perl/index.html

We developed Docear4Word[98], a Microsoft Word add-on to insert and format references directly in MS-Word, based on BibTeX. Docear4Word is open source and runs with Microsoft Word 2002 (and later) on Windows XP (and later). After the installation, Docear4Word is accessible in the "Reference" ribbon when Microsoft Word 2007 or later is used (Figure 81). In Word 2002 and 2003 a separate toolbar is installed. Docear4Word was primarily intended for users of our literature management tool Docear [33], but it can be used with any BibTeX file from any reference manager. In contrast to BibTeX4Word, Docear4Word is more user-friendly and uses the citation style language (more details in the following section).

The remainder of this chapter provides a detailed overview of Docear4Word.

## E.2 Maintaining a BibTeX database

BibTeX is a text-based format. Accordingly, a BibTeX file can be created and edited with any text editor. However, there are several tools offering a graphical user interface to create BibTeX files, for instance JabRef and our own reference manager Docear. Figure 82 shows a screenshot of Docear. Docear provides a graphical user interface for specifying title, authors, and other bibliographic data of academic literature. Based on this data Docear automatically creates a BibTeX entry. Instead of Docear, or a text editor, any reference management tool can be used that uses the BibTeX format or that may export its proprietary format to BibTeX.

## E.3 Inserting references in Microsoft Word

Figure 83 shows the dialog to search and insert references. The dialog allows selecting several references at once and specifying individual page numbers. Once, the references are selected and the "Add References" button is clicked, references are added in the document and formatted accordingly to the selected citation style (Figure 85).

[98] http://www.docear.org/software/add-ons/docear4word/overview/

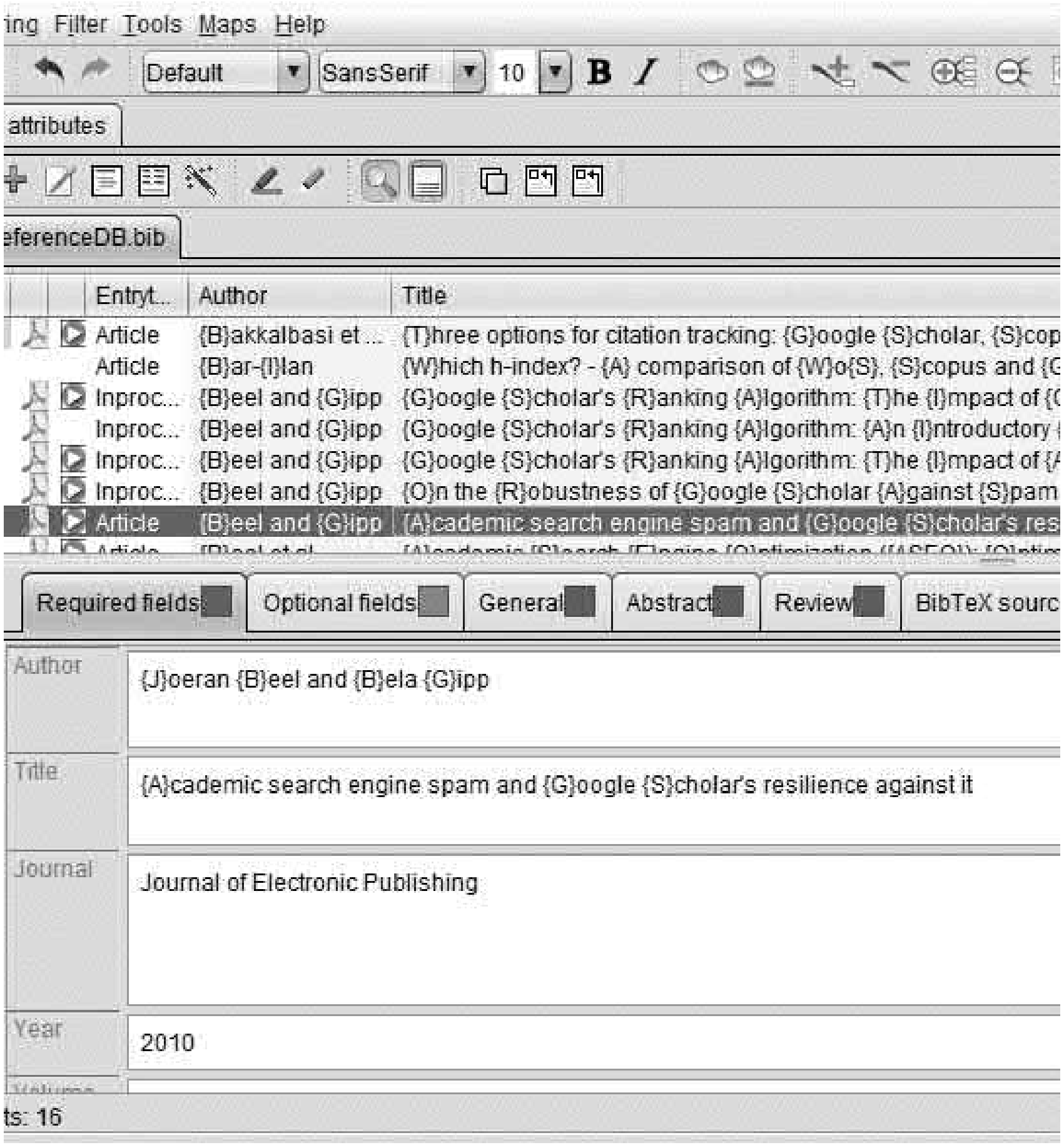

Figure 82: Maintaining the BibTeX database

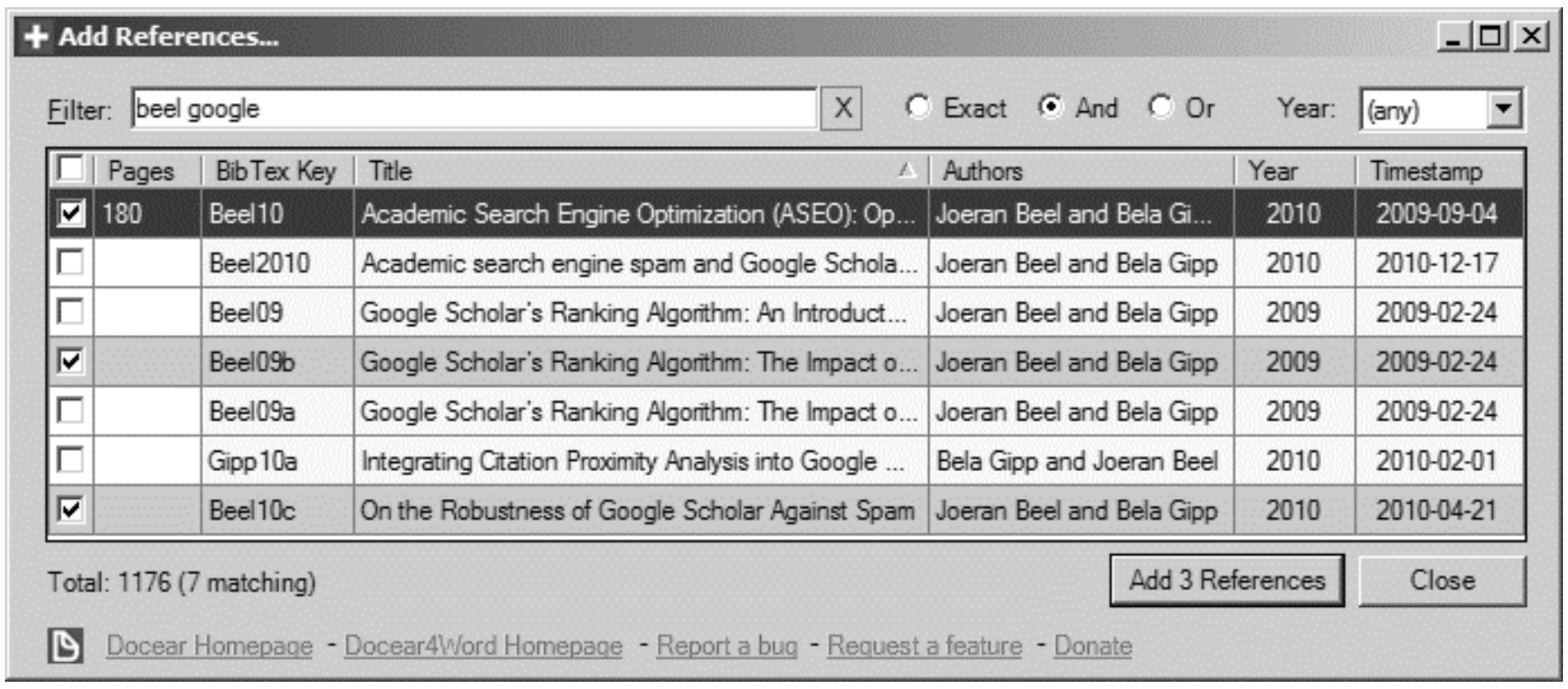

Figure 83: Selecting and inserting a reference

## E.4 Changing the Citation Style

Docear4Word uses the citation style language [430] and citeproc-js citation processor [120] to format references. The citation style language supports more than 1,700 citation styles such as IEEE, Harvard, MLA, and ACM in several variations. Docear4Word users can select a style in the style box (Figure 84) and install new styles from the style repository. When a new style is selected, all references are formatted accordingly.

## E.5 Insert a Bibliography

Docear4Word automatically creates a bibliography based on the references in the body of the document (Figure 85). The user can choose where to insert the bibliography and the bibliography is automatically updated when new references are inserted.

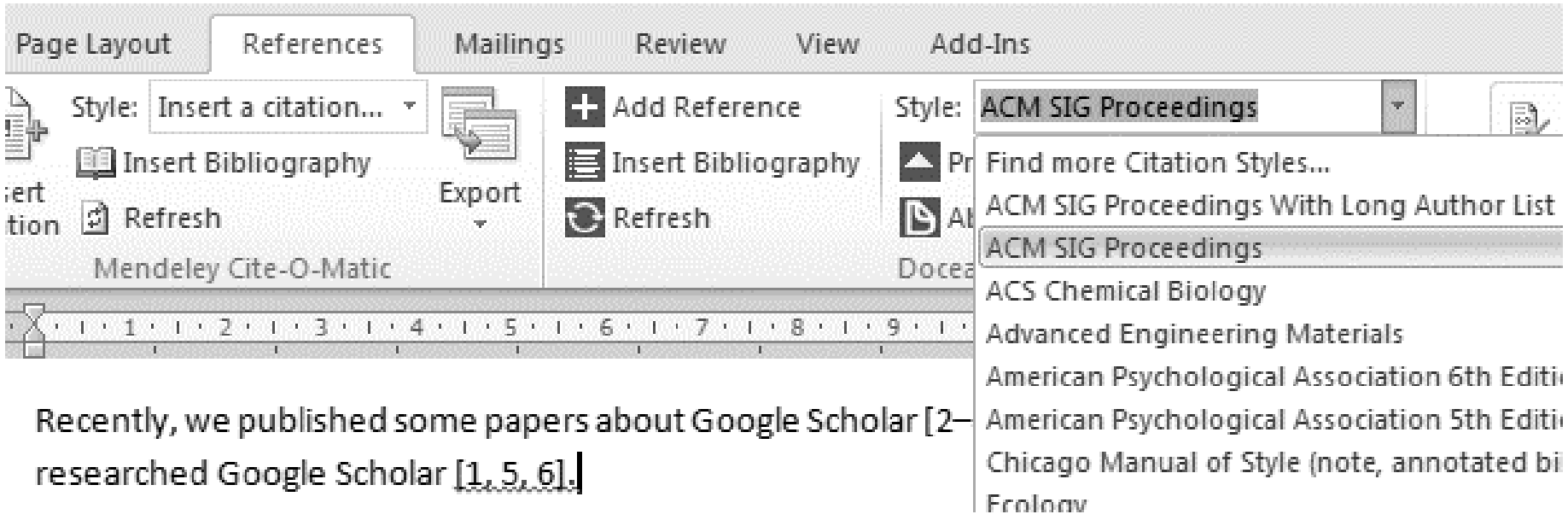

Figure 84: Style chooser

## E.6 Outlook

Docear4Word was released as final and stable version 1.0 on http://docear.org. However, we will continue to improve Docear4Word. Among others, it is planned to offer a version for Microsoft Word on MacOS; implement support for footnotes; enable suppressing author and/or year in a reference; implement an installer for new citation styles; and allow using multiple BibTeX files at the same time. As Docear4Word is available as open source, we sincerely invite other researchers to join the development.

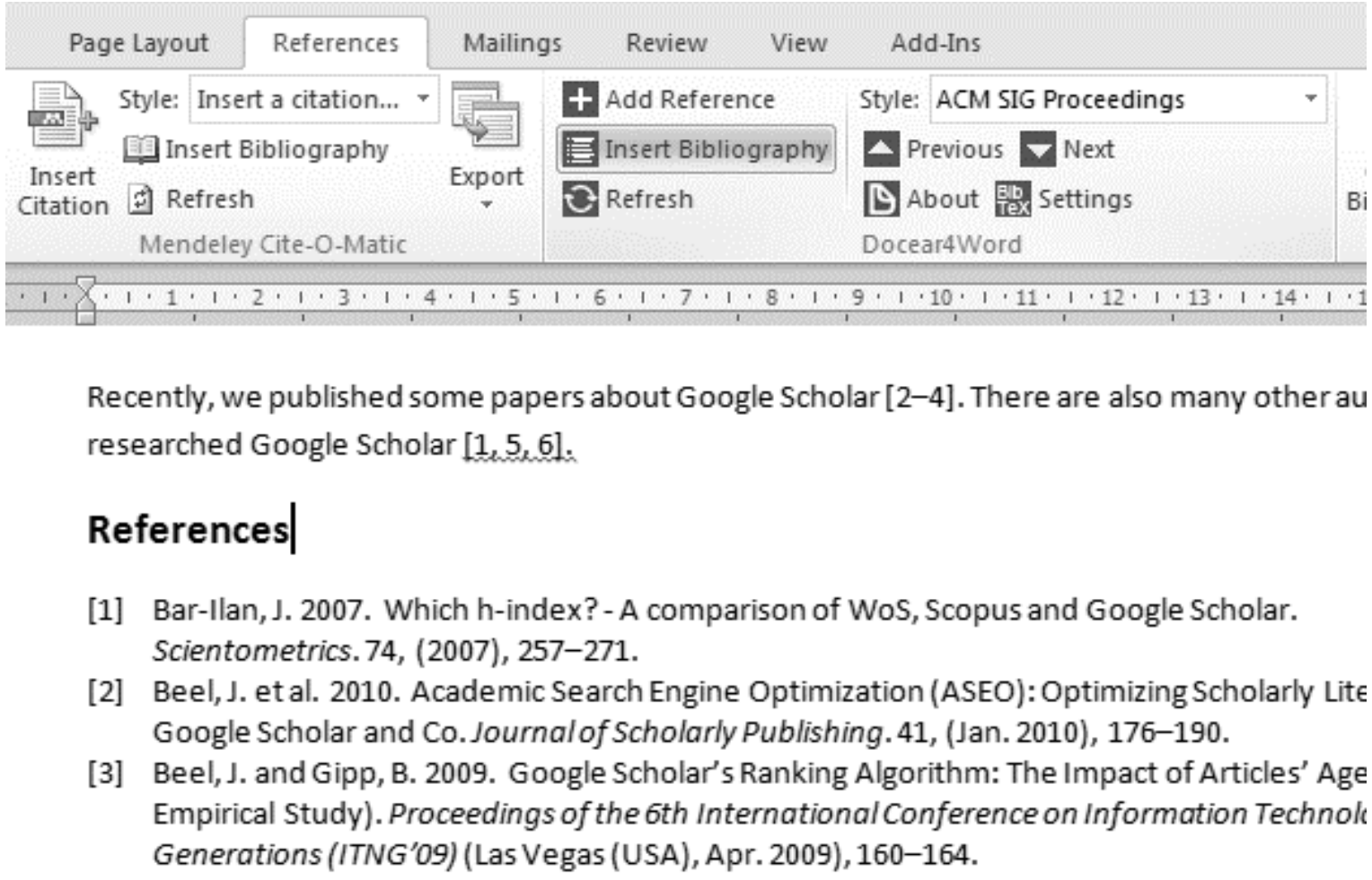

Figure 85: Automatically created bibliography

# F Review of the Recommendation Approaches

Our main goal was to provide a general overview and discussion of the research field. However, for the interested reader, we provide a short summary of the individual approaches in the following.

## F.1 Content-based Filtering Approaches

### F.1.1 CiteSeer(x) and CC-IDF

In 1998 C. Lee Giles, Kurt D. Bollacker, and Steve Lawrence introduced CiteSeer, a web-based digital library [59, 139]. With CiteSeer, users could search for research papers, and for each search result, CiteSeer offered a link for retrieving a list of "related documents". This might not be considered a "real" recommender system but it was the first step towards it, and Giles et al. also called it a recommender system. Document relatedness was calculated based on three factors. First, text similarity was calculated based on the top-20 words per document, which were determined with TF-IDF. Second, CiteSeer calculated header-similarity between documents with a string edit distance (the authors interpreted as all the text before a paper's abstract as header). Thirdly, CiteSeer introduced a new similarity measure, which they called CC-IDF. CC-IDF was identical to TF-IDF but instead of terms, citations were used. The underlying idea is that the more citations two documents share, and the less other documents contain these citations, the more similar the two documents are. Both papers [59, 139] provide identical information about the recommender system and are rather sparse of details. An evaluation of the recommendation approaches is missing.

One year later, the three researchers presented the very first 'real' research paper recommender system that offered three different approaches to users [236]. First, users could create citation alerts. Whenever a paper, that the user had to specify manually, was cited by a new paper, the user was informed. Second, users could specify constraints. For instance, users could specify keywords and whenever CiteSeer indexed a new document matching the keywords, the user was informed. Thirdly, users could specify documents they liked and whenever CiteSeer found new documents related to the liked documents, users received a recommendation. Document relatedness was based on TF-IDF and CC-IDF (header similarity was not mentioned any more). User profiles could directly be maintained by the users and recommendations were sent by email or shown on CiteSeer's website. The paper, again, contained no evaluation.

In the following years, CiteSeer was continuously improved which resulted in a new version called CiteSeer*X* [94, 247, 323, 324]. Some of CiteSeer's functionality was also patented by NEC Laboratories, the employer of Giles, Bollacker, and Lawrence [238, 239]. Today, CiteSeer(x) does not provide any recommendation functionality, although Giles et al. published several papers more about research paper recommendations, which are covered later in our review. Recommendations nowadays are given by RefSeer, which was built on top of CiteSeer and takes a text snippet or PDF file as input [342] (p. 233).

### F.1.2 Quickstep & Foxtrot

Middleton et al. published five papers between 2001 and 2009 about their recommender systems *Quickstep* and *Foxtrot* [282–286] (the fifth paper from 2009 [283] is a summary of their papers published between 2001 and 2004).

In the first paper, Quickstep was introduced and an evaluation was conducted whether flat topic lists or hierarchical ontologies are more effective for recommendations [284]. Based on machine learning, Quickstep classified each research paper with one topic, derived from DMOZ's computer science categories. When a user browsed a research paper, the paper's topic was added to the user's user model with a "topic interest" value. This interest value differed, based on how often users browsed papers with that topic, the number of days being passed since the last browsing, and some other factors. Quickstep recommended those papers whose topics correlated best with the user's topics of interests. Additionally, Middleton et al. experimented with adding the topics' parents to the user model. In DMOZ, the classification is not flat but hierarchical. This means, when a paper was assigned to one category, all parent categories were also assigned to that paper, and user models respectively. In a small-scale evaluation, the authors found that using a hierarchical ontology was slightly better than a flat list of topics in terms of click-through rate, and 7-15% better in terms of user-satisfaction.

In 2002, Middleton et al. published a poster introducing Foxtrot [285], an enhanced version of Quickstep. In Foxtrot, users' profiles were visualized and users could edit their profiles. The poster contains only brief information about Foxtrot and no new recommendation approaches or evaluation. In 2004, Foxtrot was presented in more detail in a journal article [286]. The article consisted of three parts. The first part contained the information and experiment from the first Quickstep paper (some details from the first paper were missing, and some new ones added) [284]. The second part was about bootstrapping a recommender system based on an external ontology. This part was also published as pre-print on arXiv.org in 2002 [282]. The third part was about Foxtrot, comparing user relevance feedback with profile feedback (Section 3.3.2.7). Middleton et al. also provided some statistics on the usage of their recommender system and differences between user groups (Sections 3.3.2.4 and 3.3.2.5). Another difference between Foxtrot and Quickstep was that Foxtrot used the CORA classification [273] instead of DMOZ categories.

### F.1.3 Topic Sensitive Similarity Propagation (TSSP)

Huang et al. were first to apply citation-context analysis for generating research paper recommendations [179]. Citation context is the text that surrounds a citation. For instance, the citation contexts of Huang et al.'s paper, based on our paper, would be "`Huang et al. were first to apply citation context analysis for generating research paper recommendations`", which is the sentence that includes the reference to their paper. Huang et al. calculated two similarity values separately, both being based on TF-IDF in the vector space model. One similarity value was calculated for citation contexts, and one was based on documents' body text. Then, both values were combined. The highest precision was achieved when content and citation similarity received the same weight. Huang et al. called their approach Topic Sensitive Similarity Propagation (TSSP) and evaluated it with a small user study based on 28 papers and an unknown number of participants. TSSP outperformed the baselines co-citation, SimRank [195], content-only (without citation context), and some variations. More precisely, TSSP performed twice as good as content-only (p@10 of 0.52 vs. 0.25). Also a linear combination of content and SimRank performed well (p@10=0.41). In contrast, SimRank (0.18) alone and co-citation (0.19) performed poorly. However, due to the small study size it is questionable how representative these results are.

### F.1.4 Mixed-membership model / Link-LDA

In 2004, Erosheva et al. modeled citations and terms with Latent Dirichlet Allocation (LDA) which is similar to Probabilistic Latent Semantic Analysis (PLSA) [111]. They utilized the terms contained in the document's abstracts and the references. For each document, two models were created. Terms were weighted based on simple term counts, references were weighted binary (present or not present). The paper itself has no sound evaluation, as the approach was not tested against any baseline. However, their work influenced several other papers in the field of citation recommendations [202, 298], which are covered later in this survey. The mixed membership model is later referred to as 'Link-LDA'.

### F.1.5 Papits

In 2002, Papits was introduced as a peer-to-peer system that supported researchers in finding literature, keeping a research diary, and sharing papers [309, 310][99]. Between 2004 and 2006, the authors published one paper about document classification [312] and two about integrating a research paper recommender system in Papits [311, 409]. The authors distinguished between short- and long-time interests of their users [409]. Their way of representing user models is unique in the domain of research-paper recommender systems. Each user was represented as a graph. The graph's nodes contained the terms of the users' papers. The graphs' edges were based on the frequency of co-occurrences of terms. Their approach outperformed standard CBF with a precision of 0.57 vs. 0.27. In their last paper, the authors compared a kNN approach with SVM to classify research papers [311]. Both approaches performed about the same. The papers are in some parts difficult to understand, have several typos and some important information is missing about the proposed approach and evaluation. This might be the reason why the papers received little attention (citations counts of their papers are between one and four). Nevertheless, especially the paper from 2005 is worth reading [409]. The idea of representing user models as graph is unique and the evaluation indicates an excellent precision.

### F.1.6 Trust-based Scientific Paper Recommender (SPRec)

Between 2006 and 2007, Claudia Hess developed a research paper recommender system as part of her PhD thesis [172]. Hess focused on trust between researchers and reviewers respectively. The basic idea is that if a user A trusts another user B, B's article reviews are (more) important to user A than other reviews. Hess considers the citation graph between papers as a trust network, expressing which authors trust each other. With respect to trust-based recommender systems, the thesis might be an interesting read, but the recommender system itself is only described superficially. Hess also published two papers [171, 173] but the thesis contains the more interesting information.

### F.1.7 PubMed Related Articles (PRMA)

In 2007, PubMed, a large academic search engine for biomedical literature, introduced a recommender system that is still available today [251]. When users browses a detail-page of an article, related articles are displayed. To find related articles, PubMed uses what they call PRMA (PubMed Related Articles), an algorithm similar to BM25. PRMA is based on terms contained in the documents' titles and abstracts. One fundamental difference between BM25 and PRMA is that PRMA was developed for using MeSH terms, i.e. a controlled vocabulary available only for biomedical literature. PRMA outperforms BM25 with p@5 of 0.399 vs. 0.383 (statistically significant). The authors stress that PRMA is less complex than BM25, which makes the parameter optimization process less difficult. The authors also report that 20% of users who browse an article's detail page follow at least one recommendation. Although the recommender system is active since 2007, no more papers were published about PRMA, to the best of our knowledge.

[99] Both papers are essentially the same

### F.1.8 Recommending Citations

Strohman et al. combined classic CBF and ranked papers with a combination of several metrics [369]. Strohman et al. expected a user to provide a manuscript containing text and that the user wanted recommendations for papers to cite. Based on the input manuscript, the system determined 100 recommendation candidates. The candidates were determined based on text similarity, whereas text similarity was calculated with multinomial diffusion kernel [225]. The original 100 candidates were extended by all papers being cited by the 100 candidates. Extending the candidate set increased MAP by around 10%. Strohman et al. also considered including another level of cited papers (those papers being cited by the papers being cited by the original candidates) but initial experiments showed that this did not improve the effectiveness. The papers in the final candidate set (typically 1,000-3,000) were ranked based on different combinations of publication year, text similarity, co-citation strength, same author, citation count, and the Katz measure. An offline evaluation showed that especially the Katz measure strongly improved precision. All variations including Katz were about twice as good as those variations without. For a 2-page poster, the article provides many interesting information. However, caused by the space restrictions, the authors had to omit many details, for instance how exactly the different factors were used for the ranking[100]. This limits the value of their results, and makes re-implementing their approach impossible. Strohman et al. state that full details can be found in an extended version of their paper (a technical report), but we were not able to find this technical report. Other authors [164] also criticize that Strohman et al. used a biased ground-truth to calculate the Katz measure, which would favor the Katz measure.

[100] They report to have used „coordinate ascent" to learn the weights, but do not provide a reference to this concept (of which we have never heard of).

### F.1.9 Concept-Based Recommender System

Susan Gauch et al. used the ACM classification tree to generate recommendations [80, 209, 228]. Papers were represented as trees, namely the ACM classification concepts, including parent nodes. Users were represented by the concept trees of the papers they authored. To match users and papers, Gauch et al. used cosine similarity for the concepts, and the tree-edit distance for the concept trees. Both approaches outperformed a classic keyword based cosine matching, and the tree-edit distance outperformed cosine-based concept similarity. The paper corpus was crawled from CiteSeer, and papers were assigned to the ACM classifications through machine learning. In their first article, the authors presented the basic approach to compare documents based on tree-edit-distance [228]. The second article focused on the user modeling part [80]. The third article focused on the general architecture and how the system could be integrated into CiteSeer [209]. It remains unclear whether the system was integrated into CiteSeer and whether CiteSeer supported the authors actively. The evaluations in all three articles were only small-scale with seven to nine participants.

### F.1.10 Pairwise Link-LDA & Link-PLSA-LDA

Nallapati et al. addressed the problem of jointly modeling citations and text [298]. They proposed two approaches called 'Link-PLSA-LDA' and 'Pairwise Link-LDA'. The later one is based on Erosheva et al.'s Link-LDA (p. 221) and combines it with the 'Mixed Membership Stochastic Block' [4] model, a probabilistic model originally developed to find related proteins. Although Pairwise Link-LDA achieves a better precision than Link-LDA, it is about 100 times slower and not feasible to apply with larger collections. The second approach, Link-PLSA-LDA, combined PLSA and LDA into a single graphical model and outperformed LDA as well with regard to precision as computing time.

### F.1.11 Cite-LDA & cite-PLSA-LDA

In 2010, Link-LDA and Link-PLSA-LDA were adopted by Kataria et al. who proposed Cite-LDA and Cite-PLSA-LDA [202]. The approaches utilize citations, the context of citations and the content of documents. In an evaluation, Cite-PLSA-LDA outperformed all other models on two tested datasets (CiteSeer and Webkb). The evaluation confirms, similar to TSSP (p. 221), that citation context is a highly effective source for retrieving terms. As many of the previous authors,

Kataria et al. focused only on calculating paper similarities, and they neglected the user modeling process.

### F.1.12 User's Recent Research Interests

Sugiyama and Kan used a classic CBF approach in which users were represented by the papers they published [372]. The authors experimented with several variations of building user models and weighting papers. Among others, they extended the list of papers utilized for user modeling by adding all papers that were cited or did cite the original input papers. Sugiyama and Kan tried four approaches to weight these neighbor papers. First, papers were weighted equally. Second, neighbored papers were weighted the stronger the higher their contently similarity with the original input papers was (cosine similarity). Third, papers were weighted the stronger the closer their publication year was to the publication year of the original publications. Fourth, for senior researchers with several papers being published, papers that were more recent were weighted stronger than older papers. Sugiyama and Kan incorporated PageRank to weight papers but using PageRank decreased the effectiveness. The authors assumed this was because PageRank favors older papers but users are more interested in recent papers. The authors also extended candidate models by incorporating cited and citing papers but this increased the accuracy only slightly. While the authors used plain term frequency for user modeling, they used TF-IDF for the candidate papers. Their argument was that the small number of publication utilized for user modeling would negatively affect the IDF calculation. However, we know of no evidence supporting this assumption. The authors found that it makes sense to include only papers being published in the past three years in the user modelling process.

### F.1.13 Social Tag Based Recommender System

Choochaiwattana et al. published two articles about using social tags for research paper recommendations [86, 197]. Papers were represented by the social tags that users had added. User were represented by the tags of those papers that the users had in their collections. The matching was done by calculating cosine similarity in the vector space. Sadly, both articles do not compare their approach against a baseline. The content of the two articles is nearly identical, only that the first has an evaluation with three participants [197], while the second article has an evaluation with 15 participants [86].

### F.1.14 Context Aware Relevance Model (CRM)

He et al. introduced the context-aware relevance model (CRM), which suggests papers that users could cite in a specific sentence of their manuscript [164]. He et al. expected a user to provide either a manuscript as input, or a citation context, i.e. a few sentences that needed citations. CRM searched for documents that contained citation contexts being similar to the input sentences or manuscript. These documents and the documents being cited were taken as recommendation candidates and ranked based on Gleason's Theorem, but CRM uses an approximate measure instead of maximum likelihood. He et al. differentiate between "global citations" for the bibliography, and "local citations" for each placeholder in the text. He et al. evaluated CRM against many different baselines and combinations, among others HITs, Katz, simple citation count, and simple text similarity. According to the paper, CRM is applied by CiteSeer. However, as mentioned previously, CiteSeer does not offer any recommender system (any more).

In 2011, He et al. extended their idea of a citation recommender so the recommender would analyze a manuscript and automatically recommend where exactly which citation was needed [163]. While the original approach is just a different view on recommender systems (there is no fundamental difference between a 'citation context' being provided by the user and a search query or abstract), this latter paper could be seen as the first true citation recommender because it addresses the problem of autonomously finding locations to add citations.

#### F.1.15 SVM-MAP Approach / Who Should I Cite?

Bethard and Jurafski used CBF and ranked the candidates based on a large number of factors [54]. The factors included text similarity; citation metrics such as citation count, PageRank, paper's venue citation count and h-index, and authors citation count; recency (older articles received less weight); social habits such as self-citation rate, co-authorship, boost for venues the user has previously published in, and several more. The article presents were several interesting findings. For instance, h-index had a negative impact on the algorithms' accuracy but plain citation count had a very positive impact. Bethard and Jurafski also found that authors like to cite papers they cited before. In addition, Bethard and Jurafski created an approach that learned to weight the factors with a support vector machine (SVM)[101]. In an offline evaluation, the "SVM-MAP" approach achieved an MAP of up to 28.7%, while a simple text comparison achieved only 15.9%.

[101] The authors also tried a logistic classifier but SVM performed better.

The results are interesting because they show how some simple metrics (mostly adopted from Scientometrics) can strongly enhance content-based filtering. The authors also published a simple prototype[102].

### F.1.16 Keyphrase-based recommender / Pirates Framework

Ferrara et al. created user models based on the content of papers a user had tagged [118]. Based on the terms contained in the tagged papers, they created three lists, one with uni-grams, one with bi-grams, and one with tri-grams. All n-grams were weighted based on their frequency, a part-of-speech value, phrase depth (based on the part of the document in which the term occurred first, e.g. title or abstract), phrase last occurrence (more weight to terms occurring in the conclusion), and phrase lifespan (portion of text being covered by the term). Papers were represented the same way. The cosine similarity was calculated for each of the three lists separately. The final similarity score was based on a linear combination. The approach performed better than using a uni-gram list alone. Unfortunately, the authors did not research how a different weighting (e.g. frequency only) or the individual n-gram lists (e.g. bi-grams only) performed. As such, it remains unclear how sensible it is to consider phrase depth, life span etc. for user modeling. However, the ideas are interesting but, to the best of our knowledge, the article received only little attention in the community.

### F.1.17 Source Independent Framework

A time-consuming task in building a research paper recommender system is harvesting a large-enough collection of research papers to recommend. Nascimento et al. bypassed this problem by creating brief user models which could be sent as search query to external information sources such as ACM Digital Library, IEEE XPlore, and Science Direct [301]. The search results were taken as initial candidate set and ranked by several Cosine variations. Based on the ranking, the top-ten search results were returned to the users as recommendation. The user-modeling approach itself was not novel. Nascimento et al. created user models based on terms in the title (weight=3), abstract (weight=2), and body (weight=1) of the users' papers. Results for the different search engines are interesting. Overall, ACM performed better than IEEE and Science Direct, but a combination of all three performed best. Other findings include that creating user models based on abstracts performs better than based on titles only and slightly better than based on body text. In addition, bi-grams performed better than noun-phrases. However, the evaluation was based on ten computer science students only.

[102] http://nlp.stanford.edu:8080/citation-retrieval/

### F.1.18 ResearchGate

ResearchGate is a social network for scientists and offers a recommender system, or more precisely an advanced search interface [334]. Users may enter search terms or an abstract and ResearchGate will show related papers that are determined with a classic content-based text comparison. We could find no more details on their recommender system.

### F.1.19 Docear[103]

Docear is a literature management software that allows managing PDF files, annotations, and references in mind maps [33]. In 2012, a content-based recommender system was integrated in Docear, and presented in a poster at JCDL 2013 [47]. The recommender system was based on Docear's predecessor, SciPlore MindMapping [20, 35], and utilized the users' mind maps for user-modeling. On average, click-through rate was around 6%. When users explicitly requested recommendations, CTR increased to 8.35%. Beel et al. also analyzed the impact of the user-model size. Small user models, containing five or less terms, only achieved CTRs of 1.80% on average. Larger user models with more than 11 terms, achieved CTRs of 6.67% on average. The best results were achieved when user models contained between 100 and 250 words. The authors also published several other papers about different aspects of research-paper recommender systems [42, 48, 52]. Beel et al. also applied stereotypes [51]. They assumed that all users of their reference management software *Docear* are researchers or students. Hence, papers and books were recommended that were potentially interesting for any student and researcher (e.g. a paper about optimizing scholarly literature for Google Scholar [38]). Beel et al. used stereotypes only as fallback model when other recommendation approaches could not deliver recommendations, for instance for very new users. They report mediocre performance of the stereotype approach with click-through rates (CTR) around 4%.

### F.1.20 Osusume

Osusume is the first Japanese paper recommender, according to Uchiyama et al. [395]. It distinguishes between different "viewpoints", i.e. requirements of the users. For instance, the authors assumed that some users are more interested in state-of-the-art papers and others in authoritative papers. Uchiyama et al. also distinguish between novice and expert users. To provide diversity in recommendations Osusume selects five recommendations randomly out of the top100 candidates. The authors claim that Osusume combines CF and CBF, but no details are provided in the articles. To us the approach appears like a classic CBF with some additional filtering criteria (e.g. to provide state-of-the-art paper, results were ordered by publication date). Uchiyama et al. did not evaluate their recommender system. They only asked 16 study participants which viewpoints were most relevant to them. Most users were interested in state-of-the-art and international papers (all users were Japanese and apparently interested in receiving recommendations for English articles). Only few users were interested in papers being just similar to the input paper. This finding is interesting because the main assumption of CBF is that users are interested in papers being similar to the ones they already know.

[103] Docear is included in this review, because the review was ment to be published as separate research paper.

### F.1.21 Translation Model

In 2011, Lu et al. proposed a unique and interesting view on the problem of giving citation recommendations [258]. The authors considered terms in a citation's context to be of a different language than terms contained in the cited document. Hence, they argued, the different languages needed a translation. To accomplish the translation, Lu et al. adopted the Translation Model that was introduced in 1999 by Berger and Lafferty and that is usually used for cross language search [53]. The translation model requires a training data set to 'learn' the language. Lu et al. used the citation contexts (three sentences around a citation) and the abstracts and full-texts of the recommendation candidates for the learning, and compared their approach against the context aware relevance model (p. 225), and the language model. The new approach outperformed both baselines. Interestingly, the language model performed better on the body-text than the abstract, and the translation model performed better on the abstract than on the body-text.

In 2012, He et al. enhanced their approach through position alignment, which enhanced the learning process by dividing a document into passages and considering their positions in the translation [162]. Based on an offline evaluation, the authors report a small, but statistically significant, improvement: mean average precision increased from 0.5829 for the standard translation model to 0.5919 for the position-aligned translation model. This is an improvement of around 1.5%. Somewhat interesting is the effectiveness of the baseline, i.e. the language model. In the first article, the language model achieved a mean average precision between 0.122 and 0.211, while the translation model achieved an MAP around 0.5, i.e.

more than twice as high [258]. In the second paper, the language model achieved an MAP of 0.4938, while the translation model achieved 0.5829 [162]. This is ‘only’ an increase of 18%.

### F.1.22 Citation Translation Model (CTM)

In 2012, Huang et al. adopted the idea of using the translation model for recommending citations [180]. In contrast to Lu et al., Huang et al. consider cited papers as entirely new words and give them unique IDs. In addition, they used “inverse citation context frequency” (ICF), which they adopted from the standard IDF measure. Finally, Huang et al. included co-citation data in their approach. They assigned all terms in a citation context to all references of that citing paper. Based on an offline evaluation, CTM outperforms TM and other baselines (Cite-PLSA-LDA, Link-PLSA-LDA, and CRM). Despite the excellent performance, Huang et al. acknowledge one significant problem of CTM, namely that only papers can be recommended that have been cited previously.

### F.1.23 Problem vs. Solution

Jiang et al. propose that are two types of relevancies, one problem-oriented relevance, and one solution-oriented relevance [196]. Accordingly, they try to find recommendation candidates that are most relevant to an input’s paper problem, or most relevant to its presented solution. To do so, Jiang et al. split the papers’ abstracts into a solution and a problem part. They did this manually for 200 papers (71% of the abstracts contained a clear distinction between problem and solution). Each paper then was represented by two vector space models, one containing the terms of the abstract’s problem section, and one with the terms of the abstract’s solution section. For both representations, separate recommendations were generated. According to their user study, with an unknown number of participants, their approach achieved higher user satisfaction than providing a single list of recommendations based on a combined vector space model. In addition to terms, Jiang et al. also experimented with topics and concepts. Topics were based on latent dirichlet allocation, concepts were based comparing n-gram terms with social tags from CiteULike. In many scenarios, topics and concepts performed better than single terms. The authors acknowledge that their approach requires a lot of runtime. To reduce runtime, user models (i.e. a single input paper), were only compared against those papers that were cited by the input paper or cited by the cited papers, or that the input paper was citing itself or that the cited papers were citing.

### F.1.24 Scholar Update

Scholar Update is a recommender system by Google Scholar [152]. On Google Scholar, researchers may create a profile listing their publications. Based on these publications, Scholar Update finds related articles and recommends these to the user. Related articles are based on content, the citation graph, and the authors the user works with and cites. Scholar Update also reports to consider concept drift. However, there are no further details available on the exact implementation of the recommendation approach.

### F.1.25 Mendeley Related Papers

Mendeley [165] offers two types of recommendations, namely content-based and collaborative filtering based [187]. Their content-based recommender system is a "related paper function" that shows recommendations based on a set of input papers [187]. Similarity is measured with cosine based on TF-IDF weighting and implemented with Lucene. Mendeley experimented with different document-fields for their recommender system (title, abstract, social tags, mesh-terms, author provided keywords, author name, general keywords) and combinations of the fields. Their evaluation shows that social tags outperform all other fields (p@5: 0.45 vs. e.g. abstracts p@5: 0.27). Even the best performing combination of several fields only achieved a precision of 0.36. There is no information provided about the evaluation except that it was an offline cross-validation based on a ground truth. However, since Mendeley has access to data of millions of articles and users, it seems likely that the results have some significance. Their collaborative filtering approach is reviewed later (p. 236).

### F.1.26 SemCir

In 2012/13, Zarrinkalam and Kahani introduced an approach to calculate paper similarity based on relational features [427] and built the recommender system SemCir (Semantic Citation Recommendation System) based on top of it [428][104]. SemCir indexes papers based on titles, abstracts and citation context. Users provide some text, which serves as user model. The initial candidate set is generated by selecting the *n* most similar documents based on content-similarity.

---

[104] The SemCir paper contains most of the first paper's information and reading it should be sufficient for most researchers.

All neighbor papers are included in the candidate set (cited papers, citing papers, papers from the same venue, same author, co-cited, bibliographic coupled). Then, each paper in the candidate set is ranked based on the multiplication of text and relational similarity. Relational similarity is based on the same factors as being used for extending the candidate set (citing and cited papers, number of co-authors, etc.). The weights of the factors were learned with a genetic algorithm. Results show that using the relational factors in addition to text similarity roughly doubled the effectiveness compared to text similarity alone. The optimal size, i.e. best trade-off for recall and computing time, for the initial candidate set was 25 (resulting in a total candidate set size around 3,000 wit SD=1,300). The downside of SemCir's approach is that it needs 38 times as much calculating time than a text-only comparison. The authors also self-criticize that they used a citation-based ground-truth which probably favored the citation based ranking factors. So far, the paper only received one citation. However, we believe that this paper is worth reading. The paper is well written, very detailed, and the approach seems promising.

### F.1.27 Clapper

In 2013, Wang et al. developed a system to recommend "classical papers" to researchers being new to a research field, so these researchers could easily find the most relevant standard literature in that field [408]. Wang et al. used two main factors to rank papers retrieved via a normal search query. The first factor was "download persistence" which describes how constantly papers accumulate download counts over the years. They defined that classical papers are those papers with high download persistence. The second factor was the "principle of citation approaching" (CAF). Wang et al. observed that papers, which cite a classical paper, tend to cite those papers that the classical paper is citing as well. Papers with a high CAF were ranked lower than papers with a low CAF. The authors claim that in a user study with 50 professors, and download counts retrieved from ACM Digital Library, their approach could recommend papers that were considered by the professors as suitable papers for beginners. Unfortunately, the approach was not compared against any baseline. It would have been particularly interesting to see whether "classical papers", with constantly high download counts, are preferred over papers with high download count that were not accumulated constantly over a longer period of time. In addition, the user study and the recommendation approach are described only superficially, which reduces the significance of the results.

### F.1.28 RefSeer

In 2013, Rokach et al. introduced RefSeer, which used machine learning based on the Citation Translation Model (p. 230) with a number of global relevance features [342]. Papers were ranked higher, the more citations the papers, the papers' authors, venues, or affiliations had accumulated over the past 12 months. Rokach et al. also considered title length, number of co-authors, number of affiliations, and venues types for the ranking. Papers were also ranked higher if one of its authors had co-authored with the current user, or if the user had cited the recommendation candidate or its authors previously. Finally, textual similarity between titles and venue names was considered. All these features were used to train a "Full Machine Learning Method" that combined several machine learning approaches such as LibSVM, Random Forests, and AdaBoost.

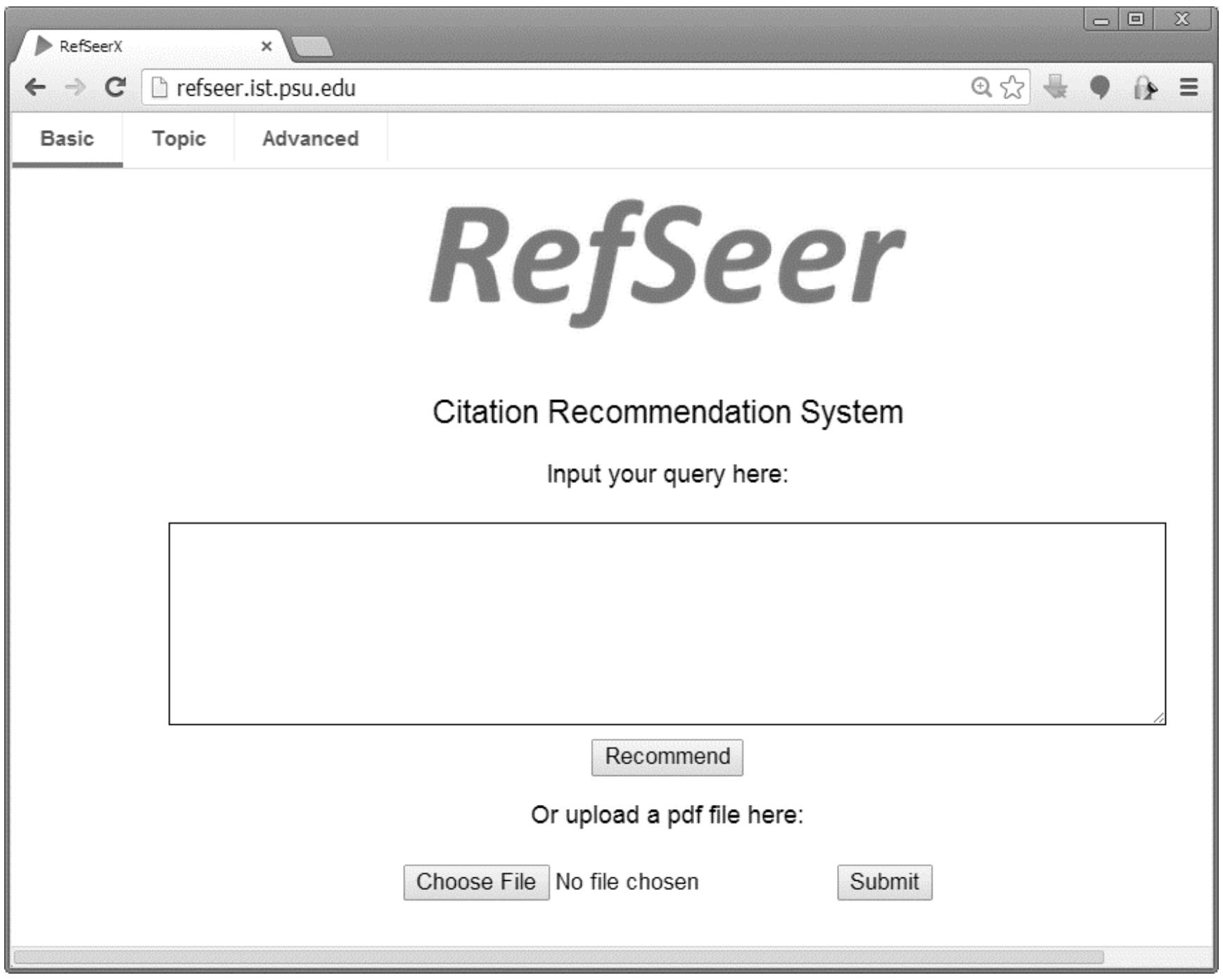

Figure 86: RefSeer website

The results showed that combining the mentioned features with the CTM lead to a twice as high precision, and f-measure, than using CTM alone. However, runtimes dramatically increased from 248ms (CTM) to 4.6 seconds (Full Machine Learning), on a small CiteSeer dataset with 3,312 papers. On a CiteULike dataset with 14,418 papers, runtime even increased to 49 seconds, while CTM required 390ms. To improve runtimes, Rokach et al. used a "Lite" machine learning which pre-filtered 500 recommendation candidates based on CTM only, and then applied the machine learning to rank the candidates. Consequently, runtime decreased to less than a second, and precision remained almost as high as with the full machine learning model. The paper impressively shows that simple metrics adopted from Scientometrics can significantly improve the recommendation process.

Whether the approach is applied by RefSeer remains unclear. On RefSeer's website (Figure 86), users have three different modes to submit PDF files, or queries (i.e. abstracts). It is not stated what exactly the differences between the modes are, and which approaches are applied to generate recommendations.

## F.2 Collaborative Filtering Approaches

### F.2.1 Personality Diagnosis

"Personality Diagnosis" was introduced in 2000 and is a collaborative filtering approach by D. M. Pennock and some of the CiteSeer authors, including C. Lee Giles [320]. Their main contribution was to assume that users were rating items with Gaussian noise and when removing the noise the 'true' rating became visible. To find the true ratings, and similar users, Pennock et al. apply probability theory. They evaluated their approach against classic collaborative filtering with Pearson correlation and a vector-similarity based CF. The evaluation showed that vector-similarity CF outperforms Pearson correlation CF, and that personality diagnosis outperformed both of the baselines. The evaluation was based on the Eachmovie dataset and a dataset from CiteSeer in which implicit ratings were inferred from users' actions such as downloading documents or viewing document details. The CiteSeer dataset contained only 1,575 documents from originally 270,000 articles because the authors removed all documents with less than 15 implicit ratings. The authors explain that their approach should be integrated into CiteSeer soon. However, none of the later CiteSeer papers mentioned 'Personality Diagnosis'. Hence, we assume the plan was never realized. Also within the research paper recommender community, Personality Diagnosis had not much impact. To the best of our knowledge, no other authors adopted their approach. Outside the research paper recommender community, the article was highly influential and was cited more than 400 times.

### F.2.2 CF Based Citation Recommender

In 2002, McNee et al. wanted to apply collaborative filtering to research paper recommendations [274]. To overcome the cold start problem, McNee at al. presented an interesting idea. They considered papers to be users and a paper's citations to be votes for other papers. This way they could fill the rating-matrix, and applied user-item and item-item CF. They compared these two approaches against four baselines. First, with co-citation matching those papers were recommended that were most often co-cited with those papers contained in the input paper's bibliography. Second, a naïve Bayesian classifier was used to find related papers. Third, the title of the input paper was send as search query to Google, and Google's results were recommended. Fourth, content similarity (title and abstract) was calculated between the input paper and all papers citing the papers in input paper's bibliography, being cited, or being co-cited. In an offline experiment, both CF variations were two or even three times as good as the alternatives (user-item a little bit better than item-item). However, in an online experiment, with real users, the Google baseline performed best.

McNee et al. also showed that the way users were asked to evaluate papers, influences their answers. For the online experiment, they asked two questions, "Would recommendations such as these be helpful in finding *related work*" and "Would recommendations such as these be helpful in finding *papers to read*" [105]. Results differed significantly although we would consider both questions as being very similar. McNee et al also showed that asking users for 'quality' and 'novelty' judgments made a difference. They concluded that there is no single-best algorithm and a recommender system should consider the usage scenario. The paper received more than 200 citations, and is interesting to read.

### F.2.3 CiteULike

CiteULike is an online reference manager providing literature recommendations since 2009 [90], whereas the algorithms are based on research from 2008 [57]. The authors compared two variations of item-based CF with user-based CF in an offline experiment [57]. User-based CF performed around twice as good as the item-based CF. Bogers and van den Bosch also found that the optimal neighborhood size for user-based CF lies between four and eight, and for item-based CF around 40, though precision still slightly increases for neighborhood sizes up to 500. Unfortunately, the paper provides hardly any detail about the algorithms, and is difficult to read. Today, on the CiteULike platform, both an item-based and a user-based algorithm are offered separately and the user decides which algorithm to use to receive recommendations [89]. CiteULike reported click-through rates in their live system of 18.96% [88]. We are not sure if the item-based approach is the item-based CF approach presented in their paper, of if it is a new approach that is based on the co-occurrence concept (see next section).

---

[105] Emphasis was made by us

### F.2.4 CARES

In 2009, Yang et al. developed a recommender system for the China American Digital Academic Library (CADAL), called CARES (CADAL Recommender System) [417]. They used collaborative filtering and inferred implicit ratings based on users' access logs, and the number of pages they read. The authors would have liked to use explicit ratings but their users were "too lazy to provide [explicit] ratings for books". Yang et al. experimented with two different ranking strategies for recommendation candidates, namely a greedy and a random walk algorithm, after similar users were determined with AP Correlation. Their results show that a random walk based ranking performs better than a simple greedy ranking algorithm. In addition, they report that the optimal neighborhood size was 20-30. It remains unclear whether CARES ever was actually integrated into CADAL. The approach was evaluated in an offline experiment, against no baseline.

### F.2.5 Synthese & Sarkanto

In 2007, Vellino and Zeber proposed a "hybrid, multidimensional recommender system" for research articles [401]. The 4-page paper contain primarily a literature survey on recommender systems and some ideas how a hybrid multidimensional recommender could look like. The paper was later referred to by Vellino as the paper in which his recommender system "Synthese" was introduced (though the term Synthese does not occur in that paper). In 2010, Vellino compared Synthese against the bx recommender (p. 238). Instead of the hybrid multidimensional approach, Synthese was now supposed to use the same approach as TechLens (p. 243), i.e. CF with papers interpreted as users [399]. Results in the two-page poster include that bx' approach, based on co-downloads, has a higher coverage than Synthese, but semantic diversity was lower. Since the overlap between the approaches, in terms of recommended papers, was low, Vellino concluded that ideally both approaches should be combined. The accuracy of the two approaches was no evaluated. A prototype of Synthese, renamed to Sarkanto, was available until recently[106], but in the past few months, the website was not available.

### F.2.6 Mendeley Suggest

*Mendeley Suggest* was introduced in 2012 by Kris Jack in several presentations [185–187]. Mendeley Suggest uses *Apache Mahout* for implementing item-based collaborative filtering, and is only available to Mendeley's premium users. Since

[106] http//www.lab.cisti-icist.nrc-cnrc.gc.ca/Sarkanto/

Mendeley Suggest is using a (slightly modified) out-of-the-box solution, it cannot shine with novel recommendation approaches but some interesting insights about running a large-scale recommender system for research papers. Among others, Jack reports that precision increased over time (0.025 in the beginning, 0.4 after six months); precision strongly depended on a user's library size (p@10=0.08 for 20 articles, p=0.40 for 140 articles), and precision depended on the similarity metric being used (1st: co-occurrence; 2nd: Log Likelihood; 3rd: Tanimoto coefficient; 4th: Cosine; 5th: Euclidian Distance; 6th: City Block). Kris was also first who reported about the monetary costs required to run a recommender system – which are surprisingly low (cf. 3.4.2, p. 52). The slides also include detailed information about the general architecture and implementation of *Mendeley Suggest*.

### F.2.7 Can't See the Forest for the Trees

In 2013, Caragea, Lee Giles, et al. used singular value decomposition (SVD) on the citation graph, and evaluated their approach against several CF variations [75]. In their 2-page poster, the authors described their approach in a single paragraph, which leaves the reader with only a rudimentary idea of the approach. Since the authors use a citation graph, containing papers and citations, it remains also unclear how the collaborative filtering approaches and user similarities respectively were computed. In addition, the test collection from CiteSeer was strongly reduced which makes it difficult to judge the validity of the results. The authors removed papers having less than ten and more than 100 citations from the collection, as well as papers citing less than 15 and more than 50 papers. Therefore, from 1.3 million citing papers, only 16 thousand papers remained in the test collection. We doubt that such a pruned collection may produce representative results.

## F.3 Co-occurrence Approaches

### F.3.1 BibTiP

BibTip was originally developed by the University of Karlsruhe, Germany and uses co-views for determining related papers [291]. The authors adopt the 'Repeat-Buying Theory' which was developed by Andrew Ehrenberg in the 1950's to explain consumer behavior [130]. Since BibTip uses co-views, no true user model is built. Instead, document similarities are calculated offline, and when a user looks at a paper, papers being previously co-viewed with that paper are recommended. The earliest BibTip papers from 2001 present only some general ideas [131, 133]. In 2002, more details were published with a first evaluation [129, 130] and an overview of BibTip's architecture [132]. Several papers more

followed in the next years [58, 121, 128, 135–138, 140, 156, 290, 291, 304]. Although the BibTip team published many papers, their work is surprisingly little acknowledged in the community. Most of their papers have only few citations. This might be because BibTip uses rather a simple approach and as such, the published papers are not particularly groundbreaking. However, it should be noted that the BibTip team around Andreas Geyer-Schulz were probably first to apply the concept of co-occurrences to research-paper recommender systems. BibTip is also one of the few recommender systems that is applied on a large scale. Today, BibTip is a commercial system available via a Web Service that can be purchased by digital libraries who want to provide literature recommendations to their visitors. The viewing behavior is collected over all libraries using BibTip. This results in observing more than one million downloads per day [213]. The authors c is clicked out of the delivered recommendation lists (containing up to 13 recommendations) [140, 213]. Interestingly, this value is similar to the one reported by PubMed which is using a completely different approach (p. 222)

### F.3.2 National Sun Yat-sen University

In 2003, the National Sun Yat-sen University in Taiwan experimented with co-occurrences and tried different methods for building user models [183]. The initial situation was that a user submitted a search query to the university's search engine. In one approach, all papers contained in the search result were utilized and those papers that most often co-occurred with those in the search results were recommended. In another approach, only those papers whose detail-page a user browsed during one session were utilized. The latter one performed best. The authors used association rules and a 'hypergraph' approach to determine relevant papers. Their association rule is a simple co-occurrence measure normalized by time. The "hypergraph" approach compares user sessions with each other based on cosine similarity. The authors conclude that the hypergraph approach performs better than the simple association rule approach. The approach does not seem too spectacular. However, it shows that also co-occurrence based approaches can utilize user models that contain more than one single input paper.

### F.3.3 bx by Exlibris

bx is a recommender system run by ExLibris and similar to BibTip. As BibTip, bx is a commercial recommender system available via a Web Service, and, as BibTip, bx utilizes co-views of research papers. Sadly, there is only little detail on the exact algorithms, which apparently are patent-pending [113]. In the paper from which bx originated, only superficial information can be found [61], but it seems that bx is applying a simple count of co-occurrences to provide recommendations.

Some presentation slides from 2011 provide information about the effectiveness of bx [390]. Click-through rates are between 3% and 10% depending on the institution in which recommendations are show (bx is providing more than 1,000 institutions with recommendations) [113]. This is an interesting result, because it shows that the same recommendation algorithm may lead to different results when they are applied at different universities, possibly in different webpage layouts, and possibly to students with different backgrounds.

### F.3.4 Co-Citations vs. Co-Downloads

For his Master's thesis, Stefan Pohl evaluated in 2007 whether recommendations based on co-citations or co-downloads were more effective [326]. He was motivated by the fact that extracting citations from articles is time consuming and error-prone and it may take years before articles are cited and become available as recommendation candidates. His most interesting finding was that after around 26 months, recommendations based on co–citations became more effective than those based on co-downloads. Pohl also pointed out that about two thirds of all papers had no co-citation at all and those who had usually had only one or two of them. In contrast, "almost all" papers had at least one co-download. Pohl concluded that co-citation approaches make only sense for few papers after a long time after publication. The most important findings of Pohl's thesis are summarized in a two-page poster [327].

### F.3.5 Scienstein and Citation Proximity Analysis

In 2009, Gipp et al. presented Scienstein, a concept for a research paper recommender system with several ideas how recommendations could be made [145]. The poster also introduced a concept called 'citation proximity analysis' (CPA) which was later presented in detail by Gipp and Beel in another paper [142]. CPA is an extension of co-citation analysis taking into account the distance of two citations in a document. For instance, if two papers are cited in the same sentence, their relatedness is assumed to be higher than that of two papers being cited in two different paragraphs. Based on a small user study, the authors report that twice as many users liked CPA-based recommendations than recommendations based on classic co-citation analysis.

## F.4 Graph Based Approaches

### F.4.1 Spreading activation in intra-book recommendations

In 2000, Woodruff et al. tackled the problem that a researcher reads a book, which contains several articles, but does not know which articles of the book to read [413]. Woodruff et al. imagined that users picked at least one article they like, and related articles in the book, or related articles being cited by articles in the book, would be recommended. They built a graph, in which papers were connected by citations, bibliographic coupling strength, co-citation strength, text similarity and several combinations. Given a single input paper, recommendation candidates were found through spreading activation in the graph. Namely, they applied the Leaky Capacitor Model, which was already introduced in 1984 [8]. Based on an evaluation with only three participants and a single book (containing 43 articles, and 676 cited articles) Woodruff et al. claimed that spreading activation achieved the best results, compared to a standard CBF. Results were particularly effective when the graph was based on papers' text similarity *and* citation based metrics.

### F.4.2 A two-layer graph approach

Usually, recommender systems use 'flat' graphs (if they are using graphs at all). Huang et al. proposed a two-layer graph model for a Chinese book recommender in a digital library [182]. In the user-layer, users are modeled and their similarities to each other based on demographic data such as age, education, gender, and number of children. In the book-layer, similarities between books are modeled, based on the books' content and attributes. For content similarity, title, keywords, foreword, and introduction were analyzed. Utilized attributes included the number of pages, layout information, publisher, weight, size, and several more. How exactly demographics and book attributes were utilized to calculate similarities was not explained. Both layers were connected by the purchases of books the users made. With this model, Huang et al. were able to apply content-based filtering, collaborative filtering, and a combination of both. For content-based filtering, those books were recommended that were similar to the books purchased by a user. For collaborative filtering, those books were recommended that were purchased by similar users. In a hybrid approach they calculated user-book similarities over three degrees in the graph. Additionally, they applied spreading activation based on Hopfield's Net algorithm. Overall, the hybrid approach performed best and no statistically significant difference was found between CF and CBF.

### F.4.3 PaperRank

In 2006, Gori and Pucci applied a *PageRank*-like *Random-Walk* algorithm to the citation network and called this approach *PaperRank* [153]. They expected a user to provide a manuscript containing already some citation to papers. The citations were taken as starting point in the citation graph. Gori and Pucci claim a precision@20 of 100% (which seems questionable to us), and did not evaluate PaperRank against any baseline. This shortcoming was made up for in 2012 when Küçüktunç et al. tested PaperRank against several baselines [219]. They confirmed a good – for some scenarios even the best – precision compared against Katz, co-citation strength, bibliographic coupling strength, CC-IDF (p. 219), DaKatz (p. 243), and DaRWR (p. 243).

### F.4.4 Multiple Graphs

In 2008 Zhou, Lee Giles, et al. used machine learning (label propagation) on several graphs to generate research paper recommendations [437]. Utilized graphs were the citation graph (papers->papers), the author graph (researchers->papers) which included papers an author had authored and cited, and the venue graph (venues->papers). Zhou et al. evaluated their approach against a SVD based CF (on the author-document matrix) and a simple graph Laplacian. They report 3-5 times better results than with Laplacian and 2.5 times better results than with SVD. Zhou at al consider their approach to be an item-based CF approach, we would rather classify it as a graph based approach. The paper is very mathematical and provides many details about graph calculations.

### F.4.5 Curated Citation Networks & Path Ranking Algorithm

Andrew Arnold and William W. Cohen are from the biological sciences and focused in 2009 on the problem of predicting genes and proteins a researcher would write about next [9]. They modeled authors, papers and genes in a graph with various connections. Authors were connected to the papers they had authored and to their co-authors. Papers were connected to papers they cited or were cited by, and to genes they mentioned. Genes were additionally connected to each other when they were related (information about relatedness was retrieved from special gene databases). Once a user specified one or several papers of interests, a random walk was performed in the graph to determine the most relevant genes to recommend. Arnold and Cohen experimented with different variations of their graph and the findings are interesting not only for researchers in the field of biology. Among others, they found that performing a query based on a paper's first author is more effective than on the paper's last author. Leaving out some connections in the graph (e.g. co-authorship and related genes) also increased accuracy. This shows that it is not always optimal to utilize all available information.

One year later, Cohen and Lao proposed the Path Ranking Algorithm (PRA) to recommend papers, venues, experts (i.e. authors), and genes [235]. In contrast to their previous work about curated citation networks from 2009, the graph also contained title-words, venues, and publication years. They used machine learning to learn the weights of the edges in the graph. Compared to an untrained RWR (with edge-weight=1) PRA performed significantly better for most of the tasks. Interesting to note is that the same algorithm, i.e. PRA, performs differently on different recommendation tasks. For instance, mean average precision for expert recommendation was 7.2% but 16.0% for papers. The original paper dates to 2010 [235]. In 2012, the PhD thesis of Lao, which was supervised by Cohen, was published with more details on the approach [231]. There are also a few more papers and posters, some being unpublished [232–234].

### F.4.6 Local and Global Relation Strength

Liang at al. proposed "Local Relation Strength" (LRS) and "Global Relation Strength" (GRS) in a citation graph to determine relatedness of papers [250]. LRS expresses how strong citing and cited papers are related. The strength is based on the "importance" of the citation, the "surrounding citation environment", and the "temporal distance". How exactly these values are calculated, is described only vaguely. GRS is essentially the Katz measure, only that Katz typically assumes a weight of 1 between two nodes in a graph, and GRS uses LRS as weight of the edges. Both LRS and GRS perform better than several baselines (CC-IDF, co-citation strength, bibliographic coupling strength, HITS, and Katz).

### F.4.7 Network-Aware Popularity

Popularity-measures such as PageRank, or a simple citation count, typically are calculated based on all papers in the citation graph. Baez et al. proposed to calculate a 'network-aware' popularity. This measure calculates e.g. citation counts only based on the citations from researchers being in the personal network related to the user (the personal network could be populated e.g. by co-authorship) [12]. Baez et al. present several, rather trivial, ideas for measuring such a network-aware popularity. For instance, one metric defines a paper as being the more popular the more researchers in a user's network authored or cited the paper. Baez et al. propose different types of graphs, for instance the venue, co-authorship, and topic graph. In their evaluation, they only seem to consider the co-authorship network. Their evaluation shows that their network-aware popularity performs better than an overall popularity. However, the authors evaluated their approach only against a single overall popularity baseline (absolute citation and author count). We also see a major problem with network-aware popularity: Citation network usually are already sparse. Reducing the citation network to related authors or papers will increase sparsity even further.

### F.4.8 TheAdvisor with direction aware Katz and RWR (daKatz & daRWR)

TheAdvisor[107] was recently developed (2012/13) and is one of the few recommender systems being publicly available without prior registration. It allows users to upload a BibTeX file with a set of references, and specifying whether recommended papers should be more recent or more traditional. TheAdvisor was initially released in January 2012 and the first paper was published as pre-print on arXiv.org [219]. An ‘official’ and briefer version followed some months later [217]. It is not clear how recommendations eventually are generated in TheAdvisor, but in their papers, Küçüktunç et al. proposed two approaches. One is based on Katz, the other one on PaperRank, i.e. RWR. They modified both approaches so they consider the direction of a citation and hence called their approaches direction aware Katz and RWR (daKatz and daRWR). They conducted an extensive evaluation with different scenarios (i.e. different weight on recent and traditional papers) and compared daKatz and daRWR against Katz, PaperRank (RWR), co-citation and bibliographic coupling strength, and CC-IDF. In most scenarios, daKatz and daRWR outperformed the baselines.

Küçüktunç et al. also published detailed information about the technical infrastructure and run-times of TheAdvisor in a conference paper [215] and an extensive pre-print of a forthcoming journal paper [216]. An approach for diversifying results, based on daRWR, is briefly presented in a poster [220], and in more detail in a pre-print of a journal article [218]. Küçüktunç et al. also report to have experimented with user feedback but results are omitted in their paper [219], and we could not find a paper that present those results.

## F.5 Hybrid Recommendation Approaches

### F.5.1 TechLens

In 2004, Torres, McNee, and three others introduced TechLens and ten different algorithms to generate research paper recommendations [393]. The algorithms were mainly adopted from Robin Burke [70] and consisted of three CBF variations, two CF variations, and five hybrid approaches.

> Content-Based Filtering: *Pure-CBF* served as baseline, being the standard CBF in which a term-based user model – in case of TechLens, terms from a single input paper – is compared with the recommendation candidates. In

[107] http://theadvisor.osu.edu/

*CBF-Separated*, for each paper being cited by the input paper, similar papers are determined separately and at the end the different recommendation lists are merged and presented to the user. In *CBF-Combined*, terms of the input paper and terms of all papers being cited by the input paper are combined in the user model. Then, those papers being most similar to this user model were recommended.

Collaborative Filtering: *Pure-CF* served as another baseline and represented the collaborative filtering approach from McNee et al., in which papers were interpreted as users, and citations as votes [274]. In *Denser-CF*, citations of the input paper were additionally included in the user model.

Hybrid: With *Pure-CF->CBF Separated*, recommendations were first created with Pure-CF. These recommendations were then used as input documents for CBF-Separated. In a similar way *Pure-CF->CBF Combined*, *CBF Separated->Pure-CF*, and *CBF-Combined->Pure-CF* were used to generate recommendations. *Fusion* created recommendations with both, CBF, and CF independently, and then merged both recommendation lists.

Torres et al. report to have evaluated all ten approaches, but results were only presented for the top-5 approaches. In an offline evaluation, the CBF approaches performed worst (CBF-Separated better than CBF-Combined), and Pure-CF performed best (even better than the hybrid approaches). The approaches were also evaluated with a user study and Torres et al. distinguished between different reading purposes (novel, survey, authority, introductory). As in 2002, no single-best algorithm could be found. For instance, Pure-CF performed best for authoritative and novel papers. For survey papers and introductory papers, CF performed worst and CBF-Separated was best. In terms of overall user satisfaction, CF delivered unsatisfactory results (second worst out of the five algorithms for which details were provided), and CBF-Separated performed best. Unfortunately, the Google-baseline, which performed best in the 2002-paper, was not used in this evaluation.

In 2006, McNee et al. criticized that researchers were concentrating too much on recommender's accuracy but ignoring users' actual needs [275]. In a user study with around 138 participants[108], they evaluated whether different algorithms performed differently for different recommendation tasks. The study is similar to the two previous papers but while the two previous papers covered users' needs only superficially, the 2006-paper investigates this issue in detail. McNee at al. compared their citation based CF approach, CBF and a PLSI and Naïve Bayesian approach with each other. The latter two approaches performed quite poor in general. However, for CF and CBF it was shown that depending on the task, user satisfaction differed for the two algorithms. Interestingly, the overall difference in satisfaction for CF and CBF was rather small. This result contradicts the previous finding, in which users were not very satisfied with CF.

In 2010, some of the TechLens authors re-evaluated some of the TechLens approaches and some newly developed approaches with an offline evaluation [110]. The novel idea was to weight implicit ratings for item-based CF, which were inferred from the citation network, based on PageRank, SALSA, and HITS. In addition, Ekstrand et al. evaluated several CBF approaches for which they used PageRank, SALSA, and HITS in the ranking, and some hybrid approaches. They combined various factors and evaluated 177 algorithms in total. Using SALSA and PageRank for CF improved performance, compared to plain CF, while HITS did not increase performance. Interestingly, for CBF, results were just the opposite. Here, a HITS enhanced ranking achieved the highest performance. Overall, CF (with whatever weighting) outperformed CBF, and the hybrid approaches. As such, the results of this offline evaluation confirmed the results presented in the 2004 paper. In an additional user study with 19 participants, Ekstrand et al. evaluated three of the approaches, namely PageRank weighted CF, CBF with HITS weighting, and a CBF-CF hybrid approach. Similar to the offline experiment, CF outperformed the other approaches, and CBF performed worst. It is interesting to note that results from the user study are exactly the opposite as in the 2004 user-study. Although it was the same evaluation scenario (creating introductory reading lists), CBF performed best in the 2004 paper, and CF performed worst. We were also confused that in the 2010 paper, item-based CF and CBF-Combined were primarily used, although in 2004, user-CF (slightly) outperformed item-CF, and CBF-Separated outperformed CBF-Combined. Again, the Google baseline from 2002 was missing.

---

[108] The exact number of study participants remains unclear. Once McNee et al write there had been 138 participants, and once they write there had been 117 professors, 18 students, and 7 others, which adds up to 142.

Interesting is also a paper published by Dong et al. [107] who are not affiliated with TechLens. In 2009, Dong et al. evaluated seven of the ten algorithms presented by TechLens in 2004. In Dong et al.'s offline-evaluation, Pure-CF is only fifth best (out of seven), while Pure-CF was best in the TechLens offline evaluations. In Dong et al.'s offline evaluation, all three CBF approaches performed better than CF or hybrid approaches, and among the CBF approaches, CBF-Combined performed better than CBF-Separated. These results contradict the results from Torres et al. [393]. There is no obvious reason why results would differ so much. The approaches seem identical, the way the offline experiment was conducted seems to be similar and both experiments were conducted on a CiteSeer dataset. The only difference we found is that Torres et al. removed papers with less than three citations from the corpus, and Dong et al. removed papers with less than two citations from the corpus[109]. However, this could only explain why rankings for CF and CBF differed, but not why in Torres et al. evaluation CBF-Separated performs better than CBF-Combined, and Dong et al. report the opposite.

### F.5.2 Papyres

Papyres is a software tool to support researchers in managing their literature [296], developed by Amine Naak as part of his Master's thesis [294]. In 2009, Papyres integrated a research paper recommender system [295]. Naak's work was motivated by the idea that most researchers were only interested in certain parts of a paper. Therefore, Naak et al. allowed users providing different explicit ratings per paper for the paper's contribution, originality, readability, technical quality, etc. These ratings were used for collaborative filtering in combination with content-based filtering. Because Papyres had not enough users, Naak et al. randomly created artificial users, randomly assigned papers to them, and randomly created ratings to evaluate their CF variations based on multiple ratings. Besides this questionable evaluation technique, the paper provides little detail on how CF and CBF are combined. In an additional user study with 83 participants, Papyrus was evaluated and achieved an average rating of 4.43 (out of 5) [296]. However, Papyrus was not evaluated against any baseline. As such, the average rating was not very meaningful.

---

[109] It should be noted that the paper from Dong et al. is not very well written, and when reading the paper, the impression occurs that Dong et al. had invented the presented algorithms and not TechLens (some might argue Dong et al. plagiarized)

# G PDF Title Extraction

## G.1 SciPlore Xtract[110]

### G.1.1 Introduction

Extracting the title from PDF documents is one of the prerequisites for many tasks in information retrieval. Among others, (academic) search engines need to identify PDF files found on the Web. One possibility to identify a PDF file is extracting the title directly from the PDF's metadata. However, often the PDF metadata is incorrect or missing. Therefore, what is often tried is to extract the title from the PDFs' full text.

Usually, machine-learning approaches such as Support Vector Machines (SVM), Hidden Markov Models and Conditional Random Fields are used for extracting titles from a document's full text. According to studies, the existing approaches achieve excellent accuracy, significantly above 90%, sometimes close to 100% [160, 178, 319]. However, all existing approaches for extracting titles from PDF files have two shortcomings. First, they are expensive in terms of runtime. Second, they usually convert PDF files to plain text and lose all style information such as font size.

For our academic search engine SciPlore.org we developed *SciPlore Xtract*, a tool applying rule based heuristics to extract titles from PDF files. In this chapter, we present this tool, the applied heuristics, and an evaluation.

### G.1.2 SciPlore Xtract

SciPlore Xtract is an open source Java program that is based on pdftohtml[111] and runs on Windows, Linux, and MacOS. The basic idea is to identify a title based on the rule that it will be the largest font on the upper first third on the first page.

In the first step, SciPlore Xtract converts the entire PDF to an XML file. In contrast to many other converters, SciPlore Xtract keeps all layout information regarding text size and text position. Figure 88 shows an example XML output file of the PDF showed in Figure 87. Lines 6 to 12 of the XML file show all font sizes that are used in the entire document (in this case it is all “Times” in a size between 7 and 22 points). Below this, each line of the original PDF file is stated including layout information such as the exact position in which the line starts, and which font is used.

[110] This chapter has been published as: Beel, Joeran, Bela Gipp, Ammar Shaker, and Nick Friedrich. "SciPlore Xtract: Extracting Titles from Scientific PDF Documents by Analyzing Style Information (Font Size)." In Research and Advanced Technology for Digital Libraries, Proceedings of the 14th European Conference on Digital Libraries (ECDL'10), edited by M. Lalmas, J. Jose, A. Rauber, F. Sebastiani, and I. Frommholz, 6273:413–416. Lecture Notes of Computer Science (LNCS). Glasgow (UK): Springer, 2010.

[111] http://www.pdftohtml.sourceforge.net

Google Scholar’s Ranking Algorithm: An
Introductory Overview

Jöran Beel
Otto-von-Guericke University
Department of Computer Science

Bela Gipp
Otto-von-Guericke University
Department of Computer Science

Figure 87: Example PDF

SciPlore Xtract now simply needs to identify the largest font type (in the example the font with the ID=0). Which text uses this font type on the first page is then identified and to assumed to be the title.

```
5  <page number="1" position="absolute" top="0" left="0" height="1262" width="893">
6      <fontspec id="0" size="22" family="Times" color="#000000"/>
7      <fontspec id="1" size="16" family="Times" color="#000000"/>
8      <fontspec id="2" size="9" family="Times" color="#000000"/>
9      <fontspec id="3" size="7" family="Times" color="#000000"/>
10     <fontspec id="4" size="13" family="Times" color="#000000"/>
11     <fontspec id="5" size="13" family="Times" color="#000000"/>
12     <fontspec id="6" size="16" family="Times" color="#000000"/>
13 <text top="106" left="245" width="415" height="27" font="0">Google Scholars Ranking Algorithm: An  </text
14 <text top="134" left="336" width="233" height="27" font="0">Introductory Overview  </text>
15 <text top="189" left="348" width="77" height="20" font="1">Jöran Beel</text>
16 <text top="186" left="424" width="6" height="13" font="2">1</text>
17 <text top="189" left="430" width="108" height="20" font="1"> and Bela Gipp</text>
18 <text top="186" left="538" width="6" height="13" font="2">2</text>
19 <text top="189" left="545" width="5" height="20" font="1"> </text>
20 <text top="225" left="340" width="7" height="11" font="3"><i>1 </i></text>
21 <text top="227" left="347" width="106" height="17" font="4"><i>joeran@beel.org </i></text>
22 <text top="225" left="453" width="5" height="11" font="3"><i>2</i></text>
23 <text top="227" left="458" width="98" height="17" font="4"><i>bela@gipp.com </i></text>
24 <text top="244" left="201" width="496" height="17" font="5">Otto-von-Guericke University, Dept. of Comput
    Science, Magdeburg, Germany </text>
25 <text top="289" left="106" width="71" height="20" font="6"><b>Abstract </b></text>
26 <text top="310" left="106" width="684" height="17" font="5">Google Scholar is one of the major academic
   search engines but its ranking algorithm for academic articles is </text>
```

Figure 88: Example XML Output

### G.1.3 Methodology

In an experiment, titles of 1000 PDF files were extracted with SciPlore Xtract. Then, titles from the same PDFs were extracted with a Support Vector Machine from CiteSeer [160] to compare results. CiteSeer’s tool is written in Perl and based on SVM Light[112] which is written in C. As CiteSeer’s SVM needs plain text, the PDFs were converted once with PDFBox[113] and once with pdftotext[114] as these are the tools recommended by CiteSeer. It was then checked for each PDF if the title was correctly extracted by SciPlore Xtract and CiteSeer's SVM (for both the pdftohtml text file and the PDFBox text file). If the title contained slight errors the title was still considered as being identified correctly. 'Slight errors' include wrongly encoded special characters or, for instance, the inclusion of single characters such as '*' at the end of the title.

---

[112] http://svmlight.joachims.org/

The PDFs analyzed were a random sample from our SciPlore.org database, a scientific (web based) search engine. A title was seen as being correctly extracted when either the main title or both the main title and the sub-title (if existent) were correctly extracted. The analyzed PDFs were not always scientific. It occurred that PDFs represented other kind of documents such as websites or PowerPoint presentations. However, we consider the collection to be realistic for an academic search engine scenario.

#### G.1.4 Results

From 1000 PDFs, 307 could not be processed by SciPlore Xtract. Apparently, SciPlore Xtract (respectively pdftohtml) struggles with PDFs that consist of scanned images on which OCR has been applied. For further analysis only the remaining 693 PDFs were used. We consider this legitimate as the purpose of our experiment was not to evaluate SciPlore Xtract, but the applied rule based heuristic.

For 54 of the 693 PDFs (7.8%), titles could neither be extracted correctly by SciPlore Xtract nor CiteSeer's SVM. Only 160 (23.1%) of the titles were correctly identified by all three approaches. Overall, SciPlore Xtract extracted titles of 540 PDFs correctly (77.9%). CiteSeer's SVM applied to pdftotext identified 481 titles correctly (69.4%). CiteSeer's SVM applied to PDFBox extracted 448 titles correctly (64.6%). Table 1 shows all these results in an overview.

When only completely correct titles are compared, SciPlore Xtract performs even better. It extracted 528 (76.2%) titles completely correct, while CiteSeer's SVM extracted only 406 (58.6%) respectively 370 (53.4%) completely correct.

---

[113] http://pdfbox.apache.org/

[114] http://www.foolabs.com/xpdf/download.html

Table 18: Title extraction of 693 PDF files

| | Correct | | Slight Errors | | Total | |
|---|---|---|---|---|---|---|
| **SciPlore Xtract** | 528 | 76.2% | 12 | 1.7% | 540 | 77.9% |
| **CiteSeer SVM + pdftotext** | 406 | 58.6% | 75 | 10.8% | 481 | 69.4% |
| **CiteSeer SVM + PDFBox** | 370 | 53.4% | 78 | 11.3% | 448 | 64.6% |

SciPlore Xtract required 8:19 minutes for extracting the titles. SVM needed 57:26 minutes for extracting the titles from the plain text files (this does not include the time to convert the PDFs to text), which is 6.9 times longer. However, we need to emphasize that these numbers are only comparable to a limited extent. CiteSeer's SVM extracts not only the title but also other header data such as the authors and CiteSeer's SVM is written in C and Perl while SciPlore Xtract is written in Java.

### G.1.5 Discussion & Summary

All three tests show significantly worse results than the often claimed close-to-100% accuracies. Our tests showed (1) that style information such as font size is suitable in many cases to extract titles from PDF files (in our experiment in 77.9%). Surprisingly, our simple rule based heuristic performed better than a support vector machine. However, it could be that with other text to PDF converters, better results may be obtained by the SVM. CiteSeer states to use a commercial tool to convert PDFs to text and recommends PDFBox and pdftotext only as secondary choice. Our tests also showed (2) that runtime of the rule based heuristic was better (8:19 min) than SVM (57:26). However, these numbers are only limitedly comparable due to various reasons.

In next steps, we will analyze why many PDFs could not be converted (30.7%) and in which cases the heuristics could not identify titles correctly. The rule based heuristic also needs to be compared to other approaches such as Conditional Random Fields and Hidden Markov Models. We also intend to take a closer look at the other studies and investigate why they achieve accuracies of around 90%, while in our test the SVM achieved significantly lower accuracies. In the long run, machine learning algorithms probably should be combined with our rule based heuristic. We assume that this will deliver the best results. It also needs to be investigated how different approaches with different languages. Existing machine learning approaches mostly are trained with English documents. It might be that our approach will outperform machine learning approaches even more significantly with non-English documents as style information is language-independent (at least for western languages).

Summarized, despite the issue that many PDFs could not be converted, the rule-based heuristic we introduced, delivers good results in extracting titles from scientific PDFs (77.9% accuracy). Surprisingly, this simple rule based heuristic performs better than a Support Vector Machine based approach.

Our dataset (PDFs, software, results) is available upon request so that other researchers can evaluate our heuristics and do further research.

## G.2 Docear's PDF Inspector[115]

### G.2.1 Introduction

Several applications in the field of Academia require extracting titles from PDF files. For instance, academic search engines identify PDFs found on the Web, and reference managers such as Mendeley and Zotero extract titles (and other metadata) from PDFs to help users creating bibliographies. In the ideal case, a PDF's title is stored in the PDF's metadata and can easily be retrieved with standard PDF libraries (e.g. PDFBox, jPod, or iText). However, often a title is not available via the PDF's metadata. To retrieve a title anyway, the full-text of a PDF must be analyzed.

In the past years, several tools used machine learning to identify titles from PDFs [95, 160, 178, 319], some of them being open source. However, the recently developed "SciPlore Xtract" [36] showed that a simple heuristic outperformed machine learning approaches. SciPlore Xtract extracted the largest font from the first page of a PDF and assumed this to be the title. Although researchers often claim accuracies of around 90% for title extraction [160, 178, 319], we recently showed that under "real-world" conditions, accuracies are rather between 50% to 70% [36].

All solutions have some shortcomings. Either they are proprietary solutions being not freely available (Mendeley), have problems in processing PDF files that do not comply 100% to the PDF standard (SciPlore Xtract), don't process PDFs at all and require third party tools (ParsCit), are rather slow and achieve low accuracies (ParsCit), are not available for all operating systems, or are available only as stand-alone tools which cannot be easily integrated into other applications.

[115] This chapter has been published as: Beel, Joeran, Stefan Langer, Marcel Genzmehr, and Christoph Müller. "Docears PDF Inspector: Title Extraction from PDF files." In Proceedings of the 13th ACM/IEEE-CS Joint Conference on Digital Libraries (JCDL'13), 443–444. ACM, 2013.

### G.2.2 Docear's PDF Inspector

We developed "*Docear's PDF Inspector*" which identifies titles from (academic) PDF files and does not suffer from the aforementioned shortcomings. Namely, Docear's PDF Inspector (a) achieves good accuracies with excellent run times (see next section for details) (b) can be used as library by other JAVA applications which means other tools can easily integrate Docear's PDF Inspector (c) can be used as a stand-alone application that returns a PDF's title on the command line or stores the data into a CSV file (Figure 89) (d) can process several PDFs in a batch (e) can process all PDF files of all PDF versions, including those with minor deviations from the PDF standard. In the rare cases that a PDF cannot be parsed the title from a PDFs metadata is returned (if available) (f) is written 100% in JAVA 1.6 which means Docear's PDF Inspector runs on any major operating system, including Windows, Linux, and MacOS, without any other tools required (besides the JAVA runtime environment, of course) (g) is released under the GNU General Public License (GPL) 2 or later, which means it is completely free to use and its source code can be downloaded and modified by anyone. Both source code and compiled library can be found at *http://www.docear.org*.

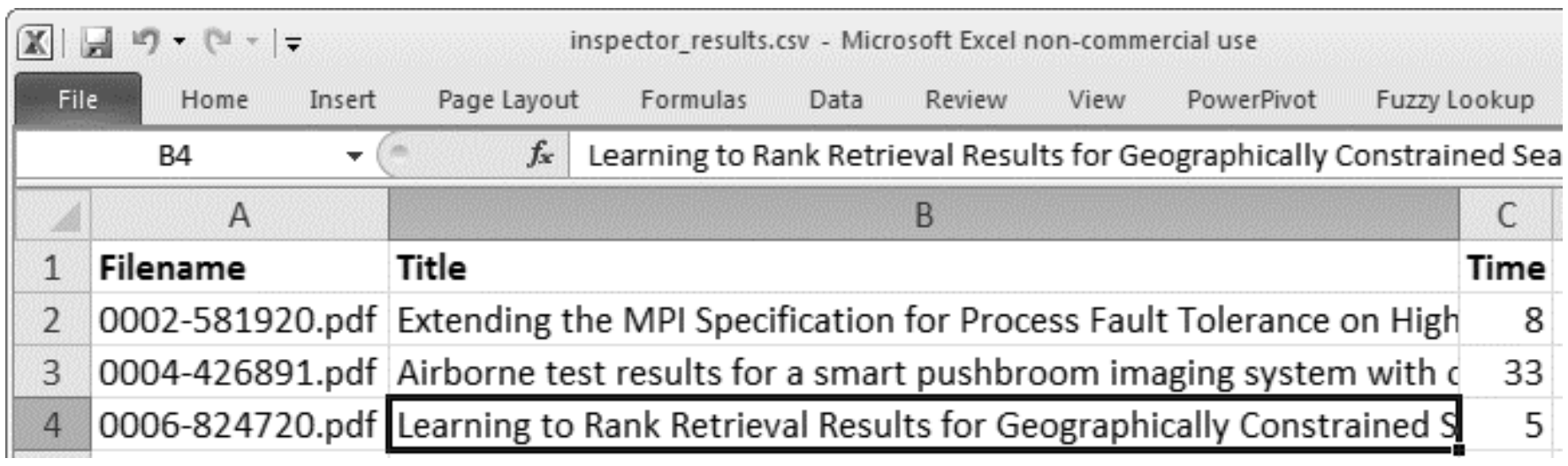

Figure 89: Output CSV opened in Microsoft Excel

Via command line, Docear's PDF Inspector is started with `java -jar PdfInspector.jar [OPTION][FILE]` and both options and files can be specified multiple times. Available options are `'header'` which includes a PDF's header in the output, `'name'` which includes the file name, `'time'` includes the time required for processing the PDF, `'out <arg>'` specifies the file to write to, `'outappend'` appends the output to an existing file instead of overwriting it, and `'delimiter'` specifies how fields are separated in the CSV file. The title extraction is performed in the same way as *SciPlore Xtract* does [36]: the largest font on the first page that is not exceeding eight lines is assumed to be the title. For processing PDF files the PDF library *jPod* is used.

### G.2.3 Methodology

To evaluate the performance of Docear's PDF Inspector we created a test collection of 500 PDF files. To have a PDF collection that contains various formats of academic articles we sent 500 search queries to Google Scholar and from the result pages (each with 100 entries) we randomly downloaded one paper. 57 PDFs were removed from the collection because they had no title or were no academic articles at all, i.e. 443 articles remained for the evaluation. The search queries were randomly generated from words contained in the mind maps of the users of our literature management software Docear [33]. We did not conduct a detailed analysis of the downloaded papers but it appeared to us that most papers were written in English, and some in German, French and Spanish. Papers were from various disciplines (computer science, psychology, biology, social sciences, business, etc.) and there was a very high variety of different formats of the articles. The collection of 500 PDFs is available upon request, so other researchers can use this PDF collection for their research and making their results comparable to ours. We also publish our research data, i.e. the extracted titles and charts we created, on http://labs.docear.org.

We evaluated Docear's PDF Inspector against SciPlore Xtract and ParsCit to have a comparison of how good the achieved results are. Because ParsCit cannot process PDF files by its own, we converted PDFs to plain text with PDFBox and jPod and run ParsCit on both text sets. If an extracted title was identical to the actual title, we classified the result as "exact match". If the extracted title was a substring of the actual title we classified the result as "partly match". Such a partly match occurred, for instance, when a tool failed to extract a PDF's sub-title. For both, exact and partly match comparisons, we ignored spaces and special characters.

Some PDFs caused parsing errors probably because they did not comply 100% with the PDF standard. For SciPlore Xtract and PDFBox (and hence ParsCit) this problem was most apparent: 35.21% (SciPlore) and 20.77% (PDFBox) of the 443 PDFs could not be parsed at all, for jPod the error was only 5.19%. While we consider the original test collection to be representative for a real-world scenario that applications such as academic search engines or reference managers face, we also wanted to have a test collection that could be processed by all tools, to evaluate the effectiveness of the title extraction algorithms (ignoring any PDF parsing problems). Therefore, we inferred a 'reduced test collection' by removing all PDFs from the original test collection which couldn't be processed by at least one of the tools. This resulted in a subset of 278 PDFs.

### G.2.4 Results

The results we present in this section also show how often titles from Google Scholar were accurate. We need to emphasize that accuracies from Google Scholar are not comparable with results from the other tools evaluated because Google Scholar often receives metadata directly from the publishers. That means, Google Scholar does not always extract metadata from PDFs. We provided these results only to show that even Google Scholar seems to have problems with extracting titles in some cases.

Docear's achieves the highest accuracies (Figure 90). For our standard test collection Docear's PDF Inspector outperforms the second best tool (SciPlore Xtract) notably. Docear extracts 65.01% of the titles exactly, i.e. without any errors, while SciPlore Xtract extracts only 50.34% accurately. ParsCit performs worst with an accuracy of 37.25% (PDFBox) and 36.79% (jPod). Docear also performs best measured by 'partly matches' with an accuracy of 74.04% (SciPlore 52.14%; ParsCit 38.83% and 36.79%).

Looking at the reduced test collection the picture slightly changes. Now, Docear and SciPlore perform about the same. Docear extracts 73.38% of the titles flawlessly, SciPlore 77.70%. Based on 'partly matches' Docear extracts 82.01% of the titles correctly, SciPlore 80.58% (differences are statistically not significant). ParsCit still performs far worse with accuracies around 50%.

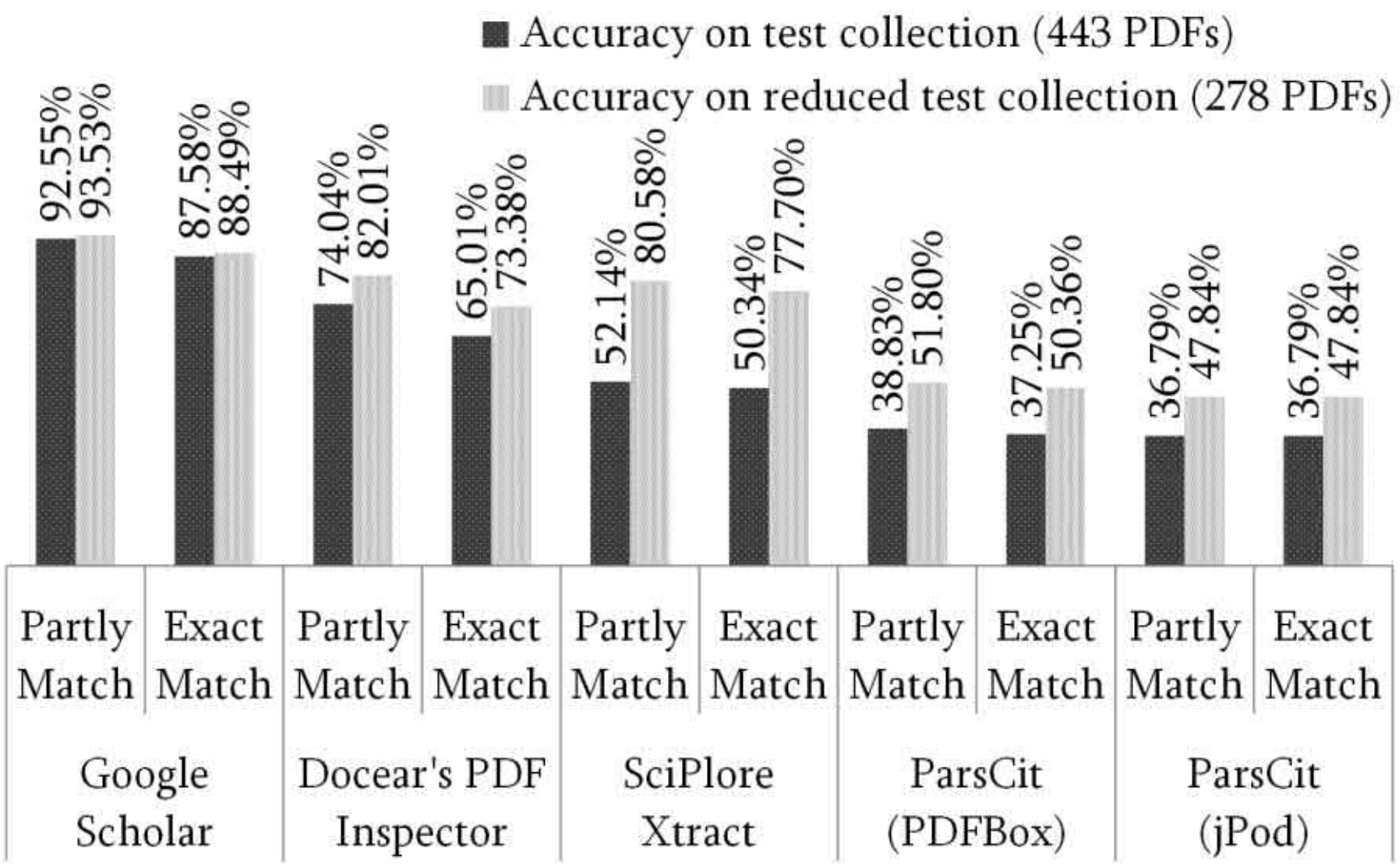

Figure 90: Accuracies of the tools on the two test collections

Docear's PDF Inspector also performs best in terms of runtime. On average (mean), Docear's PDF Inspector needs 50ms to extract a title from a PDF while SciPlore Xtract needs 428ms and ParsCit 2965ms with the PDFBox library and 1786ms with jPod (Table 19). The comparison is not completely fair because ParsCit does not only extract the title (as Docear does) but also other metadata such as authors. However, for those users being only interested in the title, Docear's PDF Inspector identifies a title definitely fastest.

Table 19: Average runtimes (in milliseconds) per PDF

| | Docear | SciPlore | ParsCit (PDFBox) | ParsCit (jPod) |
|---|---|---|---|---|
| Mean | 50 | 428 | 2965 | 1786 |
| Std. Dev. | 61 | 611 | 1383 | 1332 |
| Median | 23 | 352 | 2706 | 1394 |
| Max | 475 | 17667 | 15131 | 17585 |

Summarized, from a user perspective, Docear's PDF Inspector is the most effective tool. It is about 50% more effective than SciPlore Xtract and almost twice as effective as ParsCit for a PDF collection we consider representative for real-world scenarios. In addition, Docear's PDF Inspector is around 40 to 100 times faster than ParsCit and eight times as fast as SciPlore Xtract which uses basically the same heuristic. From a research perspective (i.e. on the reduced data set), the simple heuristic applied by Docear and SciPlore is around 50% more effective than the machine learning approach applied by ParsCit.

Final note: A recent study showed very good results for some tools which we were not aware of at the time of our evaluation [252]. Otherwise, we would have tested them against Docear's PDF Inspector.

# H Impact of User Demographics[116]

## H.1 Introduction

There are more than one hundred research articles on research paper recommender systems, and even more on recommender systems in general. Many of them report on new recommendation approaches and their effectiveness. For instance, *Papyrus* is supposed to have a precision around 20% [295]; Quickstep's approach is supposed to have a precision around 10% [286]; and Jomsri et al. claim an accuracy of 91.66% for their research paper recommender system [197]. Unfortunately, results cannot be compared with each other because researchers used different evaluation methods, metrics, and data sets.

We believe there is another factor influencing the comparability which has received too little attention: users' demographics and characteristics. In other disciplines it is well known that results from one study cannot be used to draw conclusions for a population if the study's user sample differs too much from that population. For instance, in marketing you cannot draw reliable conclusions about how elderly people in Germany will react to a product if a study about that product was conducted in France with university students. Evaluations of recommender systems widely ignored differences in user samples. Some studies report to have asked their participants for demographic data, but they do not report on them in their papers [62]. Another paper reports that age and gender had no impact on the accuracy of recommendations but test subjects were all students [316]. With students typically being all in the same age-range, it is no surprise that the study could not find any differences between different ages.

We analyzed empirical data collected with Docear's research paper recommender system [47] to find out whether users' demographics and characteristics influence the outcome of the recommender system evaluation.

[116] This chapter has been published as: Beel, Joeran, Stefan Langer, Marcel Genzmehr, and Andreas Nürnberger. "Persistence in Recommender Systems: Giving the Same Recommendations to the Same Users Multiple Times." In Proceedings of the 17th International Conference on Theory and Practice of Digital Libraries (TPDL 2013), edited by Trond Aalberg, Milena Dobreva, Christos Papatheodorou, Giannis Tsakonas, and Charles Farrugia, 8092:390–394. Lecture Notes of Computer Science (LNCS). Valletta, Malta: Springer, 2013.

## H.2 Methodology

Docear users can register an account and provide demographic information such as year of birth and gender if they like. They may also opt-in for receiving research paper recommendations (even without registration). Recommendations are shown on request or automatically every three days of use, ten at a time. During March and Mai 2013 1,028 users received 37,572 recommendations. Details on the recommendation process may be found in [47]. For the evaluation we used click-through rate (CTR) which expresses how many out of the displayed recommendations were clicked. For instance, when 37,572 recommendations were shown, and 2,361 were clicked, CTR is 6.28%. CTR is a common measure in online advertisement and equivalent to "precision" in information retrieval.

## H.3 Results

From a total of 1,028 users who received recommendations, 38.62% did not register and 61.38% registered. 21.79% registered but did not provide information about their gender, 33.17% registered and were males, and 6.42% registered and were females (Figure 91, left pie). Looking only at those users who specified their gender, 83.79% were male, and 16.22% were female (Figure 91, right pie). Among the genders there is only a marginal difference in CTR with 6.88% for males and 6.67% for females (Figure 92). However, there is a significant difference between registered users (6.95%) and unregistered users (4.97%). Interestingly, those users who registered and did not specify their gender have the highest CTR with 7.14%. Another interesting difference between genders relates to the willingness of accepting recommendations. From all male users, 38.09% activated recommendations while only 34.74% of women did and even less (28.72%) of the users who did not specify their gender during registration (Table 20). This might indicate that these users are concerned about privacy issues when receiving recommendations [365].

From the registered users, 39.62% did not specify their age. From those who did, around one quarter (24.15%) were 25 to 29 years of age (Figure 93, bar chart). 11.29% were between 20 and 24 years and only two users were younger than 20, namely 17 and 18. The vast majority (88.19%) was older than 25 years. 4.46% of the users were 60 or older. The mean age was 36.56 years, the median was 33. Of course, it might be that some users did not provide their correct age and the true ages slightly differ from the ones presented.

Looking at click-through rate by age shows that the older a user is the higher CTR becomes (Figure 93, dotted line). While younger users (20-24 years) have the lowest CTR of only 2.73% on average, CTR for users older than 60 is the highest with 9.92%. Overall, a clear linear trend is recognizable (Figure 93, dotted line). CTR for users who registered but did not provide their age was 7.66% on average (not shown in Figure 93).

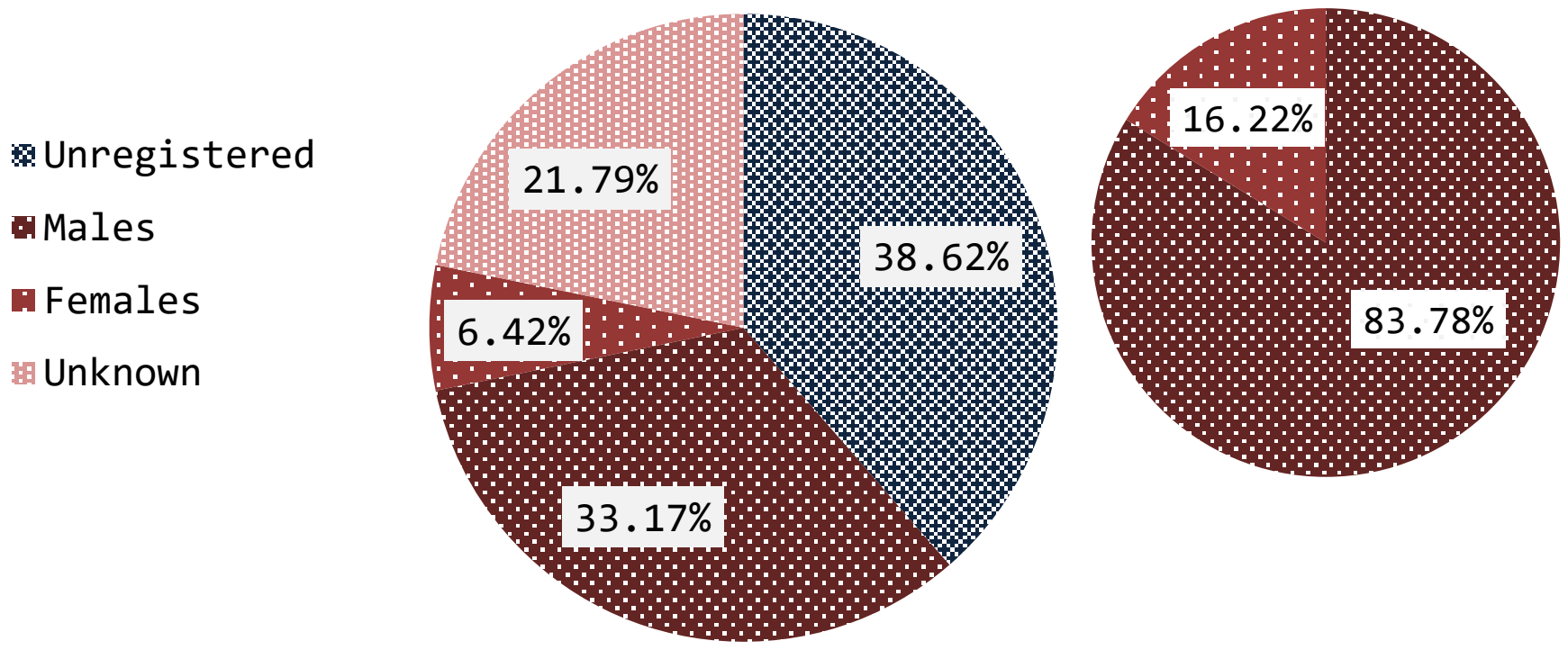

Figure 91: Gender and user type (registered/unregistered) distribution

Table 20: Percentage of activated recommendations by gender

| | Male | Female | n/a |
|---|---|---|---|
| **Recs. Activated** | 38.09% | 34.74% | 28.72% |
| **Recs. Deactivated** | 61.91% | 65.26% | 71.28% |

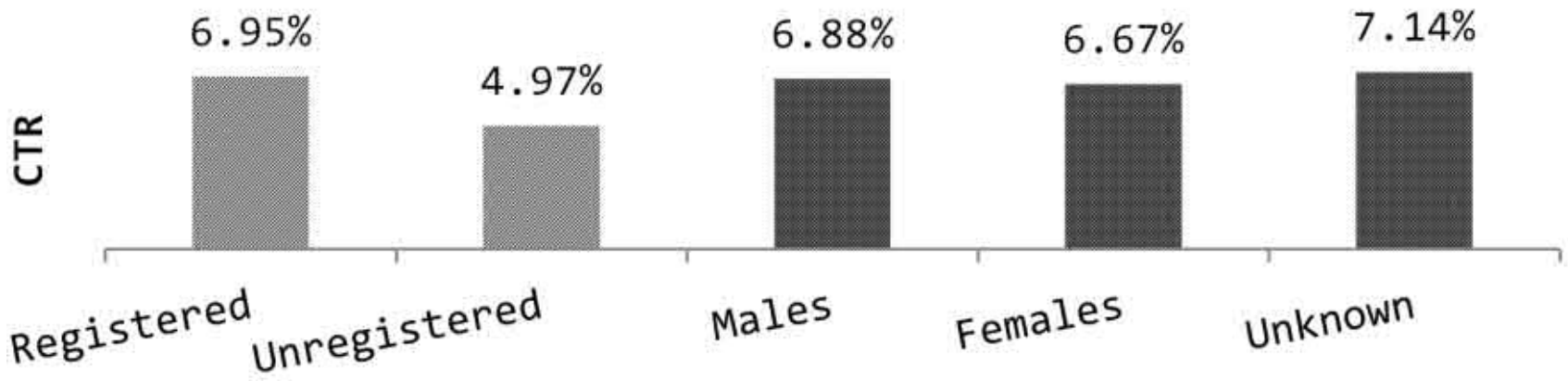

Figure 92: Click-through rate (CTR) by user type and gender

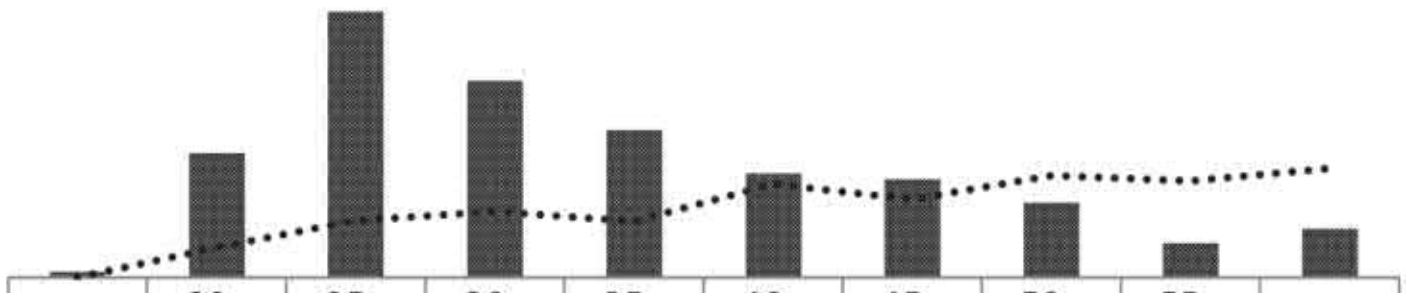

| | 0-19 | 20-24 | 25-29 | 30-34 | 35-39 | 40-44 | 45-49 | 50-54 | 55-59 | 60+ |
|---|---|---|---|---|---|---|---|---|---|---|
| # of users | 0.52% | 11.29% | 24.15% | 17.85% | 13.39% | 9.45% | 8.92% | 6.82% | 3.15% | 4.46% |
| CTR | 0.00% | 2.73% | 5.17% | 5.99% | 5.10% | 8.46% | 7.10% | 9.26% | 8.77% | 9.92% |

Age

Figure 93: Age distribution and click-through rate (CTR) by age

The analysis also indicates that the number of days on which a user started Docear impacts CTR (Figure 94). For the first 20 times a user starts Docear, CTR increases. For instance, users who started Docear on one to five days had a CTR of 5.62% on average while users having started Docear on 11-20 days had a CTR of 7.30% on average. This is not surprising assuming that the more often users start Docear, the more information they enter, the better the user models become, and hence the recommendations. However, for users having started Docear on more than 20 days, CTR decreased. For instance, users having started Docear on more than 100 days achieve a CTR of 4.92% on average.

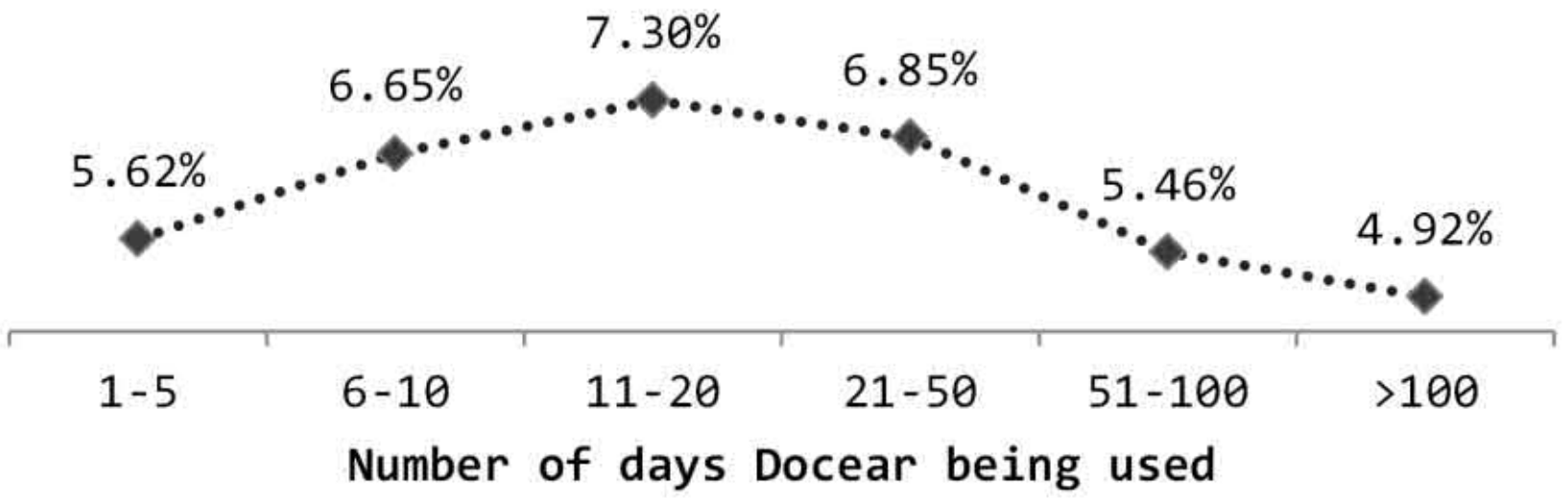

Figure 94: Click-through rate by the number of days Docear being used

Another analysis brings even more confusion. We analyzed how CTR changes based on the number of recommendations a user received. Based on the above results we assumed that the more recommendations a user received, the lower the CTR would become because users starting Docear often also receive more recommendations. Our assumption was not correct. There is a trend that the more recommendations users see, the higher the CTR becomes (Figure 95, dotted line). Users who received only one recommendation set (i.e. typically ten recommendations) had a CTR of 4.13% while users who saw 21-50 sets had a CTR of 9.91% on average.

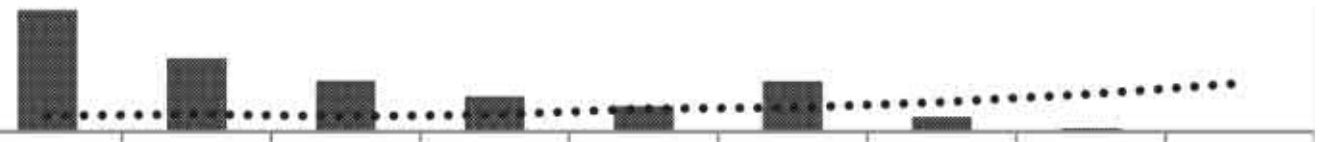

| | 1 | 2 | 3 | 4 | 5 | 6-10 | 11-20 | 21-50 | 51-100 |
|---|---|---|---|---|---|---|---|---|---|
| # of users | 32.00% | 19.46% | 13.52% | 9.34% | 6.81% | 13.42% | 3.99% | 1.17% | 0.29% |
| CTR | 4.13% | 4.56% | 4.03% | 4.39% | 6.07% | 6.32% | 7.74% | 9.91% | 12.67% |

Number of delivered recommendation sets (10 recs per set)

Figure 95: User distribution and CTR by number of recommendation sets

## H.4 Conclusion

The analysis showed that demographics and user-characteristics may have a significant impact on click-through rates on (research-paper) recommender systems. Although gender had only a marginal impact, age impacted CTR strongly. It made also a difference for CTR whether users were registered or not, how many recommendations they had seen before and how often users had started Docear. However, to fully understand the effects and correlations between the last two factors, more research is required.

We suggest that future evaluations should report on their users' demographics and characteristics in order to create valid and comparable results of recommender systems. Some of these are registered vs. unregistered; intensity of the software being used; and amount of previously shown recommendations. There are certainly further demographics and characteristics that might impact an evaluation such as nationality, field of research, and profession, whose impact should be researched.

# I Persistence in Recommender Systems[117]

## I.1 Introduction

Recommender systems became popular in many domains during the past decades and content-based and collaborative filtering became the two most dominant approaches. Some researchers in the field of collaborative filtering analyzed the effect of letting users re-rate items. They found that correlation between original ratings and new ratings was low and only 60% of users gave the same rating as before [93]. Amatriain et al. showed that it might be better to letting users re-rate items than showing new ones. By doing so accuracy of recommender systems increased by around 5% [7].

We wonder whether re-showing recommendations might make sense in general. For instance, a user might miss a recommendation the first time, simply because he was in a hurry and did not pay attention to the recommendation. In this case it would make sense for a recommender to be persistent and to display the same recommendation again. To the best of our knowledge 'recommendation persistence' has not been studied so far.

## I.2 Research Objective & Methodology

Our goal was to find out if and how often it makes sense to display the same recommendations to the same users. To answer this question we analyzed empirical data from the literature management software Docear [33] which features a research paper recommender system [47]. The recommender system recommends research papers to users regardless of whether papers were previously recommended to the users or not. We analyzed how click-through rates (CTR) between recommendations shown only once and CTR of recommendations shown multiple times differed. CTR expresses how much percent of the delivered recommendations were clicked. For instance, if 12 recommendations were clicked out of 1,000 delivered ones, CTR would be 1.2%. CTR basically measures the 'precision' of the recommendation algorithm under the assumption that a clicked recommendation is a ‘good’, i.e. useful, recommendation. For further details on Docear and its recommender system (e.g. how recommendations are generated and displayed) see [33, 47].

[117] This chapter has been published as: Beel, Joeran, Stefan Langer, Marcel Genzmehr, and Andreas Nürnberger. "Persistence in Recommender Systems: Giving the Same Recommendations to the Same Users Multiple Times." In Proceedings of the 17th International Conference on Theory and Practice of Digital Libraries (TPDL 2013), edited by Trond Aalberg, Milena Dobreva, Christos Papatheodorou, Giannis Tsakonas, and Charles Farrugia, 8092:390–394. Lecture Notes of Computer Science (LNCS). Valletta, Malta: Springer, 2013.

## I.3 Results

31,942 recommendations were shown to 1,155 users for the first time and from the 31,942 recommendations 1,677 were clicked, which equals a click-through rate of 5.25% (Table 21). From the 31,942 recommendations 2,466 were shown a second time to 375 distinct users and 154 recommendations were clicked (CTR 6.24%). From the 2,466 recommendations 574 were displayed a third time and CTR was 6.97%. Also for the fourth iteration CTR was still rather high (6.55%). Based on these results one might conclude that it could make sense to display recommendations at least two or three times because for these reiterations CTR was significantly higher than for the first one ($p<0.05$).

Table 21: Reiterations and click-through rate

| | | Reiteration | | | | | | | | | |
|---|---|---|---|---|---|---|---|---|---|---|---|
| | | 1 | 2 | 3 | 4 | 5 | 6 | ... | 11 | ... | 21 |
| | Users | 1,155 | 375 | 97 | 38 | 12 | 6 | | - | | 1 |
| | Impressions | 31,942 | 2,466 | 574 | 229 | 112 | 71 | | 2 | | 1 |
| | No clicks | 30,265 | 2,312 | 534 | 214 | 100 | 68 | | 2 | | 1 |
| | Clicks | 1,677 | 154 | 40 | 15 | 12 | 3 | | - | | - |
| | CTR, overall | 5.25% | 6.24% | 6.97% | 6.55% | 10.71% | 4.23% | | 0.00% | | 0.00% |
| | 1st click | 1,677 | 97 | 14 | 8 | 7 | - | | - | | - |
| Obliv.-clicks | 2nd click | - | 57 | 13 | 1 | 2 | 1 | | - | | - |
| Obliv.-clicks | 3rd click | - | - | 13 | 3 | 2 | 1 | | - | | - |
| Obliv.-clicks | 4th click | - | - | - | 3 | - | - | | - | | - |
| Obliv.-clicks | 5th click | - | - | - | - | 1 | - | | - | | - |
| Obliv.-clicks | 6th click | - | - | - | - | - | 1 | | - | | - |
| | Σ Obliv. clicks | - | 57 | 26 | 7 | 5 | 3 | | - | | - |
| | % Obliv. clicks | 0% | 37% | 65% | 47% | 42% | 100% | | - | | - |
| | CTR, 1st click | 5.25% | 3.93% | 2.44% | 3.49% | 6.25% | 0.00% | | 0.00% | | 0.00% |

The picture changes when looking at more detail into the data: around 50% of all clicks on reshown recommendations are ‘oblivious-clicks’ (Table 21, lower part). We define an ‘oblivious click’ as a click on a recommendation that the user should know already, because he clicked it previously. For instance, 574 recommendations were shown three times. 40 of these recommendations were clicked which equals a CTR of 6.97%. However, only 14 were clicked for the first time – the other 26 (2x13) were clicked for the second or even third time. In one case a recommendation was even shown six times to the same user and the user clicked it each time. Ignoring the oblivious-clicks, i.e. considering only $1^{st}$ clicks,

CTR decreases the more often recommendations are shown. Therefore, results may indicate that CTR increases when showing recommendations multiple times but only because users sometimes clicked on recommendations they have clicked before.

In addition, CTR increased in general the more recommendations were shown previously to a user (Figure 96). For instance, CTR did not only increase for reshown recommendations but also for ‘fresh’ recommendations, i.e. recommendations being displayed to a user for the very first time. This is not surprising because users who receive many recommendations probably are using the software for a longer time than users receiving their first recommendations. And for users using the software for a longer time, better user models can be created and hence better recommendations can be given (although this is not always the case as shown in [42].

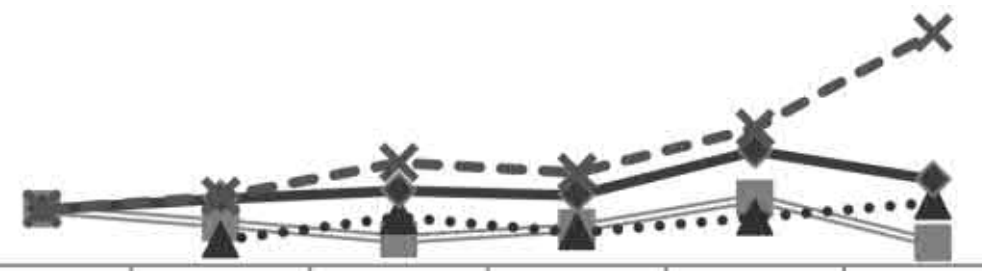

| CTR | 1 | 2 | 3 | 4 | 5 | >5 |
|---|---|---|---|---|---|---|
| Rshwn; new | 5.25% | 3.93% | 2.44% | 3.49% | 6.25% | 2.11% |
| Rshwn; obl. | | 2.31% | 4.53% | 3.06% | 4.46% | 5.91% |
| Rshwn; all | 5.25% | 6.24% | 6.97% | 6.55% | 10.71% | 8.02% |
| Fresh | 5.25% | 6.44% | 9.59% | 8.67% | 12.83% | 21.74% |

**Number of Reiterations (no delay)**

Figure 96: Redisplayed recommendations vs. fresh ones

To get a better understanding of how good re-shown recommendations performed, we compared their CTR with CTR of fresh recommendations. If a recommendation was shown the second time, it received a CTR of 6.24% on average – a CTR of 3.93% for reshown recommendations not being clicked before and a CTR of 2.31% for reshown recommendations being clicked before (Figure 96). In contrast, fresh recommendations being displayed at the same time achieved a CTR of 6.44% and hence performed better than the reshown recommendations. This is true for all iterations: fresh recommendations always performed better than reshown recommendations at the same time (including oblivious-clicks). Considering only new clicks on reshown recommendations (i.e. ignoring oblivious clicks), fresh recommendations performed even two to three times as good.

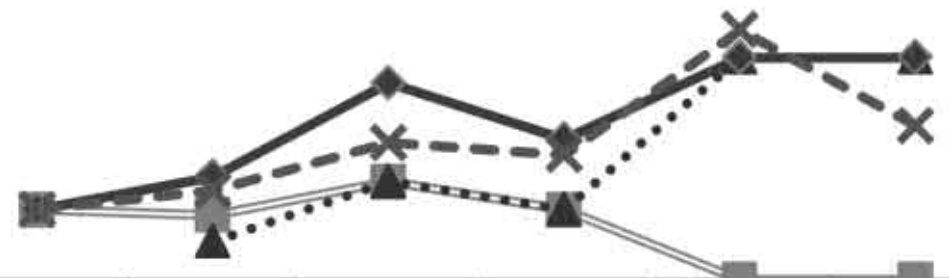

| CTR | 1 | 2 | 3 | 4 | 5 | >5 |
|---|---|---|---|---|---|---|
| Rshwn; new | 5.25% | 4.85% | 7.37% | 5.26% | 0.00% | 0.00% |
| Rshwn; obl. | | 2.88% | 7.37% | 5.26% | 16.67% | 16.67% |
| Rshwn; all | 5.25% | 7.72% | 14.75% | 10.53% | 16.67% | 16.67% |
| Fresh | 5.25% | 6.69% | 10.27% | 9.38% | 18.75% | 11.54% |

Number of Reiterations (1 day delay)

Figure 97: Fresh recommendations vs. redisplayed ones (one-day delay)

Based on the presented numbers one could conclude that reshowing recommendations would never make sense. However, we did the same analysis for recommendations that were reshown with at least one day delay (Figure 97). That means we ignored all recommendations in the analysis that were reshown to the same user within 24 hours. In this case, CTR of reshown recommendations is often better than for fresh recommendations (with oblivious-clicks included). For instance, for the second iteration CTR for fresh recommendations was 6.69% but for reshown recommendations 7.72%. However, when ignoring oblivious-clicks again fresh recommendations always perform better than reshown recommendations. We also conducted the same analysis with a longer delay (three, seven, and fourteen days). Results were similar to the ones presented. Due to space restrictions we omit further details.

## I.4 Interpretation and Outlook

Our results indicate that it makes no sense to generally display recommendations multiple times to the same users – fresh recommendations usually perform better. Nevertheless, about 2-3 % of recommendations shown the second or third time were clicked by the users for the first time. By showing recommendations only once, researchers would miss this 2-3% of interesting articles. In further research it should be studied why users sometimes click recommendations only when they were shown multiple times and whether users eventually found those recommendations useful or not. If they found the recommendations useful, then it should be studied how to find out which recommendations to show multiple times and how often. For instance, it might be that the interest of a user has changed – maybe even due to the recommendations he has seen – and on first display the recommendation simply was not relevant for him. That means if a strong concept drift was determined by the recommender system, recommendations shown previously (before the concept drift) might be given again.

In addition, it should be studied why users click several times on the same recommendations. We assumed that users were just oblivious. In this case it probably would be of little benefit for the user to see the same recommendations several times. But maybe obliviousness is not the only reason for clicking recommendations multiple times.

It is also quite interesting that it made a difference whether a recommendation was reshown before or after 24 hours of a previous impression. In latter case (delay of one day or more), click through rates were significantly higher than for recommendations being re-shown within 24 hours and CTR of the reshown recommendations was even higher than for fresh recommendations. Under the assumption that oblivious clicks are desirable, reshowing recommendations could make sense. It might also make sense to transfer this finding to collaborative filtering and study how long to set a delay before letting users re-rate their items.

# J Impact of Labels[118]

## J.1 Introduction

In the Web community there is lots of discussion about organic and sponsored search. ‘Organic search’ is the classic search where users enter search terms and search engines return a list of relevant web pages. ‘Sponsored search’ describes additional ‘results’ that are often shown beside the organic results. Usually these results are related to the search terms but companies pay for them to be displayed (in other words, ‘sponsored search’ is a nice paraphrase for personalized advertisement). While typical online advertisement has click-through rates (CTR) around 0.5% [262], sponsored search achieves CTRs around 2% and sometimes even more than 30% [351]. CTR is a common performance measure in online advertisement. It describes how many ads were clicked relative to the delivered ones. For instance, if 1,000 ads were delivered, and users clicked 61 of them, CTR was 6.1%. The higher the CTR the better is the algorithm behind the search results.

In academia, there are several academic recommender systems which typically only show organic recommendations [153, 248]. However, we were interested which CTR was to expect for sponsored recommendations in academia and more importantly, how much, or how little, users would like recommendations in general that were displayed for profit-making.

## J.2 Methodology

Our academic literature management software ‘*Docear*’ [33] features a research paper recommender system [47]. Every third start Docear displays ten recommendations that can be freely downloaded (Figure 98). We modified Docear’s recommender system and analyzed the effects of the modifications on click-through rates (overall, 22,452 recommendations were delivered to 587 users). Modifications were related to a label describing the nature of the recommendations (organic or commercial) and the way of presenting

---

[118] This chapter has been published as: Beel, Joeran, Stefan Langer, and Marcel Genzmehr. “Sponsored vs. Organic (Research Paper) Recommendations and the Impact of Labeling.” In Proceedings of the 17th International Conference on Theory and Practice of Digital Libraries (TPDL 2013), edited by Trond Aalberg, Milena Dobreva, Christos Papatheodorou, Giannis Tsakonas, and Charles Farrugia, 395–399. Valletta, Malta, 2013.

recommendations (Figure 98). More information on the recommender system can be found in [33, 47].

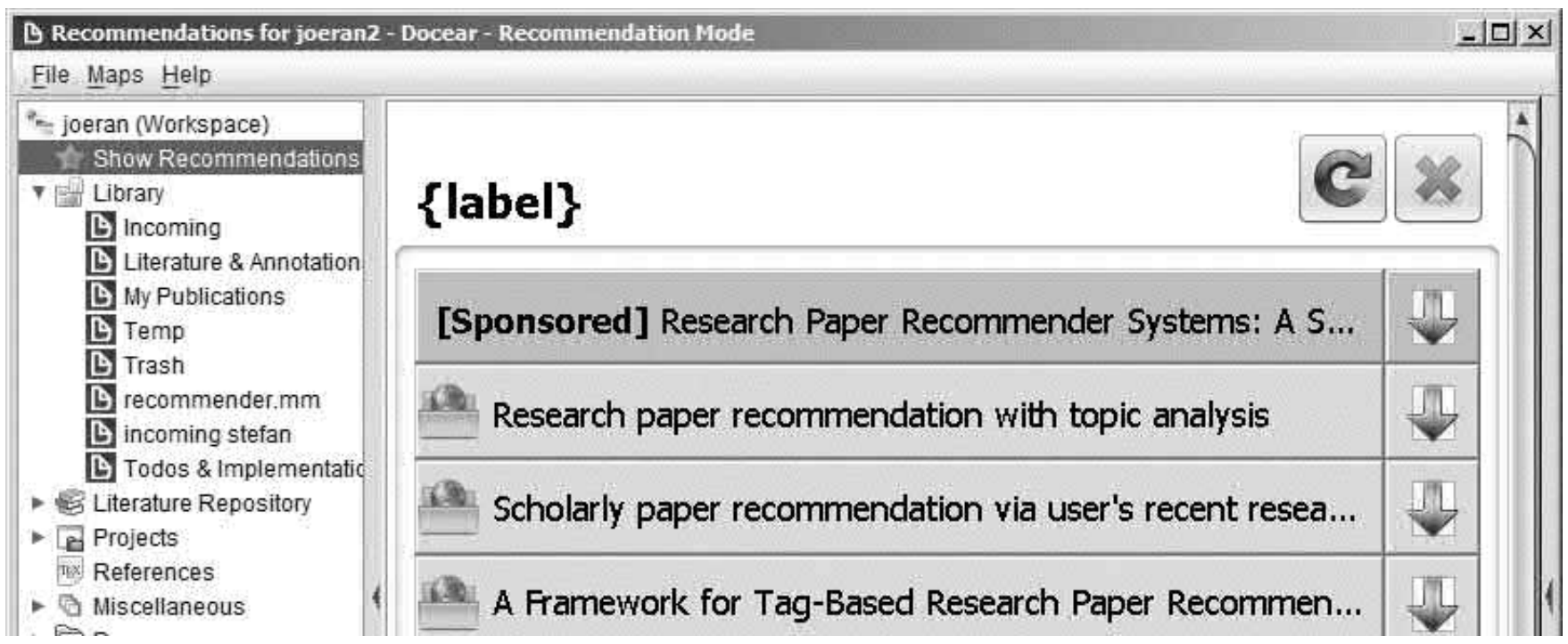

Figure 98: Recommendations in Docear with labels

Recommendations in Docear were ‘labeled’ to explain the ‘nature’ of the recommendations (Figure 98). The ‘basic’ label was ‘Research Papers’. We modified this label for each user by randomly choosing whether to add a prefix such as ‘Free’ or ‘Free Full-text’ (Table 22) or a suffix such as ‘(Advertisement)’ or ‘(Sponsored)’ which resulted in labels like ‘Free Research Papers’, ‘Research Papers from our partners’, or ‘Free Full-text Research Papers (Sponsored)’. When a suffix was chosen, user must have assumed that the recommendations had a commercial background. When no suffix was chosen, users must have assumed that recommendations were organic. In addition, when no suffix was chosen it was randomly chosen whether to mark the first recommendation as ‘[Sponsored]’ and whether to highlight this recommendation or not (Figure 98). Whatever label was displayed, recommendations were always calculated with the same algorithms and always linked to freely downloadable PDFs.

Table 22: Labels for the recommendations

| Prefix | | | | Suffix | | |
|---|---|---|---|---|---|---|
| Free | Free Full-text | Full-text | None | (Sponsored) | (Advertisement) | From our partners |

We selected two metrics to measure the effectiveness of recommendations and determine differences between the labels. With click-through rate (CTR) we measured how many recommendations out of the displayed ones were clicked overall. For instance, if 1,000 recommendations with a certain label were shown and 50 were clicked, CTR was 5%. If CTR for recommendations with another label was, for instance, 3.2%, the first label performed better. CTR is a common measure on advertisement but it suffers from one problem, especially when recommendations of only a few users are analyzed. In this case, a few users could spoil the results. For instance, one user receiving and clicking many recommendations would strongly increase overall CTR, although maybe all other users hardly clicked on any recommendations. Therefore, we also used mean average precision (MAP) for users' click-through rates. That means, for each user we calculated his average CTR and then we calculated the mean CTR over all users. For instance, if one user had seen 50 recommendations and clicked all of them, and 95 other users had each seen 10 recommendations but clicked none, CTR for the first user was 100% but CTR for the 95 others were 0% each. Hence, MAP was $\frac{100\%+0\%+0\%+\cdots+0\%}{96} = 1.04\%$.

## J.3 Results

Based on CTR organic recommendations clearly outperform commercial ones with a CTR of 8.86% vs. 5.86% (Figure 99, blue line). This is probably what most people would expect. However, it is still interesting to have it quantified that only because recommendations are labeled as some kind of commercial, users are far less likely to click on them. Based on CTR, recommendations with the first recommendation being labeled as '[Sponsored]', but not highlighted, also clearly outperform those being highlighted (8.38% vs. 5.16%). However, the evaluation based on MAP shows a different picture (Figure 99, beige line). Here, organic (MAP=5.21%) and commercial recommendations (4.91%) perform very much alike. In addition, recommendations with the first one being labeled as sponsored *and* being highlighted (MAP=7.47%) outperform those being not highlighted (5.25%). What is evident with both metrics is that completely unlabeled recommendations performed better than all other label variations (CTR=9.87%; MAP=8.76%).

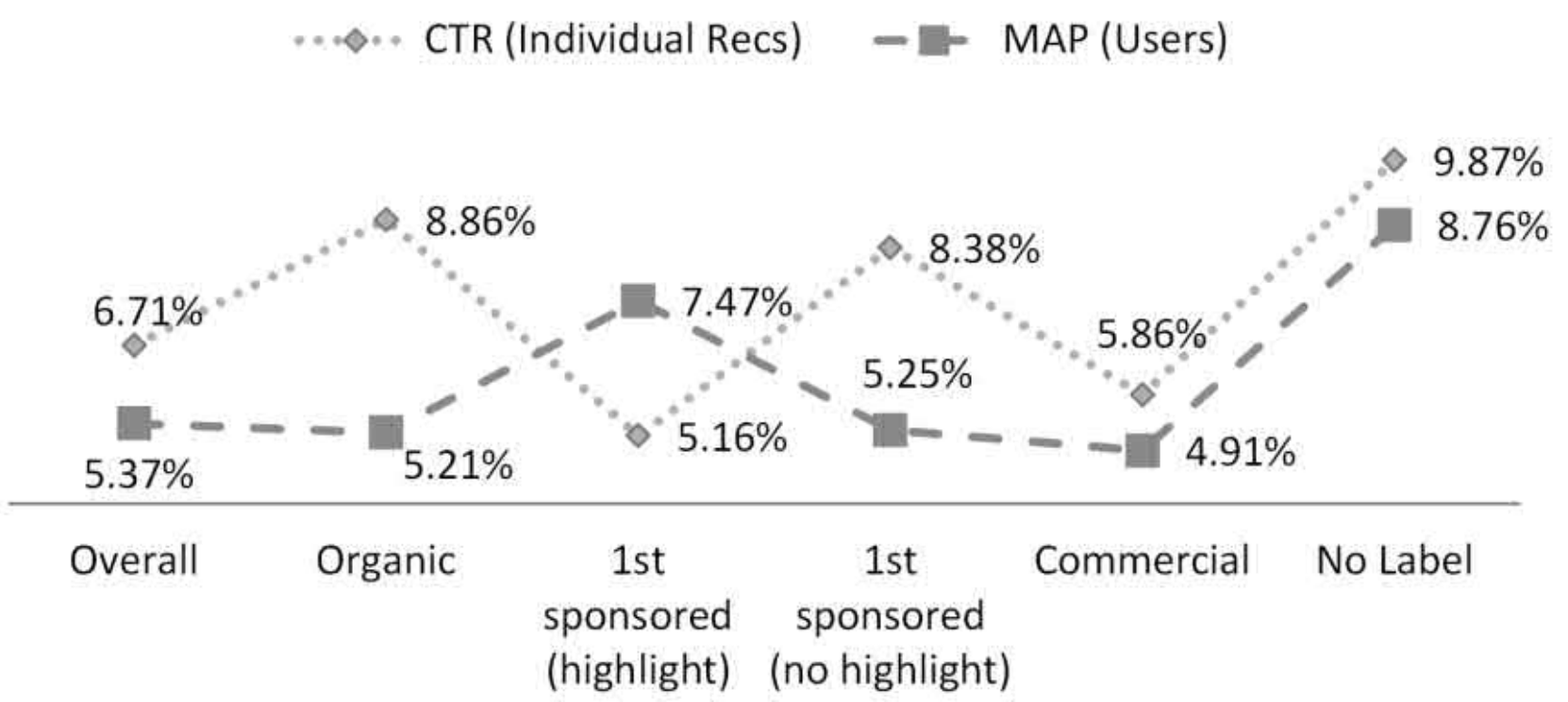

Figure 99: CTR and MAP of different labels

For organic recommendations, the 'free' and 'free full-text' labels clearly outperformed those labels not indicating that the recommended papers were free to download (Figure 100). This is true for both metrics CTR and MAP[119]. However, for commercial recommendations results differed. Here, using no suffix at all (MAP=6.51%; CTR=7.26%) performed better than any of the suffixes. We cannot explain this difference. For suffixes, both CTR and MAP indicate that 'Advertisement' leads to the lowest performance (Figure 101). Based on MAP 'Sponsored' recommendations (5.95%) performed better than 'partner' recommendations (4.85%). Based on CTR, 'partner' recommendations performed better (6.79%) than 'sponsored' ones (5.93%).

Summarized, the most surprising result was that recommendations with no label at all performed best, and that based on MAP commercial and organic recommendations performed about alike. Our study also showed that click-rates on recommendations varied strongly based on how they were labeled (although they were all based on the same algorithms). In particular recommendations labeled as 'advertisement' were least liked by the users. Results based on CTR often contradicted those based on MAP and also using certain prefixes had different effects on commercial and organic recommendations. More research is needed to clarify these contradictions. In some cases a small sample size might have caused the contradictions. For instance, for some labels (e.g. 'Free Research Papers') results were only based on twelve users. However, other results were based on larger samples and still contradict each other.

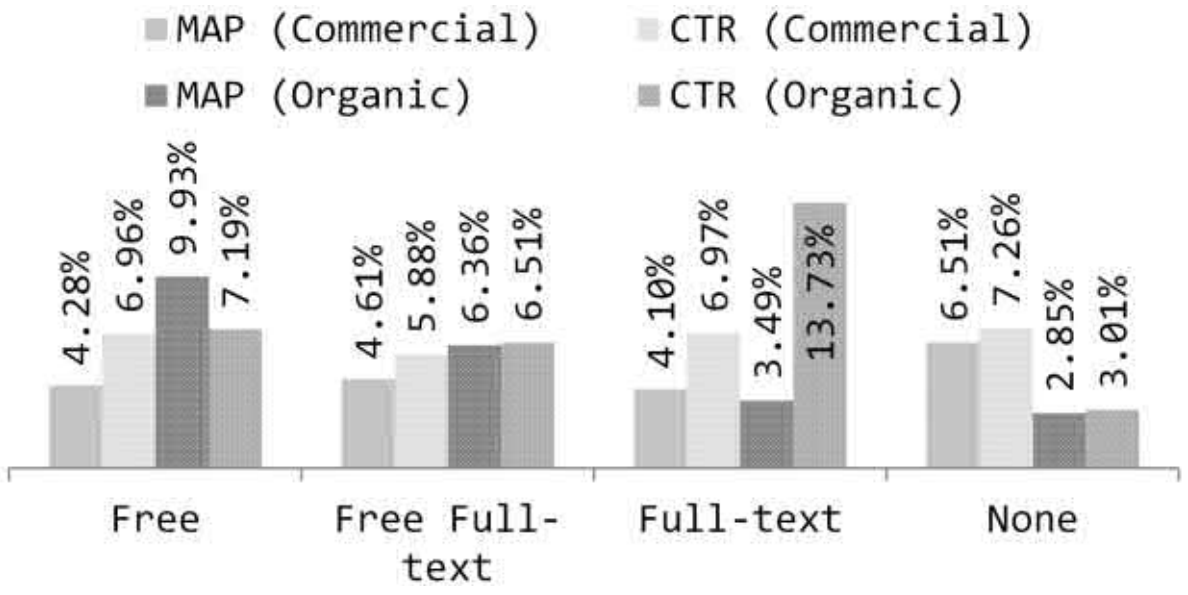

Figure 100: MAP and CTR for prefixes (commercial and organic)

[119] For 'full-text' CTR is an outlier. We investigated the result and found that in this case few users had extremely high CTRs based on few received recommendations they almost all clicked.

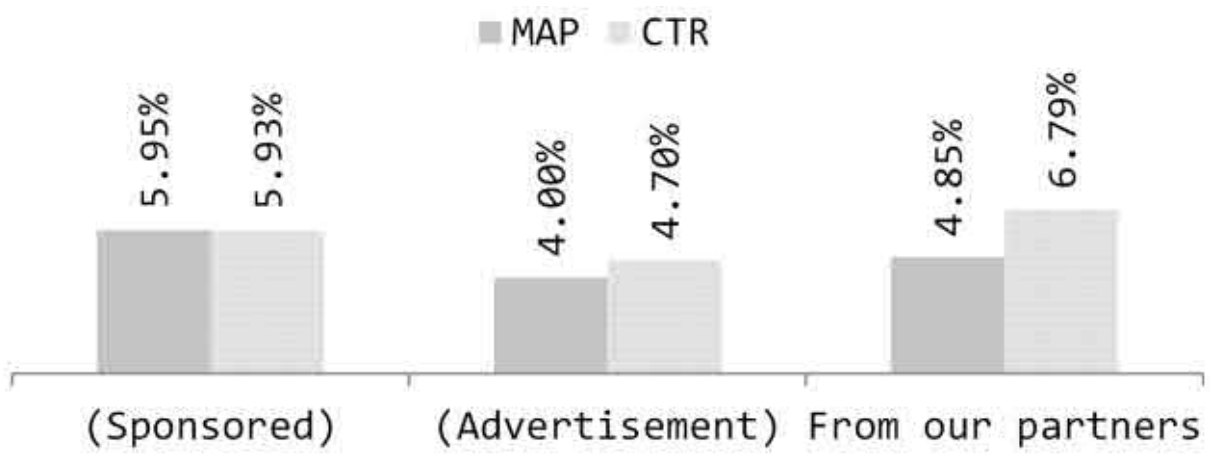

Figure 101: MAP and CTR for suffixes (commercial only)

# K Patent Application

As part of my work, I filed an international patent application (PCT) in 2011 for Docear's mind-map-specific user modeling approach (PCT/EP2011/070873). The following text is a copy of the application in German. A machine translation to English is available via *Google Patents*[120].

## K.1 Patentbeschreibung

### Verfahren und System zum Erstellen von Nutzermodellen

### Gebiet der Erfindung

Die Erfindung betrifft ein Verfahren und ein System zum Erstellen von Nutzermodellen und darauf basierenden Empfehlungen, vorzugsweise durch Analyse von baumförmigen Datenstrukturen.

### Hintergrund der Erfindung und Stand der Technik

Nutzer von Computersystemen unterscheiden sich in vielerlei Hinsicht, unter anderem hinsichtlich ihrer Interessen, ihres Wissens und ihrer demographischen Daten. Viele Computersysteme versuchen diesen Unterschieden gerecht zu werden, indem sie abhängig etwa vom Wissen und den Interessen des Nutzers, individuelle Informationen oder Benutzeroberflächen zur Anzeige bringen. Um Softwaresysteme, etwa Computerprogramme oder Internet-basierte Anwendungen individuell an seine Nutzer anpassen zu können, benötigen die Softwaresysteme Zugriff etwa auf die Interessen der Nutzer. Diese Daten können entweder manuell vom Nutzer angegeben, oder automatisch vom System erzeugt werden. In jedem Fall werden die Informationen über die Nutzer in so genannten Nutzermodellen gespeichert.

Häufig werden solche Nutzermodelle von Empfehlungsdiensten verwendet. Abhängig von den Interessen eines Nutzers zeigen diese Empfehlungsdienste individuelle Empfehlungen beispielsweise für Filme, Bücher, Musik, oder auch auf den Nutzer abgestimmte Werbung an. Bei einem Empfehlungsdienst handelt es sich immer um ein sogenanntes „User-Item Matching Problem“: Die Frage bei diesem Problem ist, welche kleine Auswahl an relevanten Items (z.B. Musikstücke, Webseiten, Bücher, etc.) aus einer großen Menge von verfügbaren Items einem Nutzer empfohlen werden soll. Dieser aus dem Stand der Technik bekannte Ansatz ist in Fig. 1 dargestellt. Gezeigt ist in Fig. 1 eine Menge von Nutzer n (User 1 bis User 3) und eine Menge von Items (Item 1 bis Item 3). Mit entsprechenden Verfahren wird die Relevanz von Nutzern und Items zueinander berechnet. Danach können alle Items die einen bestimmten Schwellenwert bzgl. der Relevanz überschreiten den entsprechenden Nutzern empfohlen werden.

[120] http://www.google.com/patents/WO2013075745A1?cl=en&hl=de

Aus dem Stand der Technik bekannte Empfehlungsdienste nutzen zwei grundsätzliche Verfahren, um Nutzermodelle zu erzeugen bzw. Empfehlungen zu geben. Diese Verfahren sind bekannt als "Content Based Filtering" bzw. "Collaborative Filtering".

Beim Content Based Filtering (CBF) nimmt das Computersystem an, dass der Inhalt (Content) der Items mit denen ein Nutzer in Verbindung steht, die Interessen und/oder das Wissen des Nutzers wiederspiegelt. Dieser aus dem Stand der Technik bekannte Ansatz ist in Fig.2 gezeigt.

Ein Nutzer (User 1) steht mit einer Anzahl von Items (Item 1 bis Item j) in Verbindung. Ein Item kann jedes mögliche Objekt sein. Ein Item kann etwa ein Dokument (Bücher, Webseiten, Emails, etc.), ein Multimediaobjekt (Filme, Musik, Fotos), eine Personen oder ein Ort sein. Items können aber auch Menüeinträge einer Computeranwendung oder Komponenten grafischer Benutzeroberflächen sein.

In Verbindung steht ein Nutzer mit einem Item, wenn irgendein Bezug zwischen ihnen besteht. Das heißt, wenn der Nutzer beispielsweise ein Buch gelesen, gekauft oder auch nur kurz betrachtet hat, eine Person kennt oder einen Film geschaut, heruntergeladen oder auf einem Filmportal bewertet hat, steht der Nutzer mit dem Buch, der Person, bzw. dem Film in Verbindung. Die Verbindung kann dabei unterschiedlich stark gewichtet werden, je nach Art der Verbindung. Beispielsweise könnte das Kaufen eines Buches stärker gewichtet werden als das bloße Betrachten des Buchcovers. Oder eine Verbindung zu einem Item kann umso stärker gewichtet werden, je öfter das Item genutzt wurde.

Beim Content Based Filtering wird der Inhalt der verbundenen Objekte genutzt um ein Nutzermodell zu erstellen. In der Regel wird dieses Verfahren bei textuellen Items, also Dokumenten, angewandt, da der Inhalt von Dokumenten (also der Text) gut von Computern verarbeitet werden kann im Gegensatz, z. B. zu Bildern. Um den Inhalt der Items zu nutzen, wird für jedes Item ein Modell erstellt. Bei Dokumenten wird häufig das sogenannte "Vector Space Model" genutzt, ein Modell, welches Dokumente als Vektor ihrer Terme darstellt. Jeder Vektor drückt durch seine Länge aus, wie gut der entsprechende Term das eigentliche Dokument beschreibt. Diese Gewichtung kann mit verschiedenen Verfahren errechnet werden. Ein gängiges Verfahren ist das sogenannte TF-IDF Verfahren. Hierbei ist das Gewicht eines Terms für ein Dokument umso größer je öfter der Term in dem Dokument vorkommt und je weniger Dokumente in der gesamten Kollektion es mit diesem Term gibt.

Das Nutzermodell wird dann aus den Modellen der verschiedenen verbundenen Items erzeugt. Dies bedeutet, wenn ein Nutzer viele Bücher besitzt die den Term „Recommender" mit hohem Gewicht enthalten, dann bekommt auch das Nutzermodell diesen Term mit einem hohem Gewicht zugeordnet. Die verschiedenen Item-Modelle können dabei mit unterschiedlicher Gewichtung in das Nutzermodell einfließen. Üblich ist es beispielsweise, Items die vor kurzem genutzt wurden, stärker zu gewichten als Items deren Nutzung bereits längere Zeit zurückliegt. Üblicherweise wird das Nutzermodell in dem gleichen Format gespeichert, wie die Item-Modelle – also beispielsweise als Vector Space Model.

Von den Items, die später dem Nutzer gegebenenfalls empfohlen werden sollen, wird ebenfalls ein Modell erstellt, beispielsweise wieder mit TF-IDF-Verfahren und dem Vector Space Model. Diese Items müssen nicht notwendigerweise die gleichen Items sein, die mit Nutzern in Verbindung stehen. Beispielsweise wäre es möglich, ein Nutzermodell zu erstellen basierend auf Webseiten die ein Nutzer besucht hat, und ihm basierend auf diesem Modell Bücher zu empfehlen, oder auch personalisierte Werbung anzuzeigen. Das Matching der Nutzermodelle mit den zu empfehlenden Items basiert vorzugsweise auf Ähnlichkeitsvergleichen zwischen den Nutzer- und den Item-Modellen. Bei textbasierten Items bedeutet das, wenn das Nutzermodell die gleichen Terme mit hohem Gewicht enthält wie die Item-Modelle der zu empfehlenden Items, dann ist die Ähnlichkeit groß und das Item wird dem Nutzer empfohlen. Ein übliches Ähnlichkeitsmaß im Vector Space Model ist etwa die Cosine Similarity.

Beim sogenannten Collaborative Filtering (CF), welches in **Fig. 3** gezeigt ist, spielt der Inhalt von Items keine Rolle. Es wird lediglich die Information verwendet, welche Items mit welchen Nutzern (wie stark) in Verbindung stehen. Die Gewichtung wird entweder direkt vom Nutzer angegeben, indem der Nutzer ein Item bewertet, oder indirekt, indem das System die Nutzung des Items überwacht.

Wie beim Content Based Filtering kann die Gewichtung beispielsweise umso stärker sein, je öfter ein Item genutzt wird. Oder ein Item, welches gekauft wurde, wird stärker gewichtet als ein Item welches kostenlos heruntergeladen wurde. Berechnet wird dann auch nicht die Ähnlichkeit zwischen Nutzer-Modellen und Item-Modellen sondern ausschließlich die Ähnlichkeit von Nutzermodellen zueinander (User-User Matching). Hier gibt es wieder viele bekannte Verfahren. Im Wesentlichen wird bei allen Verfahren geprüft, welche Nutzermodelle möglichst viele Items in ähnlicher Gewichtung gemeinsam haben. Wurden nun ähnliche Nutzermodelle identifiziert, werden dem Nutzer 1 die Items empfohlen die mit dem ähnlichen Nutzer 2 in starker Verbindung stehen (und die Nutzer 1 gegebenenfalls noch nicht kennt).

Es ist auch möglich, die Nutzermodelle wie beim Content Based Filtering zu erzeugen und basierend auf diesen Nutzermodellen gleiche Nutzer zu identifizieren. Dieser Ansatz ist in **Fig. 4** gezeigt.

Zumindest beim Content Based Filtering (CBF) ist das Erstellen der Item-Modelle ein zentraler Bestandteil, da alles andere, d.h., das Erstellen der Nutzermodelle und das Matching von Nutzern und Items, hierauf basiert. Wie vorstehend erwähnt, ist ein häufig verwendetes Modell das Vector Space Model, welches ein textuelles Item als Vektor seiner Terme speichert, wobei die Länge des Vektors die Gewichtung des jeweiligen Terms in Bezug auf das Item repräsentiert. Um das Gewicht der Terme zu bestimmen, gibt es zahlreiche Verfahren. Nachteilig ist hierbei jedoch, dass mit diesen aus dem Stand der Technik bekannten Verfahren nur die Modellierung von "normalen" textuellen Items, also Dokumenten, wie Emails, Webseiten, Büchern, News Artikel, wissenschaftliche Artikel, etc., möglich ist. Eine Abbildung von Termen in baumförmigen Strukturen, etwa Mind Maps und Verzeichnisstrukturen, im Vector Space Model, um hierauf basierend Nutzermodelle und Empfehlungsdienste zu realisieren ist mit den aus dem Stand der Technik bekannten Verfahren nicht möglich.

## Aufgabe der Erfindung

Aufgabe der Erfindung ist es daher, ein Verfahren und ein System bereitzustellen, welche es auf einfache Weise erlauben Item-Modelle bzw. Nutzermodelle auch für hierarchische, d.h. baumförmige Strukturen zu erzeugen, um basierend hierauf Nutzermodelle und Empfehlungen zu erstellen.

## Erfindungsgemäße Lösung

Diese Aufgabe wird durch ein Verfahren und ein System gemäß den unabhängigen Ansprüchen gelöst. Vorteilhafte Ausgestaltungen der Erfindung sind in den jeweiligen abhängigen Ansprüchen angegeben.

Bereit gestellt wird demnach ein Verfahren zum Erzeugen eines Nutzermodells, insbesondere für einen Empfehlungsdienst, aus zumindest einer baumförmigen Datenstruktur, wobei das Nutzermodell Informationen über einen Nutzer umfasst, wobei die zumindest eine baumförmige Datenstruktur dem Nutzer zuordenbar ist, wobei die baumförmige Datenstruktur einen Wurzelknoten und eine Anzahl von Kinderknoten umfasst, welche über Kanten mit dem Wurzelknoten oder mit einem Kinderknoten verbunden sind, wobei zumindest einem Knoten zumindest ein Element zugeordnet ist, und wobei

- die den Knoten zugeordneten Elemente ermittelt werden, wobei die Elemente einen Inhalt des jeweiligen Knoten repräsentieren,
- die ermittelten Elemente gewichtet werden und jedem Element eine Elementgewichtung zugeordnet wird, und
- ein Nutzermodell generiert wird, wobei das generierte Nutzermodell die ermittelten Elemente und die dem jeweiligen Element zugeordnete Elementgewichtung umfasst.

Die Knoten der baumförmigen Datenstruktur können gewichtet werden und jedem Knoten kann eine Knotengewichtung zugeordnet werden.

In einem Initialisierungsschritt kann jedem Element eine vorbestimmte Elementgewichtung oder die Knotengewichtung des zugeordneten Knotens zugeordnet werden.

Das Verfahren kann ferner einen Vorverarbeitungsschritt umfassen, bei dem
- Knoten, denen keine Elemente zugeordnet sind, gelöscht werden, und/oder
- Knoten gelöscht werden, denen ein vorbestimmtes Element zugeordnet oder nicht zugeordnet ist, und/oder
- Knoten gelöscht werden, welche vorbestimmte Attribute aufweisen oder nicht aufweisen, und/oder
- Knoten und/oder Elemente der Knoten gelöscht werden, welche nicht direkt dem Nutzer zugeordnet sind.

Das Gewichten der Knoten kann eine statische Knotengewichtung und/oder eine dynamische Knotengewichtung umfassen, wobei

- bei der statischen Knotengewichtung die Anzahl der dem jeweiligen Knoten zugeordneten Kinderknoten, die Anzahl der jeweiligen Geschwisterknoten, die Tiefe des jeweiligen Knotens in der baumförmigen Datenstruktur, die Sichtbarkeit des Knotens, oder eine Kombination hiervon berücksichtigt werden, und
- bei der dynamischen Knotengewichtung für jeden Knoten das Alter, der Zeitpunkt der letzen Änderung, die Anzahl der Änderungen, die Anzahl der Verschiebungen innerhalb der baumförmigen Datenstruktur, die Anzahl der Markierungen, die Sichtbarkeit des Knotens, ein Dämpfungsfaktor, oder eine Kombination hiervon berücksichtigt werden.

Das Ermitteln der den Knoten zugeordneten Elemente kann ein Vorverarbeiten der ermittelten Elemente umfassen, wobei beim Vorverarbeiten der Elemente Text in Token und/oder Terme zerlegt wird, sofern das Element ein Textelement ist, und/oder Verweise verarbeitet werden, sofern das Element ein Verweiselement ist.

Die in dem Initialisierungsschritt den Elementen zugeordneten Elementgewichtungen können angepasst werden, wobei beim Anpassen der jeweiligen Elementgewichtung der Elementtyp, Attributsausprägungen der von dem Element zugeordneten Attribute, eine Häufigkeit des Elements innerhalb der baumförmigen Datenstruktur, die Anzahl der baumförmigen Datenstrukturen in einer Kollektion von baumförmigen Datenstrukturen in denen das Element vorkommt, eine Häufigkeit des Elements innerhalb einer Kollektion von baumförmigen Datenstrukturen, die Größe der baumförmigen Datenstruktur im Verhältnis zu anderen baumförmigen Datenstrukturen in einer Kollektion von baumförmigen Datenstrukturen, die Position des Elementes innerhalb des Knotens, die Sprache des Elementes, die Anzahl der Elemente innerhalb des Knotens, der Abstand des Elementes zu gleichartigen Elementen anderer Knoten, Häufigkeit des Elementes in dem Pfad zwischen dem Knoten und dem Wurzelknoten, das Alter des Elements, der Zeitpunkt der letzen Änderung, die Anzahl der Änderungen, die Anzahl der Markierungen, die Sichtbarkeit des Elements, ein Dämpfungsfaktor, oder eine Kombination hiervon berücksichtigt werden.

Bei der statischen Knotengewichtung und/oder bei der dynamischen Knotengewichtung oder nach der statischen Knotengewichtung und/oder nach der dynamischen Knotengewichtung und/oder bei oder nach der Elementgewichtung kann eine Vererbung des Knotengewichts bzw. des Elementgewichts berücksichtigt werden.

Das generierte Nutzermodell kann in einer Speichereinrichtung gespeichert werden, um dem Empfehlungsdienst zur Verfügung gestellt zu werden.
Alle Elemente können zusammen mit den jeweiligen Elementgewichtungen als Nutzermodell gespeichert werden, oder für jeden Elementtyp kann ein eigenes Nutzermodell gespeichert werden, wobei die Nutzermodelle der verschiedenen Elementtypen ein Gesamtnutzermodell bilden.

Bei mehreren dem Nutzer zuordenbaren baumförmigen Datenstrukturen kann für jede baumförmige Datenstruktur eine Anzahl von Nutzermodellen generiert werden, welche zusammen ein dem Nutzer zugeordnetes Gesamtnutzermodell bilden.

Jeder baumförmigen Datenstruktur kann eine Baumgewichtung zugeordnet werden.

Für eine neue dem Nutzer zuordenbare baumförmige Datenstruktur kann das dem Nutzer zugeordnete Nutzermodell angepasst werden.

Von der baumförmigen Datenstruktur referenzierte Elemente können in das Nutzermodell eingefügt werden und wie Elemente der baumförmigen Datenstruktur behandelt werden.

Einem generierten Nutzermodell kann eine Information über den Nutzermodelltyp zugeordnet werden.

Das Verfahren kann ferner ein Auswählen von Objekten anhand vorbestimmter Auswahlkriterien umfassen, wobei ein Objekt ein Nutzermodell oder ein Itemmodell umfasst.

Die Auswahlkriterien können umfassen:
- Objekte eines vorbestimmten Typs, und/oder

- Objekte die eine vorbestimmte Ähnlichkeit zu dem Nutzermodell und/oder Itemmodell aufweisen, wobei vor dem Auswählen Ähnlichkeitswerte zwischen dem generierten Nutzermodell und/oder Itemmodell und den Objekten ermittelt werden.

Durch ein Itemmodell kann ein Förderprogramm repräsentiert werden, wobei das das Förderprogramm repräsentierende Itemmodell ausgewählt wird, wenn das Nutzermodell eine vorbestimmte Ähnlichkeit zu dem Itemmodell aufweist.

Bereit gestellt wird ferner ein System zum Erzeugen eines Nutzermodells, insbesondere für einen Empfehlungsdienst, aus zumindest einer baumförmigen Datenstruktur, wobei das Nutzermodell Informationen über einen Nutzer umfasst, wobei die baumförmige Datenstruktur dem Nutzer zuordenbar ist, wobei die baumförmige Datenstruktur einen Wurzelknoten und eine Anzahl von Kinderknoten umfasst, welche über Kanten mit dem Wurzelknoten oder mit einem Kinderknoten verbunden sind, und wobei zumindest einem Knoten zumindest ein Element zugeordnet ist, wobei das System aufweist:

- wenigstens eine Speichereinrichtung zum Speichern wenigstens einer baumförmigen Datenstruktur, und
- eine Verarbeitungseinrichtung, welche mit der Speichereinrichtung gekoppelt ist und welche angepasst ist ein Verfahren nach einem der vorhergehenden Ansprüche auszuführen, um ein Nutzermodell zu generieren und das generierte Nutzermodell in der Speichereinrichtung abzuspeichern und einem Empfehlungsdienst zur Verfügung zu stellen.

Des Weiteren wird ein Datenträgerprodukt bereit gestellt, mit einem darauf gespeicherten Programmcode, welcher in einen Computer und / oder in ein Computernetzwerk ladbar ist und angepasst ist, ein erfindungsgemäßes Verfahren auszuführen.

## Kurzbeschreibung der Figuren

Die Erfindung wird anhand eines Ausführungsbeispiels und der Zeichnung näher erläutert. In der Zeichnung zeigt:

Fig. 1 einen aus dem Stand der Technik bekannten Ansatz für ein sogenanntes "User-Item Matching";

Fig. 2 ein aus dem Stand der Technik bekanntes "Content Based Filtering" Verfahren;

Fig. 3 ein sogenanntes "Collaborative Filtering" Verfahren, wie es aus dem Stand der Technik bekannt ist;

Fig. 4 eine Abwandlung des aus dem Stand der Technik bekannten "Collaborative Filtering" Verfahrens;

Fig. 5 eine baumförmige (hierarchische) Datenstruktur;

Fig. 6a, 6b zwei baumförmige Datenstrukturen, welche im Sinne der Erfindung die gleiche Aussage haben; und

Fig. 7a, 7b ein Ablaufdiagramm eines erfindungsgemäßen Verfahrens.

## Detaillierte Beschreibung der Erfindung

### Definitionen

- Baumförmige Datenstruktur - Als baumförmige Datenstruktur (im Folgenden BD) wird eine Datenstruktur bezeichnet mit der sich eine Monohierarchie abbilden lässt. Dabei sind in der Datenstruktur Knoten mittels Kanten baumförmig verbunden. Es gibt genau einen Wurzelknoten, der beliebig viele Kinderknoten haben kann. Jeder Kinderknoten kann wiederum beliebig viele Kinderknoten haben.

  Beispiele für BD im Sinne der Erfindung sind vor allem, aber nicht ausschließlich, Verzeichnisstrukturen und/oder Dateisysteme auf einer Festplatte (Ordner und Dateien) oder sogenannte Mind Maps. Handelt es sich bei der BD um ein Dateisystem entsprechen die "Blätter" (das sind die jeweils letzten Knoten eines Pfades in einer BD) Dateien oder Dateiverknüpfungen und alle anderen Knoten entsprechen Verzeichnissen bzw. Ordnern.

  Knoten einer BD enthalten in der Regel ein oder mehrere Elemente. Diese Elemente können unterschiedlichen Typs sein. Übliche Elemente bzw.

Elementtypen sind: Text (im Falle eines Dateisystems wäre der Knotentext der Datei- oder Verzeichnisname), zusätzliche Notizen, Tabellen, Termine, Multimediaobjekte (Musik, Film, Bild), Icons, Formeln, Verweise (in der Regel auf externe Items), Zahlen, und / oder Binärcode (insbesondere falls es sich bei der BD um eine Verzeichnisstruktur handelt und der Knoten eine Datei ist). Ein Verweis kann eine eindeutige URI (Uniform Resource Identifier) sein, z.B. Hyperlink, lokaler Link/Verknüpfung auf eine Datei auf einem Speichermedium (z.B. Festplatte). Ein Verweis kann aber auch eine nicht eindeutige Beschreibung sein, die ein Item identifiziert (z.B. Titel eines Dokumentes, Autorenname, Foto, BibTeX Key, Name eines Ortes oder Produktes).

Jedes der Elemente kann eine Anzahl von Attributen besitzen. So kann Text unterschiedlich formatiert sein, also hinsichtlich z.B. Größe und Farbe unterschiedliche Werte annehmen. Auch Knoten selbst können Attribute besitzen, insbesondere um die Anzeige der Knoten zu formatieren oder dem Knoten bestimmte Funktionen zuzuordnen. Beispielsweise können Knoten mittels Attributen als "eingeklappt" oder "ausgeklappt" dargestellt werden, das heißt für den Nutzer sichtbar oder unsichtbar sein. Genauso wie Knoten können einzelne Elemente eines Knotens sichtbar oder unsichtbar für den Nutzer sein.

Kanten in einer BD sind in der Regel ungerichtet und enthalten üblicherweise keine textuellen Informationen. Kanten können aber auch gerichtet sein.

- Item - Items sind beliebige Objekte, d.h., zum Beispiel Dokumente (Bücher, Webseiten, wissenschaftliche Artikel), Dateien, Werbeanzeigen (in Bild, Text, Ton), Personen, Musikstücke oder Musikalben, Filme, Produkte, geographische Orte, etc. oder deren digitale Repräsentation (d.h. nicht zwangsweise ein physisches Buch, sondern z.B. die digitale Kopie/Repräsentation des Buches in verschiedensten Formaten).

- Nutzer - Ein Nutzer ist eine Person die das erfindungsgemäße System anwendet bzw. nutzt. Ein Nutzer kann auch ein sogenannter Agent, eine Art elektronische Person bzw. ein System, welches das Verhalten einer realen Person simuliert.

- Nutzermodell - Ein Nutzermodell umfasst die Interessen, das Wissen oder andere Informationen über die Person, üblicherweise in maschinenlesbarer

Form. Im Folgenden werden die Begriffe Interessen bzw. Wissen eines Nutzers bzw. Information über den Nutzer synonym verwendet.

- Verbindung zwischen BD und Nutzer - Eine BD steht mit einem Nutzer in Verbindung bzw. ist dem Nutzer zuordenbar, wenn dieser Nutzer z.B. die BD erstellt, editiert, heruntergeladen, oder geöffnet hat oder sich die BD im Besitz des Nutzers befindet oder befand (z.B. auf der Festplatte des Nutzers gespeichert ist bzw. war).

- Kollektion - Eine Kollektion ist die Menge aller BD auf die das erfindungsgemäße System Zugriff hat.

Erfindungsgemäß werden baumförmige Datenstrukturen BD analysiert, die mit dem Nutzer in Verbindung stehen bzw. einem Nutzer zuordenbar sind, um ein Modell des Nutzers, d.h. ein Nutzermodell, zu erstellen. Ein Nutzermodell umfasst insbesondere, aber nicht ausschließlich, Informationen über die Interessen und das Wissen des Nutzers.

Fig. 5 zeigt eine erfindungsgemäße baumförmige Datenstruktur BD. Eine baumförmige Datenstruktur BD umfasst eine Anzahl von Knoten, wobei ein spezieller Knoten den Root-Knoten bzw. den Wurzelknoten repräsentiert. Die anderen Knoten werden als Kinderknoten bezeichnet, wobei die Kinderknoten über Kanten mit dem Wurzelknoten oder mit einem Kinderknoten verbunden sind. Knoten, welche keine Kinderknoten enthalten, werden als "Blätter" bezeichnet. Jeder Knoten kann einen oder mehrere Verweise auf externe Items enthalten. In Fig. 5 besitzt der Knoten 2.i einen solchen Verweis auf ein Item.

Erfindungsgemäß beschreiben der Inhalt einer baumförmige Datenstruktur und gegebenenfalls die Items die aus einer baumförmige Datenstruktur verlinkt werden bzw. die Inhalte der verlinkten Items die Interessen des Nutzers und können zur Generierung eines Nutzermodells verwendet werden. Vereinfacht gesagt bedeutet dies, wenn die Knoten einer baumförmigen Datenstruktur häufig das Wort "Patent" enthalten, kann daraus geschlossen werden, dass der Nutzer der baumförmige Datenstruktur bzw. der Nutzer, dem die baumförmige Datenstruktur zuordenbar ist sich für Patente interessiert oder Kenntnisse auf diesem Gebiet hat.

Der gleiche Schluss kann auch gezogen, wenn das Wort nicht in der baumförmige Datenstruktur selbst vorkommt, aber viele Dokumente (z.B. Patentschriften oder Webseiten) in der baumförmige Datenstruktur verlinkt sind, die das Wort "Patent" enthalten.

**Fig. 7a** und **Fig. 7b** zeigen ein Ablaufdiagramm eines erfindungsgemäßen Verfahrens zum Erzeugen eines Nutzermodells aus zumindest einer baumförmigen Datenstruktur.

In einem ersten Schritt findet eine Vorverarbeitung statt, bei der die baumförmigen Datenstrukturen für eine weitere Verarbeitung angepasst bzw. aufbereitet werden. Der Schritt der Vorverarbeitung ist ein optionaler Schritt und muss nicht notwendigerweise durchgeführt werden, etwa wenn die baumförmigen Datenstrukturen bereits das für die weitere Verarbeitung notwendige Format aufweisen.

Die Vorverarbeitung kann ein Konvertieren der baumförmigen Datenstrukturen in ein für das System lesbares Format umfassen. Ferner kann die Vorverarbeitung ein Löschen von Knoten aus den baumförmigen Datenstrukturen umfassen, d.h., es können bestimmte Knoten gelöscht werden, die für das Erzeugen eines Nutzermodells nicht relevant sind. Löschen eines Knoten heißt, dass dieser aus der baumförmige Datenstruktur entfernt wird und die Kinderknoten des zu löschenden Knotens entweder ebenfalls entfernt werden oder die Kinderknoten dem Elternknoten des zu löschenden Knotens zugeordnet werden. Ein Knoten kann etwa gelöscht werden, wenn der Knoten eines oder mehrere der folgenden Kriterien erfüllt:

- Der Knoten ist leer;
- Der Knoten enthält ein bestimmtes Element (nicht), wie z.B. Text oder Verweis;
- Der Knoten besitzt ein bestimmtes Attribut (nicht);
- Der Knoten oder Elemente des Knotens stehen nicht direkt mit dem Nutzer in Verbindung. Dies kann der Fall sein, wenn ein Knoten nicht vom Nutzer selbst erzeugt wurde (z.B. bei Knoten einer Mind Map die auf eine Datei verlinken und wo der Text des Knotens gleich dem Dateinamen der verlinkten Datei ist kann angenommen werden, dass der Knoten automatisch, z.B. durch "Drag & Drop" entstanden ist, also der Text des Knotens nicht vom Nutzer erzeugt wurde, und deshalb keine oder nur eine geringe Aussagekraft besitzt).

Selbstverständlich können beim Löschen eines Knotens auch noch weitere Kriterien berücksichtigt werden.

Nach der (optionalen) Vorverarbeitung wird in einem nächsten Schritt basierend auf der baumförmigen Datenstruktur ein Nutzermodell erzeugt. Hierbei werden die Knoten einer baumförmigen Datenstruktur bzw. die Elemente der Knoten analysiert, um die Interessen etc. des Nutzers zu identifizieren und in einem Nutzermodell zu speichern, welches dem Nutzer zugeordnet wird. Dies geschieht in den folgenden Schritten:

A) Gewichten der Knoten

Dem Gewichten der Knoten liegt die Annahme zugrunde, dass einige Knoten bzw. ihre Elemente aussagekräftiger sind um die Interessen des Nutzers zu beschreiben als andere Knoten bzw. ihre Elemente. Bei der Gewichtung der Knoten können zwei Teilgewichte berechnet werden. Es ist aber auch möglich nur eines der beiden Teilgewichte zu berechnen und dieses eine Teilgewicht als Knotengewicht eines Knotens zu betrachten. Die beiden Teilgewichte umfassen die statische Knotengewichtung und die dynamische Knotengewichtung. Selbstverständlich können auch noch weitere hier nicht genannte Teilgewichte berechnet werden. Die Kombination der berechneten Teilgewichte ergibt das Knotengewicht eines Knotens.

Bei der statischen Knotengewichtung können folgende Kriterien berücksichtigt werden:

- Anzahl der Kinderknoten: Ein Knoten wird abhängig von der Anzahl der Kinderknoten gewichtet, z.B. je mehr Kinder der Knoten hat desto mehr Gewicht erhält dieser Knoten.
- Anzahl der Geschwisterknoten: Geschwisterknoten eines Knotens sind jene Knoten, die denselben Elternknoten wie der betrachtete Knoten haben. Hier wird der Knoten abhängig von der Anzahl der Geschwisterknoten des Knotens gewichtet, z.B. je mehr Geschwister ein Knoten hat, desto weniger Gewicht bekommt er.
- Knoten-Tiefe: Je weiter oben (also je näher zum Wurzelknoten) in einer baumförmige Datenstruktur ein Knoten ist, desto mehr Gewicht erhält er. Der Wurzelknoten bekommt also viel Gewicht, die Blattknoten weniger.

- Sichtbarkeit des Knotens: Sichtbare Knoten erhalten mehr Gewicht als unsichtbare Knoten.
- Attribute: Ist der Knoten durch ein bestimmtes Attribut hervorgehoben, z.B. durch farbige Markierung oder Unterstreichung, so erhält er mehr Gewicht. Ist er durch bestimmte Attribute abgeschwächt, z.B. indem er ausgegraut oder durchgestrichen wurde, erhält er weniger Gewicht.

Bei der statischen Knotengewichtung wird die baumförmige Datenstruktur nur zu einem bestimmten Zeitpunkt betrachtet. Erfindungsgemäß können die Knoten aber auch dynamisch gewichtet, d.h., Veränderungen und Nutzungsintensivität der baumförmigen Datenstruktur über die Zeit können in die Knotengewichtung einfließen. Wenn beispielsweise ein Knoten intensiver genutzt wird, bzw. in der Vergangenheit intensiver genutzt wurde als andere Knoten, so kann dieser Knoten ein höheres Gewicht erhalten als die anderen Knoten. Die Gewichtung kann sich unter anderem ergeben aus:

- Alter des Knotens: Ältere Knoten können mehr oder weniger Gewicht erhalten als jüngere Knoten. Vorzugsweise erhalten jüngere Knoten ein höheres Gewicht. Es kann auch ein Schwellenwert vorgesehen sein, beispielsweise der Art "Knoten die mindestens 12 Stunden und maximal 5 Tage alt sind". Knoten, die das Schwellenwertkriterium erfüllen erhalten ein höheres Gewicht.
- Zeitpunkt der letzten Bearbeitung (z.B. Editierung): Knoten die kürzlich editiert wurden erhalten ein höheres Gewicht.
- Anzahl der Bearbeitungen: Vorzugsweise kann ein Knoten, der öfter editiert wurde als andere Knoten ein höheres Gewicht erhalten.
- Anzahl der Verschiebungen: Je öfter ein Knoten verschoben (ausgeschnitten und wieder eingefügt) wurde, desto mehr Gewicht erhält er.
- Anzahl der Markierungen: Je öfter ein Knoten ausgewählt/markiert wurde desto mehr Gewicht erhält er.
- Sichtbarkeitsdauer: Je länger ein Knoten sichtbar war, desto mehr Gewicht erhält er.
- Anzahl der Sichtbarkeiten: Je öfter ein Knoten unsichtbar und wieder sichtbar gemacht wurde, das heißt ein- und ausgeklappt wurde, desto stärker ist sein Gewicht.
- Anzahl der verfolgten Verweise: Je öfter ein Verweis eines Knoten geöffnet wurde, desto größer ist das Gewicht des Knoten.

Jede der vorstehend genannten (dynamischen) Gewichtungen kann mittels eines Zeitparameters geschwächt oder verstärkt werden. Hierzu ein Beispiel: Ein Knoten wird doppelt so stark gewichtet wenn er wenigstens zwei Mal editiert wurde. Liegt die letzte Editierung aber schon länger als X Wochen zurück, wird die Gewichtung durch einen dämpfenden Zeitparameter nur 1,5 Mal so stark gewichtet.

Zusätzlich zur vorstehend genannten statischen und/oder dynamischen Knotengewichtung kann eine Vererbung von Gewichten vorgesehen sein. Die Vererbung wird vorzugsweise dann durchgeführt, nachdem die statische bzw. dynamische Gewichtung durchgeführt worden ist. Bei der Vererbung von Gewichten können Knoten Gewichtungen von ihren umliegenden Knoten "erben". Hat etwa ein Elternknoten ein sehr hohes Gewicht (weil er z.B. oft ausgewählt wurde), kann auch der Kindsknoten ein höheres Gewicht bekommen als wenn er nur für sich betrachtet würde. Bevorzugt bekommen alle Kinderknoten und deren Knoten, alle Geschwisterknoten und alle Elternknoten und deren Eltern bis zur Wurzel ein höheres Gewicht, wobei das zusätzliche Gewicht schwächer wird, je weiter entfernt von dem vererbenden Knoten sich der erbende Knoten befindet. Zudem kann vorgesehen sein, dass nur Knoten, die einen Schwellenwert (z.B. Gewichtung fünf Mal größer als normal) überschreiten das Gewicht an umliegende Knoten vererben können. Gehören mehrere Knoten einer baumförmige Datenstruktur einer bestimmten "Gruppe" an, kann die Gewichtung der einzelnen Knoten der Gruppe aneinander angeglichen bzw. vererbt werden. Gruppen können visuell in der baumförmige Datenstruktur erkennbar sein oder sich durch bestimmte Attribute bzw. Elementtypen auszeichnen. Beispiel: Eine baumförmige Datenstruktur enthält einige Knoten, die Verweise haben. Alle diese Knoten sind der Gruppe "Verweis-Knoten" zugeordnet. Obwohl nur 95% dieser Knoten eine sehr hohe Gewichtung haben, vergibt das System an alle Knoten (also auch an die restlichen 5%) eine sehr hohe Gewichtung. Die schwachen Knoten einer Gruppe erben quasi von ihren anderen Gruppenknoten.

B) Identifizieren der Elemente in den Knoten

In einem weiteren Schritt werden die Elemente in den Knoten identifiziert und ggf. einer Vorverarbeitung zugeführt. Hierbei werden zunächst die in jedem Knoten enthaltenden Elemente und deren Attribute identifiziert bzw. ermittelt.

Handelt es sich bei dem ermittelten Element um ein Textelement, so wird dieser Text weiter zerlegt und zwar in Token bzw. Terme (Begriffe). Häufig kann als Term ein einzelnes Wort gelten, manchmal aber auch zusammengesetzte Wörter wie "Mind Map". Im Folgenden gilt jeder Term als eigenständiges Element vom Typ Text.

Die Terme können weiter verarbeitet werden. Hierfür können aus dem Stand der Technik bekannte Verfahren, etwa aus dem Bereich Information Retrieval herangezogen werden. Beispiele für solche Verfahren sind etwa

- Stemming: Wörter werden auf ihre Stämme reduziert. Beispielsweise würde das Wort "Stämme" auf "Stamm" gestemmt bzw. reduziert.
- Stop Word Removal: Sehr häufig vorkommende Wörter mit wenig Aussagekraft (beispielsweise der, die, das, wo, wer, weshalb, schon, so, darum, …) werden entfernt.
- Latent Semantic Indexing (LSI): Mit Latent Semantic Indexing werden Synonyme von Wörtern zusammengefasst bzw. berücksichtigt.
- Translation: die Worte werden in eine Referenzsprache, z.B. Englisch, übersetzt.
- Spelling Correction: Rechtschreibfehler werden erkannt und korrigiert oder gelöscht.

Bei dem ermittelten Element kann es sich auch um einen Verweis handeln. Verweise können ebenfalls vorverarbeitet werden, indem beispielsweise für jeden Verweis die URI (Uniform Ressource Identifier) und/oder die Sonderzeichen auf ein einheitliches Format konvertiert werden oder falls es sich um keinen eindeutigen Verweis handelt (z.B. lediglich der Titel eines Dokumentes), versucht wird einen eindeutigen Identifikator zu finden (im Falle eines Dokumentes beispielsweise die ISBN).

C) Gewichten der Elemente

Ähnlich wie die Knoten kann auch jedes Element eines Knotens gewichtet werden. Besonders Text- und Verweiselemente sind wichtig für die Erstellung des erfindungsgemäßen Nutzermodells. Vorzugsweise erhält jedes Element zunächst eine vorbestimmte Gewichtung (Initialgewichtung), etwa die Gewichtung 1 oder die Gewichtung seines zugehörigen Knotens. Dies kann etwa in einem Initialisierungsschritt erfolgen, bei dem alle Elemente mit einer Initialgewichtung versehen werden.

Die Initialgewichtung eines Elementes kann verstärkt bzw. geschwächt werden, vorzugsweise basierend auf einen oder mehreren der folgenden Faktoren:

- Element-Typ: Elemente bestimmter Typen können unterschiedliche Basisgewichtungen erhalten. Beispielsweise kann ein Text-Element welches den allgemeinen Knotentext darstellt eine höhere Gewichtung erhalten als ein Text-Element welches eine zusätzliche Notiz darstellt.
- Attribute: Abhängig von den Attributen können Elemente eine stärkere oder schwächere Gewichtung bekommen. Beispielsweise kann ein Text-Element, welches fett formatiert ist stärker gewichtet werden als ein Text-Element ohne Formatierung.
- Element Frequenz: Je öfter ein Element in der baumförmigen Datenstruktur vorkommt, desto stärker kann seine Gewichtung sein.
- BD Frequenz: In je weniger baumförmigen Datenstrukturen der gesamten Kollektion ein Element vorkommt, desto stärker wird es gewichtet. Dies beruht auf der Annahme, dass ein Element, welches nur wenige Male in allen baumförmigen Datenstrukturen vorkommt, aussagekräftiger ist als ein Element, das in fast jeder baumförmigen Datenstruktur vorkommt. Enthält beispielsweise in einer Kollektion von 100 baumförmigen Datenstrukturen nur eine einzige baumförmige Datenstruktur den Term "Baum", dann würde dieser Term stärker gewichtet in Bezug auf die baumförmige Datenstruktur als wenn 90 weitere baumförmige Datenstrukturen diesen Term ebenfalls enthalten.
- Kollektionsfrequenz: Je seltener das Element in der Gesamtmenge aller Elemente der gesamten Kollektion vorkommt, desto stärker wird es gewichtet. Dies ist sehr ähnlich zur BD Frequenz, mit dem Unterschied, dass bei der BD Frequenz die Anzahl der baumförmigen Datenstrukturen gezählt wird in denen das Element vorkommt und bei er Kollektionsfrequenz die Gesamtanzahl der Elemente selbst.
- BD Größe: Je größer die baumförmige Datenstruktur, desto weniger stark wird das Element gewichtet. Dies unterliegt der Annahme, dass große baumförmige Datenstrukturen tendenziell mehr Elemente enthalten aber nicht gegenüber kleinen baumförmige Datenstrukturen bevorzugt werden sollen. Die Größe einer baumförmigen Datenstruktur kann angegeben werden durch die Anzahl der Knoten einer baumförmigen Datenstruktur oder durch die Anzahl der Elemente in einer baumförmigen Datenstruktur.
- Position im Knoten: Elemente die vorne im Knoten stehen werden anders gewichtet als Elemente weiter hinten im Knoten. Enthält ein Knoten beispielsweise 100 Terme, dann kann vorgesehen sein, dass nur die ersten 10

Terme berücksichtigt werden. Ferner kann vorgesehen sein, dass die weiteren Terme (z.B. die nächsten 10 Terme) mit weniger Gewicht berücksichtigt werden.

- Sprache (falls der Knoten ein Text-Elemente enthält): Im Gegensatz zu Dokumenten, wie Webseiten, kommt es bei baumförmigen Datenstrukturen häufig vor, dass Terme in verschiedenen Sprachen enthalten sind. Die Elemente eines Knotens können abhängig von der Sprache unterschiedlich stark gewichtet werden. Das heißt auch, dass wenn z.B. der Text eines Knotens in einer bestimmten Sprache ist, die anderen Elemente des Knotens (zum Beispiel ein Verweis) weniger oder mehr gewichtet werden.
- Knotenlänge: Elemente werden abhängig von der Knotenlänge gewichtet. Je weniger Elemente ein Knoten enthält, desto stärker können seine Elemente gewichtet werden.
- Abstand zu gleichartigen Elementen: Je weniger gleichartige Elemente es in der Nähe eines Knotens gibt zu dem das Element gehört, desto mehr Gewicht bekommt das Element. Zum Beispiel: Hat ein Knoten einen Verweis auf ein Item und die umliegenden Knoten (z.B. alle Kinder, Geschwister und Elternknoten) enthalten keine Verweise, dann könnte dieser Verweis ein besonders hohes Gewicht bekommen, da die Vermutung nahe liegt, dass sich der Verweis auch auf die umliegenden Knoten bezieht. Haben hingegen Geschwisterknoten ebenfalls Verweise, bekommt dieser Verweis kein besonders hohes Gewicht.
- Element-Wiederholung: Baumförmige Datenstrukturen können sehr benutzerspezifisch erstellt werden. Beispielsweise kann es vorkommen, dass ein Nutzer Elemente in den Knoten oft wiederholt, ein anderer Nutzer aber nicht. Hier kann es vorteilhaft sein, das Gewicht eines Elementes zu verringern, je öfter einer der Elternknoten (bis hoch zum Wurzelknoten) oder Geschwisterknoten dieses Element bereits enthält. Fig. 6a und Fig. 6b verdeutlichen diesen Fall. **Fig. 6a** und **Fig. 6b** zeigen jeweils eine baumförmige Datenstruktur mit der gleichen Aussage von zwei Nutzern, wobei die baumförmigen Datenstrukturen dennoch unterschiedlich aussehen. In Fig. 6a wiederholt sich der Term "Recommender" mehrfach, in Fig. 6b hingegen nicht. Trotzdem wäre der Term "Recommender" für beide baumförmigen Datenstrukturen bzw. Nutzer gleichermaßen zutreffend und sollte gleichermaßen gewichtet werden.

Die Gewichtung der Elemente kann auch mit den gleichen Verfahren stattfinden mit dem die Knoten gewichtet werden. Beispielsweise können ältere Elemente schwächer gewichtet werden als neuere und auch bei den Elementgewichtungen kann eine Vererbung stattfinden.

D) Speichern des Nutzermodells

In einer vorteilhaften Ausgestaltung der Erfindung kann es vorteilhaft sein, das generierte Nutzermodell zu speichern, um es etwa einem Empfehlungsdienst zur Verfügung zu stellen. Alternativ kann ein Nutzermodell aber auch auf Anforderung erstellt werden, ohne es zu speichern.

Erfindungsgemäß können mindestens zwei verschiedene Ansätze genutzt werden, um ein Nutzermodell zu speichern. Die beiden hier gezeigten Ansätze sind das Typ-Neutrale Speichern und das Typ-Abhängige Speichern eines Nutzermodells.

Bei der Typ-Neutralen Speicherung werden alle Elemente mit ihrer Gewichtung gespeichert. Das heißt, Terme, Links/Verweise, Bilder, etc. werden alle gemeinsam in dem Model gespeichert. In einer konkreten Ausgestaltung des erfindungsgemäßen Verfahrens kann hierfür das eingangs beschriebene Vector Space Model genutzt werden, welches durch die Erfindung so erweitert wird, dass nicht nur Terme mit einer Gewichtung gespeichert werden können, sondern vielmehr beliebige Elemente verschiedenen Typs mit ihrer Gewichtung und ihrem Typ.

Bei der Typ-Abhängigen Speicherung kann für jeden Elementtyp ein separates Nutzermodell erzeugt werden, welche zusammen ein Gesamtnutzermodell bilden. Ein Nutzermodell umfasst dann beispielsweise ein Text-Modell und ein Verweis-Modell. Hierfür können Standardverfahren aus dem Information Retrieval Bereich bzw. User Modelling Bereich genutzt werden. Ein Standardmodell für ein Text-basiertes Modell wäre beispielsweise wieder das genannte Vector Space Model in dem die einzelnen Terme entsprechend des oben beschriebenen Verfahrens gewichtet sind. Verweise können auch in anderen Modellen gespeichert werden, die beispielsweise auch die Reihenfolge der Elemente in der baumförmigen Datenstruktur berücksichtigen.

E) Weitere Schritte

Die nachfolgenden Schritte können optional zu den vorstehend genannten Schritten ausgeführt werden.

Hat ein Nutzer Beziehungen zu mehreren baumförmigen Datenstrukturen, kann für jede baumförmige Datenstruktur ein (bzw. mehrere) Modelle erzeugt werden, wie vorstehend beschrieben, und die verschiedenen Modelle am Ende zu einem Gesamtmodell zusammen gefügt werden. Hierbei können unterschiedliche baumförmige Datenstrukturen mit unterschiedlicher Gewichtung versehen werden. Die Gewichtung erfolgt nach ähnlichen Prinzipien wie die Gewichtung der Knoten oder der Elemente. Beispielsweise kann eine neuere baumförmige Datenstruktur oder baumförmige Datenstrukturen, die häufiger geöffnet oder editiert wurden, stärker gewichtet werden.

Entsteht eine neue Beziehung zwischen einer baumförmigen Datenstruktur und einem Nutzer zu dem bereits ein Nutzermodell existiert, kann das bestehende Nutzermodell um die Elemente der neuen baumförmigen Datenstruktur erweitert werden.

Enthält eine baumförmige Datenstruktur Verweise auf Items, können diese Items ebenfalls für die Generierung eines Nutzermodells genutzt werden. Das heißt, die Elemente in dem verlinkten Item werden in das Nutzermodell eingefügt und zwar auf vergleichbare Weise wie Elemente der baumförmige Datenstruktur selbst. Diese Items können mit einer niedrigeren Gewichtung versehen werden. Ist das verlinkte Item eine baumförmige Datenstruktur, werden deren Elemente mit dem vorstehend beschriebenen Verfahren gewichtet. Ist das verlinkte Element z.B. eine Webseite, dann kann die Gewichtung mit Standardverfahren, wie dem TF-IDF durchgeführt werden.

Die vorstehend genannten Typ-Neutralen und Typ-Abhängigen Modelle können in Untermodelle unterteilt sein, beispielsweise in

- Modelle für Kurzzeitinteressen: Dieses Modell würde beispielsweise nur Daten aus einer Session bzw. der zuletzt editierten baumförmigen Datenstruktur enthalten (oder Daten der baumförmigen Datenstruktur, die in einem bestimmten Zeitraum editiert wurden).
- Modelle für Langzeitinteressen: Dieses Modell würde Interessen basierend auf allen oder zumindest mehreren baumförmigen Datenstrukturen enthalten.
- Modelle für verschiedene Interessen: Es ist denkbar, dass Nutzer verschiedene baumförmige Datenstrukturen erstellen für z.B. verschiedene Projekte. Das heißt, eine baumförmige Datenstruktur (oder auch mehrere) werden genutzt für Projekt A und eine andere baumförmige Datenstruktur (oder auch mehrere) für ein anderes Projekt B. Erfindungsgemäß können baumförmige

Datenstrukturen, die sehr unterschiedlich sind für die Erstellung unterschiedlicher Modelle genutzt werden (die ggf. auch wieder unterteilt werden in Langzeit und Kurzzeitinteressen). Die Identifizierung von zusammengehörigen baumförmigen Datenstrukturen kann folgendermaßen erfolgen:

- Inhaltliche Analyse: Hier werden mit den Verfahren zur Gewichtung von Termen oder Verweisen ähnliche baumförmige Datenstrukturen ermittelt. Unterschreiten die baumförmigen Datenstrukturen einen bestimmten Ähnlichkeits-Schwellenwert, werden sie für unterschiedliche Nutzermodelle genutzt.
- Zeitliche Analyse: Baumförmige Datenstrukturen, die selten oder nie zur gleichen Zeit genutzt, geöffnet, etc. werden, werden für unterschiedliche Nutzermodelle genutzt.

Baumförmige Datenstrukturen können für unterschiedliche Arten von Anwendungen genutzt werden, zum Beispiel Dateiverwaltung, Brainstorming, Dokumentenmanagement, Projektplanung etc. Die Art der Anwendung wird im Nutzermodell vermerkt. Wenn ein Nutzer verschiedene baumförmige Datenstrukturen für verschiedene Arten von Anwendungen erstellt, werden wieder jeweils verschiedene Nutzermodelle erstellt. Die Art der Anwendung kann wie folgt festgestellt werden:

- Über die Anwendung, mit der die baumförmige Datenstruktur erstellt wurde. Beispielsweise kann pauschal angenommen werden, dass eine baumförmige Datenstruktur, mit dem Windows Explorer erstellt worden ist zur Dateiverwaltung dient.
- Durch manuelle Angabe des Nutzers: Der Nutzer kann in der Anwendung zum Erstellen der baumförmigen Datenstruktur angeben für welchen Zweck er die baumförmige Datenstruktur erstellen will (z.B. Brainstorming, Projektplanung, etc.).
- Automatische Analyse: Das System analysiert die baumförmige Datenstruktur und schließt, z.B. an Hand ihres Aufbaus, ihrer Nutzung oder ihres Quellformates, automatisch auf die Art der Anwendung.

Bei der automatischen Analyse können folgende Regeln angewandt werden:

- Enthalten baumförmige Datenstrukturen viele Verweise, ist die primäre Anwendung Dateiverwaltung, Webseitenverwaltung bzw. Dokumentenverwaltung.

- Werden erst sehr schnell sehr viele Knoten in der baumförmigen Datenstruktur erstellt, diese dann verschoben und editiert und die baumförmige Datenstruktur danach nie oder nur selten geöffnet, wurde sie für Brainstorming erstellt.
- Wächst die baumförmige Datenstruktur langsam und kontinuierlich ist es keine baumförmige Datenstruktur für Brainstorming.

Bei der automatischen Analyse können Faktoren wie Wachstumsrate, Größe, Nutzungsdauer, Art der Nutzung, Anwendung zum Erstellen der baumförmigen Datenstruktur und/oder weitere Faktoren eine Rolle spielen.

Mit dem vorstehend beschriebenen erfindungsgemäßen Verfahren sind für eine Anzahl von Nutzer jeweils ein oder auch mehrere Nutzermodelle erzeugt worden. Diese Nutzermodelle werden erfindungsgemäß genutzt, um einem Nutzer Empfehlungen für Items zu geben. Das heißt, basierend auf den Nutzermodellen können Items identifiziert werden, die der Nutzer wahrscheinlich als interessant/relevant empfindet. Um Items zu empfehlen, wird gemäß der Erfindung ein Verfahren vorgeschlagen, welches sowohl Item-Modelle als auch Nutzermodelle als gleich ansieht. Im Folgenden werden Items und Nutzer zusammengefasst als Objekt bezeichnet, wobei jedes Objekt einen Typ haben kann (Typ = Nutzer; Typ = Webseite; Typ = Email; Typ = Wissenschaftlicher Artikel; etc.).

Das erfindungsgemäße Verfahren zum Vorschlagen von Objekten kann zumindest die nachfolgend genannten Schritte umfassen:

Die Nutzermodelle wurden bereits mit vorstehen genannten erfindungsgemäßen Verfahren erstellt und gespeichert. Nun wird von allen Items die potentiell empfohlen werden können, ein Item-Modell mit aus dem Stand der Technik bekannten Verfahren erzeugt. Handelt es sich bei dem Item beispielsweise um eine Webseite könnten Terme mittels TF-IDF gewichtet und als Vector Space Model gespeichert werden. Handelt es sich bei dem Item um eine wissenschaftliche Arbeit könnte ebenfalls TF-IDF genutzt werden, aber auch andere Verfahren, wie etwa Citation Proximity Analysis um das Item abzubilden.

Es kann auch das vorstehend genannte Typ-Neutrale Verfahren genutzt werden, um ein entsprechendes Modell von Items zu erstellen die, wie baumförmige Datenstrukturen, mehrere Elementtypen enthalten. Dies trifft zum Beispiel auf wissenschaftliche Artikel zu. Eine wissenschaftliche Arbeit enthält in der Regel Text und Verweise (Referenzen auf andere wissenschaftlichen Arbeiten), vergleichbar zu einer baumförmigen Datenstruktur, die ebenfalls Text und Verweise (z.B. auf Dateien) enthält. Darum können beide relativ leicht mit kompatiblen Modelltypen abgebildet werden. Wichtig ist, dass erfindungsgemäß alle Objekte in dem gleichen bzw. einem kompatiblen Modell abgebildet werden, um diese später miteinander vergleichen zu können.

Nachdem von allen Objekten Modelle erstellt und gespeichert wurden, werden diese auf Ähnlichkeit verglichen. Enthalten die Objektmodelle mehrere Untermodelle (beispielsweise für Kurz- und Langzeitinteressen oder für verschiedene Elementtypen), wird jedes dieser Untermodelle mit den anderen Objektmodellen verglichen. Der Vergleich kann mit Standardverfahren stattfinden, wie Cosine für Vergleiche im Vector Space Model oder Ähnlichkeitsmaßen wie Greedy Citation Tiling für verweisbasierte Modelle. Das heißt letztlich, dass durch das Objekt-Objekt Matching die Ähnlichkeit der Nutzer zueinander, die Ähnlichkeit der Items zueinander und die Ähnlichkeit der Items und Nutzer zueinander in einem Schritt berechnet werden. Dies hat den Vorteil, dass einem Nutzer später sehr flexible Empfehlungen gegeben werden können.

Soll nun einem Nutzer eine Empfehlung gegeben werden, werden basierend auf seinem Objekt bzw. Nutzermodell:

- alle Objekte empfohlen, die einen bestimmten Typ haben (z.B. Webseite bzw. allgemein „Item“) oder ungleich eines bestimmten Typs sind (z.B. Nutzer) und einen gewissen Ähnlichkeitswert überschreiten;
- alle Objekte empfohlen, auf die in einem der im vorherigen Schritt bestimmten Objekte mit hohem Gewicht verwiesen wird. Wurde z.B. im vorherigen Schritt ein zum Nutzermodell ähnliches Objekt vom Typ Webseite bestimmt, werden dem Nutzer die Webseiten empfohlen die häufig auf der ähnlichen Webseite verlinkt sind. Oder, wurde ein ähnliches Objekt vom Typ "Nutzer" bestimmt werden die Items empfohlen, auf die häufig in dem Nutzermodell des ähnlichen Nutzers verwiesen wird bzw. die in enger Verbindung zu dem Nutzer stehen; und/oder
- alle Objekte empfohlen, die ähnlich zu den Objekten sind, auf die im Nutzermodell des Nutzers verwiesen wird. Das heißt: Hat ein Nutzer in seinem Nutzermodell einen Verweis auf eine Webseite X, werden die Objekte empfohlen die ähnlich zu dieser Webseite X sind.

Zusätzlich bzw. alternativ zu dem vorstehend beschriebenen Ansatz kann ein Verfahren basierend auf Machine Learning genutzt werden. Die Verweise eines

Nutzermodells können mit Machine Learning Verfahren genutzt werden, um die Präferenzen von Nutzern zu lernen. Hierbei gilt jeder Verweis auf ein Item als positive Assoziation, welche das System lernt und darauf basierend Empfehlungen für neue Items gibt.

In einer speziellen Ausführung des erfindungsgemäßen Verfahrens kann dieses verwendet werden, um Empfehlungen für Fördergelder/Förderprogramme zu geben. Hierzu wird zunächst ein aus einer baumförmigen Datenstruktur ein Nutzermodell erzeugt, wie vorstehend beschrieben. Das Förderprogramm selbst wird als Item betrachtet. Das Item wiederum wird repräsentiert durch einen Text welcher das Förderprogramm beschreibt. Dieser Text kann eine Webseite sein, eine Broschüre im PDF Format, Social Tags, etc. Enthält beispielsweise ein Nutzermodell den stark gewichteten Term "Recommender Systems" und gibt es ein Förderprogramm, dessen Webseite ebenfalls diesen Term oft beinhaltet, würde dieses Förderprogramm dem Nutzer empfohlen.

Durch die erfindungsgemäßen Nutzermodelle und Empfehlungen kann der Nutzen vieler Softwareprogramme für den Anwender gesteigert werden, da sie interessante Empfehlungen erhalten.

## K.2 Ansprüche

Computer-implementiertes Verfahren zum Erzeugen eines Nutzermodells, insbesondere für einen Empfehlungsdienst, aus zumindest einer baumförmigen Datenstruktur, wobei das Nutzermodell Informationen über einen Nutzer umfasst, wobei die zumindest eine baumförmige Datenstruktur dem Nutzer zuordenbar ist, wobei die baumförmige Datenstruktur einen Wurzelknoten und eine Anzahl von Kinderknoten umfasst, welche über Kanten mit dem Wurzelknoten oder mit einem Kinderknoten verbunden sind, wobei zumindest einem Knoten zumindest ein Element zugeordnet ist, und wobei

- die den Knoten zugeordneten Elemente ermittelt werden, wobei die Elemente einen Inhalt des jeweiligen Knoten repräsentieren,
- die ermittelten Elemente gewichtet werden und jedem Element eine Elementgewichtung zugeordnet wird, und
- ein Nutzermodell generiert wird, wobei das generierte Nutzermodell die ermittelten Elemente und die dem jeweiligen Element zugeordnete Elementgewichtung umfasst.

1. Verfahren nach Anspruch 1, wobei die Knoten der baumförmigen Datenstruktur gewichtet werden und jedem Knoten eine Knotengewichtung zugeordnet wird.

2. Verfahren nach einem der vorhergehenden Ansprüche, wobei in einem Initialisierungsschritt jedem Element eine vorbestimmte Elementgewichtung oder die Knotengewichtung des zugeordneten Knotens zugeordnet wird.

3. Verfahren nach einem der vorhergehenden Ansprüche, wobei das Verfahren ferner einen Vorverarbeitungsschritt umfasst, bei dem
   - Knoten, denen keine Elemente zugeordnet sind, gelöscht werden, und/oder
   - Knoten gelöscht werden, denen ein vorbestimmtes Element zugeordnet oder nicht zugeordnet ist, und/oder
   - Knoten gelöscht werden, welche vorbestimmte Attribute aufweisen oder nicht aufweisen, und/oder
   - Knoten und/oder Elemente der Knoten gelöscht werden, welche nicht direkt dem Nutzer zugeordnet sind.

4. Verfahren nach einem der vorhergehenden Ansprüche, wobei das Gewichten der Knoten eine statische Knotengewichtung und/oder eine dynamische Knotengewichtung umfasst, wobei
   - bei der statischen Knotengewichtung die Anzahl der dem jeweiligen Knoten zugeordneten Kinderknoten, die Anzahl der jeweiligen Geschwisterknoten, die Tiefe des jeweiligen Knotens in der baumförmigen Datenstruktur, die Sichtbarkeit des Knotens, oder eine Kombination hiervon berücksichtigt werden, und
   - bei der dynamischen Knotengewichtung für jeden Knoten das Alter, der Zeitpunkt der letzen Änderung, die Anzahl der Änderungen, die Anzahl der Verschiebungen innerhalb der baumförmigen Datenstruktur, die Anzahl der Markierungen, die Sichtbarkeit des Knotens, ein Dämpfungsfaktor, oder eine Kombination hiervon berücksichtigt werden.

5. Verfahren nach einem der vorhergehenden Ansprüche, wobei das Ermitteln der den Knoten zugeordneten Elemente ein Vorverarbeiten der ermittelten Elemente umfasst, wobei beim Vorverarbeiten der Elemente Text in Token und/oder Terme zerlegt wird, sofern das Element ein Textelement ist, und/oder Verweise verarbeitet werden, sofern das Element ein Verweiselement ist.

6. Verfahren nach einem der vorhergehenden Ansprüche, wobei die in dem Initialisierungsschritt den Elementen zugeordneten Elementgewichtungen angepasst werden, wobei beim Anpassen der jeweiligen Elementgewichtung der Elementtyp, Attributsausprägungen der von dem Element zugeordneten Attribute, eine Häufigkeit des Elements innerhalb der baumförmigen Datenstruktur, die Anzahl der baumförmigen Datenstrukturen in einer Kollektion von baumförmigen Datenstrukturen in denen das Element vorkommt, eine Häufigkeit des Elements innerhalb einer Kollektion von baumförmigen Datenstrukturen, die Größe der baumförmigen Datenstruktur im Verhältnis zu anderen baumförmigen Datenstrukturen in einer Kollektion von baumförmigen Datenstrukturen, die Position des Elementes innerhalb des Knotens, die Sprache des Elementes, die Anzahl der Elemente innerhalb des Knotens, der Abstand des Elementes zu gleichartigen Elementen anderer Knoten, Häufigkeit des Elementes in dem Pfad zwischen dem Knoten und dem Wurzelknoten, das Alter des Elements, der Zeitpunkt der letzen Änderung, die Anzahl der Änderungen, die Anzahl der Markierungen, die Sichtbarkeit des Elements, ein Dämpfungsfaktor, oder eine Kombination hiervon berücksichtigt werden.

Verfahren nach einem der Ansprüche 5 bis 7, wobei bei der statischen Knotengewichtung und/oder bei der dynamischen Knotengewichtung oder nach der statischen Knotengewichtung und/oder nach der dynamischen Knotengewichtung und/oder bei oder nach der Elementgewichtung die Vererbung des Knotengewichts bzw. des Elementgewichts berücksichtigt werden.

7. Verfahren nach einem der vorhergehenden Ansprüche, wobei das generierte Nutzermodell in einer Speichereinrichtung gespeichert wird, um dem Empfehlungsdienst zur Verfügung gestellt zu werden.

8. Verfahren nach Anspruch 9, wobei alle Elemente zusammen mit den jeweiligen Elementgewichtungen als Nutzermodell gespeichert werden, oder wobei für jeden Elementtyp ein eigenes Nutzermodell gespeichert wird, wobei die Nutzermodelle der verschiedenen Elementtypen ein Gesamtnutzermodell bilden.

9. Verfahren nach einem der vorhergehenden Ansprüche, wobei bei mehreren dem Nutzer zuordenbaren baumförmigen Datenstrukturen für jede baumförmige Datenstruktur eine Anzahl von Nutzermodellen generiert wird, welche zusammen ein dem Nutzer zugeordnetes Gesamtnutzermodell bilden.

10. Verfahren nach Anspruch 11, wobei jeder baumförmigen Datenstruktur eine Baumgewichtung zugeordnet wird.

11. Verfahren nach Anspruch 11 oder 12, wobei für eine neue dem Nutzer zuordenbare baumförmige Datenstruktur das dem Nutzer zugeordnete Nutzermodell angepasst wird.

12. Verfahren nach einem der vorhergehenden Ansprüche, wobei von der baumförmigen Datenstruktur referenzierte Elemente in das Nutzermodell eingefügt werden und wie Elemente der baumförmigen Datenstruktur behandelt werden.

13. Verfahren nach einem der vorhergehenden Ansprüche, wobei einem generierten Nutzermodell eine Information über den Nutzermodelltyp zugeordnet wird.

14. Verfahren nach einem der vorhergehenden Ansprüche, wobei das Verfahren ferner ein Auswählen von Objekten anhand vorbestimmter Auswahlkriterien umfasst, wobei ein Objekt ein Nutzermodell oder ein Itemmodell umfasst.

15. Verfahren nach Anspruch 16, wobei die Auswahlkriterien umfassen:
    - Objekte eines vorbestimmten Typs, und/oder
    - Objekte die eine vorbestimmte Ähnlichkeit zu dem Nutzermodell und/oder Itemmodell aufweisen, wobei vor dem Auswählen Ähnlichkeitswerte

zwischen dem generierten Nutzermodell und/oder Itemmodell und den Objekten ermittelt werden.

16. Verfahren nach einem der vorhergehenden Ansprüche, wobei durch ein Itemmodell ein Förderprogramm repräsentiert wird, und wobei das das Förderprogramm repräsentierende Itemmodell ausgewählt wird, wenn das Nutzermodell eine vorbestimmte Ähnlichkeit zu dem Itemmodell aufweist.

17. System zum Erzeugen eines Nutzermodells, insbesondere für einen Empfehlungsdienst, aus zumindest einer baumförmigen Datenstruktur, wobei das Nutzermodell Informationen über einen Nutzer umfasst, wobei die baumförmige Datenstruktur dem Nutzer zuordenbar ist, wobei die baumförmige Datenstruktur einen Wurzelknoten und eine Anzahl von Kinderknoten umfasst, welche über Kanten mit dem Wurzelknoten oder mit einem Kinderknoten verbunden sind, und wobei zumindest einem Knoten zumindest ein Element zugeordnet ist, aufweisend

    - wenigstens eine Speichereinrichtung zum Speichern wenigstens einer baumförmigen Datenstruktur, und
    - eine Verarbeitungseinrichtung, welche mit der Speichereinrichtung gekoppelt ist und welche angepasst ist ein Verfahren nach einem der vorhergehenden Ansprüche auszuführen, um ein Nutzermodell zu generieren und das generierte Nutzermodell in der Speichereinrichtung abzuspeichern und einem Empfehlungsdienst zur Verfügung zu stellen.

18. Datenträgerprodukt mit einem darauf gespeicherten Programmcode, welcher in einen Computer und / oder in ein Computernetzwerk ladbar ist und angepasst ist, ein Verfahren nach einem der Ansprüche 1 bis 18 auszuführen.

## K.3 Figuren

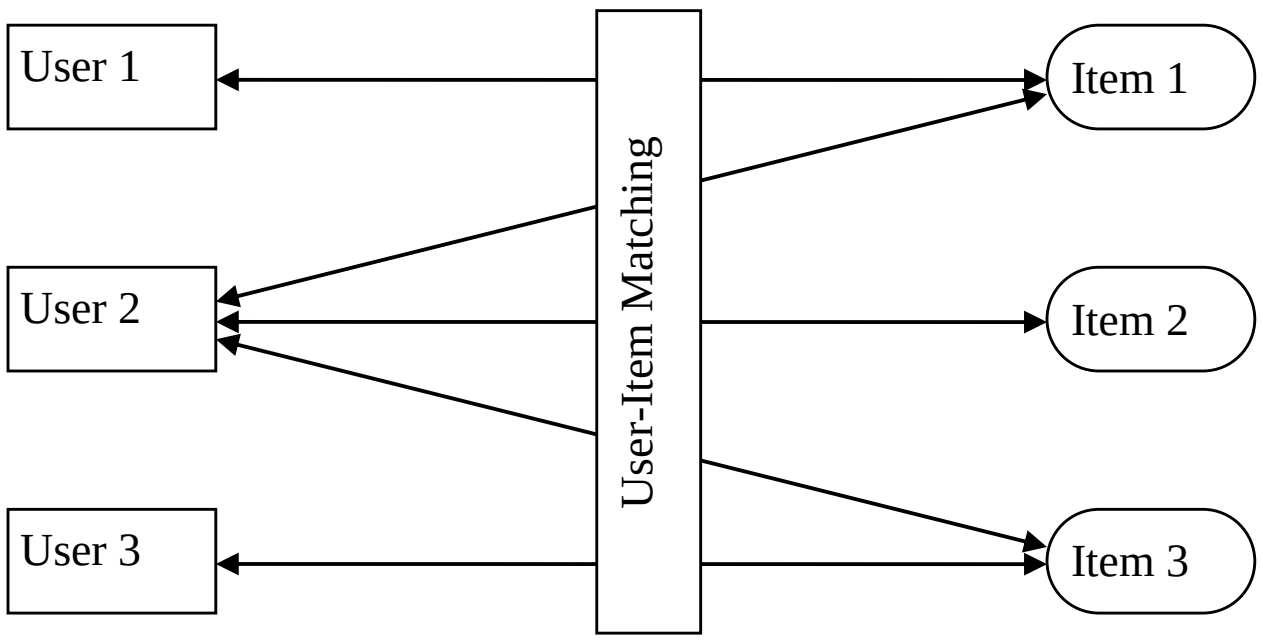

**Fig. 1**
**(Stand der Technik)**

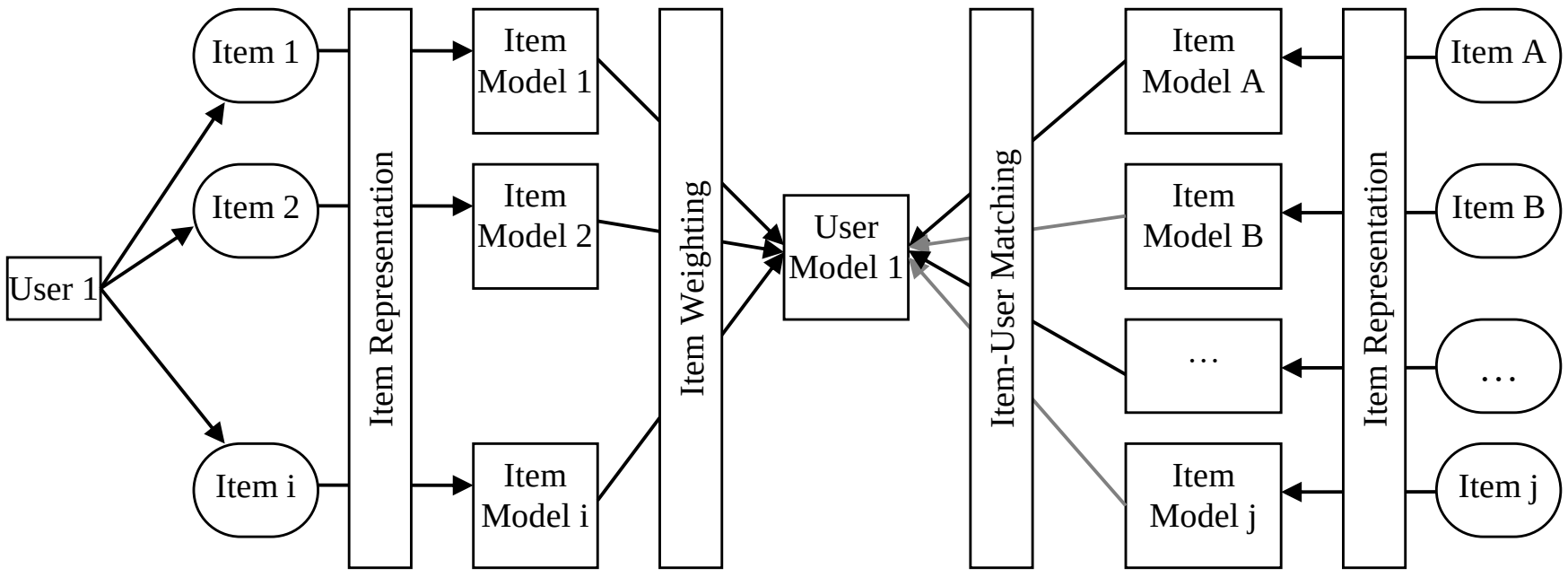

**Fig. 2**
**(Stand der Technik)**

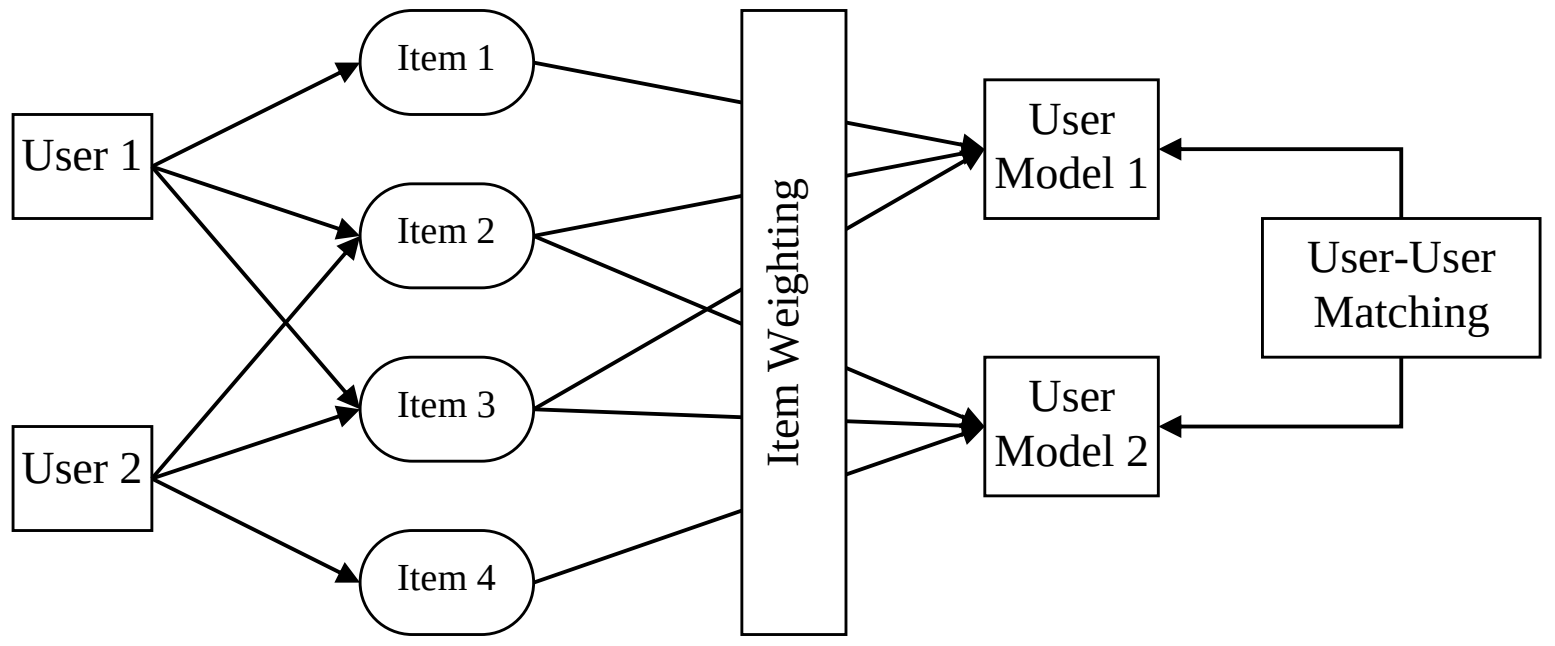

**Fig. 3**
**(Stand der Technik)**

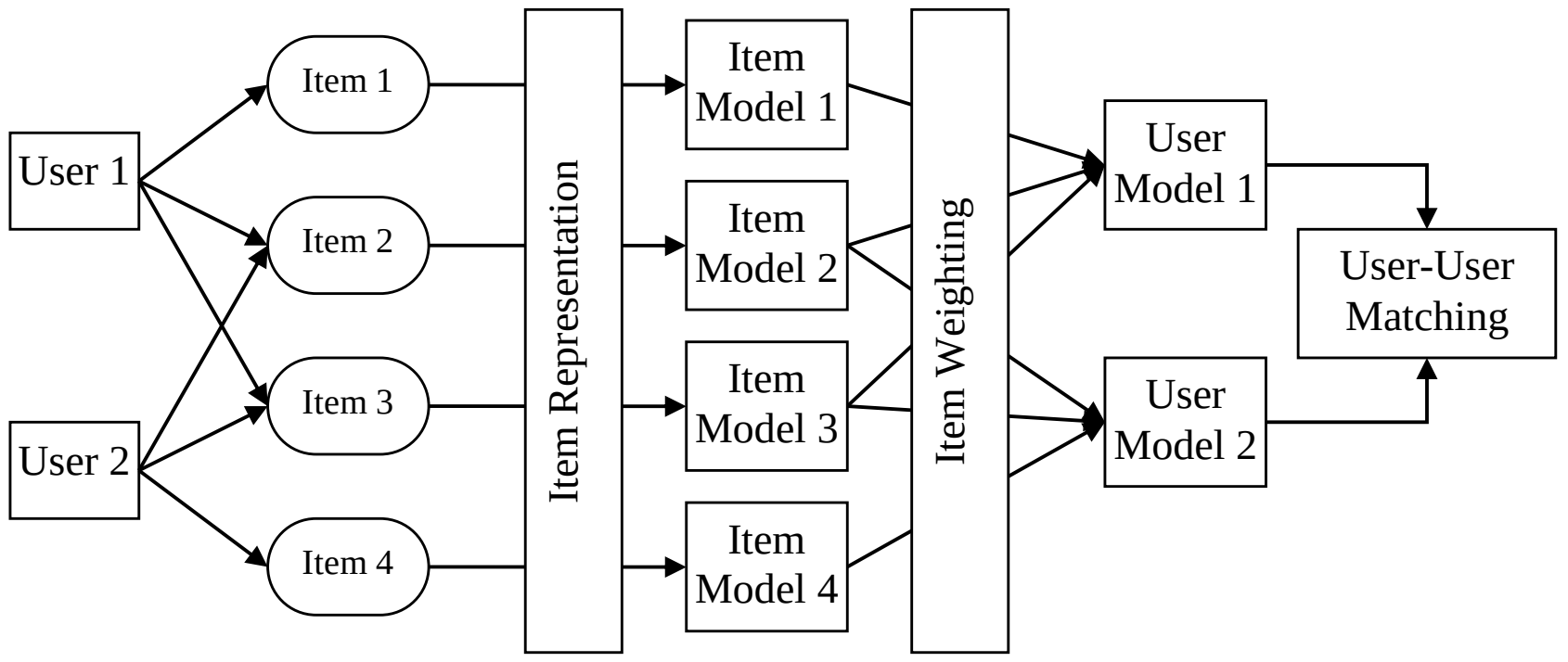

**Fig. 4**
**(Stand der Technik)**

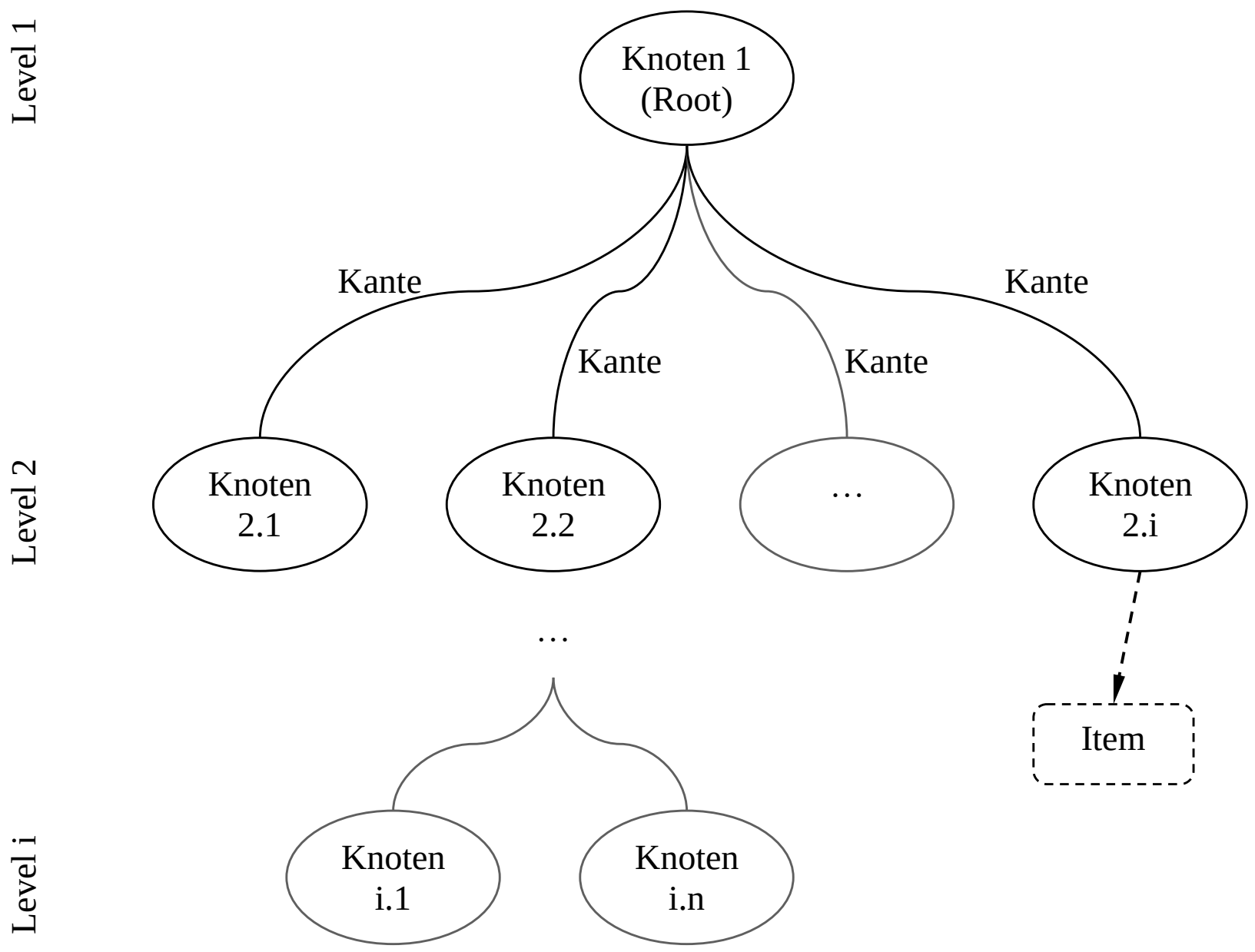

**Fig. 5**

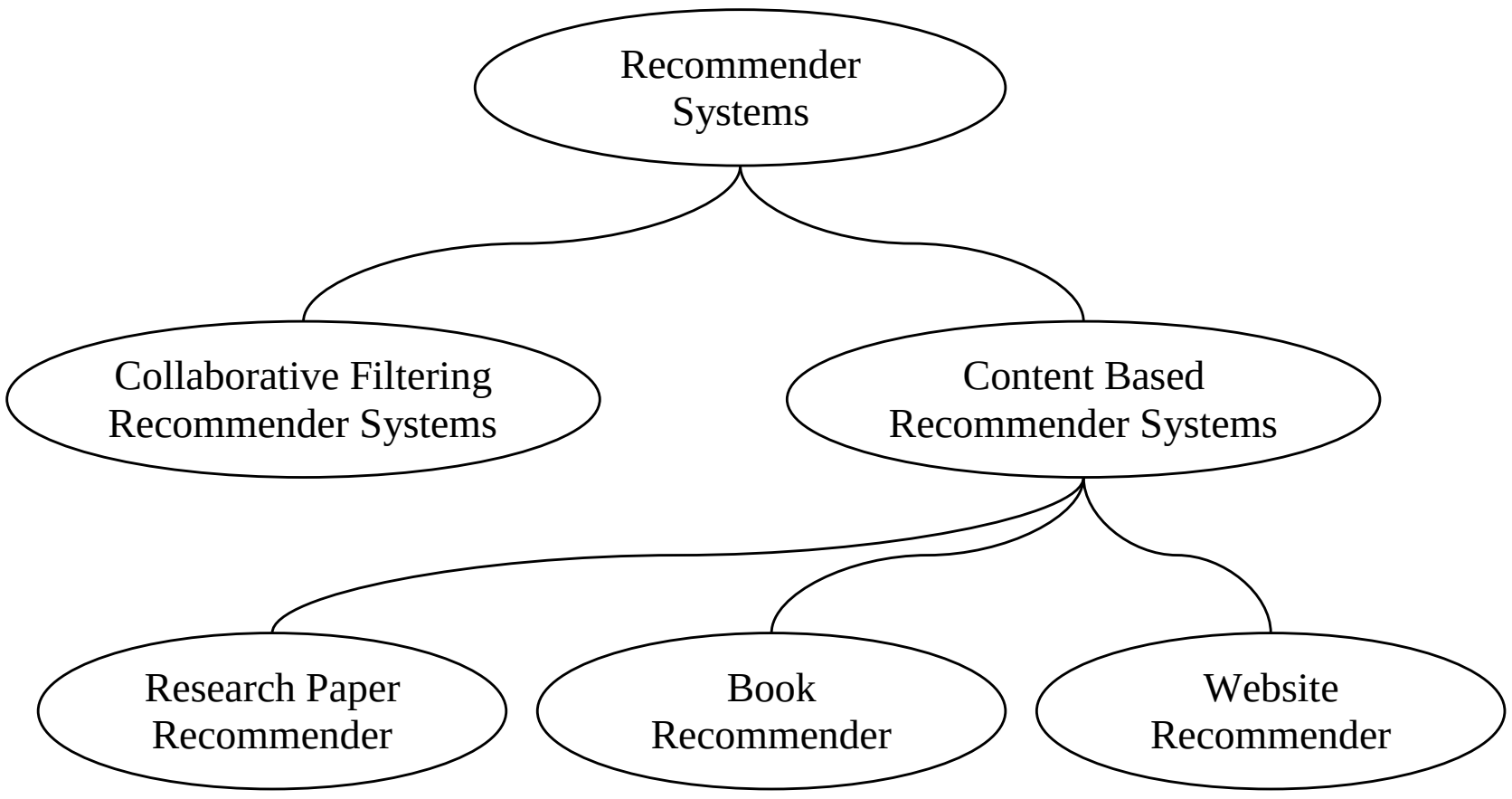
Recommender Systems
Collaborative Filtering Recommender Systems
Content Based Recommender Systems
Research Paper Recommender
Book Recommender
Website Recommender

**Fig. 6a**

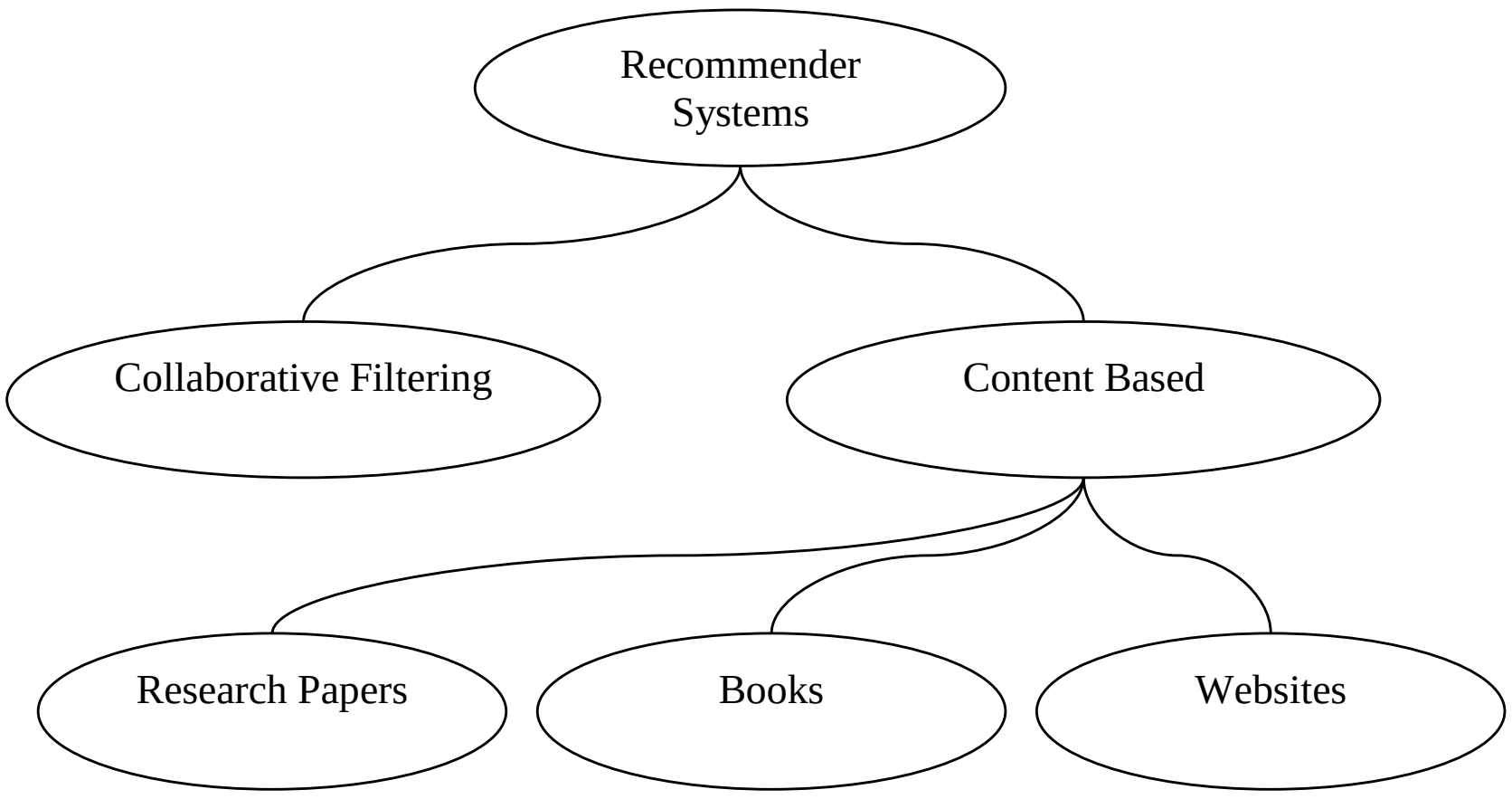

**Fig. 6b**

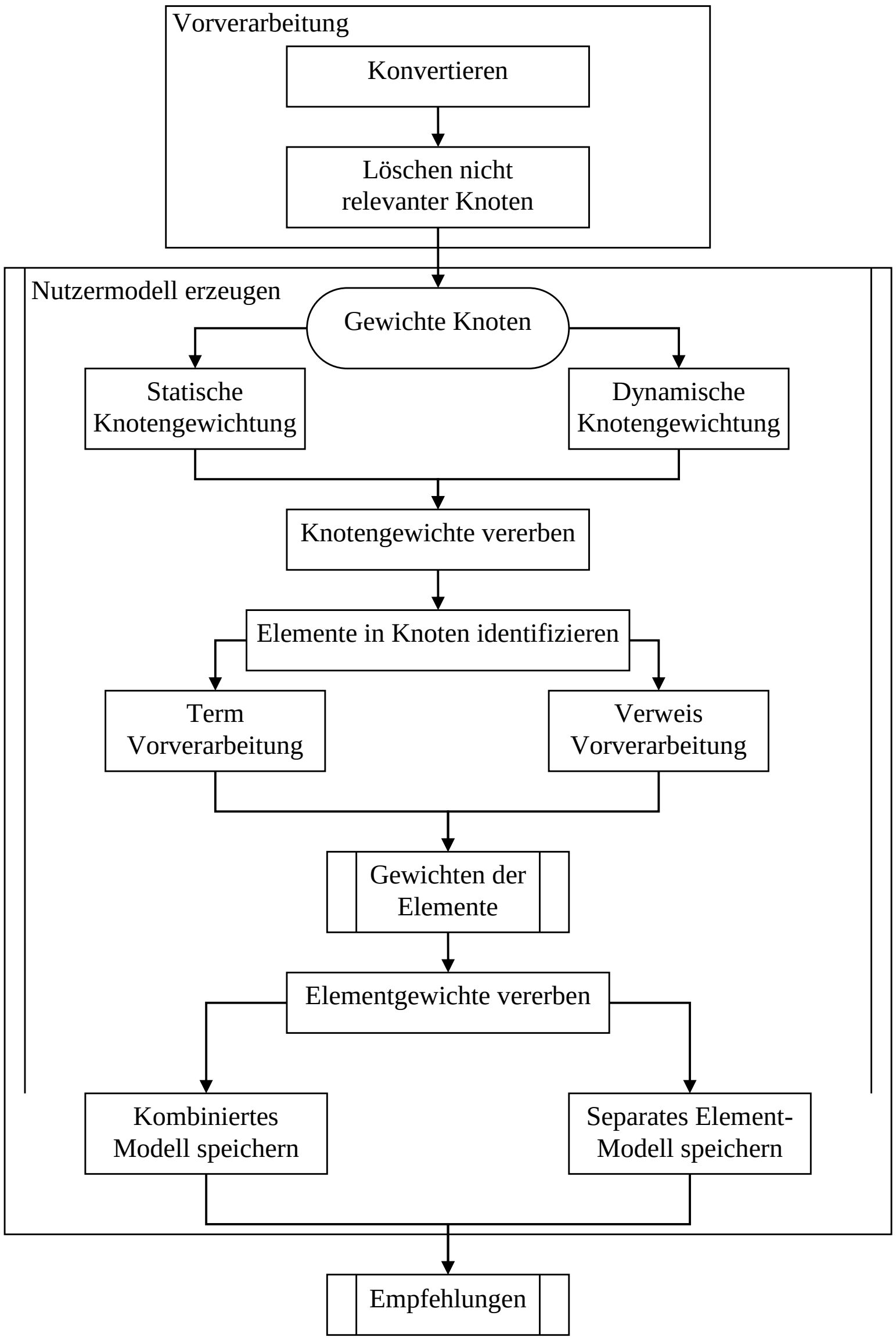

**Fig. 7a**

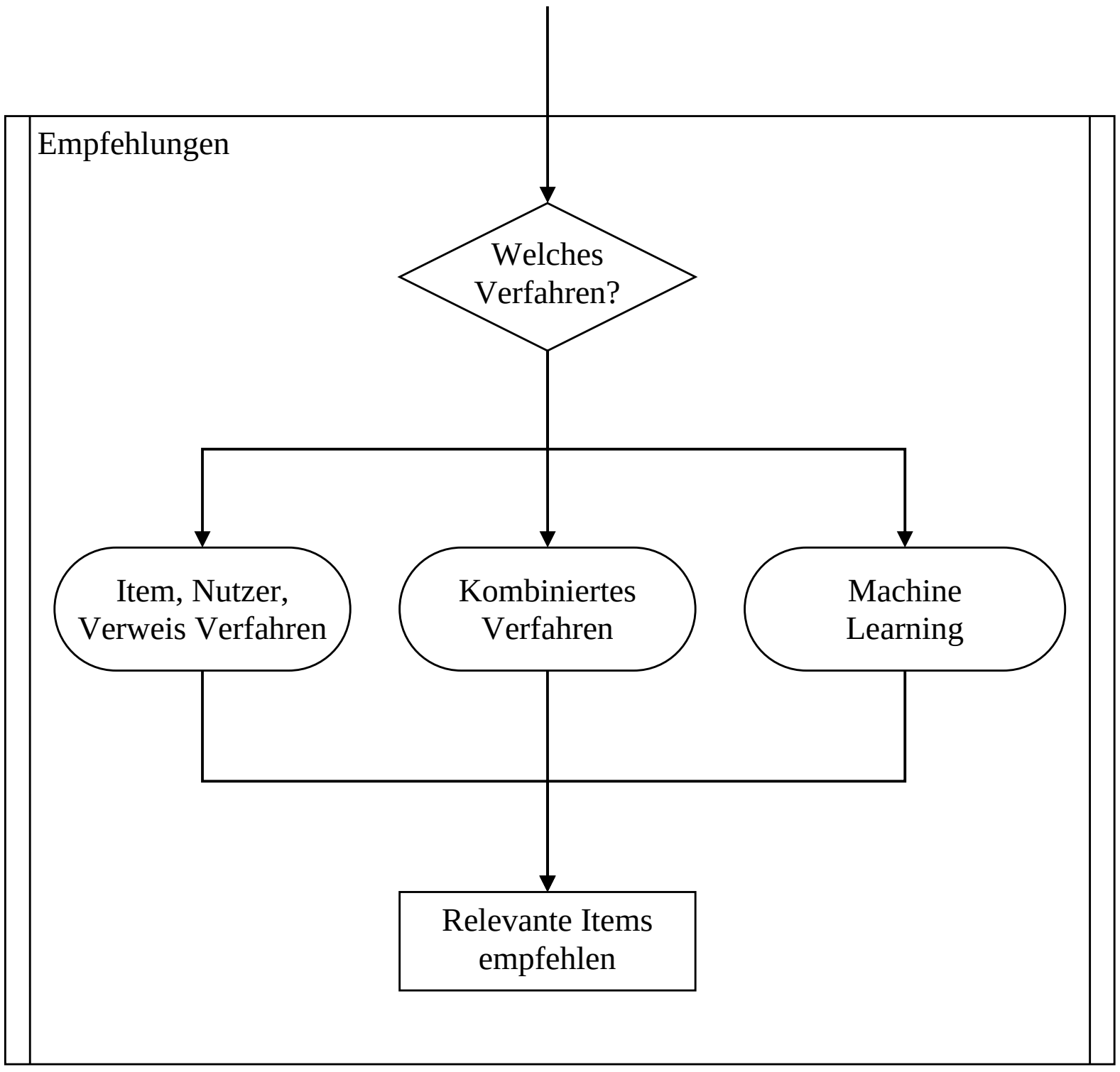

**Fig. 7b**

# Ehrenerklärung

Ich versichere hiermit, dass ich die vorliegende Arbeit ohne unzulässige Hilfe Dritter und ohne Benutzung anderer als der angegebenen Hilfsmittel angefertigt habe; verwendete fremde und eigene Quellen sind als solche kenntlich gemacht. Insbesondere habe ich nicht die Hilfe eines kommerziellen Promotionsberaters in Anspruch genommen. Dritte haben von mir weder unmittelbar noch mittelbar geldwerte Leistungen für Arbeiten erhalten, die im Zusammenhang mit dem Inhalt der vorgelegten Dissertation stehen. Ich habe insbesondere nicht wissentlich:

- Ergebnisse erfunden oder widersprüchliche Ergebnisse verschwiegen,
- statistische Verfahren absichtlich missbraucht, um Daten in ungerechtfertigter Weise zu interpretieren,
- fremde Ergebnisse oder Veröffentlichungen plagiiert,
- fremde Forschungsergebnisse verzerrt wiedergegeben.

Mir ist bekannt, dass Verstöße gegen das Urheberrecht Unterlassungs- und Schadensersatzansprüche des Urhebers sowie eine strafrechtliche Ahndung durch die Strafverfolgungsbehörden begründen kann. Die Arbeit wurde bisher weder im Inland noch im Ausland in gleicher oder ähnlicher Form als Dissertation eingereicht und ist als Ganzes auch noch nicht veröffentlicht.

Jöran Beel, Magdeburg, Deutschland, 22. Oktober 2014